\documentclass[11pt,a4paper]{article}
\usepackage[colorlinks=true, linkcolor=black!50!blue, urlcolor=blue, citecolor=blue, anchorcolor=blue]{hyperref}
\usepackage[font=small,labelfont=bf,margin=0mm,labelsep=period,tableposition=top]{caption}
\usepackage[a4paper,top=3cm,bottom=2.5cm,left=2.5cm,right=2.5cm,bindingoffset=0mm]{geometry}

\usepackage{graphicx,placeins}
\usepackage{float}
\usepackage{afterpage}
\usepackage{epsfig,cite}
\usepackage{amssymb}
\usepackage{amsmath}
\usepackage{dsfont}
\usepackage{multirow}
\usepackage{url}
\usepackage{xcolor,colortbl}
\usepackage{float}
\usepackage{afterpage}
\usepackage{url}
\usepackage{hyperref}
\usepackage{booktabs}
\usepackage{mathrsfs}

\usepackage{tikz}
\usepackage{tikz-3dplot}
\usepackage[compat=1.0.0]{tikz-feynman}

\usepackage{enumitem}
\usepackage{hyperref}
\usepackage{cite}

\usepackage{pifont}

\usetikzlibrary{shapes, arrows}
\usetikzlibrary{decorations.pathreplacing}
\usetikzlibrary{positioning, calc}
\tikzstyle{fitted} = [rectangle, minimum width=5cm, minimum height=1cm, text centered, draw=black, fill=red!30]
\tikzstyle{operations} = [rectangle, rounded corners, minimum width=2cm,text centered, draw=black, fill=red!30]
\tikzstyle{roundtext} = [rectangle, rounded corners, minimum width=2cm, minimum height=0.8cm, text centered, draw=black, fill=red!30]
\tikzstyle{n3py} = [rectangle, rounded corners, minimum width=3cm, minimum height=1cm, text centered, draw=black, fill=green!30]
\tikzstyle{myarrow} = [thick,->,>=stealth]
\tikzstyle{line} =[draw, -latex']
\tikzstyle{decision} = [diamond, draw, fill=red!20, text width=7.5em, text centered,  inner sep=0pt, minimum height=2em, aspect=4]
\tikzstyle{cloud} = [draw, ellipse,fill=green!20, minimum height=2em]
\tikzstyle{inout} = [rectangle, draw, fill=green!20, text width=9.5em, text centered, rounded corners, minimum height=2em, minimum width=10em]
\tikzstyle{block}=[rectangle, draw, fill=blue!20, text width=9.5em, 
                   text centered, rounded corners, minimum height=2em, 
                   minimum width=10em]

\definecolor{darkgreen}{rgb}{0.0, 0.5, 0.13}

\newcommand{\be}{\begin{equation}}
\newcommand{\ee}{\end{equation}}
\newcommand{\bea}{\begin{eqnarray}}
\newcommand{\eea}{\end{eqnarray}}
\newcommand{\bi}{\begin{itemize}}
\newcommand{\ei}{\end{itemize}}
\newcommand{\ben}{\begin{enumerate}}
\newcommand{\een}{\end{enumerate}}

\newcommand{\lp}{\left(}
\newcommand{\rp}{\right)}

\def\gsim{\mathrel{\rlap{\lower4pt\hbox{\hskip1pt$\sim$}}
    \raise1pt\hbox{$>$}}}         
\def\lsim{\mathrel{\rlap{\lower4pt\hbox{\hskip1pt$\sim$}}
    \raise1pt\hbox{$<$}}}         

\newcommand{\draft}[1]{}

\def\beq{\begin{equation}}
\def\eeq{\end{equation}}

\def\lapprox{\lower .7ex\hbox{$\;\stackrel{\textstyle <}{\sim}\;$}}
\def\gapprox{\lower .7ex\hbox{$\;\stackrel{\textstyle >}{\sim}\;$}}

\def\d{{\rm d}}

\numberwithin{equation}{section}
\numberwithin{figure}{section}
\numberwithin{table}{section}

\usepackage{tabularx}
\newcolumntype{C}[1]{>{\centering\arraybackslash}p{#1}}

\usepackage{amsmath}
\usepackage{amsfonts}
\usepackage{amssymb}
\usepackage{dsfont}
\usepackage{pifont}
\usepackage{booktabs}
\usepackage{graphicx}
\usepackage{epstopdf}
\usepackage{epsfig}
\usepackage{framed}
\usepackage{makeidx}
\usepackage{siunitx}
\usepackage[capitalise]{cleveref}
\usepackage{hyperref}
\usepackage{placeins}
\usepackage[font=small,labelfont=bf]{caption}

\RequirePackage{csquotes}
\usepackage{bbm}

\begin{document}
\newgeometry{top=1.5cm,bottom=1.5cm,left=1.5cm,right=1.5cm,bindingoffset=0mm}

\vspace{-2.0cm}
\begin{flushright}
Nikhef-2023-020\\
TIF-UNIMI-2023-23\\
Edinburgh 2023/29\\
CERN-TH-2024-033\\
\end{flushright}
\vspace{0.3cm}

\begin{center}
  {\Large \bf The Path to  N$^3$LO Parton Distributions}
  \vspace{1.1cm}

   {\bf The NNPDF Collaboration}: \\[0.1cm]
   Richard D. Ball$^1$,
   Andrea Barontini$^2$,
   Alessandro Candido$^{2,3}$,
   Stefano Carrazza$^2$,
   Juan Cruz-Martinez$^3$,\\[0.1cm]
   Luigi Del Debbio$^1$,
   Stefano Forte$^2$,
   Tommaso Giani$^{4,5}$,
   Felix Hekhorn$^{2,6,7}$,
   Zahari Kassabov$^8$,\\[0.1cm]
   Niccol\`o Laurenti,$^2$
   Giacomo Magni$^{4,5}$,
   Emanuele R. Nocera$^9$,
   Tanjona R. Rabemananjara$^{4,5}$,
   Juan Rojo$^{4,5}$,\\[0.1cm]
   Christopher Schwan$^{10}$,
   Roy Stegeman$^1$, and
   Maria Ubiali$^8$

    \vspace{0.7cm}
    
    {\it \small
   
    ~$^1$The Higgs Centre for Theoretical Physics, University of Edinburgh,\\
      JCMB, KB, Mayfield Rd, Edinburgh EH9 3JZ, Scotland\\[0.1cm]
    ~$^2$Tif Lab, Dipartimento di Fisica, Universit\`a di Milano and\\
      INFN, Sezione di Milano, Via Celoria 16, I-20133 Milano, Italy\\[0.1cm]
      ~$^3$CERN, Theoretical Physics Department, CH-1211 Geneva 23, Switzerland\\[0.1cm]
       ~$^4$Department of Physics and Astronomy, Vrije Universiteit, NL-1081 HV Amsterdam\\[0.1cm]
      ~$^5$Nikhef Theory Group, Science Park 105, 1098 XG Amsterdam, The Netherlands\\[0.1cm]
   ~$^6$University of Jyvaskyla, Department of Physics, P.O. Box 35, FI-40014 University of Jyvaskyla, Finland\\[0.1cm]
   ~$^7$Helsinki Institute of Physics, P.O. Box 64, FI-00014 University of Helsinki, Finland\\[0.1cm]
      ~$^8$ DAMTP, University of Cambridge, Wilberforce Road, Cambridge, CB3 0WA, United Kingdom\\[0.1cm]
        ~$^9$ Dipartimento di Fisica, Universit\`a degli Studi di Torino and\\
      INFN, Sezione di Torino, Via Pietro Giuria 1, I-10125 Torino, Italy\\[0.1cm]
      ~$^{10}$Universit\"at W\"urzburg, Institut f\"ur Theoretische Physik und Astrophysik, 97074 W\"urzburg, Germany\\[0.1cm]

      }
  
 \vspace{0.5cm}
        {\it This paper is dedicated to the memory of Stefano Catani,\\
          Grand Master of QCD, great scientist and
          human being}
 \vspace{0.5cm}

{\bf \large Abstract}

\end{center}

We extend the existing leading (LO), next-to-leading (NLO), and
next-to-next-to-leading order (NNLO)   NNPDF4.0 sets of parton distribution functions (PDFs)
to approximate next-to-next-to-next-to-leading order (aN$^3$LO).
We construct an approximation to  the  N$^3$LO splitting functions
that includes all available 
partial information from both fixed-order computations and from
small and large $x$ resummation, and estimate the uncertainty on this
approximation by varying the set of basis functions used to construct
the approximation.
We include known  N$^3$LO  corrections to 
deep-inelastic scattering structure functions and extend the FONLL general-mass scheme to $\mathcal{O}\lp \alpha_s^3\rp$
accuracy.
We determine a set of aN$^3$LO PDFs by accounting both 
for the uncertainty  on splitting functions due to 
the incomplete knowledge of 
N$^3$LO terms, and to the uncertainty related to missing higher
corrections (MHOU), estimated by scale variation,  through a theory
covariance matrix formalism. 
We assess the perturbative stability of the resulting PDFs, we study
the impact of MHOUs on them, and we compare our results to the
aN$^3$LO PDFs from the MSHT group.  
We examine the phenomenological impact of aN$^3$LO corrections on parton
luminosities at the LHC, and give a first assessment of the impact of
aN$^3$LO PDFs on the   Higgs and Drell-Yan total production cross-sections.
We find that the aN$^3$LO NNPDF4.0 PDFs are consistent within uncertainties 
with their NNLO counterparts, that they
improve the description of the global dataset and  the perturbative
convergence of Higgs and Drell-Yan cross-sections, and that MHOUs on PDFs
decrease substantially with the increase of perturbative order.

\clearpage

\tableofcontents

\section{Introduction}
\label{sec:intro}

Calculations of hard-scattering cross-sections at fourth perturbative order in
the strong coupling, i.e. at next-to-next-to-next-to-leading order (N$^3$LO),
have been available for a long time for massless deep-inelastic
scattering (DIS)~\cite{Vermaseren:2005qc,Moch:2004xu,Moch:2007rq,Moch:2008fj},
and have more recently become available for a rapidly growing set of hadron
collider processes. These include inclusive Higgs production in
gluon-fusion~\cite{Anastasiou:2015ema,Mistlberger:2018etf},
bottom-fusion~\cite{Duhr:2019kwi}, in association with vector
bosons~\cite{Baglio:2022wzu}, and in
vector-boson-fusion~\cite{Dreyer:2016oyx},
Higgs pair production~\cite{Chen:2019lzz}, inclusive
Drell-Yan production~\cite{Duhr:2020sdp,Duhr:2021vwj}, differential Higgs
production~\cite{Dulat:2017prg,Dulat:2018bfe,Chen:2021isd,Billis:2021ecs,
  Camarda:2021ict}, and differential Drell-Yan
distributions~\cite{Chen:2021vtu,Chen:2022lwc}, see~\cite{Caola:2022ayt}
for an overview.

In order to obtain predictions for hadronic observables with this
accuracy, these  partonic cross-sections must be combined with parton
distribution functions (PDFs) determined  at the same perturbative order.
These, in turn, must be determined by comparing to experimental data theory
predictions computed at the same accuracy. The main bottleneck in
carrying out this programme is the lack of exact expressions for
the N$^3$LO splitting functions
that govern the scale dependence of the PDFs: for these only partial information
is available~\cite{Davies:2016jie,Moch:2017uml,Davies:2022ofz,Henn:2019swt,
  Duhr:2022cob,Moch:2021qrk,Soar:2009yh,Falcioni:2023luc,Falcioni:2023vqq,
  Moch:2023tdj,Falcioni:2023tzp}. This information includes a set of integer
$N$-Mellin moments, terms proportional to $n_f^k$ with $k\ge 1$, and the large-
and small-$x$ limits. By combining these partial results it is possible
to attempt an approximate determination of the N$^3$LO splitting
functions~\cite{Moch:2023tdj,McGowan:2022nag}, as was successfully done in
the past at NNLO~\cite{vanNeerven:2000wp}.

At present a global PDF determination at N$^3$LO must consequently be
based on incomplete information: the approximate knowledge of
splitting functions, and full knowledge of partonic cross-sections
only for a subset of processes. A first attempt towards achieving this
was recently made in Ref.~\cite{McGowan:2022nag}, where the missing
theoretical information
on N$^3$LO calculations was parametrized in terms of a set of nuisance
parameters, which were determined together with the PDFs from a fit to
experimental data.

Here we adopt a somewhat different strategy. Namely, we use a
theory covariance matrix formalism
in order to account for the missing
perturbative information. It was shown in
Ref.~\cite{Ball:2018twp} that nuclear
uncertainties can be included through a theory covariance matrix,
and it was further shown in
Refs.~\cite{NNPDF:2019vjt,NNPDF:2019ubu} how such a theory covariance
matrix can be constructed to account for missing
higher-order uncertainties (MHOUs), estimated through renormalization
and factorization scale variation. Here we will use the same formalism
in order to also construct a theory
covariance matrix for incomplete higher-order uncertainties (IHOUs),
namely, those related
to incomplete knowledge of the  N$^3$LO theory, specifically for the splitting
functions and for the massive DIS coefficient functions.
Equipped with such theory covariance matrices, we can
perform a determination of PDFs at ``approximate N$^3$LO''
(hereafter denoted aN$^3$LO), in
which the theory covariance matrix accounts both for incomplete
knowledge of  N$^3$LO  splitting functions and massive coefficient
functions (IHOUs), and for missing N$^3$LO corrections to 
the partonic cross-sections for hadronic processes (MHOUs).

We will thus present the  aN$^3$LO NNPDF4.0 PDF determination, to be added to
the existing LO, NLO and NNLO sets~\cite{NNPDF:2021njg}, as well as the
more recent NNPDF4.0 MHOU PDFs~\cite{NNPDF:2024dpb} that also include
MHOUs in the PDF uncertainty. Besides using a different methodology to
the MSHT20 study~\cite{McGowan:2022nag}, here we are also
able to include more recent exact results~\cite{Falcioni:2023luc,
  Falcioni:2023vqq,Moch:2023tdj,Falcioni:2023tzp} 
that stabilize the N$^3$LO splitting function parametrisation.
Our construction is implemented in the open-source 
NNPDF framework~\cite{NNPDF:2021uiq}. Specifically, our
aN$^3$LO evolution is implemented in {\sc EKO}~\cite{Candido:2022tld}
and the N$^3$LO DIS coefficient functions, including the FONLL general-mass
scheme, in {\sc YADISM}~\cite{Candido:2024rkr}.
With PDFs determined from the same global dataset and using the same
methodology at four consecutive perturbative orders it is now possible to
assess carefully perturbative stability and provide a reliable uncertainty
estimation.

The outline of this paper is as follows. In Sect.~\ref{sec:dglap} we construct
an approximation to the N$^3$LO splitting functions based on all known exact
results and limits. We compare it with the MSHT
approximation~\cite{McGowan:2022nag} as well as with the
more recent approximation of Refs.~\cite{Falcioni:2023luc,Falcioni:2023vqq,
  Moch:2023tdj}. In Sect.~\ref{sec:n3lo_coefffun} we discuss available and
approximate N$^3$LO corrections to hard cross-sections: specifically, DIS
coefficient functions, including a generalization to this order of the
FONLL~\cite{Forte:2010ta,Ball:2015dpa,Ball:2015tna} method for the inclusion of
heavy quark mass effects, and the Drell-Yan cross-section.
In Sect.~\ref{sec:results} we present the main results
of this work, namely the aN$^3$LO  NNPDF4.0 PDF set, based on the  results of
Sects.~\ref{sec:dglap} and~\ref{sec:n3lo_coefffun}. Perturbative
convergence before and after the inclusion of MHOUs is
discussed in detail, and results are compared to those of
the MSHT group~\cite{McGowan:2022nag}. 
A first assessment of the impact of aN$^3$LO PDFs on Drell-Yan and
Higgs production is presented in Sect.~\ref{sec:pheno}.
Finally, a summary and outlook on
future developments are presented in Sect.~\ref{sec:summary}.
Expressions for the anomalous dimensions parametrized in
Sects.~\ref{sec:non_singlet}-\ref{sec:singlet}
are given in Appendix~\ref{app:splitting_asy}.

\section{Approximate  N$^3$LO evolution}
\label{sec:dglap}

We proceed to the construction and implementation of aN$^3$LO evolution.
We first describe our strategy to approximate the N$^3$LO
evolution equations, the way this is used to construct aN$^3$LO
anomalous dimensions and splitting functions, and to estimate the
uncertainty in the approximation and its impact on theory predictions.
We then use this strategy to construct an approximation in the nonsinglet
sector, where accurate results have been available for a
while~\cite{Moch:2017uml}, and benchmark it against these results. We
then present our construction of aN$^3$LO singlet splitting functions,
examine  our results, their uncertainties and their
perturbative behavior, and also how they relate to NLL small-$x$
resummation. We next
describe our implementation of aN$^3$LO evolution and study the
impact of aN$^3$LO on the perturbative evolution of PDFs. Finally,
we compare our aN$^3$LO singlet splitting functions to 
those of the MSHT group and to the recent
results of~\cite{Falcioni:2023luc,Falcioni:2023vqq,Moch:2023tdj}.

\subsection{Construction of the approximation}
\label{sec:general_strategy}

We write the QCD evolution equations as
\be
    \mu^2 \frac{\partial f_i(x,\mu^2)}{\partial \mu^2} = \int_{x}^{1} \frac{dz}{z} P_{ij} (x/z,a_s(\mu^2)) f_j(z, \mu^2) \, ,
    \label{eq:dglap}
\ee
where $f_i(x,\mu^2)$ is a vector of PDFs and, with $n_f$ active quark
flavors, the $\lp 2n_f+1 \rp \times\lp  2n_f+1 \rp$ splitting function
matrix   $P_{ij}(x,a_s(\mu^2))$ is expanded perturbatively as
\be
P_{ij}(x,a_s(\mu^2)) = a_s P_{ij}^{(0)}(x)
+ a_s^2 P_{ij}^{(1)}(x)
+ a_s^3 P_{ij}^{(2)}(x)
+ a_s^4 P_{ij}^{(3)}(x)
+ \mathcal{O}\lp a_s^5\rp \, .
\label{eq:p_expansion}
\ee
in powers of the strong coupling $a_s(\mu^2) = \alpha_s(\mu^2)/4\pi$.

Defining Mellin space PDFs $ f_i(N,\mu^2)$ (denoted in a slight abuse of notation by
the same symbol as the $x$-space PDFs), and
anomalous dimensions $\gamma_{ij}(N,a_s(\mu^2))$ as minus the
Mellin transforms of splitting functions,
\begin{align}
  \label{eq:def_mellin}
    f_i(N,\mu^2) &= \mathcal{M}[f_i(x,\mu^2)](N) = \int\limits_0^1 dx\,x^{N-1}f_i(x,\mu^2) \\
    \gamma_{ij}(N,a_s(\mu^2)) &= -\mathcal{M}[P_{ij}(x,a_s(\mu^2))](N) = - \int\limits_0^1 dx\,x^{N-1}P_{ij}(x,a_s(\mu^2)) \label{eq:melp}
  \end{align}
the evolution equations become
    \be
    \mu^2 \frac{\partial f_i(N,\mu^2)}{\partial \mu^2} = -\gamma_{ij} (N,a_s(\mu^2)) f_j(N, \mu^2) \, ,
    \label{eq:dglap_mellin}
\ee
where the perturbative expansion of the anomalous dimensions is
\be
    \gamma_{ij}(N,a_s(\mu^2)) = a_s \gamma_{ij}^{(0)}(N)
    + a_s^2 \gamma_{ij}^{(1)}(N)
    + a_s^3 \gamma_{ij}^{(2)}(N)
    + a_s^4 \gamma_{ij}^{(3)}(N)
+ \mathcal{O}\lp a_s^5\rp  \, .
    \label{eq:ad_expansion}
    \ee
    
The $\lp 2n_f+1 \rp \times\lp  2n_f+1 \rp$ matrix of  anomalous dimensions
has seven independent entries (see e.g.~\cite{Vogt:2004ns}), driving the
evolution of various PDF combinations as follows:
\begin{itemize}
\item All nonsinglet combinations 
\begin{align}\label{eq:nsdef}
  q^{\pm}_{ij} &= q^\pm_i-\bar   q^\pm_j,\\
  \quad q^\pm_i&=q_i\pm\bar q_i \label{eq:qpmdef}
\end{align}
  satisfy  
    decoupled evolution equations with the same two anomalous
    dimension $\gamma_{{\rm ns},\pm}$; note that the plus and minus variants of
    $\gamma_{{\rm ns},\pm}$ start differing from each other already at NLO.
  \item The total valence combination 
\begin{equation}\label{eq:valdef}
  V=\sum_{i=1}^{n_f} q^-_i
\end{equation}
  satisfies a decoupled
      evolution equation with an anomalous dimension
\be\label{eq:valad}
    \gamma_{\rm ns,v}= \gamma_{\rm ns,s}+\gamma_{{\rm ns},-} \, ;
\ee
note that the flavor-independent ``sea'' contribution $\gamma_{\rm ns,s}$ 
starts being nonzero only at NNLO.

\item The singlet combination
  \be\label{eq:singdef}
\Sigma = \sum_{i=1}^{n_f} q_i^+ 
\ee
mixes with the gluon 
\be
   \mu^2 \frac{\partial}{\partial \mu^2} \begin{pmatrix} \Sigma(N,\mu^2) \\ g(N,\mu^2) \end{pmatrix} =
    - \begin{pmatrix} \gamma_{qq}(N,a_s(\mu^2)) & \gamma_{qg}(N,a_s(\mu^2)) \\ \gamma_{gq}(N,a_s(\mu^2)) & \gamma_{gg}(N,a_s(\mu^2)) \\ \end{pmatrix} 
    \begin{pmatrix} \Sigma(N,\mu^2) \\ g (N,\mu^2)\end{pmatrix} \, .
    \label{eq:dglap_singlet}
    \ee
   The quark-quark entry of the anomalous dimension matrix can be further
   decomposed into nonsinglet and pure singlet contributions according to
\be
   \gamma_{qq} = \gamma_{{\rm ns},+} + \gamma_{{qq},{\rm ps}} \, ,
    \label{eq:p_qq}
\ee
where the pure singlet contribution $\gamma_{{qq},{\rm ps}}$ starts at NLO.
\end{itemize}

There are thus seven independent contributions: three in the nonsinglet sector,
$\gamma_{{\rm ns},\pm}$ and $\gamma_{\rm ns,s}$, and four in the singlet
sector, $\gamma_{{qq},{\rm ps}}$, $\gamma_{{qg}}$,
$\gamma_{{gq}}$, and $\gamma_{{gg}}$.
In turn, each of these anomalous dimensions can be expanded
according to Eq.~(\ref{eq:ad_expansion}). Our goal is to determine an
approximate expression for the corresponding seven $\gamma_{ij}^{(3)}(N)$ N$^3$LO
terms.

The information that can be exploited in order to achieve this goal
comes from three different sources: (1) full analytic knowledge of
contributions to the anomalous dimensions proportional to the highest
powers of the number of flavors $n_f$; (2) large-$x$ and small-$x$ resummations
provide all-order information on terms that are logarithmically enhanced by
powers of
$\ln (1-x)$ and $\ln x$ respectively; (3) analytic knowledge of a finite
set of integer moments. We construct an approximation based on this
information by first separating off the analytically known terms
(1-2),  then expanding the remainder on a set of basis functions and
using the known moments to  determine the expansion coefficients.
Finally, we vary the set of basis functions in order to obtain an
estimate of the uncertainties. 

Schematically, we proceed as follows:
\begin{enumerate}

\item We include all terms in the expansion
  \be\label{eq:gammaf}
  \gamma_{ij}^{(3)}(N) = \gamma_{ij}^{(3,0)}(N) + n_f
  \gamma_{ij}^{(3,1)}(N) + n_f^2
  \gamma_{ij}^{(3,2)}(N) + n_f^3
  \gamma_{ij}^{(3,3)}(N) \, ,
  \ee
  of the anomalous dimension in powers of $n_f$ 
  that are fully or partially known analytically.
  We collectively denote such terms as $\gamma_{ij,n_f}^{(3)}(N)$.

\item We include all terms from large-$x$ and small-$x$ resummation,
  to the highest known logarithmic accuracy, including all known
  subleading power corrections in both limits.
  We  denote these terms as $\gamma_{ij,N\to \infty}^{(3)}(N)$
  and $\gamma_{ij,N\to 0}^{(3)}(N)$, $\gamma_{ij,N\to 1}^{(3)}(N)$ respectively.
  Possible double counting coming from the overlap of these terms with
  $\gamma_{ij,n_f}^{(3)}(N)$
  is removed.

\item We write the approximate anomalous dimension matrix element
  $\gamma_{ij}^{(3)}(N)$ as the sum of the
  terms which are known exactly and a remainder 
  $\widetilde{\gamma}_{ij}^{(3)}(N)$ according to
  \be
  \label{eq:ad_expansion_terms}
  \gamma_{ij}^{(3)}(N) = \gamma_{ij,n_f}^{(3)}(N)
  + \gamma_{ij,N\to \infty}^{(3)}(N)
  + \gamma_{ij,N\to 0}^{(3)}(N)
  + \gamma_{ij,N\to 1}^{(3)}(N)
  + \widetilde{\gamma}_{ij}^{(3)}(N) \,.
  \ee
  We determine the remainder  as a linear
  combination of a set of  $n^{ij}$
  interpolating functions $G^{ij}_\ell(N)$ (kept fixed)
  and $H^{ij}_\ell(N)$ (to be varied)
  \be
  \label{eq:gamma_residual_basis}
  \widetilde{\gamma}_{ij}^{(3)}(N) = \sum_{\ell=1}^{n^{ij }-n_H}
  b^{ij}_\ell G^{ij}_\ell(N) +\sum_{\ell=1}^{n_H}
  b^{ij}_{n^{ij} -2+\ell} H^{ij}_{\ell}(N) \, ,
  \ee
  with $n^{ij}$ equal to the number of known
  Mellin moments of 
  $\gamma_{ij}^{(3)}(N)$. We determine the coefficients
  $b^{ij}_\ell$ by equating the evaluation of
  $\widetilde{\gamma}_{ij}^{(3)}(N)$ to the known moments of the splitting
  functions.

\item In the singlet sector, we take $n_H=2$ and we make  $\widetilde{N}_{ij}$
  different choices for the two functions $H^{ij}_\ell(N)$, by selecting them
  out of a list of distinct  basis
  functions (see Sect.~\ref{sec:singlet} below). Thereby, we obtain
  $\widetilde{N}_{ij}$  expressions for the
  remainder $\widetilde{\gamma}_{ij}^{(3)}(N)$ and accordingly for the
  N$^3$LO anomalous dimension matrix element ${\gamma}_{ij}^{(3)}(N)$ through
  Eq.~(\ref{eq:ad_expansion_terms}). These are used to construct the
  approximate anomalous dimension matrix and the uncertainty on it, in
  the way discussed in Sect.~\ref{sec:ihou} below. In the nonsinglet
  sector instead, we take  $n_H=0$, i.e.\ we take a unique answer as our
  approximation, and we neglect the uncertainty on it, for reasons to
  be discussed in greater detail at the end of Sect.~\ref{sec:non_singlet}.  
\end{enumerate}

\subsection{The approximate anomalous dimension matrix and its uncertainty}
\label{sec:ihou}

The procedure described in Sect.~\ref{sec:general_strategy} provides us with an
ensemble of $\widetilde{N}_{ij}$ different
approximations to the N$^3$LO anomalous
dimension, denoted $\gamma_{ij}^{(3),\,(k)}(N)$, $k=1,\dots\widetilde{N}_{ij}$ . 
Our best estimate for the approximate anomalous dimension is then their average
\begin{equation}\label{eq:centralgij}
  \gamma_{ij}^{(3)}(N)=
  \frac{1}{\widetilde{N}_{ij}}\sum_{k=1}^{\widetilde{N}_{ij}} \gamma_{ij}^{(3),\,(k)}(N).
\end{equation}

We include  the uncertainty on the approximation, and the
ensuing uncertainty on N$^3$LO theory predictions,
using the general formalism for the treatment of theory uncertainties
developed in
Refs.~\cite{Ball:2018twp,NNPDF:2019vjt,NNPDF:2019ubu}. Namely, we
consider the uncertainty on each anomalous dimension matrix element
due to its incomplete knowledge as a source of
uncertainty on  theoretical predictions, uncorrelated from
other sources of uncertainty, and neglecting possible correlations
between our incomplete knowledge of each individual matrix element
$\gamma_{ij}^{(3)}$. This uncertainty on the incomplete higher
(N$^3$LO) order terms (incomplete higher order uncertainty, or IHOU)
is then treated in the same way as the uncertainty due to missing
higher order terms (missing higher order uncertainty, or MHOU).

Namely, we construct the shift of theory prediction for the $m$-th
data point due to replacing the central anomalous dimension matrix
element $\gamma_{ij}^{(3)}(N)$, Eq.~(\ref{eq:centralgij}), with each of
the instances $\gamma_{ij}^{(3),\,(k)}(N)$, viewed as an independent
nuisance parameter:
\begin{equation}\label{eq:deltaij}
  \Delta_m(ij,k)=  T_{m}(ij,k) - \bar T_{m},
\end{equation}
where  $\bar T_{m}$ is the prediction for the $m$-th datapoint
obtained using the best estimate Eq.~(\ref{eq:centralgij}) for the
full anomalous dimension matrix, while $T_{m}(ij,k)$ is the prediction
obtained when the the $ij$ matrix element of our best estimate is replaced
with the $k$-th instance $\gamma_{ij}^{(3),\,(k)}(N)$.

We then construct the covariance matrix over theory predictions for
individual datapoints due to the IHOU on the $ij$
N$^3$LO matrix element as the covariance of the shifts
$\Delta_m(ij,k)$ over all $\widetilde{N}_{ij}$ instances:
\be\label{eq:covihouij}
\text{cov}^{(ij)}_{mn} = \frac{1}{\widetilde{N}_{ij}-1} \sum_{k=1}^{\widetilde{N}_{ij} }\Delta_m(ij,k)\Delta_n(ij,k).
\ee
We recall that we do not associate an IHOU to the nonsinglet anomalous
dimensions and we assume conservatively that there is no correlation
between the different
singlet anomalous dimension matrix elements.
Thus we can write the total contribution to the theory covariance matrix due to IHOU as
\be\label{eq:covihou}
\text{cov}^\text{IHOU}_{mn} = \text{cov}^{(gg)}_{mn}  +\text{cov}^{(gq)}_{mn}  +\text{cov}^{(qg)}_{mn}  +\text{cov}^{(qq)}_{mn}.  
\ee

The mean square uncertainty on the anomalous dimension matrix element itself is
then determined, by viewing it as a pseudo-observable, as the variance 
\be\label{eq:sigihou}
(\sigma_{ij}(N))^2=  \frac{1}{\widetilde{N}_{ij}-1}
\sum_{k=1}^{\widetilde{N}_{ij}}
\left(\gamma_{ij}^{(3),\,(k)}(N)-\gamma_{ij}^{(3)}(N)\right)^2.
\end{equation}

\subsection{aN$^3$LO anomalous dimensions: the nonsinglet sector}
\label{sec:non_singlet}

Information on the Mellin moments of nonsinglet anomalous dimensions is
especially abundant, in that eight moments of
$\gamma_{{\rm ns},\pm}^{(3)}$ and nine moments of $\gamma_{{\rm ns,s}}^{(3)}$
are known. An approximation based on this knowledge  was 
given in Ref.~\cite{Moch:2017uml}. More recently, further information
on the small-$x$ behavior of
$\gamma_{{\rm ns},\pm}^{(3)}$ was derived in Ref.~\cite{Davies:2022ofz}.
While for $\gamma_{{\rm ns},s}^{(3)}$ we directly rely on the approximation of
Ref.~\cite{Moch:2017uml}, which already includes all the available
information, we construct an approximation to $\gamma_{{\rm ns},\pm}^{(3)}$ 
based on the procedure described in
Sect.~\ref{sec:general_strategy}, in order to include also this
more recent information, and also as a warm-up for the construction of
our approximation to the singlet sector anomalous dimension that we
present in the next section.

Contributions to $\gamma_{{\rm ns},\pm}^{(3)}$ proportional to $ n_f^2$ and $ n_f^3$
are known exactly~\cite{Davies:2016jie} (in particular the $n_f^3$
contributions to  $\gamma_{{\rm ns},\pm}^{(3)}$ coincide), while
$\mathcal{O}( n_f^0)$ and $\mathcal{O}( n_f)$ terms\footnote{The $n_f C^3_F$
  terms have also been published very recently~\cite{Gehrmann:2023iah},
  but we do not include them yet.} are known in the large-$N_c$
limit~\cite{Moch:2017uml} and we include these in
$\gamma_{{\rm ns},\pm ,n_f}^{(3)}(N)$.

Small-$x$ contributions to  $\gamma_{{\rm ns},\pm}$
are double logarithmic, i.e.\ of the form
$a_s^{n+1} \ln^{2n-k}(x)$, corresponding in Mellin space
to poles of order $2n-k+1$ in
$N=0$, i.e.\ $\frac{1}{N^{2n-k+1}}$, so at N$^3$LO we have $n=3$ and thus
\be
    P_{\rm ns,\,\pm}^{(3)}(x) = \sum_{k=1}^{6} c^k_{{\rm ns}, \,N\to 0} \ln^k(1/x)+ O(x)\, .
    \label{eq:ns_smallxx}
\ee
The coefficients $c^k_{{\rm ns},\,N\to 0}$
are known~\cite{Davies:2022ofz} exactly up to
NNLL accuracy ($k=4,\,5,\,6)$, and approximately  up to N$^6$LL ($k=1,\,2,\,3$).
Hence, we let
\begin{equation}
    \label{eq:ns_smallx}
    \gamma_{{\rm ns},\pm, \,N\to 0}^{(3)}(N)=\sum_{k=1}^{6} c^k_{{\rm ns}, \,N\to 0} (-1)^k \frac{k!}{N^{k+1}}.
\end{equation}

Large-$x$ logarithmic contributions in the $\overline{\rm MS}$
scheme only appear in coefficient
functions~\cite{Albino:2000cp},
and so the $x\to 1$ behaviour of splitting functions is provided by the
cusp anomalous dimension $\sim\frac{1}{(1-x)_+}$, corresponding to a
single $\ln(N)$ behavior in Mellin space as $N\to\infty$. This behavior
is common to the pair of anomalous dimensions $\gamma_{{\rm ns},\pm}^{(3)}(N)$.
Furthermore,
several subleading power corrections as $N\to\infty$ can also be
determined and we set
\be
\gamma_{{\rm ns},\pm,\,N\to \infty}^{(3)}(N) =A^{q}_4 S_1(N) + B_4^{q} + C_4^{q} \frac{S_1(N)}{N} + D_4^{q} \frac{1}{N} ,
\label{eq:ns_largen}
\ee
where $S_1$ denotes the harmonic sum (see Eqs.~\ref{eq:multi_idx_harmonics}-\ref{eq:harmterm}).
The coefficient of the $\ln(N)$ term  $A^{q}_4$, is the quark cusp
anomalous dimension~\cite{Henn:2019swt}. The constant coefficient
$B^{q}$ is determined by the integral of the nonsinglet splitting
function, which was originally computed in \cite{Moch:2017uml} 
in the large-$N_c$ limit and recently updated to the full color
expansion~\cite{Duhr:2022cob}  
as a result of computing different N$^3$LO cross-sections in the soft limit.
The coefficients of the terms suppressed by $1/N$ in the large-$N$ limit,
$C^{q}$ and $ D^{q}$,
can be obtained directly from lower-order anomalous dimensions
by exploiting  large-$x$ resummation techniques~\cite{Davies:2016jie}.
The explicit expressions of $\gamma_{{\rm ns}\, \pm,\,N\to \infty}^{(3)}(N)$ and
$\gamma_{{\rm ns}\, \pm,\,N\to 0}^{(3)}(N)$ are given in
Appendix~\ref{app:splitting_asy}.

\begin{table}[!t]
  \centering
  \small
  \renewcommand{\arraystretch}{1.50}
\begin{tabularx}{0.6\textwidth}{Xcc}
\midrule
$G^{{\rm ns}, \pm}_1(N)$ &   1 \\
$G^{{\rm ns}, \pm}_2(N)$ & $\mathcal{M}[(1-x)\ln(1-x)](N)$ \\
$G^{{\rm ns}, \pm}_3(N)$ & $\mathcal{M}[(1-x)\ln^2(1-x)](N)$ \\
$G^{{\rm ns}, \pm}_4(N)$ &  $\mathcal{M}[(1-x)\ln^3(1-x)](N)$ \\
$G^{{\rm ns}, \pm}_5(N)$ & $\frac{S_1(N)}{N^2}$  \\
$G^{{\rm ns}, \pm}_6(N)$ &  $\frac{1}{(N+1)^2}$  \\
$G^{{\rm ns}, \pm}_7(N)$ & $\frac{1}{(N+1)^3}$ \\
$G^{{\rm ns}, +}_8(N)$, $G^{{\rm ns}, -}_8(N)$  &  $\frac{1}{(N+2)}$, $\frac{1}{(N+3)}$ \\
\bottomrule
\end{tabularx}
\vspace{0.2cm}
\caption{The Mellin space interpolating functions $G^{{\rm ns},\pm}_\ell(N)$ 
  entering the parametrisation of the remainder term
  $\widetilde{\gamma}^{(3)}_{{\rm ns}\, \pm}(N)$
  for the nonsinglet anomalous dimension expansion
  of Eq.~(\ref{eq:gamma_residual_basis}).}
\label{tab:functions_interpolating}
\end{table}

The remainder terms, $\widetilde{\gamma}_{\rm ns, \pm}^{(3)}(N)$, are
expanded over the set of eight functions $G^{{\rm ns},\pm}_\ell(N)$ listed in
Table~\ref{tab:functions_interpolating}. The 
coefficients $b^{\rm ns,\pm}_\ell$ (defined in Eq.\eqref{eq:gamma_residual_basis})
are determined by imposing  that the values of the eight moments given in
Ref.~\cite{Moch:2017uml} be reproduced. The set of functions
$G^{{\rm ns},\pm}_\ell(N)$ is chosen to adjust the overall constant ($\ell=1$),
model the large-$N$ behavior ($2\le \ell\le 5$) and model the small-$N$ behavior
($\ell=6,7$), consistent with the general analytic
structure of fixed order anomalous dimensions~\cite{Ball:2013bra}.
Specifically, the large-$N$ functions are chosen as the
logarithmically enhanced next-to-next-to-leading power terms ($\ln^k(N)/N^2$,
$\ell=2,3,4,5$) and the small-$N$ functions are chosen as
logarithmically enhanced subleading poles ($1/(N+1)^k$, $\ell=6,7$)
and sub-subleading poles ($1/(N+2)$ or $1/(N+3)$,  $\ell=8$).
The last element, $\ell = 8$, is chosen
at a fixed distance from the lowest known moment, $N=2$ for
$\gamma_{{\rm ns},+}^{(3)}(N)$ and $N=1$ for
$\gamma_{{\rm ns},-}^{(3)}(N)$.

\begin{figure}[!t]
  \centering
  \includegraphics[width=.49\textwidth]{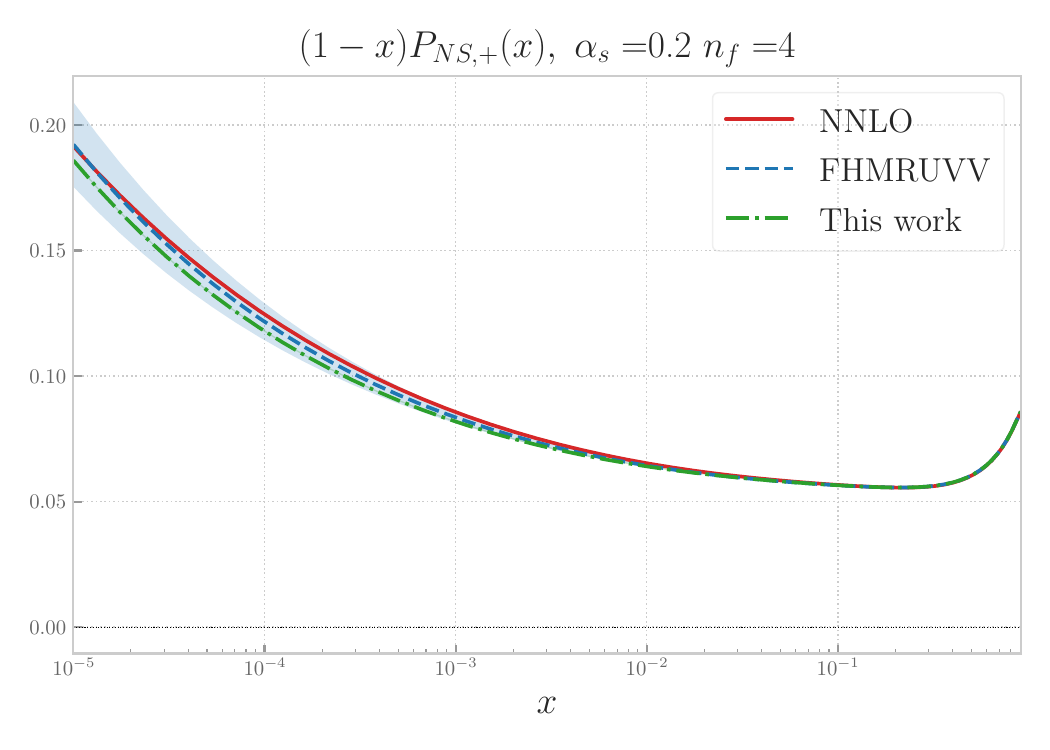}
  \includegraphics[width=.49\textwidth]{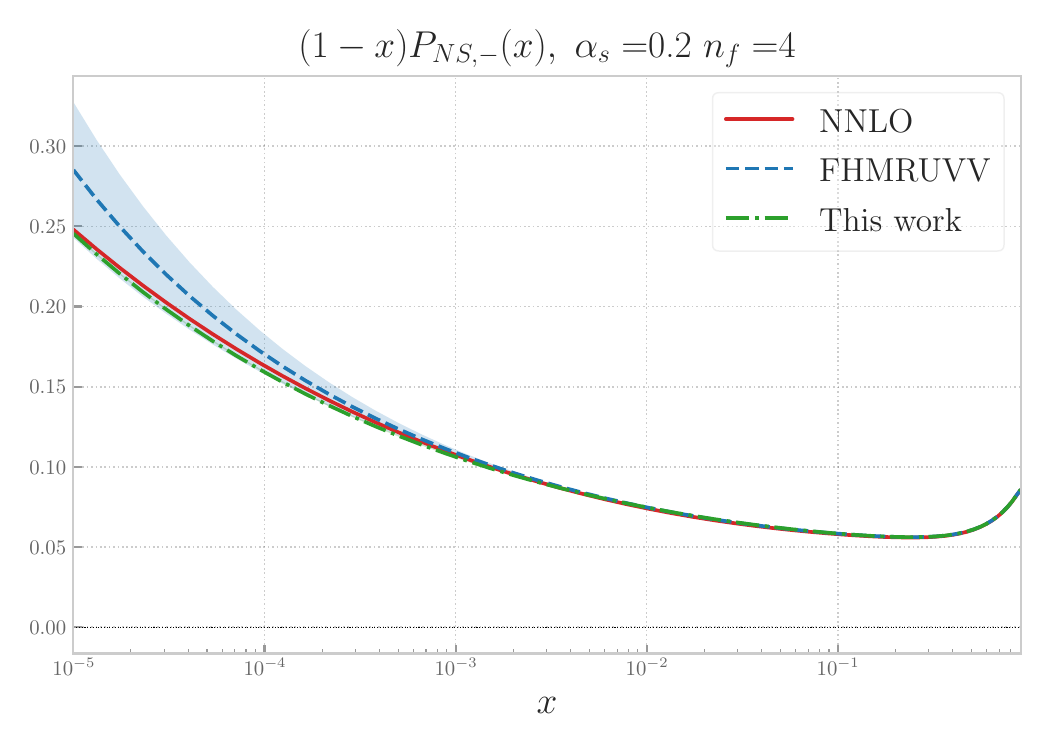}
  \caption{The aN$^3$LO nonsinglet splitting functions $(1-x)
    P_{{\rm ns},+}(x,\alpha_s)$ 
    and $(1-x) P_{{\rm ns},-}(x,\alpha_s)$, evaluated 
    as a function of $x$  for $n_f=4$ and $\alpha_s=0.2$ in our
    approximation compared to the previous  approximation of
    Ref.~\cite{Moch:2017uml} (denoted FHMRUVV), for which the approximation
    uncertainty, as estimated by its authors, is also displayed. 
    For comparison, the (exact) NNLO result is also shown.}
  \label{fig:nsgamma} 
\end{figure}

In Fig.~\ref{fig:nsgamma} we plot the resulting splitting functions
$P_{{\rm ns},\pm}^{(3)}(x)$, obtained by Mellin inversion of the anomalous
dimension. We compare our approximation to the approximation of
Ref.~\cite{Moch:2017uml}, for $\alpha_s=0.2$ and $n_f=4$, and also show 
the (exact) NNLO result for reference. Because
the splitting function is a distribution at $x=1$ we plot $(1-x)P(x)$.
The result of Ref.~\cite{Moch:2017uml} also provides an estimate
of the uncertainty related to the approximation, shown in the
figure as a band, and we observe that this uncertainty
is negligible except at very small $x$.
As we include further constraints on the small-$x$ behavior,
the uncertainty on the approximation becomes negligible, as it can be
checked by comparing results obtained by including increasingly more
information in the construction of the approximation. 
Consequently, as mentioned in Sect.~\ref{sec:general_strategy} above, we take
$n_H=0$ in Eq.~(\ref{eq:gamma_residual_basis}).

\subsection{aN$^3$LO anomalous dimensions: singlet sector}
\label{sec:singlet}

In order to determine the singlet-sector anomalous dimension
matrix entering Eq.~(\ref{eq:dglap_singlet}), we must determine
$\gamma_{{qq},{\rm ps}}$ that, together with the previously determined
nonsinglet anomalous dimension, contributes to the $qq$
entry, Eq.~(\ref{eq:p_qq}), and then also the three remaining matrix elements
$\gamma_{{qg}}$, $\gamma_{{gq}}$, and $\gamma_{{gg}}$.

For all matrix elements, the leading large-$n_f$ $\mathcal{O}(n_f^3)$
contributions in Eq.~(\ref{eq:gammaf}) are 
known analytically~\cite{Davies:2016jie}, while for
$\gamma_{{qq},{\rm ps}}$~\cite{Gehrmann:2023cqm}
and $\gamma_{{gq}}$~\cite{Falcioni:2023tzp} the $\mathcal{O}(n_f^2)$
contributions are also known and we include all of them in
$\gamma_{ij,n_f}^{(3)}(N)$.

Small-$x$ contributions in the singlet sector include, on top of the
double-logarithmic contributions $a_s^{n+1} \ln^{2n-k}(x)$ that are
present in the nonsinglet case, also single-logarithmic contributions
$a_s^{n+1} \frac{1}{x} \ln^{n}(x)$. In Mellin space, this means that
on top of order $2n-k+1$ subleading poles in
$N=0$, there are also leading poles in $N=1$ of order $n-k+1$, 
i.e.\ $\frac{1}{(N-1)^{n-k+1}}$.
The leading-power single logarithmic contributions can be extracted from the
leading~\cite{Fadin:1975cb,Kuraev:1976ge,Lipatov:1976zz,Kuraev:1977fs,
  Balitsky:1978ic}
and next-to-leading~\cite{Fadin:1996nw,Fadin:1997hr,Fadin:1997zv,Camici:1997ij,Fadin:1998py} high-energy resummation at
LL$x$~\cite{Jaroszewicz:1982gr} and NLL$x$~\cite{Ball:1995vc,Ball:1999sh,Bonvini:2018xvt}
accuracy. This allows for a determination of the coefficients of the leading
$\frac{1}{(N-1)^4}$ and next-to-leading $\frac{1}{(N-1)^3}$ contributions to
$\gamma^{(3)}_{gg}$ and of the next-to-leading $\frac{1}{(N-1)^3}$ contributions
to $\gamma^{(3)}_{qg}$. The remaining entries can be obtained from these
by using the color-charge (or Casimir scaling) relation
$\gamma_{iq}=\frac{C_F}{C_A}\gamma_{ig}$~\cite{Catani:1994sq,Bonvini:2018xvt}.
Hence, we set
\begin{align}
 \label{eq:smallxgg}
 \gamma_{gg,\,N\to 1}^{(3)}(N) & = c^4_{gg, \,N\to 1}\frac{1}{(N-1)^4}+c^3_{gg, \,N\to 1}\frac{1}{(N-1)^3}; \\
 \label{eq:smallxqg}
 \gamma_{qg,\,N\to 1}^{(3)}(N) & = c^3_{qg, \,N\to 1}\frac{1}{(N-1)^3}; \\ \label{eq:smallxiq}
 \gamma_{iq,\,N\to 1}^{(3)}(N) & = \frac{C_F}{C_A} \gamma_{ig,\,N\to 1}^{(3)}(N),\quad i= q,\,g.
\end{align}

Although only the leading pole of $\gamma_{gq}$ satisfies
Eq.~\eqref{eq:smallxiq} exactly,
at NNLO this relation is only violated at the sub-percent
level~\cite{Vogt:2004mw}, so this is likely to be an adequate
approximation also at this order: this approximation is also adopted in
Ref.~\cite{Moch:2023tdj}. An important observation is
that both NLO and NNLO coefficients of
the leading poles, $\frac{1}{(N-1)^2}$ and $\frac{1}{(N-1)^3}$
respectively, vanish accidentally. Hence, at N$^3$LO the leading poles
contribute for the first time beyond leading order.
The subleading poles can be
determined up to NNLL accuracy~\cite{Davies:2022ofz} and, thus, fix
the coefficients of the $\frac{1}{N^7}$,  $\frac{1}{N^6}$ and
$\frac{1}{N^5}$ subleading poles for all entries of the singlet
anomalous dimension matrix. All these contributions are included in
$\gamma_{ij,N\to 1}^{(3)}(N)$ and $\gamma_{ij,N\to 0}^{(3)}(N)$. 

In the singlet sector, large-$x$ contributions whose Mellin transform
is not suppressed in 
the large-$N$ limit only appear in the diagonal $qq$ and $gg$
channels. In the quark channel these are already included, through
Eq.~(\ref{eq:p_qq}) in $\gamma_{{\rm ns},+,\,N\to \infty}^{(3)}(N)$,
according to Eq.~(\ref{eq:ns_largen}), while  $\gamma^{(3)}_{qq,{\rm ps}}$
is suppressed in this limit. In the gluon-to-gluon channel they take
the same form as in the nonsinglet and diagonal quark channel. Hence,
we expand, as in Eq.~(\ref{eq:ns_largen}),
\be\label{eq:largexgg}
 \gamma_{gg,\,N\to \infty}^{(3)}(N)
= A^{g}_4 S_1(N) + B^{g}_4 +  C^{g}_4 \frac{S_1(N)}{N} + D^{g}_4 \frac{1}{N} \, .
\ee
The coefficients $A^{g}_4$, $ B^{g}_4$, $C^{g}_4$
and $D^{g}_4$ are the  counterparts of those of
\cref{eq:ns_largen}: the gluon cusp anomalous dimension was determined
in Ref.~\cite{Henn:2019swt} and the constant in
Ref.~\cite{Duhr:2022cob}, while the  $C^{g}_4$ and $D^{g}_4$
coefficients can be determined using results from
Refs.~\cite{Dokshitzer:2005bf,Moch:2023tdj}.  

Off-diagonal $qg$ and $qg$ splitting functions have logarithmically
enhanced next-to-leading power behavior at large-$x$:
\begin{equation}
      P_{ij}^{(3)}(x) = \sum_{k=0}^{6} \sum_{l=0}^{\infty} c^{k,l}_{ij,\,N\to \infty} (1-x)^l \ln^k(1-x) . 
      \label{eq:largex_expansion}
\end{equation}
For $l=0$ the coefficients of the higher logs $k=4,5,6$ can be determined from 
N$^3$LO coefficient functions, based on a conjecture~\cite{Soar:2009yh,Almasy:2010wn} 
on the large-$x$ behavior of the physical evolution kernels that give the scale
dependence of structure functions. 
The coefficient with the highest power $k=6$ cancels and thus we let
\begin{align}
      \label{eq:largexgq}
      \gamma_{gq,\,N\to \infty}^{(3)}(N) = \sum_{k=4}^5 c^{k,0}_{gq, \,N\to \infty} & \mathcal{M}\left[\ln^k(1-x)\right](N), \\
      \label{eq:largexqg}
      \gamma_{qg,\,N\to \infty}^{(3)}(N) = \sum_{k=4}^5 c^{k,0}_{qg, \,N\to \infty} \mathcal{M}\left[\ln^k(1-x)\right](N) & + c^{k,1}_{qg, \,N\to \infty} \mathcal{M}\left[(1-x)\ln^k(1-x)\right](N), 
\end{align}
where in $\gamma_{qg,\,N\to \infty}^{(3)}$ we have retained also the
$l=1$ terms
\cite{Falcioni:2023vqq}.

Finally, the pure singlet quark-to-quark
splitting function starts at next-to-next-to-leading power as $x\to1$,
i.e.\ it behaves as  $(1-x)\ln^k(1-x)$, with $k\le 4$. 
The coefficients of the higher logs $k=3,4$  can be extracted by expanding the
$x=1$ expressions from Refs.~\cite{Soar:2009yh,Falcioni:2023luc}. Hence, we let
 \be
 \label{eq:largexps}
 \gamma_{qq,{\rm ps},\,N\to \infty}^{(3)}(N)=
 \sum_{k=3}^4 \left[c^{k,1}_{qq,{\rm ps},\,N\to \infty} \mathcal{M}\left[(1-x)\ln^k(1-x)\right](N) + c^{k,2}_{qq,{\rm ps},\,N\to \infty} \mathcal{M}\left[(1-x)^2\ln^k(1-x)\right](N)\right] 
\ee
Note that for the $qq$ and $qg$ entries we also include the (known) 
next-to-leading power contributions, while we do not include them for $gq$ and
$gg$ because for these anomalous dimension matrix elements a
significantly larger number of higher Mellin moments is known, hence
there is no risk that the inclusion of these contributions could
contaminate the intermediate $x$ region where they are not
necessarily dominant. The explicit expressions of
$\gamma_{ij\,N\to \infty}^{(3)}(N)$,
$\gamma_{ij\,N\to 0}^{(3)}(N)$ and $\gamma_{ij\,N\to 1}^{(3)}(N)$ 
are all given in Appendix~\ref{app:splitting_asy}.

As discussed in Sect.~\ref{sec:general_strategy}, the remainder
contribution $\widetilde{\gamma}_{ij}^{(3)}(N)$,
Eq.~(\ref{eq:gamma_residual_basis}), is determined by expanding each
of its matrix elements  over
a set of  $n^{ij}$ basis functions, where  $n^{ij}$ is the number of
known Mellin moments,  and determining the expansion coefficients by
demanding that the known moments be reproduced.
Specifically, the known moments  are the four moments
computed in Ref.~\cite{Moch:2021qrk}, the six additional moments
for $\gamma_{qq,{\rm ps}}$ and $\gamma_{qg}$ computed in
Ref.~\cite{Falcioni:2023luc} and Ref.~\cite{Falcioni:2023vqq} respectively,
and the additional moment $N=10$ for $\gamma_{gg}$ and $\gamma_{gq}$
evaluated in Ref.~\cite{Moch:2023tdj}. 
These constraints automatically implement momentum conservation:
\begin{align}
\begin{split}
    \gamma_{qg}(N=2) + \gamma_{gg}(N=2) &= 0 \, , \\
    \gamma_{qq}(N=2) + \gamma_{gq}(N=2) &= 0 \, .
    \label{eq:singlet_scaling}
\end{split}
\end{align}

\begin{table}[!t]
  \centering
  \small
  \renewcommand{\arraystretch}{1.70}
  \begin{tabularx}{\textwidth}{Xcc}
    \toprule
    \multirow{5}{*}{$\gamma_{gg}^{(3)}(N)$} 
    & $ G^{gg}_{1}(N) $ &	$\mathcal{M}[(1-x)\ln^3(1-x)](N)$\\
    & $ G^{gg}_{2}(N) $  &	$\frac{1}{(N-1)^2} $\\
    & $ G^{gg}_{3}(N) $  &	$\frac{1}{N-1}$ \\
    & $\{ H^{gg}_{1}(N), \ H^{gg}_{2}(N) \}$ &
    $\frac{1}{N^4},\ 
    \frac{1}{N^3},\ 
    \frac{1}{N^2},\ 
    \frac{1}{N+1},\ 
    \frac{1}{N+2},\ 
    \mathcal{M}[(1-x)\ln^2(1-x)](N),\ 
    \mathcal{M}[(1-x)\ln(1-x)](N) $ \\
    \midrule
    \multirow{5}{*}{$\gamma_{gq}^{(3)}(N)$} 
    & $ G^{gq}_{1}(N) $  &	$\mathcal{M}[\ln^3(1-x)](N)$ \\ 
    & $ G^{gg}_{2}(N) $  &	$\frac{1}{(N-1)^2} $ \\
    & $ G^{gq}_{3}(N) $  &	$\frac{1}{N-1}$ \\
    & $\{ H^{gq}_{1}(N), \ H^{gq}_{2}(N) \}$ &	
    $\frac{1}{N^4},\
    \frac{1}{N^3},\ 
    \frac{1}{N^2},\ 
    \frac{1}{N+1},\ 
    \frac{1}{N+2},\ 
    \mathcal{M}[\ln^2(1-x)](N),\ 
    \mathcal{M}[\ln(1-x)](N)$ \\
    \midrule
    \multirow{6}{*}{$\gamma_{qg}^{(3)}(N)$}  
    & $ G^{qg}_{1}(N) $ &	$\mathcal{M}[\ln^3(1-x)](N)$ \\
    & $ G^{qg}_{2}(N) $ &	$\frac{1}{(N-1)^2}$ \\
    & $ G^{qg}_{3}(N) $ &	 $\frac{1}{N-1}-\frac{1}{N}$ \\
    & $ G^{qg}_{4,\dots,8}(N) $  &	
    $\frac{1}{N^4},\
    \frac{1}{N^3},\
    \frac{1}{N^2},\
    \frac{1}{N},\ 
    \mathcal{M}[\ln^2(1-x)](N)$ \\
    &  \multirow{2}{*}{$\{ H^{qg}_{1}(N), \ H^{qg}_{2}(N) \}$}  &	
    $\mathcal{M}[\ln(x) \ln(1-x)](N),\
    \mathcal{M}[\ln(1-x)](N),\
    \mathcal{M}[(1-x)\ln^3(1-x)](N)$ \\
    & & $\mathcal{M}[(1-x)\ln^2(1-x)](N),\ 
    \mathcal{M}[(1-x)\ln(1-x)](N),\ 
    \frac{1}{1+N}$ \\
    \midrule
    \multirow{6}{*}{$\gamma_{qq,{\rm ps}}^{(3)}(N)$}
    & $ G^{qq,{\rm ps}}_{1}(N) $  &	$\mathcal{M}[(1-x)\ln^2(1-x)](N)$  \\ 
    & $ G^{qq,{\rm ps}}_{2}(N) $  &	$-\frac{1}{(N-1)^2} + \frac{1}{N^2}$ \\
    & $ G^{qq,{\rm ps}}_{3}(N) $  &	$-\frac{1}{(N-1)} + \frac{1}{N}$ \\
    & \multirow{2}{*}{$ G^{qq,{\rm ps}}_{4,\dots,8}(N) $}  & 
    $\frac{1}{N^4},\
    \frac{1}{N^3},\
    \mathcal{M}[(1-x)\ln(1-x)](N)$ \\
    & & $\mathcal{M}[(1-x)^2\ln(1-x)^2](N),\ 
    \mathcal{M}[(1-x)\ln(x)](N)$ \\
    & \multirow{2}{*}{$\{ H^{qq,{\rm ps}}_{1}(N), \ H^{qq,{\rm ps}}_{2}(N) \}$} &  
    $\mathcal{M}[(1-x)(1+2x)](N),\ 
    \mathcal{M}[(1-x)x^2](N)$, \\
    & & $\mathcal{M}[(1-x) x (1+x)](N),\ 
    \mathcal{M}[(1-x)](N)$ \\
    \bottomrule
  \end{tabularx}
  \vspace{0.2cm}
  \caption{The set of basis functions $G^{ij}_\ell (N)$ and $H^{ij}_\ell (N)$
    used to parametrize the singlet sector remainder 
    anomalous dimensions matrix elements $\widetilde{\gamma}_{ij}^{(3)}(N)$
    according to Eq.~(\ref{eq:gamma_residual_basis}).}
  \label{tab:functions_interpolating_ihou}
\end{table}

The set of basis functions is chosen based on the idea of constructing
an approximation that reproduces the singularity structure of the Mellin
transform of the anomalous dimension viewed as analytic functions in
$N$ space~\cite{Ball:2013bra}, hence corresponding to the leading and
subleading (i.e.\ rightmost)
$N$-space poles with unknown coefficients as well as the leading
unknown large-$N$ behavior.  As mentioned in
Sect.~\ref{sec:general_strategy}, the uncertainty on the answer is
then estimated by varying the set of basis functions, specifically by
varying two out of the  $n^{ij}$ basis functions. The way the basis
functions are partitioned between the fixed functions  $G^{ij}$
and  the varying functions  $H^{ij}$ is by always including in the
fixed set the most leading unknown contributions, and in the  $H^{ij}$
further subleading ones. The number of varying  $H^{ij}$  is chosen to
be larger when less exact information is known.

Specifically, the functions $G^{ij}$ are chosen as follows.
\begin{enumerate}
\item The function $G^{ij}_1(N)$ reproduces the leading unknown
  contribution in the large-$N$ limit, i.e.\ the unknown term in
  Eq.~(\ref{eq:largex_expansion}) with highest $k$ and lowest $l$.
\item The functions $G^{ij}_2(N)$ and $G^{ij}_3(N)$
  reproduce the first two
  leading  unknown contributions in the small-$N$ limit, i.e.\ the unknown
  $\frac{1}{(N-1)^k}$ leading poles with highest and next-to-highest
  values of $k$, i.e.\ $k=2$ and  $k=1$. For $\gamma_{qq,{\rm ps}}$
  and  $\gamma_{qg}$ a 
  subleading small-$x$ pole 
  with the same power and opposite sign is added to the leading pole
  with respectively  $k=1,\,2$ and $k=1$, so as to leave unaffected the
  respective
  large-$x$ leading power behavior Eqs.~(\ref{eq:largexqg}-\ref{eq:largexps}).
\item For $\gamma_{qq,{\rm ps}}$
  and  $\gamma_{qg}$, for which an additional five moments are known, the
  functions   $G^{qj}_{4,\dots,8}(N)$ reproduce subleading small- and
  large-$N$ terms.
\end{enumerate}
Note that a larger number of basis functions is chosen to describe the
small-$N$ poles  rather than the large-$N$ behavior because  less
exact information  is available in the former case: so for instance
only the leading pole Eq.~(\ref{eq:smallxqg}) is known for
$\gamma_{qg}^{(3)}(N)$, while the first two logarithmically enhanced
large-$N$ contributions to it Eq.~(\ref{eq:largexqg}) are known. 

As mentioned, the functions $H^{ij}$ are chosen to reproduce further
subleading contributions:
\begin{enumerate}
\item The functions  $H^{gj}_1(N),H^{gj}_2(N)$ in the gluon sector,
  where only five moments are known exactly, are chosen to reproduce subleading
  small- and large-$N$ terms, i.e.\ similar to  $G^{qj}_{4,\dots,8}(N)$.
\item The functions  $H^{qg}_1(N),H^{qg}_2(N)$ are chosen as
  subleading and next-to-leading power large-$x$ terms and the
  remaining unknown leading small-$N$ pole.
\item The functions  $H^{qq,\,{\rm ps}}_1(N),H^{qq,\,{\rm ps}}_2(N)$
  are chosen as low-order polynomials, i.e., sub-subleading small-$x$
  poles.
\end{enumerate}
Also as mentioned, the number of basis functions is greater for anomalous
dimension matrix elements for which less exact information is
available: 7 in the gluon sector (i.e.\ $gg$ and $gq$), 6 for the $qg$ entry
and 4 for the pure singlet entry. For the $gg$ entry
two combinations are discarded as they lead to unstable
(oscillating) results and we thus end up with $\widetilde{N}_{gg}=19$,
$\widetilde{N}_{gq}=21$, $\widetilde{N}_{qg}=15$, and $\widetilde{N}_{qq}=6$
different parametrizations.
The full set of basis functions $G^{ij}$ and $H^{ij}$ is listed in 
Table~\ref{tab:functions_interpolating_ihou}. We have checked that
results are stable upon variation of these choices, so for instance
including a larger number of   $H^{ij}$ functions does not lead to
significantly larger uncertainties. 

Upon combining the exactly known contributions with the
$\widetilde{N}_{ij}$ remainder terms according to
Eq.~(\ref{eq:ad_expansion_terms}) we end up with an ensemble of
$\widetilde{N}_{ij}$ instances of $\gamma_{ij}^{(3),\,(k)}(N)$ for each singlet
anomalous dimension matrix element and the final matrix elements
$\gamma_{ij}^{(3)}(N)$ and their uncertainties $\sigma_{ij}(N)$ are computed using
Eqs.~(\ref{eq:centralgij}) and~(\ref{eq:sigihou}) respectively.

\subsection{Results: aN$^3$LO splitting functions}
\label{sec:n3lo_eko}

\begin{figure}[!t]
  \centering
  \includegraphics[width=.49\textwidth]{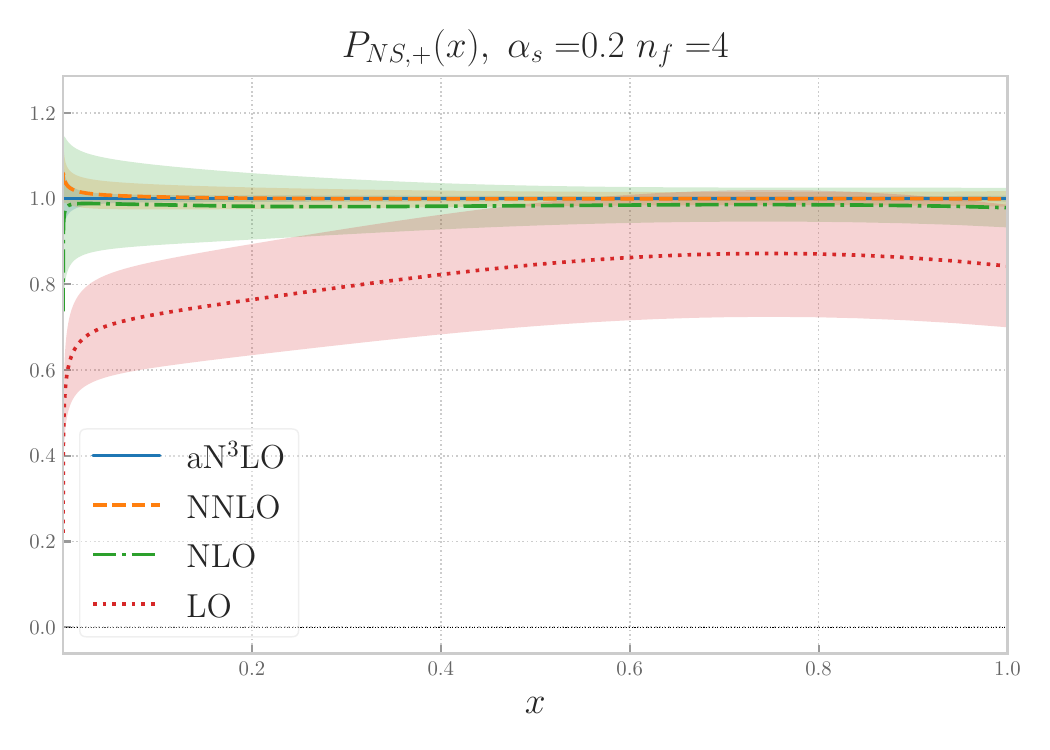}
  \includegraphics[width=.49\textwidth]{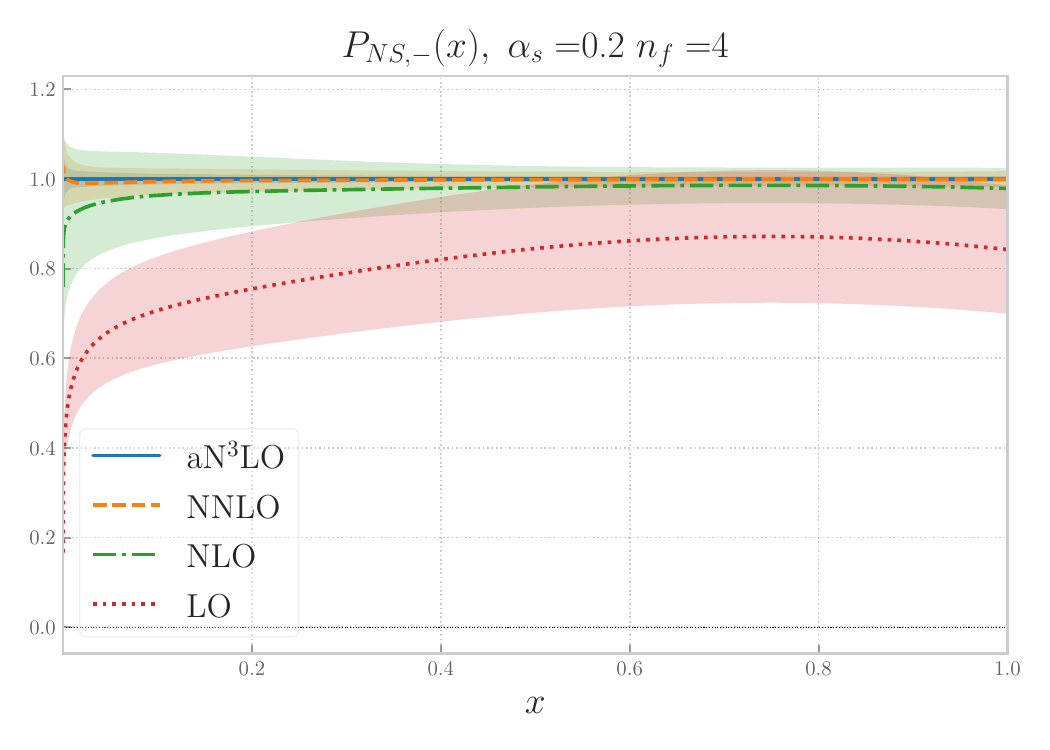}
  \caption{\small The nonsinglet splitting functions at
    LO, NLO, NNLO, and aN$^3$LO, normalized to the aN$^3$LO
    central value and with a linear scale on the $x$ axis.
    In each case we shown also the 
    uncertainty due to missing higher orders (MHOU) estimated by scale
    variation according to Refs.~\cite{NNPDF:2019vjt,NNPDF:2019ubu}.
  }
  \label{fig:splitting-functions-ns} 
\end{figure}

\begin{figure}[!t]
  \centering
  \includegraphics[width=.49\textwidth]{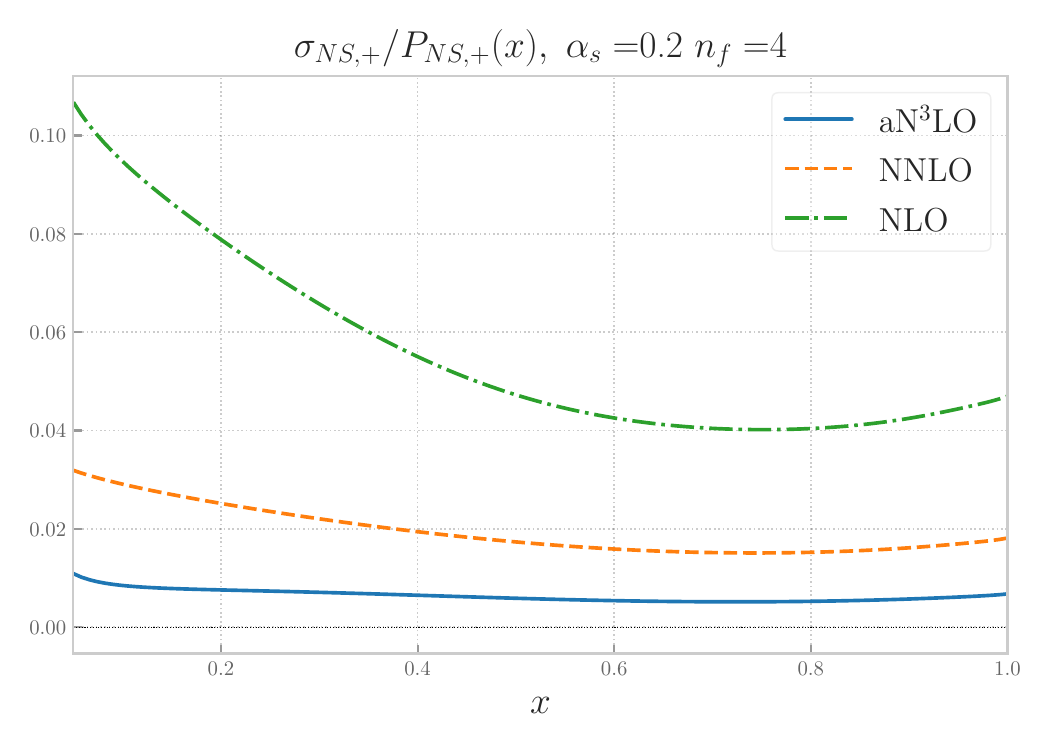}
  \includegraphics[width=.49\textwidth]{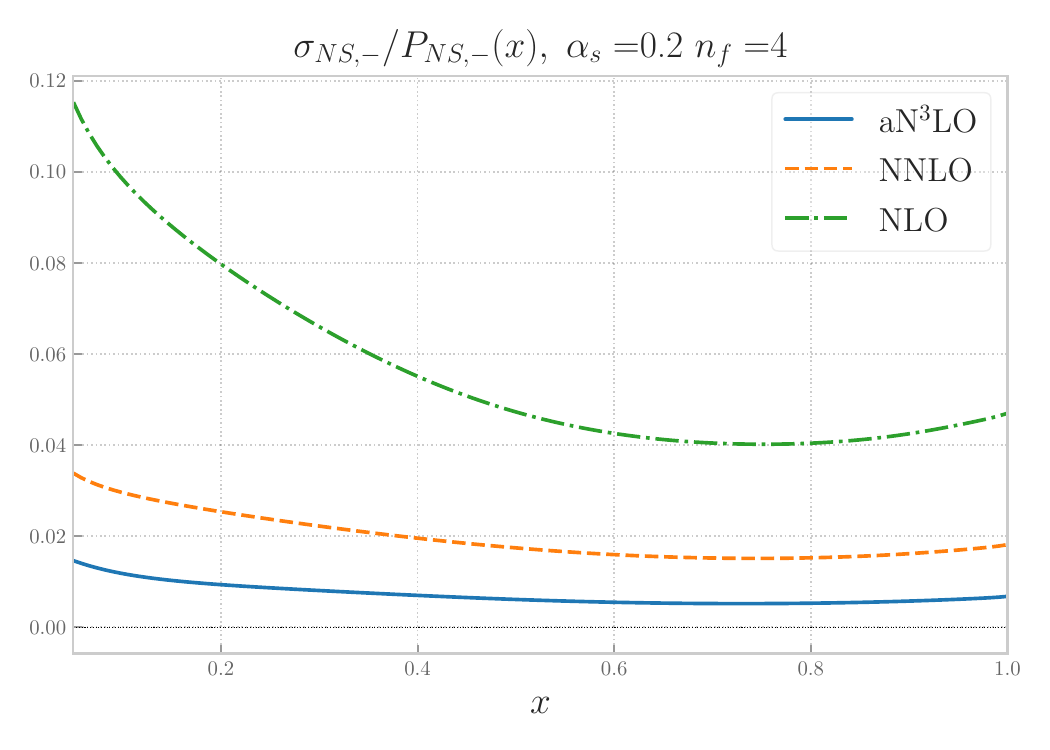}
  \caption{\small The relative size of the uncertainty due to missing higher
    orders (MHOU) on the splitting functions
    of Fig.~\ref{fig:splitting-functions-ns}.}
  \label{fig:splitting-functions-ns-unc} 
\end{figure}

\begin{figure}[!t]
  \centering
  \includegraphics[width=.49\textwidth]{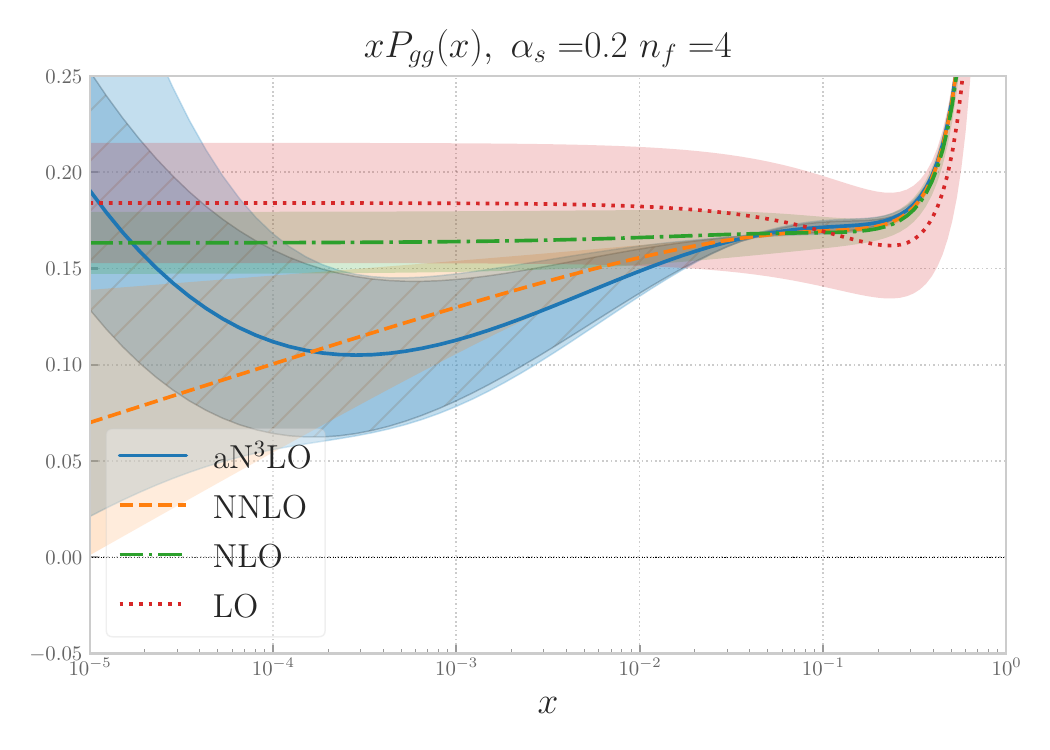}
  \includegraphics[width=.49\textwidth]{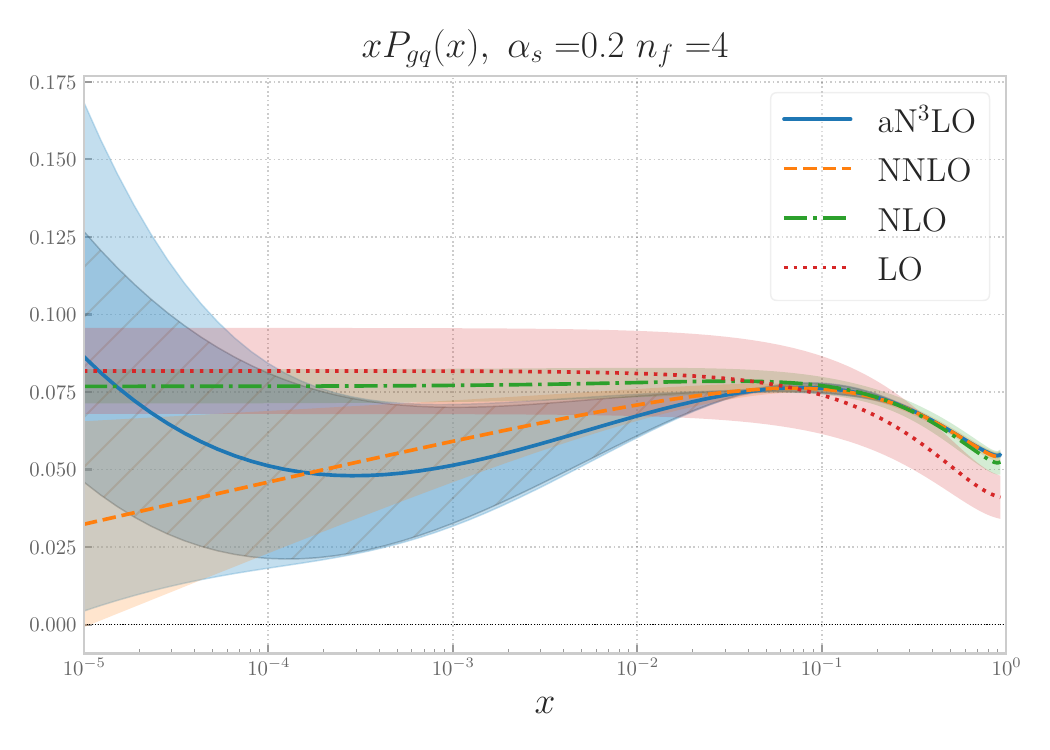} \\
  \includegraphics[width=.49\textwidth]{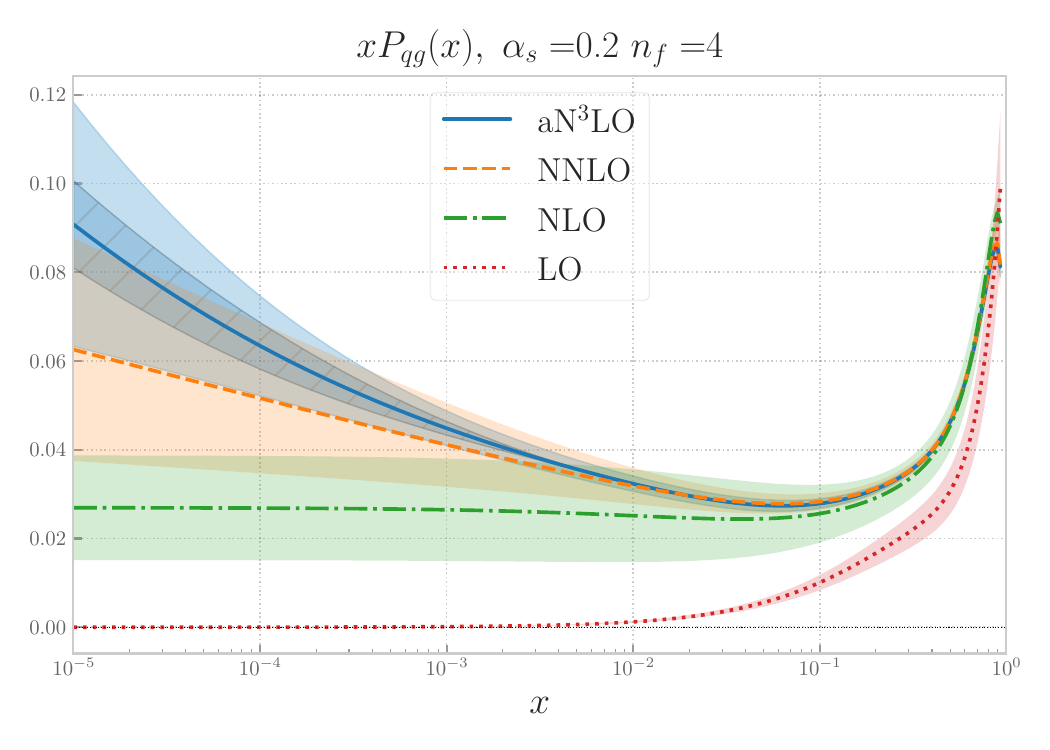}
  \includegraphics[width=.49\textwidth]{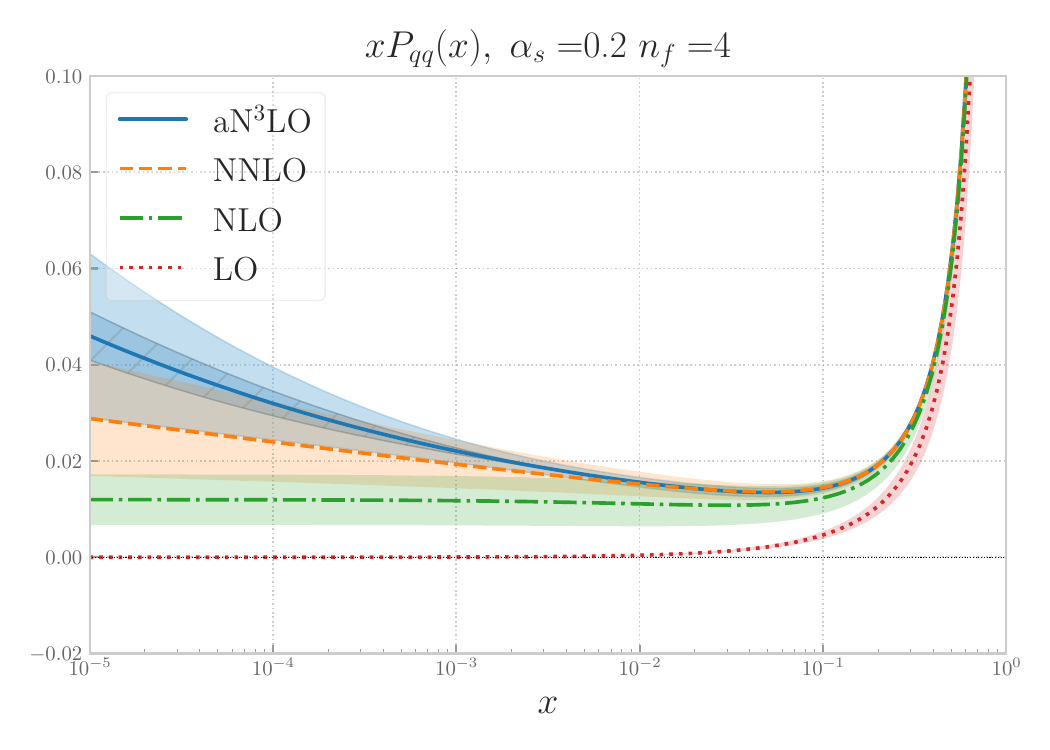}
  \caption{\small The singlet matrix of splitting functions $xP_{ij}$  at
    LO, NLO, NNLO  and aN$^3$LO. From left to
    right and from top to bottom the $gg$, $gq$, $qg$ and $qq$ entries
    are shown. The MHOU estimated by scale
    variation is shown to all orders. At aN$^3$LO
    the dark blue band corresponds to IHOU only, 
    while the light blue band is the sum in 
    quadrature of IHOU and MHOU.}
  \label{fig:splitting-functions-singlet-logx} 
\end{figure}

\begin{figure}[!t]
  \centering
  \includegraphics[width=.49\textwidth]{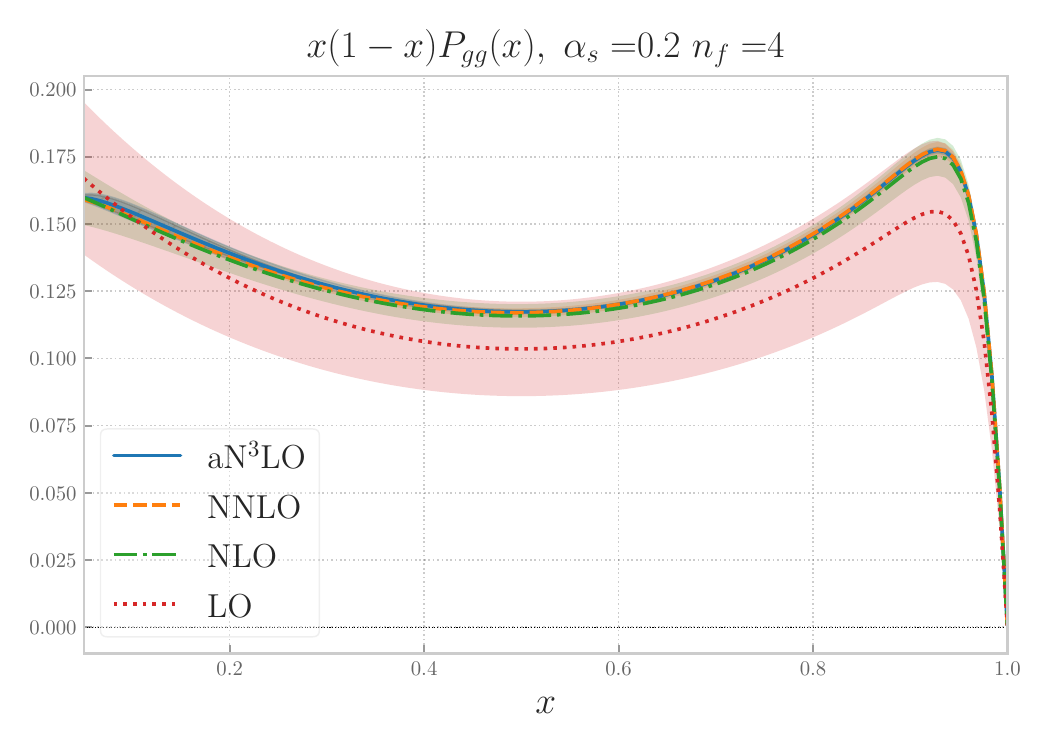}
  \includegraphics[width=.49\textwidth]{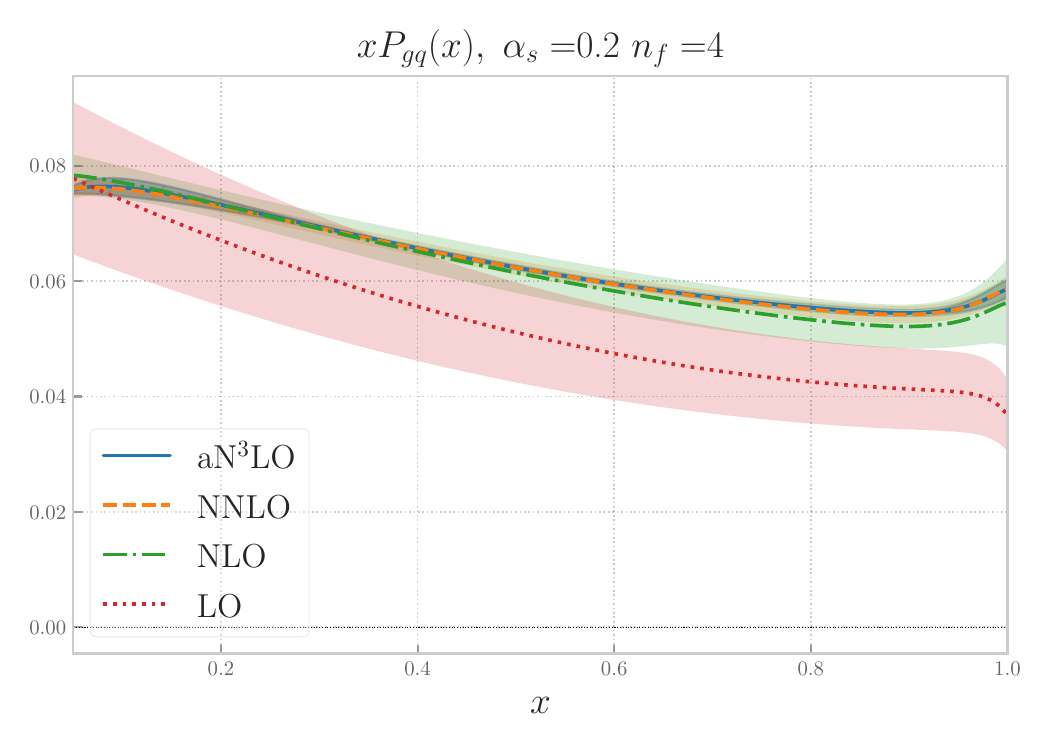} \\
  \includegraphics[width=.49\textwidth]{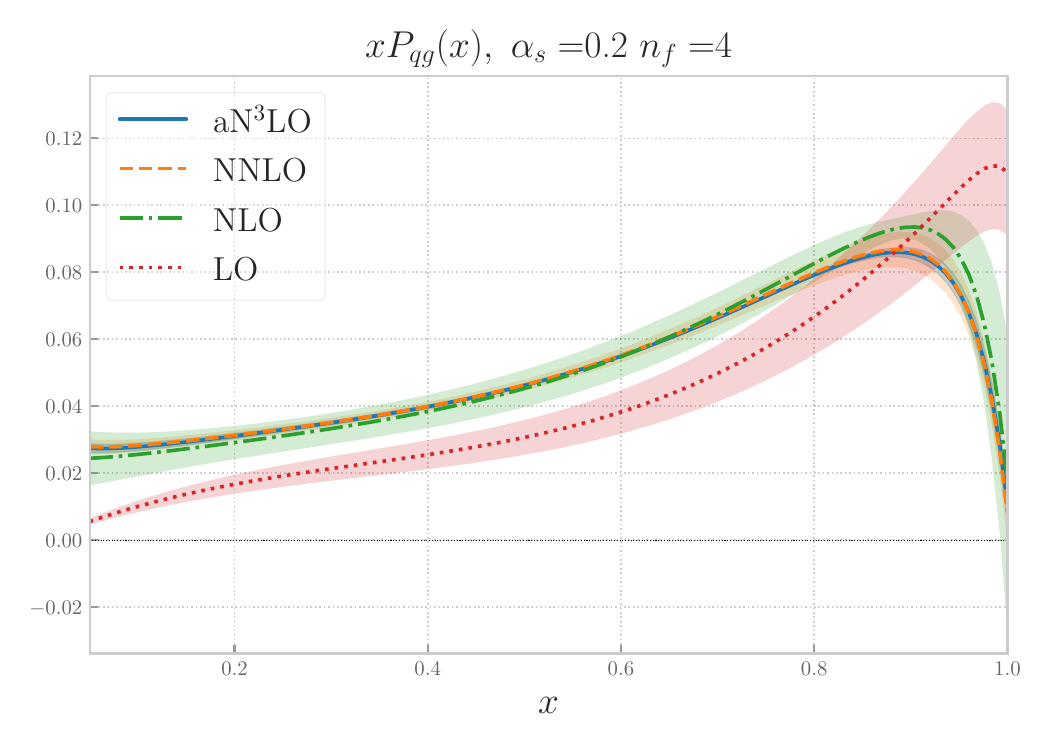}
  \includegraphics[width=.49\textwidth]{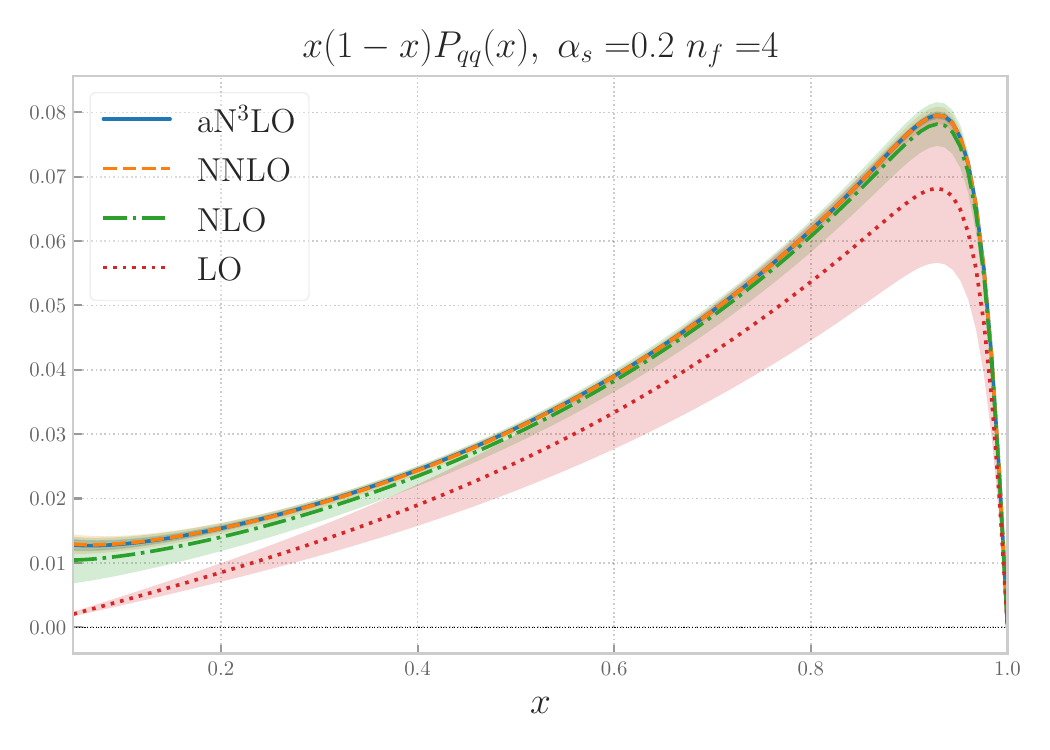}
  \caption{\small Same as
    Fig.~\ref{fig:splitting-functions-singlet-logx}
    with a  linear scale on the $x$ axis, and plotting
    $(1-x)x P_{ii}$ for diagonal entries.}
  \label{fig:splitting-functions-singlet-linx} 
\end{figure}

\begin{figure}[!t]
  \centering
  \includegraphics[width=.49\textwidth]{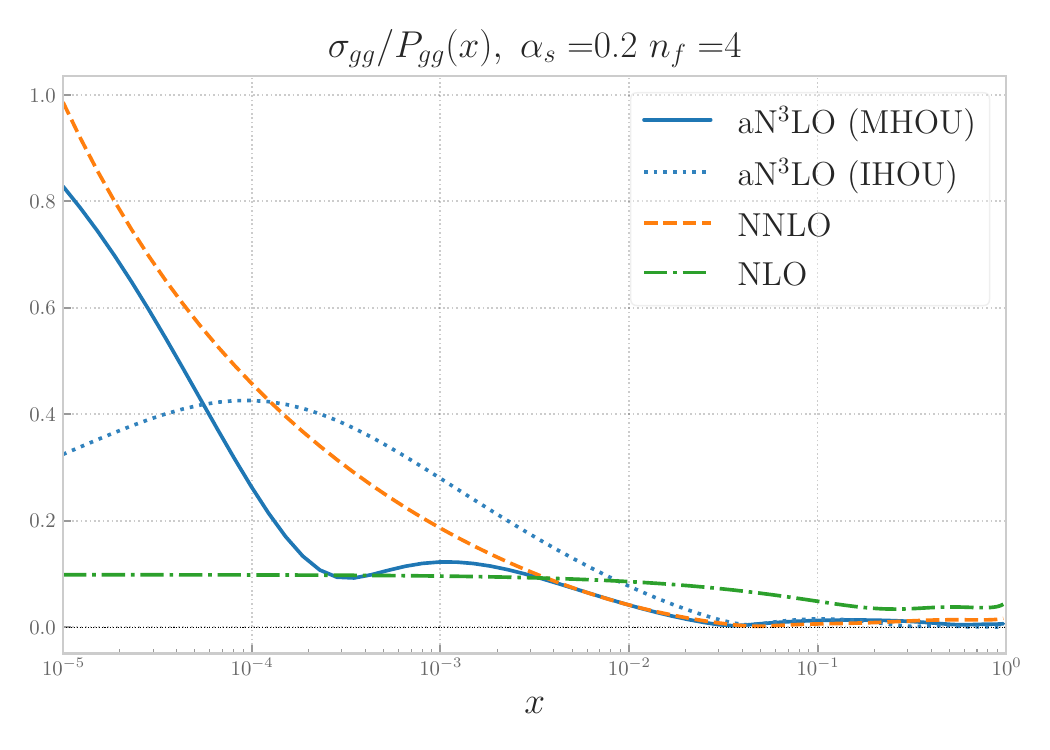}
  \includegraphics[width=.49\textwidth]{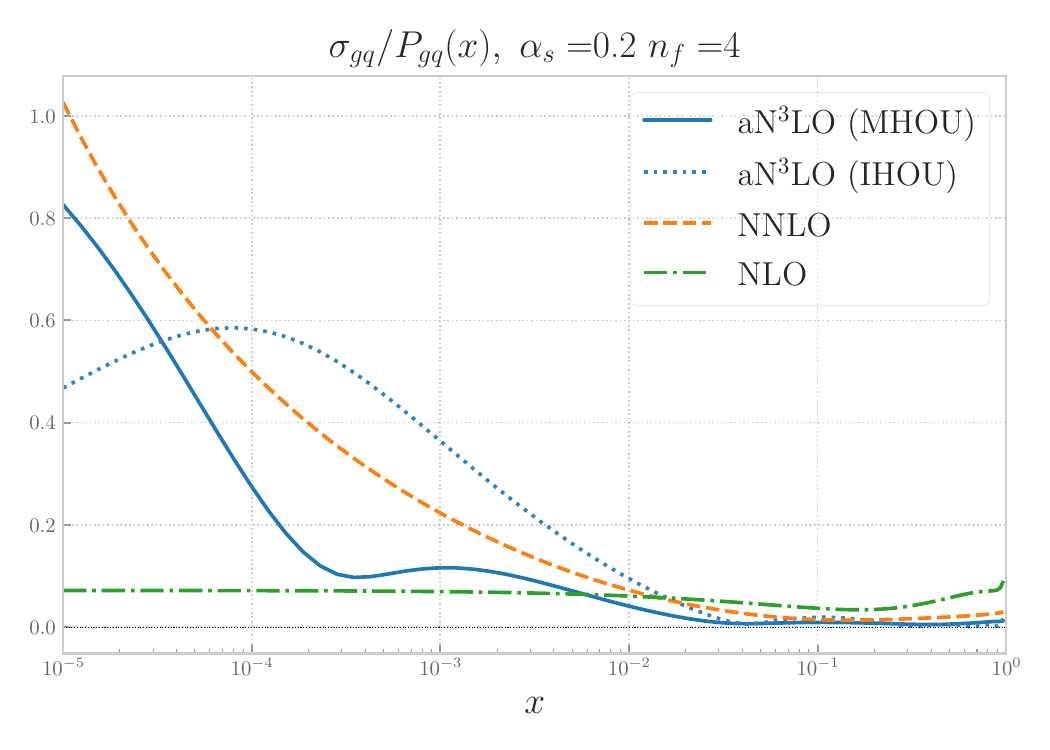} \\
  \includegraphics[width=.49\textwidth]{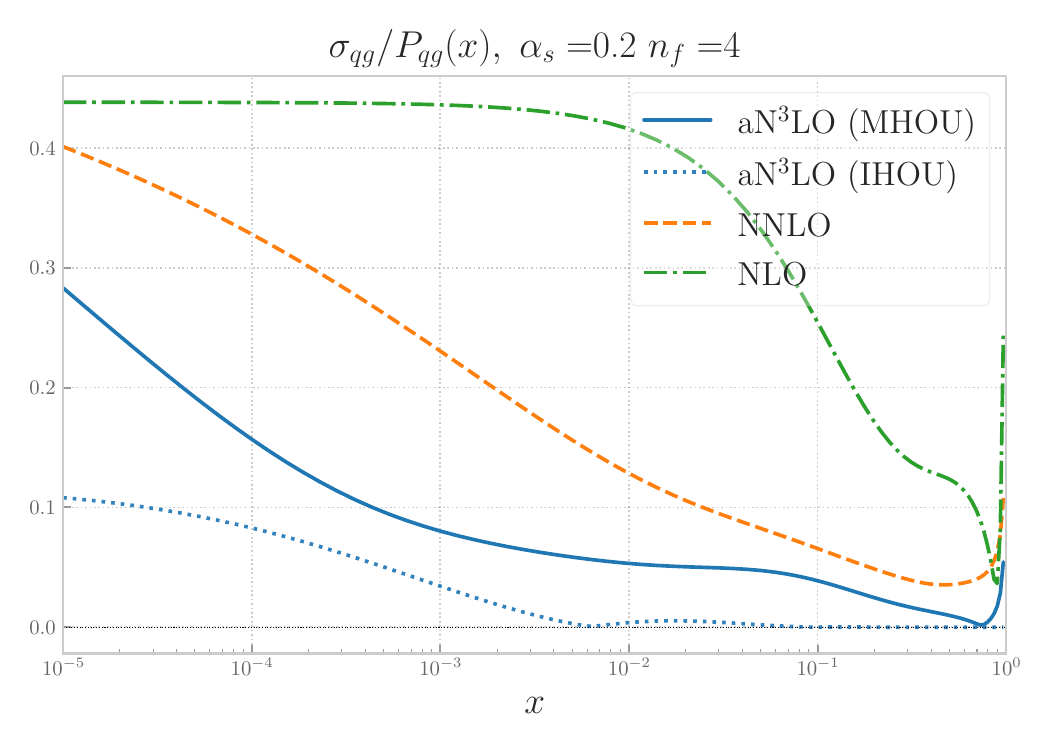}
  \includegraphics[width=.49\textwidth]{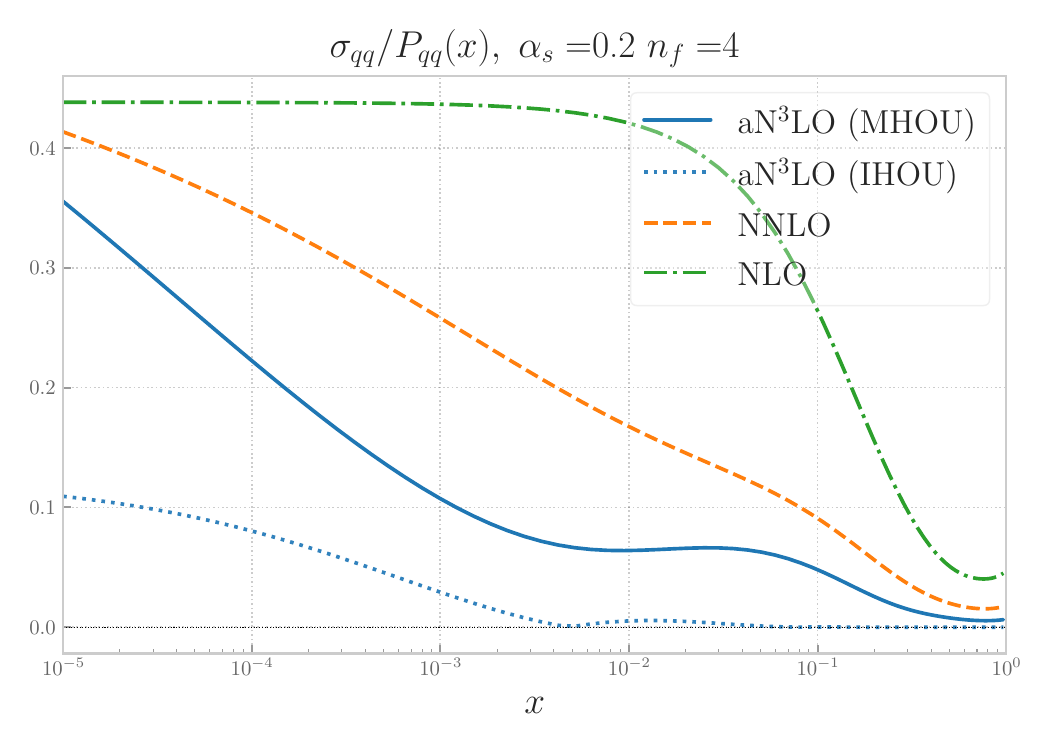}
  \caption{\small Same as Fig.~\ref{fig:splitting-functions-ns-unc}
    for the singlet splitting function matrix elements. At NLO and NNLO we show
    the MHOU, while at aN$^3$LO we also show the IHOU.}
  \label{fig:splitting-functions-singlet-unc} 
\end{figure}

\begin{figure}[!t]
  \centering
  \includegraphics[width=.49\textwidth]{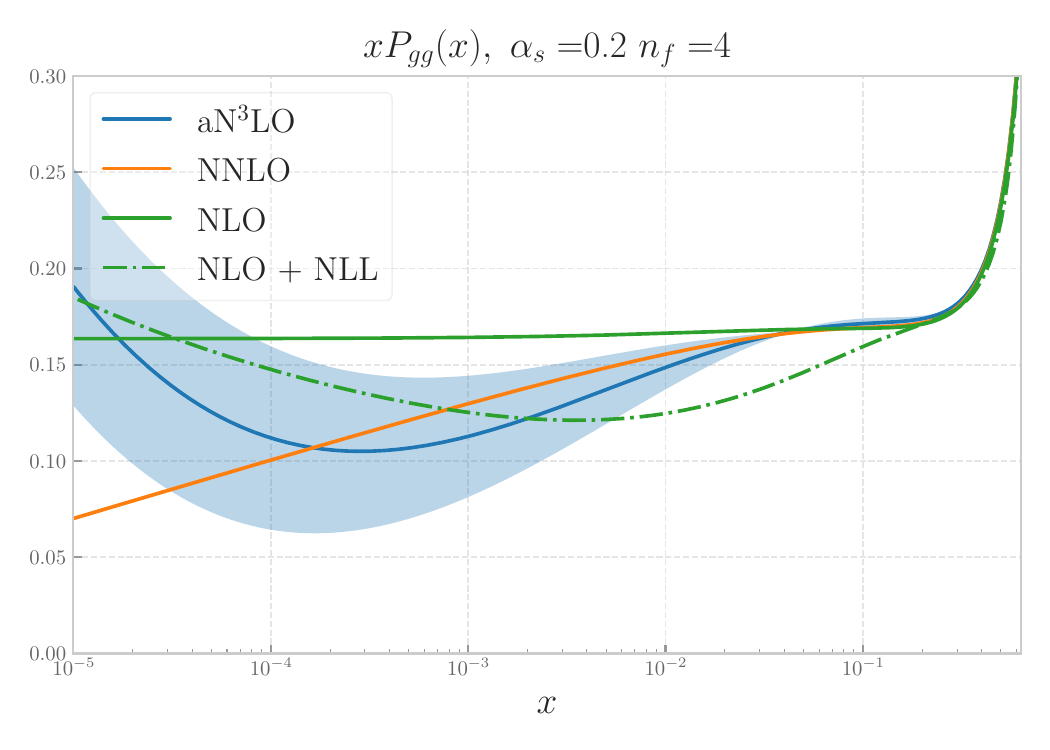}
  \includegraphics[width=.49\textwidth]{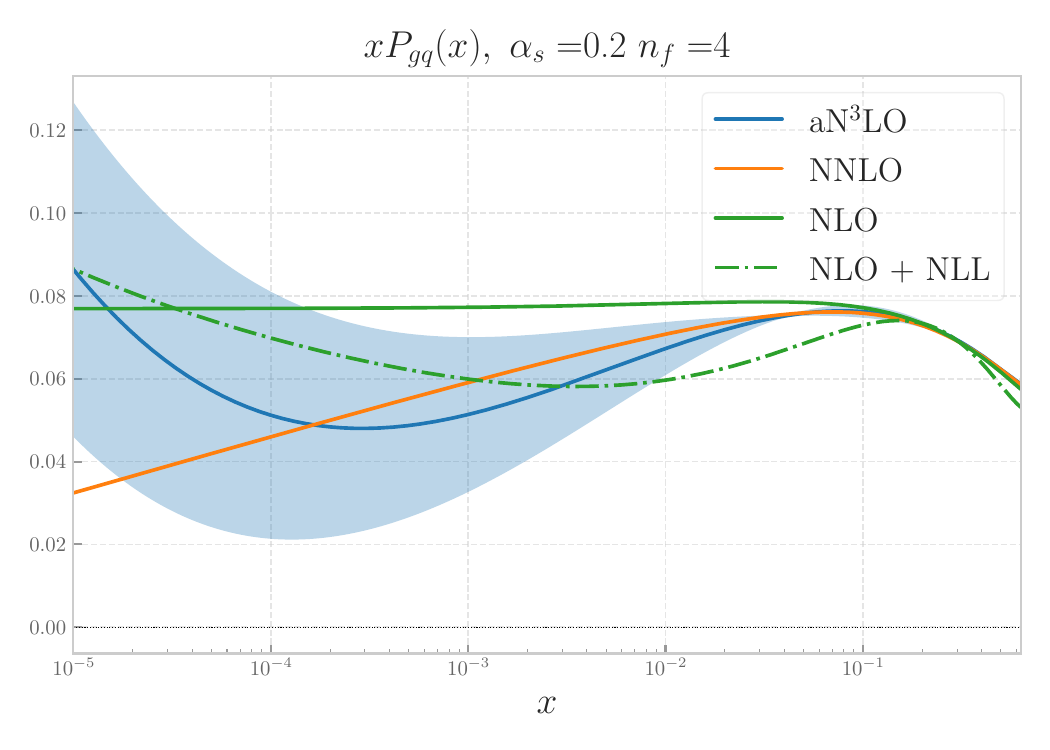}\\
  \includegraphics[width=.49\textwidth]{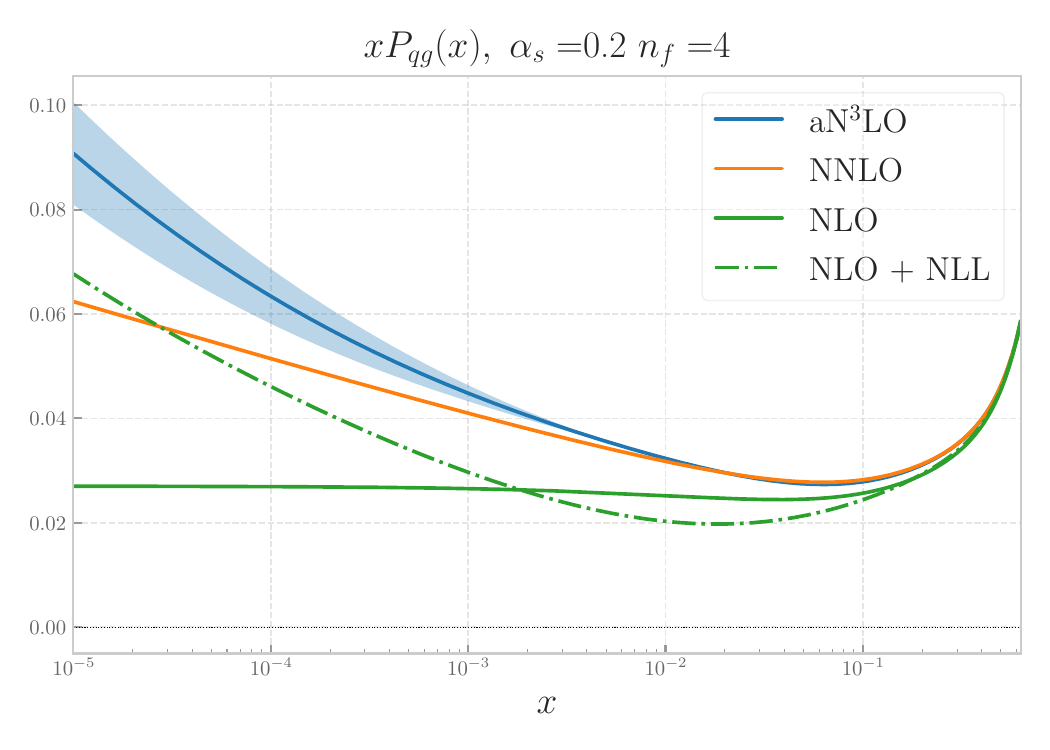}
  \includegraphics[width=.49\textwidth]{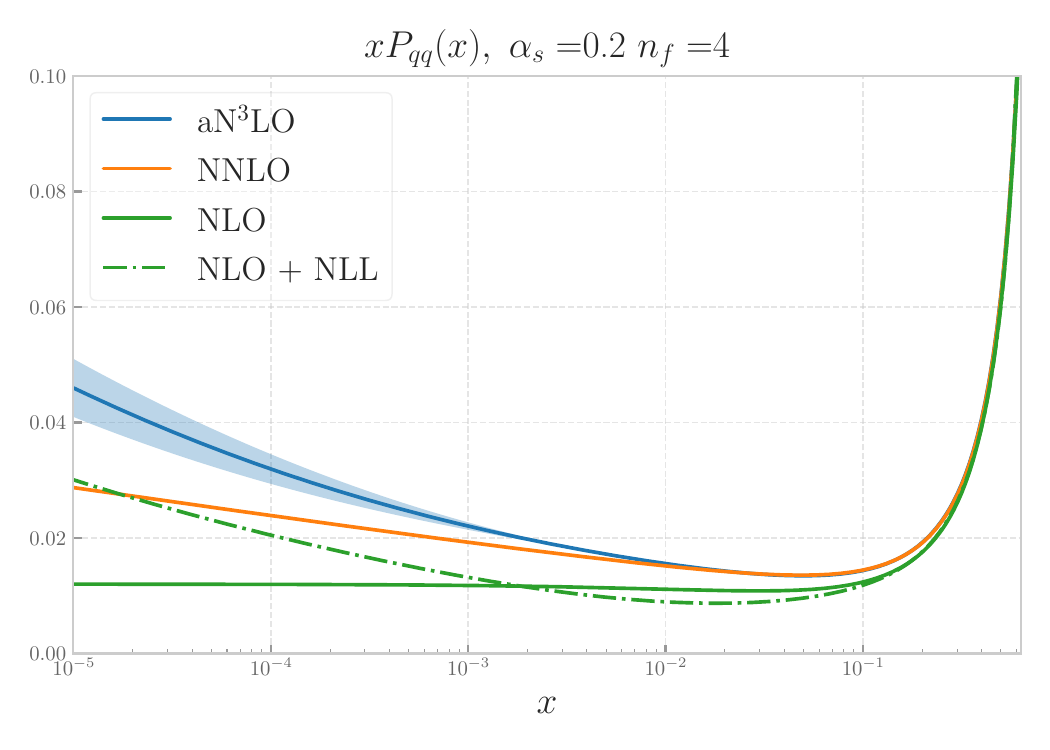}
  \caption{\small Comparison of the NLO, NNLO, and aN$^3$LO splitting functions
    (same as in  Fig.~\ref{fig:splitting-functions-singlet-logx}) to the
    small-$x$ resummed NLO+NLL result of
    Ref.~\cite{Altarelli:2008aj}. Only the IHOU on the aN$^3$LO result is
    shown.}
  \label{fig:splitting-functions-singlet-logx-res} 
\end{figure}

We now present the aN$^3$LO splitting functions constructed following the
procedure described in Sects.~\ref{sec:general_strategy}--\ref{sec:singlet}.
The nonsinglet result, already compared in  Fig.~\ref{fig:nsgamma} to
the previous approximation of  Ref.~\cite{Moch:2017uml}, is shown 
in Fig.~\ref{fig:splitting-functions-ns} at the first four
perturbative orders as a ratio to the aN$^3$LO result. 
For each order we include the MHOU determined by
scale variation according to Refs.~\cite{NNPDF:2019vjt,NNPDF:2019ubu} and recall
that there are no IHOU in the nonsinglet sector.
As the nonsinglet splitting function are subdominant at small
$x$ we only show the plot with a linear scale in $x$. The relative size of the
MHOU is shown in Fig.~\ref{fig:splitting-functions-ns-unc}.

Inspecting Figs.~\ref{fig:splitting-functions-ns} and
\ref{fig:splitting-functions-ns-unc}
reveals good perturbative convergence\footnote{Here and henceforth by
``convergence'' we mean that the size of the missing N$^4$LO corrections is
negligible compared to the target accuracy of theoretical predictions,
i.e.\ at the sub-percent level.} for all values of $x$.
Specifically, the differences between two subsequent perturbative orders are
reduced as the accuracy of the calculation increases, and, correspondingly, the
MHOUs associated to factorization scale variations decrease with the
perturbative accuracy. Indeed, the MHOU
appears to reproduce well the observed behaviour of the
higher orders, with overlapping uncertainty bands between subsequent
orders except at LO at the smallest $x$ values. Hence, the behavior of
the perturbative series suggests that the MHOU estimate based on scale
variation at N$^3$LO is reliable. 

Based on these results it is clear that in the nonsinglet
sector the N$^3$LO contribution to the splitting function is
essentially negligible except at the smallest $x$ values, as shown in
Fig.~\ref{fig:nsgamma}. Consequently, for all practical purposes we
can consider the current approximation to the nonsinglet anomalous
dimension to be essentially exact, and with negligible MHOU.

The situation in the singlet sector is more challenging.
The singlet matrix of splitting functions is shown in
Figs.~\ref{fig:splitting-functions-singlet-logx} and
\ref{fig:splitting-functions-singlet-linx}, respectively
with a logarithmic or linear scale on the $x$ axis.
Because the diagonal splitting functions are
distributions at $x=1$ in the linear scale plots we display $x(1-x)P_{ii}$.
The corresponding relative size of the MHOU is shown in
Fig.~\ref{fig:splitting-functions-singlet-unc} for the first
four perturbative orders, along with the IHOU on the aN$^3$LO result,
determined using Eq.~(\ref{eq:sigihou}).

A different behaviour is observed for the quark sector $P_{qi}$ and for
the gluon sector $P_{gi}$. In the quark sector, the MHOU decreases with
perturbative order for all $x$, but it remains sizable at aN$^3$LO
for essentially
all $x$, of order 5\% for $10^{-2}\lesssim x\lesssim10^{-1}$. 
In the gluon sector instead for $x \gtrsim 0.03$ the MHOU is negligible, 
but at smaller $x$ it grows rapidly, and in fact at very small $x$
it becomes larger than the NLO MHOU. This is due to the presence of
leading small-$x$ logarithms, Eq.~(\ref{eq:smallxgg}), which are absent at NLO.
In fact the true gluon-sector MHOU at very small $x$ is likely to be
underestimated
by scale variation, because while it generates the fourth-order leading pole
present in the N$^4$LO (the fifth-order pole vanishes), it fails to generate the sixth-order pole known to be present in the N$^5$LO splitting function.

We now turn to the IHOU and find again contrasting behaviour in the different
sectors. In the quark sector, thanks to the large
number of known Mellin moments and the copious information on the
large-$x$ limit, the IHOU are significantly smaller than the MHOU, by about a
factor three, and become negligible for $x\gtrsim 10^{-2}$. In the gluon
sector instead the IHOU, while still essentially negligible for
$x\gtrsim0.1$, is larger than the MHOU except at very small
$x\lesssim 10^{-4}$ where the MHOU dominates.

Consequently, for all matrix elements at large $x\gtrsim 0.1$ the behaviour of
the singlet is similar to the behaviour of the nonsinglet: IHOU and MHOU are
both negligible, meaning that aN$^3$LO results are essentially
exact, and the perturbative expansion has essentially converged, see
Fig.~\ref{fig:splitting-functions-singlet-linx}. At smaller $x$, while
the aN$^3$LO and NNLO results agree within uncertainties, the
uncertainties on the aN$^3$LO are sizable, dominated by MHOUs in the
quark channel and by IHOUs in the gluon channel.

In the singlet sector the most dramatic impact of the aN$^3$LO
correction is at small $x$. It is thus interesting to compare the
aN$^3$LO singlet splitting functions with those obtained by
the resummation of leading and next-to-leading order small-$x$ logarithms
of Ref.~\cite{Altarelli:2008aj}, namely the two highest powers of $\ln x$
contained in the N$^3$LO result; this comparison is shown in
Fig.~\ref{fig:splitting-functions-singlet-logx-res}.
The agreement of all four entries
$xP_{gg}$, $xP_{gq}$, $xP_{qg}$ and $xP_{qq}$ is remarkably good and well within
the uncertainties in the two approaches. In particular the dip in
$xP_{gg}$ at intermediate $x$ at aN$^3$LO (albeit with significant IHOU)
is also a feature of the resummation. This is nontrivial, 
as the resummation includes only the asymptotic LL$x$ and NLL$x$ singularities
at $N=1$,  but none of the subleading results incorporated at
aN$^3$LO. Instead, it uses a symmetrization which resums collinear and
anti-collinear logarithms in the small-$x$ expansion, and the effects
of running coupling which change the nature of the small-$x$
singularity (from a fourth order pole at $N=1$ in the fixed order N$^3$LO
result to a simple pole a little further to the right on the real axis).

That both the resummed and fixed order approaches converge to
very similar results, at least in the range of $x$ relevant for
HERA and LHC, is very reassuring. It shows that in a global fit with
current data, while NLL$x$ resummation significantly improves the
quality of a fixed order NNLO fit \cite{Ball:2017otu}, the same
improvement should also be seen by adding aN$^3$LO corrections. Thus to
find evidence for small-$x$ resummation at aN$^3$LO, it will probably
be necessary to go to yet smaller values of $x$, e.g.\ below $10^{-5}$,
where the fixed order and resummed results will eventually diverge again.

\subsection{Results: aN$^3$LO evolution}
\label{sec:n3lo_evol}

The aN$^3$LO anomalous dimensions discussed in the previous sections
have been implemented in the
Mellin-space open-source evolution code {\sc\small EKO}~\cite{Candido:2022tld}
which enters the new pipeline~\cite{Barontini:2023vmr}
adopted by NNPDF in order to produce theory predictions used for PDF
determination. The  parametrization is expressed in
terms of a basis of Mellin space functions 
which are numerically efficient to evaluate.
In order to achieve full aN$^3$LO accuracy, in addition to the
anomalous dimensions,  the  four-loop
running of   the strong coupling constant $\alpha_s(Q)$
and the  N$^3$LO matching conditions
dictating the transitions between schemes with different numbers of active
quark flavor have also been implemented.

The N$^3$LO matching conditions have been presented in
Ref.~\cite{Bierenbaum:2009mv} and subsequently computed analytically in
Refs.~\cite{Bierenbaum:2008yu,Bierenbaum:2009zt,Ablinger:2010ty,
  Ablinger:2014vwa,Ablinger:2014uka,Behring:2014eya,Blumlein:2017wxd,
  Ablinger_2014,Ablinger_2015,Ablinger:2022wbb}.
The exception is the $a_{Hg}^{(3)}$ entry of the matching condition matrix, which
is still unknown\footnote{The terms recently computed in
  Ref.~\cite{Ablinger:2023ahe} are not yet included and left for future
  updates.} and which instead is parametrized using the first 5 known
moments~\cite{Bierenbaum:2009mv} and the LL$x$ contribution as done in
Ref.~\cite{Kawamura:2012cr}. Also these matching conditions are implemented in
{\sc\small EKO} and thus it is possible to assess
the impact of the inclusion of aN$^3$LO terms on perturbative evolution.

\begin{figure}[!t]
  \centering
  \includegraphics[width=0.49\textwidth]{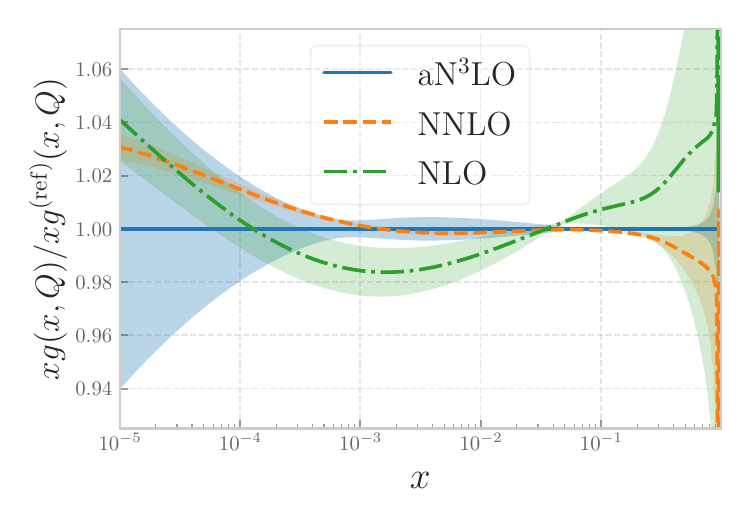}
  \includegraphics[width=0.49\textwidth]{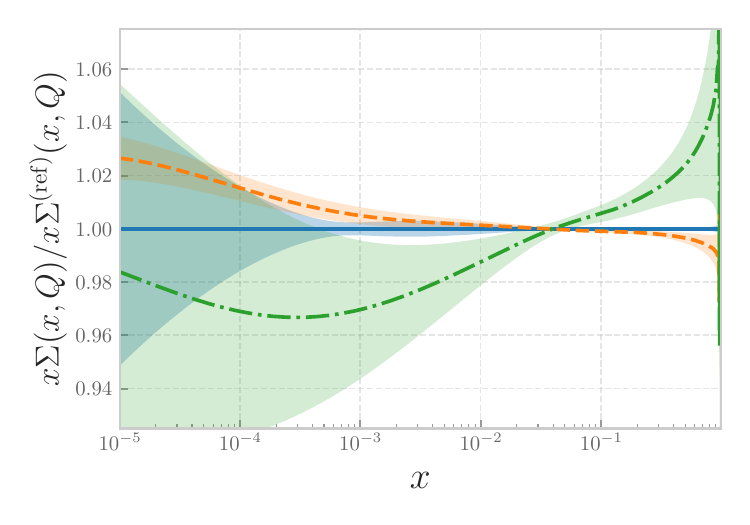}\\
  \includegraphics[width=0.49\textwidth]{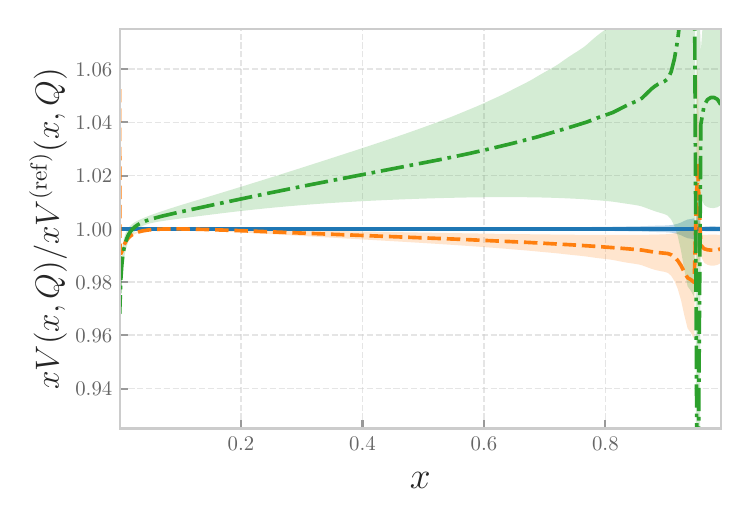}
  \includegraphics[width=0.49\textwidth]{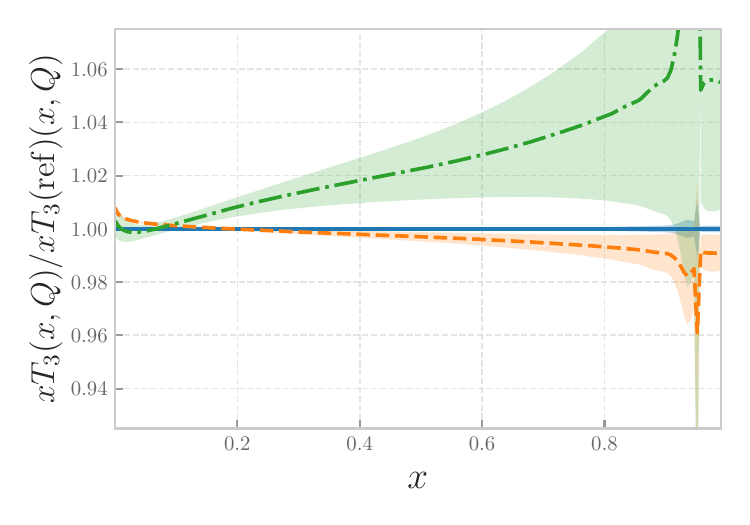}\\ 
  \includegraphics[width=0.49\textwidth]{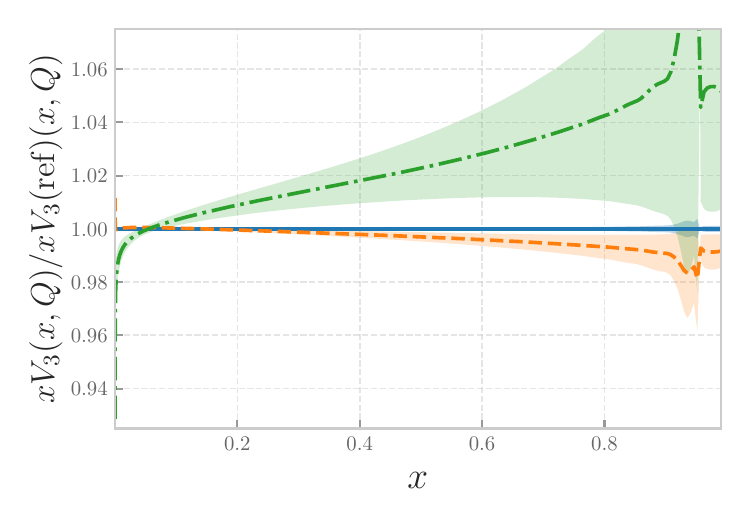} 
  \caption{Comparison of the result obtained evolving from  $Q_0=1.65$~GeV
    to $Q=100$~GeV at NLO, NNLO, and aN$^3$LO using NNPDF4.0 NNLO as fixed
    starting PDF. Results are shown as ratio to the aN$^3$LO (from left to right
    and from top to bottom) for the gluon and singlet $\Sigma$, and for the
    $V$, $V_3$ and $T_3$ quark eigenstates of perturbative evolution (see
    Sect.~\ref{sec:general_strategy}). The total theory uncertainty is
    shown in all cases, i.e.\ the MHOU at NLO and NNLO, and the sum in
    quadrature of MHOU and IHOU at aN$^3$LO.}
  \label{fig:N3LOevolution-q100gev-ratios} 
\end{figure}

\begin{figure}[!t]
  \centering
  \includegraphics[width=0.49\textwidth]{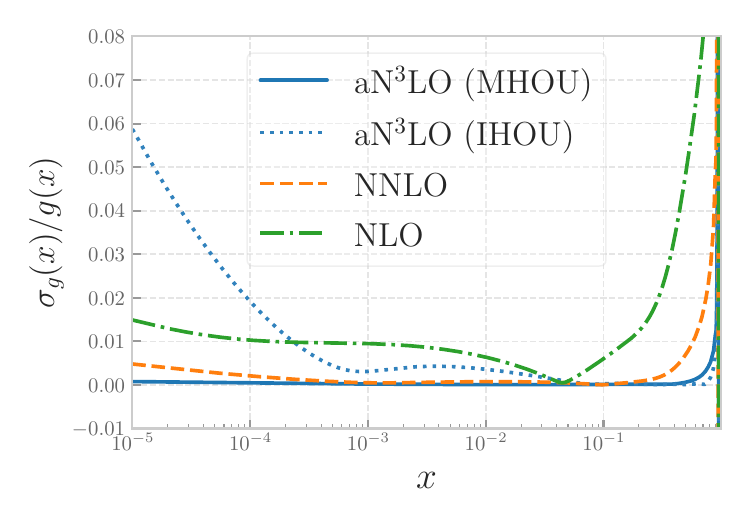}
  \includegraphics[width=0.49\textwidth]{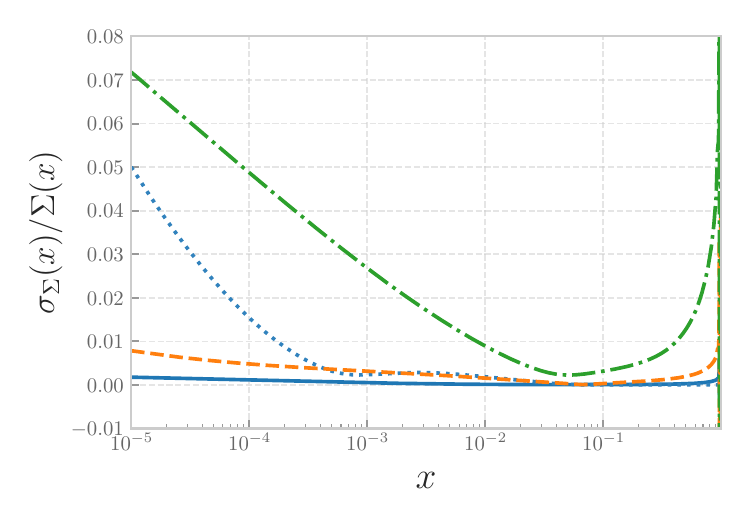}
  \caption{The relative size of the uncertainty on the gluon and
    singlet PDFs shown in Fig.~\ref{fig:N3LOevolution-q100gev-ratios}.
    The MHOU is shown in all cases, and at  aN$^3$LO the IHOU is also shown.}        \label{fig:N3LOevolution-q100gev-pull_ihou}
\end{figure}

In Fig.~\ref{fig:N3LOevolution-q100gev-ratios} we compare the result of
evolving a fixed set of PDFs  from $Q_0=1.65$ GeV up to $Q=100$ GeV at
NLO, NNLO, and aN$^3$LO. We take as input the NNPDF4.0NNLO PDF set,
and show results normalized to the aN$^3$LO evolution. Results are shown for
all the combinations that evolve differently, as discussed in
Sect.~\ref{sec:general_strategy}, namely the singlet, gluon, total
valence and nonsinglet $\pm$ combinations, with a logarithmic scale
on the $x$ axis for the singlet sector and a linear scale for the
valence and nonsinglet combinations. The relative uncertainty on the
gluon and singlet are shown in
Fig.~\ref{fig:N3LOevolution-q100gev-pull_ihou}, with MHOU and IHOU
separately displayed at N$^3$LO.

In all cases the perturbative expansion appears to have converged
everywhere, with almost no difference between NNLO and aN$^3$LO except
at small $x\lesssim10^{-3}$, where singlet evolution is weaker at
aN$^3$LO than at NNLO due to the characteristic dip seen in the
gluon sector splitting functions of
Fig.~\ref{fig:splitting-functions-singlet-logx}. Because the
gluon-driven small-$x$ rise dominates small-$x$ evolution this is a
generic feature of all quark and gluon PDFs in this small-$x$ region.
It is interesting to observe that this is an all-order feature that persists
upon small-$x$ resummation, as already discussed at the end of
Sect.~\ref{sec:n3lo_eko} and seen in
Fig.~\ref{fig:splitting-functions-singlet-logx-res}.
In fact, the total theory uncertainty at aN$^3$LO is at the
sub-percent level for all  $x\gtrsim10^{-3}$. Hence, not only has the
MHOU become negligible, but also the effect of IHOU on PDF evolution
is only significant at small $x$. 

\subsection{Comparison to other groups}
\label{sec:n3lo_comp}

We finally compare our approximation of the N$^3$LO splitting
functions to other recent results from
Refs.~\cite{McGowan:2022nag,Falcioni:2023luc,Falcioni:2023vqq,Moch:2023tdj}.
While the approach of
Refs.~\cite{Falcioni:2023luc,Falcioni:2023vqq,Moch:2023tdj}
(FHMRUVV, henceforth) is very similar to our own,
with differences only due to details of the choice of basis functions,
a rather different approach is adopted in
Ref.~\cite{McGowan:2022nag} (MSHT20, henceforth). 
There, the approximation is constructed from similar theoretical 
constraints (small-$x$, large-$x$ coefficients and Mellin moments), 
but supplementing the parametrization with additional
nuisance parameters, which control the uncertainties arising from 
unknown N$^3$LO terms. However, these approximations are taken as a prior, 
and the nuisance parameters are fitted to the
data along with the PDF parameters. The best-fit values of the
parameters determine the posterior for the splitting function, and
their uncertainties are interpreted as the final IHOU on it. A consequence
of this procedure is that the posterior can reabsorb not only N$^3$LO
corrections, but any other missing contribution, of theoretical or
experimental origin. 

\begin{figure}[!t]
  \centering
  \includegraphics[width=.49\textwidth]{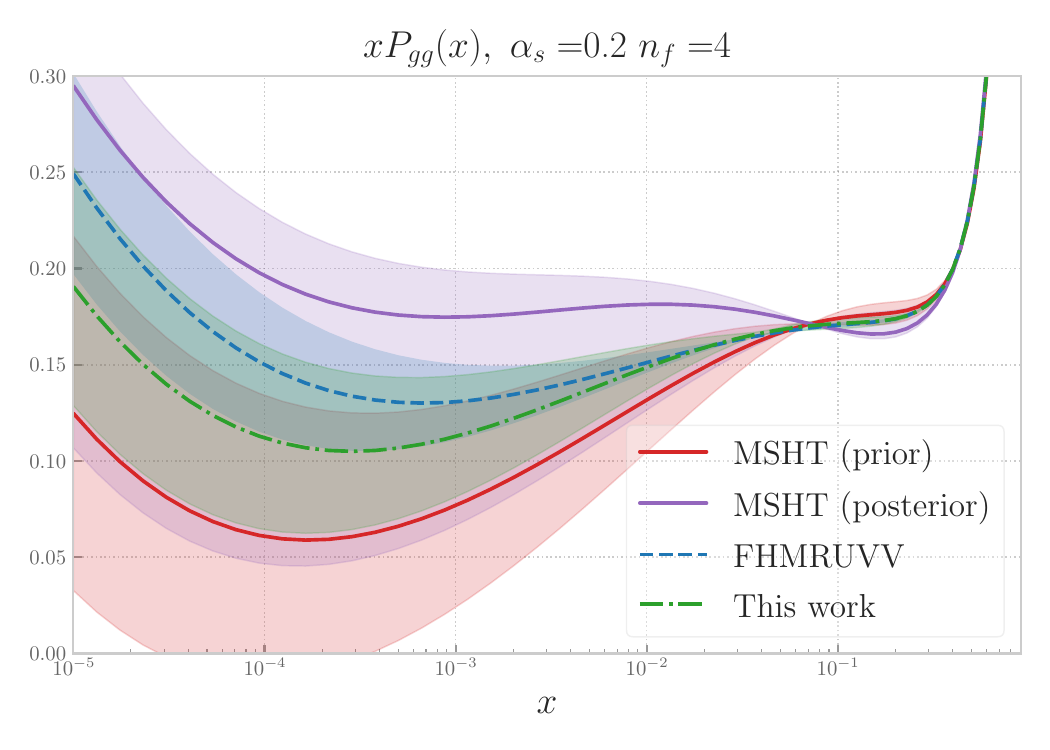}
  \includegraphics[width=.49\textwidth]{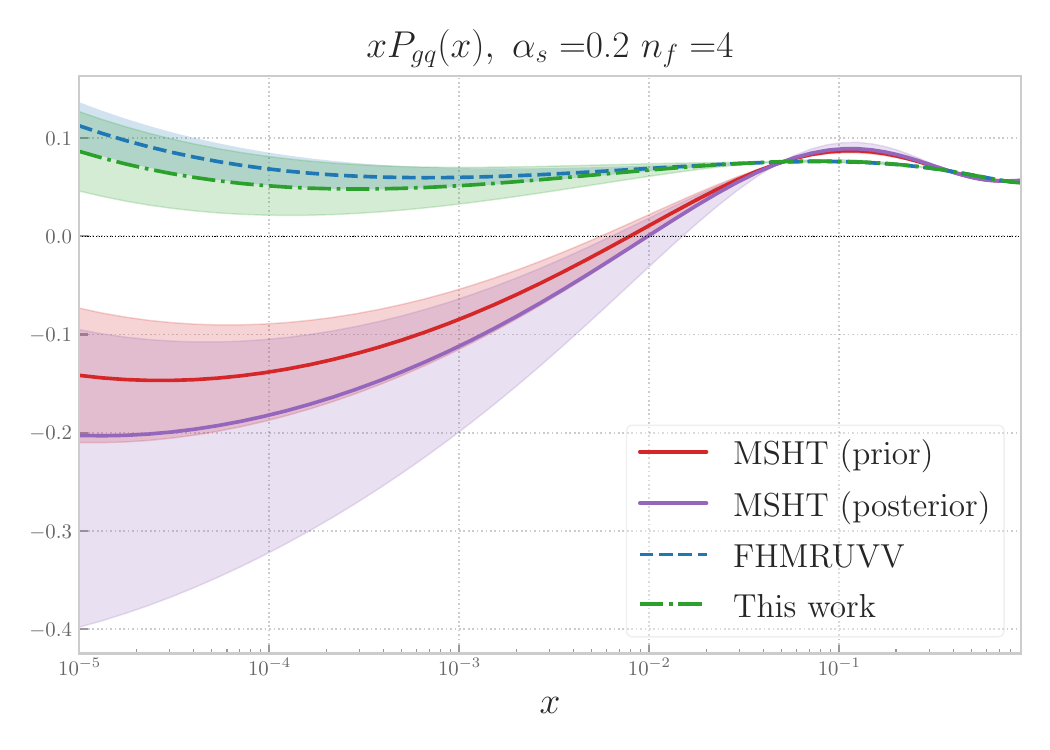} 
  \includegraphics[width=.49\textwidth]{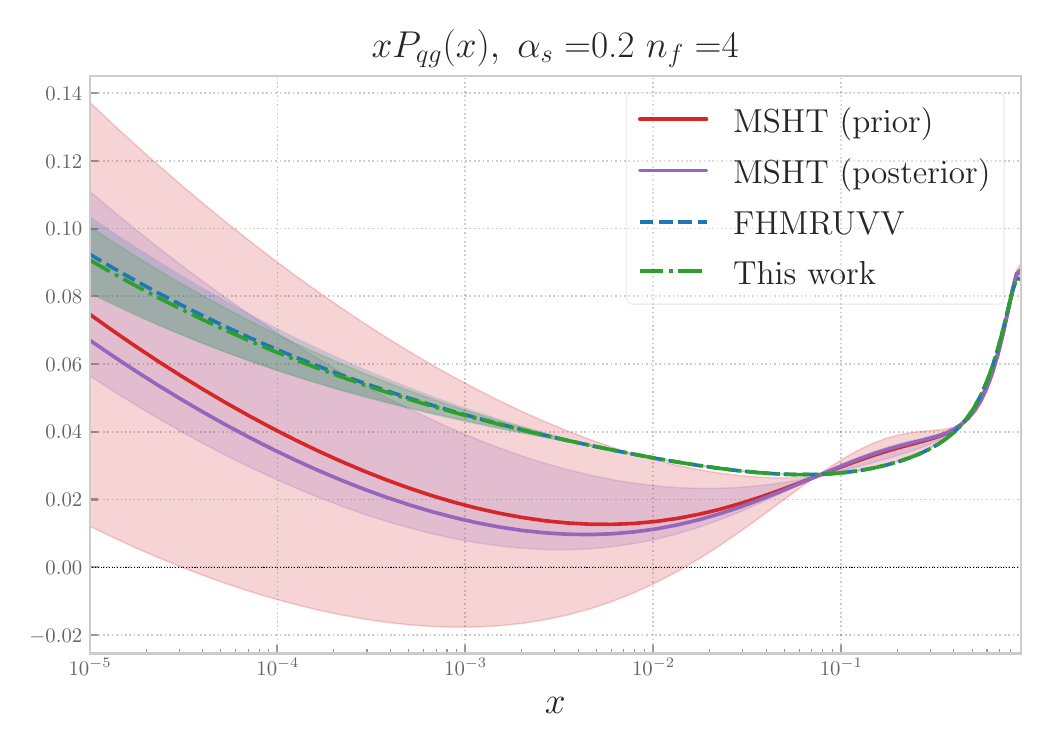}
  \includegraphics[width=.49\textwidth]{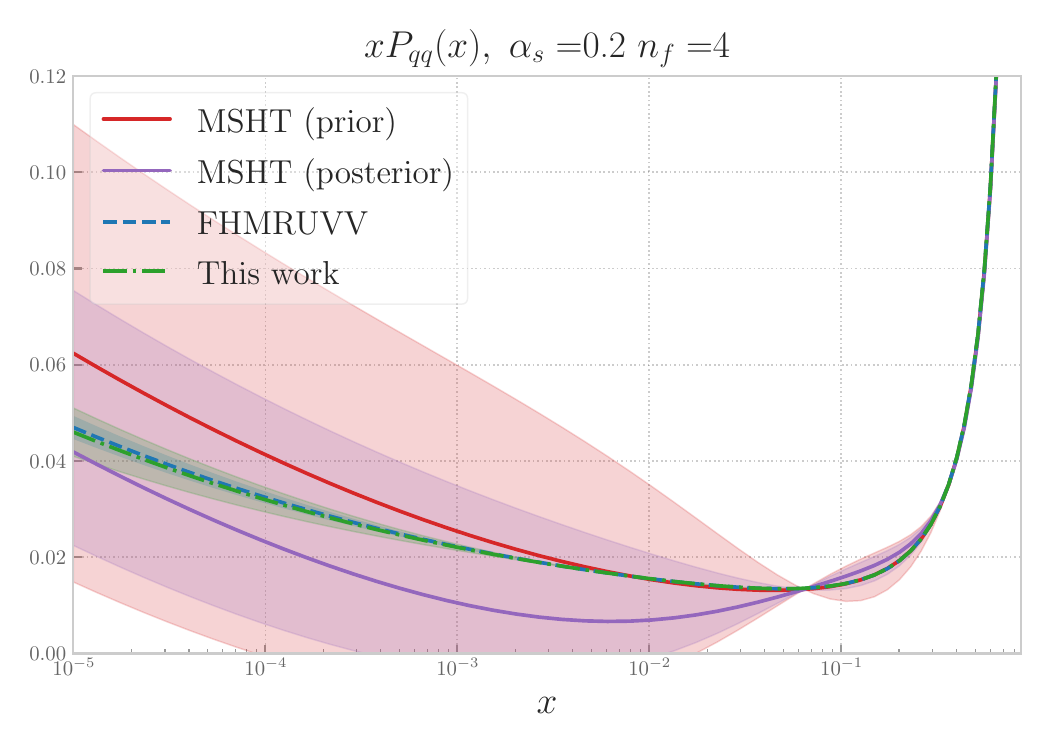}
  \caption{\small Same as
    Fig.~\ref{fig:splitting-functions-singlet-logx}, now comparing
    our aN$^3$LO result to those of Ref.~\cite{McGowan:2022nag} (MSHT20) and
    Refs.~\cite{Falcioni:2023luc,Falcioni:2023vqq,Moch:2023tdj} (FHMRUVV).
    In all cases the uncertainty
    band correspond to the IHOU as estimated by the various groups.
    For the MSHT20 results, we display both the prior and the
    posterior parametrizations (see text).}
  \label{fig:splitting-functions-mhst} 
\end{figure}

The comparison is presented in Fig.~\ref{fig:splitting-functions-mhst},
for all the four singlet splitting functions. For the MSHT20 results 
both prior and posterior are displayed. It should be noticed that even though
the uncertainty bands on the NNPDF4.0, DHMRUVV and MSHT20 prior are
all obtained by varying the set of basis functions, they are found
using somewhat different procedures, and their meaning is accordingly
somewhat different. Indeed, for NNPDF4.0 the is constructed out of the
covariance matrix according to 
Eq.~(\ref{eq:sigihou}). For FHMRUVV is instead the band between an
upper and lower estimates which are representative of the envelope of
all variations. Finally  for the MSHT20 prior it is the variance of
the probability distribution obtained assuming a  multigaussian
distribution of suitable nuisance parameters. 

As expected, excellent agreement is found with the FHMRUVV result, for  
all  splitting functions and for all $x$, especially for the $P_{qg}$
and $P_{qq}$ splitting functions, for which the highest number of
Mellin moments is known. Good qualitative agreement is also found for $P_{gq}$
and $P_{gg}$, although at small $x$ IHOUs
are larger and consequently  central values differ somewhat more,
though still in agreement within uncertainties. Uncertainties are
qualitatively similar, except at small $x$, where less exact
information is available and both central values and uncertainties are
less constrained. In this region the NNPDF4.0 is generally somewhat
more conservative, possibly due to the fact that it is obtained by
adding individual shifts in quadrature, rather than taking their envelope.

Coming now to  MSHT20 results, good agreement is found with the
prior, except for $P_{gq}$, for which MSHT20 shows a small-$x$ dip
accompanied by a large-$x$ bump. The different small-$x$ behaviour
is likely due to the fact that
MSHT20 do not enforce the color-charge relation Eq.~(\ref{eq:smallxiq})
at NLL$x$, with the large-$x$ bump then following from the constraints
Eq.~(\ref{eq:singlet_scaling}).  Also, in the quark sector the MSHT20
prior has significantly larger IHOUs due to the fact that it does not include 
the more recent information on Mellin moments from
Refs.~\cite{Davies:2022ofz,Falcioni:2023luc,Falcioni:2023vqq,Moch:2023tdj,
  Falcioni:2023tzp},
which were not available at the time of the MSHT20
analysis~\cite{McGowan:2022nag}.
At the level of posterior, however, significant differences appear
also for  $P_{gg}$, while persisting for $P_{gq}$. 
This means that the gluon evolution at aN$^3$LO is being significantly modified
by the data entering the global fit, and it is not fully determined by the
perturbative computation. Further benchmarks of aN$^3$LO splitting functions
will be presented in Ref.~\cite{lh24}.

\section{N$^3$LO partonic cross-sections}
\label{sec:n3lo_coefffun}

A PDF determination at N$^3$LO requires, in  addition to the splitting
functions discussed in Sect.~\ref{sec:dglap}, hard cross-sections at the same
perturbative order. Exact N$^3$LO massless DIS coefficient functions have been
known for several years~\cite{Vermaseren:2005qc,Moch:2004xu,Moch:2007rq,
  Moch:2008fj,Davies:2016ruz,Blumlein:2022gpp}, 
while massive coefficient functions are only available in various
approximations~\cite{Kawamura:2012cr,Laurenti:2024anf,bbl2023}.
For hadronic processes, N$^3$LO results are available for inclusive
Drell-Yan production for the total
cross-section~\cite{Duhr:2021vwj,Duhr:2020sdp,Baglio:2022wzu} as well
as for rapidity~\cite{Chen:2021vtu} and transverse momentum
distributions~\cite{Chen:2022lwc}, though neither of these is publicly
available.

We now describe the implementation of these corrections.
First, we review available results on DIS coefficient functions and summarize
the main features of the approximation that we will use for massive coefficient
functions~\cite{bbl2023,Laurenti:2024anf}.
Next we discuss how massless and massive DIS
coefficient functions are combined to 
extend the FONLL general-mass variable-flavor 
number scheme to $\mathcal{O}\lp \alpha_s^3\rp$.
Finally, we discuss 
N$^3$LO corrections for hadronic processes and different options for
their inclusion in PDF determination.

\subsection{N$^3$LO corrections to DIS structure functions}
\label{sec:n3lo_dis}

\begin{figure}[!t]
  \centering
  \includegraphics[width=.49\textwidth]{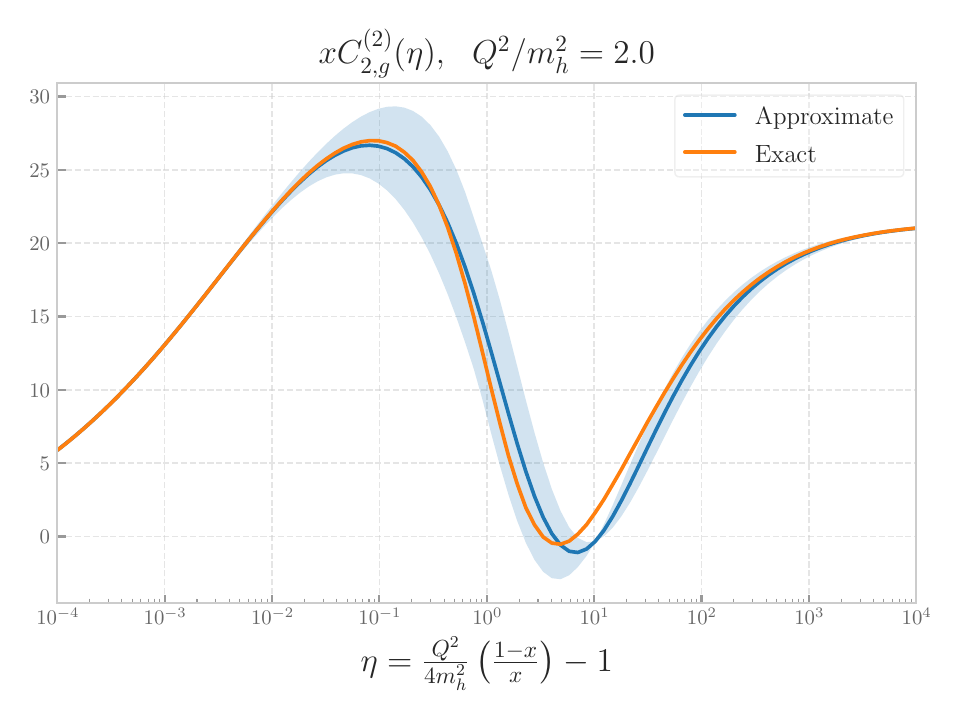}
  \includegraphics[width=.49\textwidth]{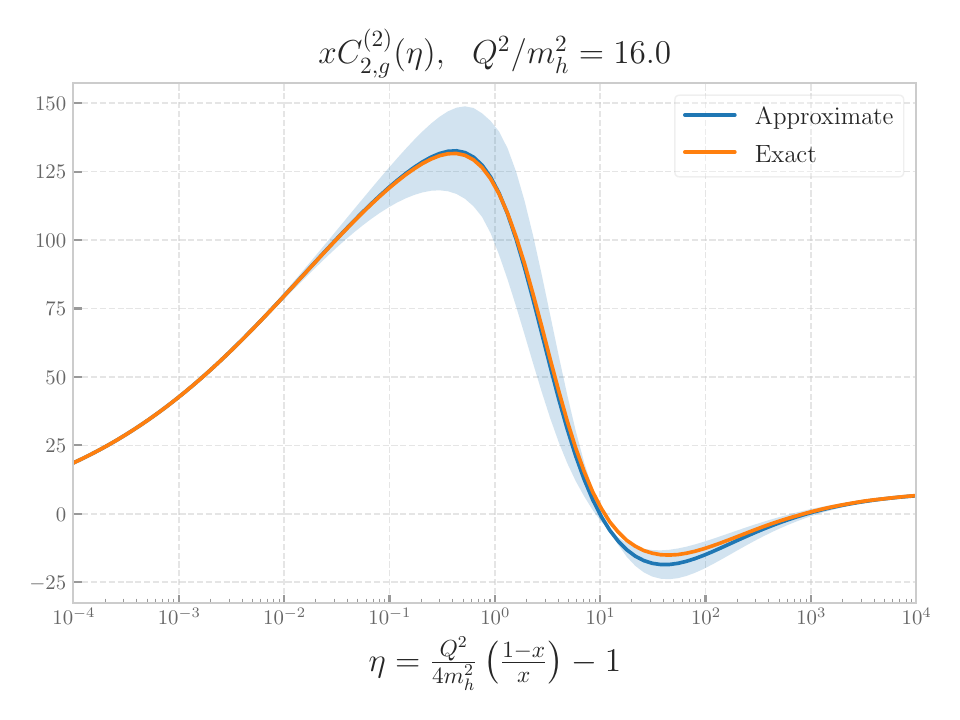}
  \caption{Comparison of the exact
    NNLO massive gluon-initiated coefficient function $x C_{2,g}^{(2)}(\eta)$
    to the approximation Eq.~(\ref{eq:massive_cf}) from
    Ref.~\cite{Laurenti:2024anf}, plotted as a function of $\eta$,
    Eq.~(\ref{eq:etadef}), for fixed  $Q^2$. 
    Results are shown for two different values of $Q^2$, one
    close to threshold $Q^2=2 m_h^2$ (left) and one at high scales $Q^2=16 m_h^2$
    (right). The uncertainty on the approximate result
    is obtained by varying the interpolating functions
    $f_1(x)$ and $f_2(x)$ in Eq.~(\ref{eq:massive_cf}).}
  \label{fig:massive_nnlo_bench} 
\end{figure}

\begin{figure}[!t]
  \centering
  \includegraphics[width=.49\textwidth]{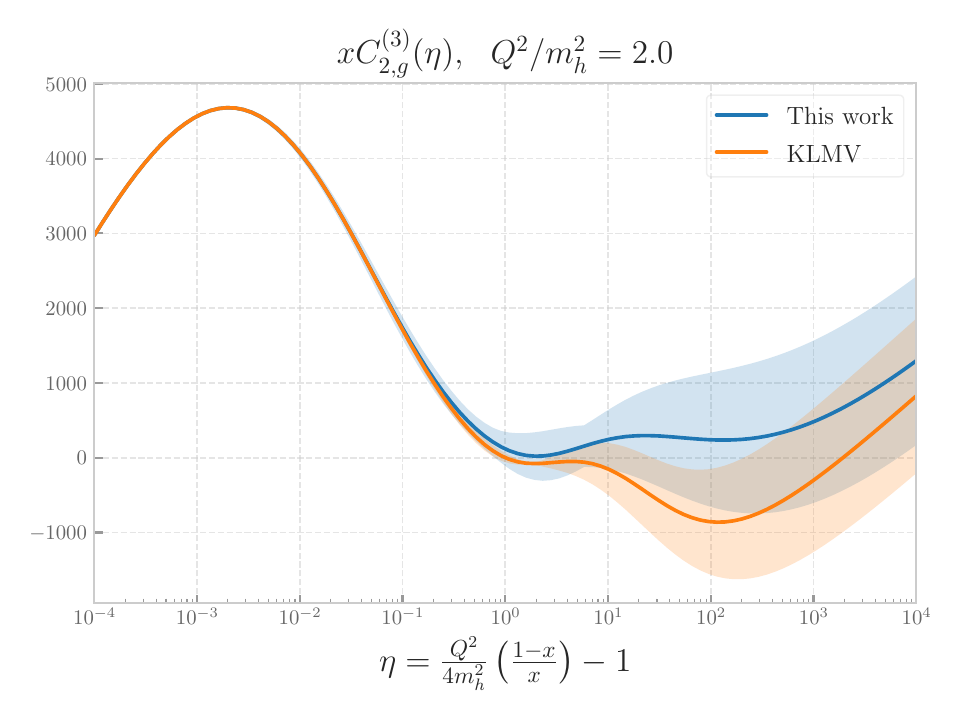}
  \includegraphics[width=.49\textwidth]{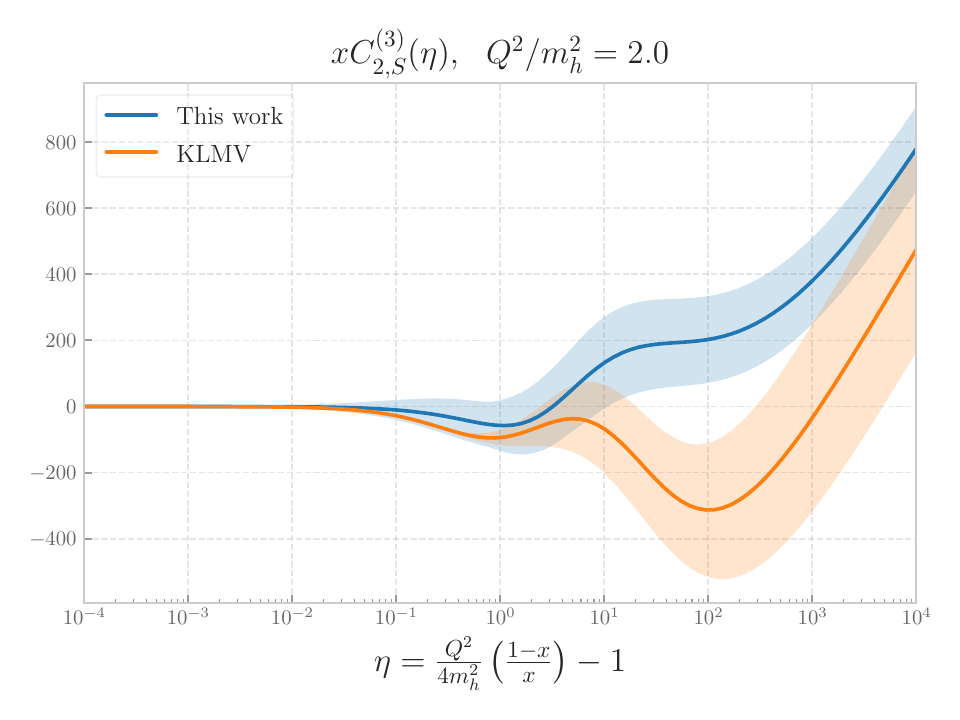}
  \caption{The approximate N$^3$LO massive gluon (left) and quark
    singlet (right) coefficient functions as a function of $\eta$
    for fixed  $Q^2=2 m_h^2$. Our result based on the approximation
    of Ref.~\cite{Laurenti:2024anf} is compared to the approximation of
    Ref.~\cite{Kawamura:2012cr} (KLMV).}
  \label{fig:massive_n3lo_bench} 
\end{figure}

\begin{figure}[!t]
  \centering
  \includegraphics[width=.85\textwidth]{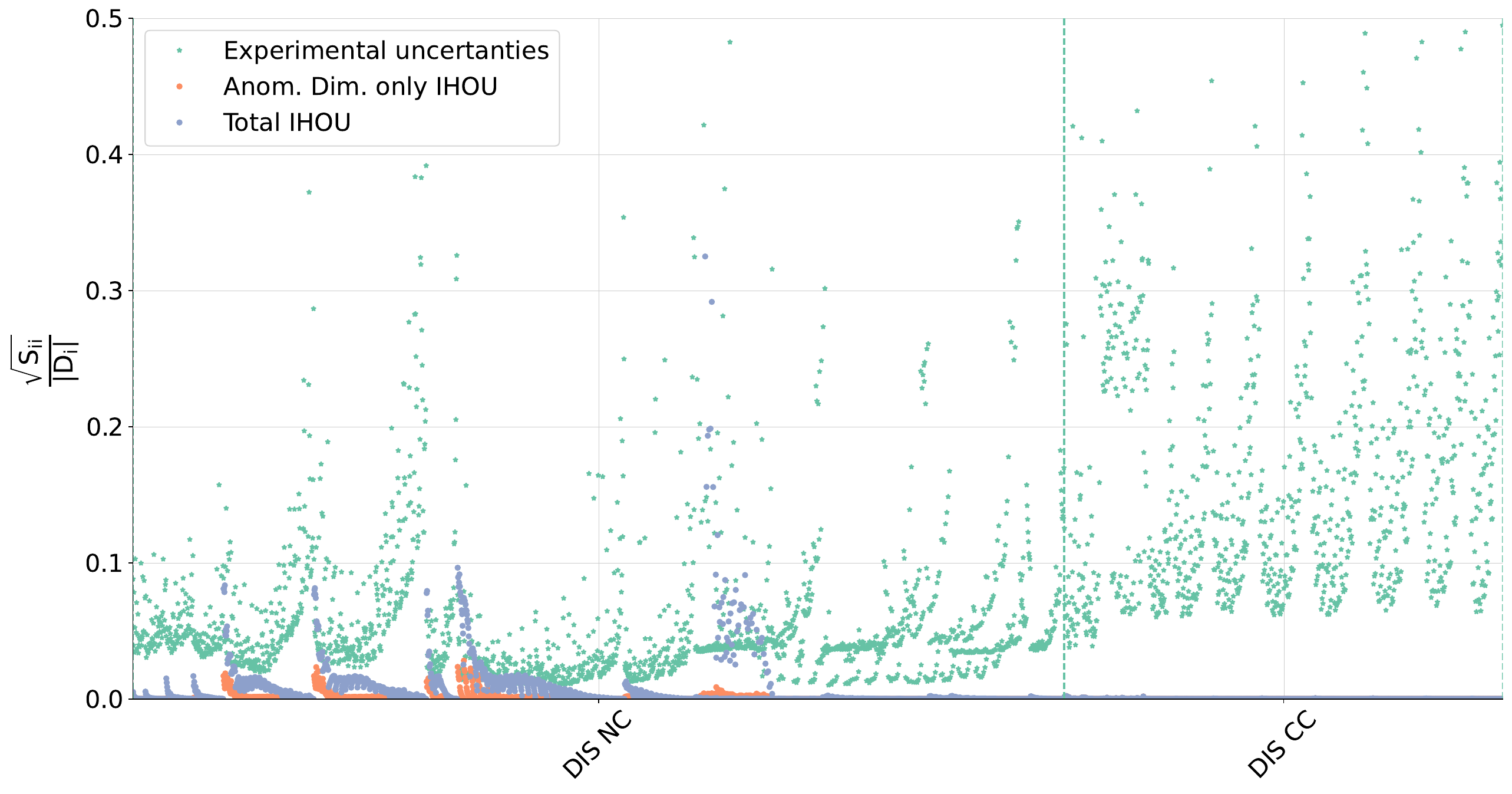} \\
  \caption{
    Square root of the diagonal entries of the IHOU
    covariance matrix for the DIS
    datasets normalized to the experimental central value $D_i$. We show the
    IHOU before and after adding to the covariance matrix 
    Eq.~(\ref{eq:covihou}) that accounts for uncertainty on anomalous
    dimensions the extra component
    Eq.~(\ref{eq:covihouC}) due to the  massive coefficient
    function. The experimental uncertainty is also shown for comparison.
  }    
  \label{fig:dis_ihou}
\end{figure}

The DIS structure functions $F_i$ are evaluated from the convolution of PDFs and
coefficient functions
\begin{align}
  \begin{split}
     F_{i}(x,Q^2) = \sum_{k=1}^{n_f} C_{i,k}(x,\alpha_s) \otimes xq^+_k(x, Q^2) & +  C_{i,g}(x,\alpha_s)  \otimes xg(x, Q^2)\, , \qquad i=\{2,L\}  \, ,\\
    x F_3(x,Q^2) = \sum_{k=1}^{n_f} C_{3,k}(x,\alpha_s) & \otimes xq^-_k(x, Q^2) \, ,
  \end{split}
\end{align}
with the coefficient functions evaluated in a perturbative QCD expansion
\be
C_{i,k}(x,\alpha_s(Q^2)) = \sum_{n=0} \alpha_s^{n}(Q^2) C_{i,k}^{(n)}(x) \, .
\ee
Coefficient functions with all quarks assumed to be massless 
were evaluated at N$^3$LO in~\cite{Vermaseren:2005qc,Moch:2004xu} for
neutral-current charged-lepton
scattering and recently independently benchmarked in~\cite{Blumlein:2022gpp}.
The corresponding results for charged-current scattering were presented
in~\cite{Moch:2007rq,Moch:2008fj,Davies:2016ruz}.

For sufficiently low scale, some or all of the heavy quark masses
cannot be neglected. Heavy quark contributions to structure functions may be
treated in a decoupling scheme~\cite{Collins:1978wz}, in which heavy quarks do
not contribute to the running of $\alpha_s$ and to PDF evolution, and 
coefficient functions acquire a dependence on
the heavy quark mass $m_h$~\cite{Buza:1996wv}:
$C_{i,k}=C_{i,k}(x,\alpha_s,m^2_h/Q^2)$ (massive coefficient functions,
henceforth). The  massive 
coefficient functions are known exactly up to
NNLO for photon~\cite{Laenen:1992zk,Alekhin:2003ev},
Z~\cite{Hekhorn:2018ywm,Hekhorn:2019nlf} and W~\cite{Gao:2017kkx}
exchange (for massless to massive transitions only)  while at  N$^3$LO only
partial results are  
available~\cite{Laenen:1998kp,Catani:1990eg,Kawamura:2012cr,bbl2023}
or in the $Q^2 \gg m_h^2$ limit~\cite{Bierenbaum:2008yu,Bierenbaum:2009mv,
  Ablinger:2010ty,Ablinger:2014vwa,Ablinger:2014nga,Behring:2014eya}.

We adopt an approximation for the N$^3$LO contribution
$C_{i,k}^{(3)}(x,\alpha_s,m^2_h/Q^2)$ to massive
coefficient functions for photon-induced DIS and neglect the axial-vector
coupling of the Z boson, while we treat heavy quarks in the massless
approximation for the W boson exchange.
Such an approximation, based on known partial results, has
been presented in Ref.~\cite{Kawamura:2012cr}, and recently revisited 
in Ref.~\cite{Laurenti:2024anf}. The approaches of these references rely on the
same known exact results, and differ in the details of the way they are
combined and interpolated. Here we will follow
Ref.~\cite{Laurenti:2024anf}, see also Ref.~\cite{bbl2023}, to which
we refer for further details. 
Exact results come from threshold resummation and high-energy
resummation, and are further combined with the asymptotic large-$Q^2$ limit,
thereby ensuring that the approximate massive coefficient function
reproduces the exact massless result in the $Q^2/m_h^2\to\infty$ limit.
In the approach of Refs.~\cite{Laurenti:2024anf,bbl2023} the massive
coefficient functions 
are written as
\be
C_{i,k}^{(3)}(x,m_h^2/Q^2) = C_{i,k}^{(3),{\rm thr}}(x,m_h^2/Q^2) f_1(x) + C_{i,k}^{(3),{\rm asy}}(x,m_h^2/Q^2) f_2(x) \, ,
\label{eq:massive_cf}
\ee
where $C_{i,k}^{(3),{\rm thr}}$ and $C_{i,k}^{(3),{\rm asy}}$
correspond to the contributions coming from differently resummations,
and  $f_1(x)$ and  $f_2(x)$ are 
two suitable matching functions.

For massive quarks the threshold limit is
$x\to x_{\rm max}$ with $x_{\rm max} = \frac{Q^2}{4m_h^2 + Q^2}$ or
$\beta \to 0$, with
$\beta\equiv \sqrt{1- \frac{4m_h^2}{s}}$ and $s=Q^2\frac{1-x}{x}$
the center-of-mass energy of the partonic cross-section.
In this limit, the coefficient
function contains logarithmically enhanced terms of the form
$\alpha_s^n \ln^m\beta$ with $m\le 2n$ due to soft gluon emission, which are
predicted by threshold
resummation~\cite{Bonciani:1998vc}.  Further contributions
of the form $\alpha_s^n \beta^{-m}\ln^l\beta$, with $m\le n$, arise from
Coulomb exchange between the heavy quark and antiquark, and can also
be resummed using  non-relativistic QCD
methods~\cite{Pineda:2006ri}. At N$^3$LO all these contributions are
known and can be extracted from available resummed
results~\cite{Kawamura:2012cr}; they are included in
$C_{i,k}^{(3),{\rm thr}}$.

In the high-energy limit, the coefficient function contains logarithmically
enhanced terms of the form 
$\alpha_s^n \ln^m x$ with $m\le n-2$, which are determined at all
orders through small-$x$
resummation at the LL
level~\cite{Catani:1990eg}, from which the N$^3$LO expansion can be
extracted~\cite{Kawamura:2012cr}. This result can be further
improved~\cite{Laurenti:2024anf,bbl2023} by including a particular
class of NLL terms related to NLL perturbative evolution and the
running of the coupling. In the approach of
Refs.~\cite{Laurenti:2024anf,bbl2023} the high-energy contributions
are combined into $C_{i,k}^{(3),{\rm asy}}$ with 
the asymptotic $Q^2 \gg m_h^2$ limit of the coefficient function  
in the decoupling
scheme~\cite{Bierenbaum:2008yu,Bierenbaum:2009mv,Ablinger:2010ty,
  Ablinger:2014vwa,Ablinger:2014nga,Behring:2014eya},
while subtracting overlap terms. This
ensures that in the $Q^2 \gg m_h^2$ limit, the structure function,
computed from  $C_{i,k}^{(3),{\rm asy}}$
combined with decoupling-scheme PDFs, coincides with the structure function
computed in the limit in which the heavy quark mass is
neglected and the heavy quark is treated as a massless parton.
However, the asymptotic limit can
only be determined approximately since in particular some of the
matching conditions are not fully known. 

The interpolating functions, used to combine the two contributions in
Eq.~(\ref{eq:massive_cf}), are chosen to  
satisfy the requirements 
\begin{align}
\begin{split}
    f_1(x) \xrightarrow[x \to 0]{} 0, & \quad  f_1(x) \xrightarrow[x \to x_{\rm max}]{} 1 \, , \\
    f_2(x) \xrightarrow[x \to 0]{} 1, & \quad  f_1(x) \xrightarrow[x \to x_{\rm max}]{} 0 \, , 
    \label{eq:dampingfunctions}
\end{split}
\end{align}
which ensure that the threshold contribution vanishes in the small-$x$
limit and conversely. This guarantees that the approximation
Eq.~(\ref{eq:massive_cf}) is  reliable in a broad kinematic range in
the $(x,Q^2)$ plane:  $C_{i,k}^{(3),{\rm asy}}$
reproduces the massless limit for large $Q^2$ values and for all values of $x$,
including
the small-$x$ limit, while  $C_{i,k}^{(3),{\rm thr}}$ describes the threshold limit,
with $x$ close to $x_{\rm max}$. An uncertainty on the approximate coefficient
function  can be constructed varying the functional form of the
interpolating functions, as well as that of terms which are not fully
known. This includes the NLL small-$x$ resummation and the matching
functions that enter the asymptotic high $Q^2$ limit.
This uncertainty vanishes in the $x\to x_{\rm max}$ limit, for which
the exact known limit is reproduced  (with a fixed choice for the
unknown constant $\beta$-independent terms), and becomes larger in the
intermediate $\eta$ region. The interpolating functions and their
uncertainties are optimized by using the same methodology at NNLO,
where the full result is known.
We refer to Ref.~\cite{Laurenti:2024anf} for a detailed discussion of this
construction. 

This optimized  approximation at NNLO is shown in
Fig.~\ref{fig:massive_nnlo_bench}, where  we compare it to
the  exact result for the
massive gluon-initiated coefficient function $x C_{2,g}^{(2)}(\eta)$,
expressed in terms of the variable
\be\label{eq:etadef}
\eta= \frac{Q^2(1-x)}{4m_h^2x}-1 \, .
\ee
Results are shown for two different values of the $Q^2/m_h^2$ ratio,
close to threshold and at higher scales.
Note that 
 $\eta \to 0$ corresponds to $x\to x_{\rm max}$ (threshold limit),
while $\eta \to \infty$ corresponds to either $Q^2/m_h^2 \to \infty$ for fixed $x$ (asymptotic limit),
or $x\to 0$ for fixed $Q^2$ (high-energy limit). In this case the
uncertainty band is obtained by varying the interpolating functions only.

The results found using the same procedure for the gluon and quark singlet
coefficient functions at N$^3$LO are displayed in
Fig.~\ref{fig:massive_n3lo_bench}, compared to the
approximation of Ref.~\cite{Kawamura:2012cr}, each shown with the
respective uncertainty estimate.
Good agreement between the different approximations is found,
especially for  the dominant  gluon 
coefficient function. The approximations agree in the asymptotic
$\eta\to 0$ and $\eta\to \infty$ limits and in most of the $\eta$
range,  but differ somewhat in the subasymptotic large $\eta$ region
at fixed $Q^2$, which corresponds to the small $x$ limit at fixed
$Q^2$. These differences can be traced to the aforementioned inclusion
in the procedure of Ref.~\cite{Laurenti:2024anf,bbl2023} of a  particular
class of NLL terms related.

The uncertainty involved in the approximation can be included as a
further IHOU, alongside that discussed in Sect.~\ref{sec:ihou},
through an additional contribution to the theory  
covariance matrix. Namely, we define
\be\label{eq:covihouC}
\text{cov}^{C}_{mn} = \frac{1}{2} \left(\Delta_m(+)\Delta_n(+)+\Delta_m(-)\Delta_n(-)\right).
\ee
Here $\Delta_m(\pm)$ is the shift in the prediction for the $m$-th
DIS data point obtained by replacing the central approximation to the
massive coefficient function with the upper or lower edge of the
uncertainty range determined in Ref.~\cite{Laurenti:2024anf} and shown
as an uncertainty band in Fig.~\ref{fig:massive_n3lo_bench}. Note that
unlike in Eq.~(\ref{eq:covihou}), we divide by the number of
independent variations, without decreasing it by one, because the
central value is not the average of the variations, and thus is independent.
The contribution Eq.~(\ref{eq:covihouC}) is then added to the IHOU covariance
matrix as a further term on the right-hand side of Eq.~(\ref{eq:covihou}).

The impact of this contribution to the IHOU is assessed in
Fig.~\ref{fig:dis_ihou}, where 
the square root of the diagonal component of the covariance matrix is shown 
for all the DIS data points in our dataset, comparing the IHOU before
and after adding to Eq.~(\ref{eq:covihou}) the extra component
Eq.~(\ref{eq:covihouC}) due to
the IHOU on the massive coefficient function. It is clear that the
impact of IHOUs due to perturbative evolution is generally negligible,
in agreement with the results discussed in Sect.~\ref{sec:n3lo_evol}
and shown in Fig.~\ref{fig:N3LOevolution-q100gev-ratios}: IHOUs on
splitting functions  are
only significant at small $x$, but available small-$x$ data are at
relatively low scale where the evolution length is small. The impact
of IHOUs on massive coefficient functions is relevant for data on
tagged bottom and charm structure functions, but otherwise moderate
and only significant for structure function data close to the heavy
quark production thresholds. 

\subsection{A general-mass variable flavor number scheme at N$^3$LO}
\label{sec:fonll_d}

\begin{figure}[!t]
    \centering
    \includegraphics[width=\textwidth]{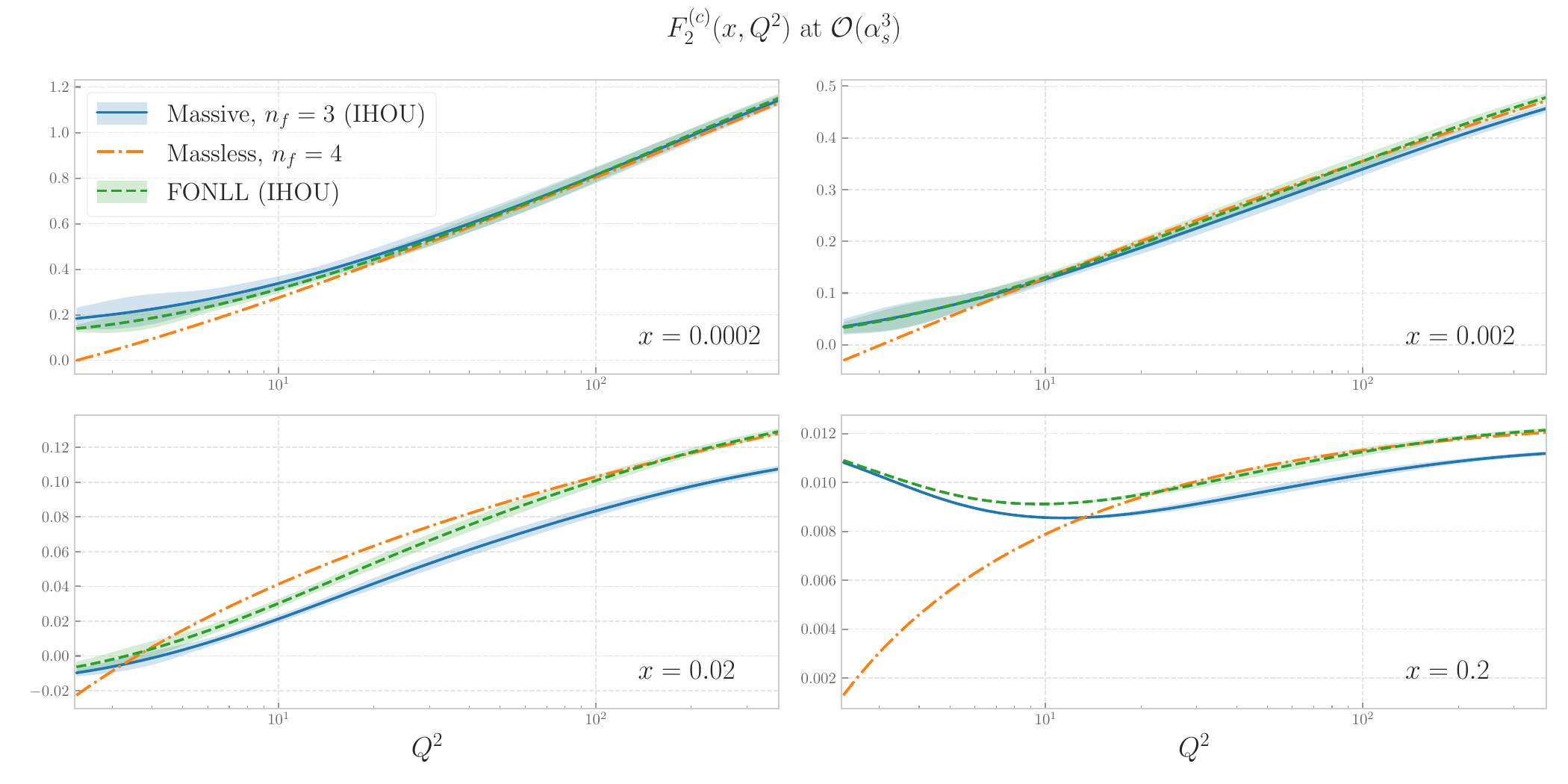}
    \caption{
      The charm structure function $F^{(c)}_{2}(x,Q^2,m_c^2)$ in the
      FONLL-E scheme, compared to the massive and massless scheme
      results (see text). Results are shown as a function of  $Q^2$
      for $x=2\times 10^{-4}$ (top left), $x=2\times 10^{-3}$ (top right),  
      $x=2\times 10^{-2}$ (bottom left), and $x=2\times 10^{-1}$ (bottom
      right). The uncertainty shown on the FONLL and massive curves is
      the IHOU on the heavy quark coefficient functions Eq.~(\ref{eq:covihouC}).}
    \label{fig:f2_charm_n3lo} 
\end{figure}
\begin{figure}[!t]
    \centering
    \includegraphics[width=\textwidth]{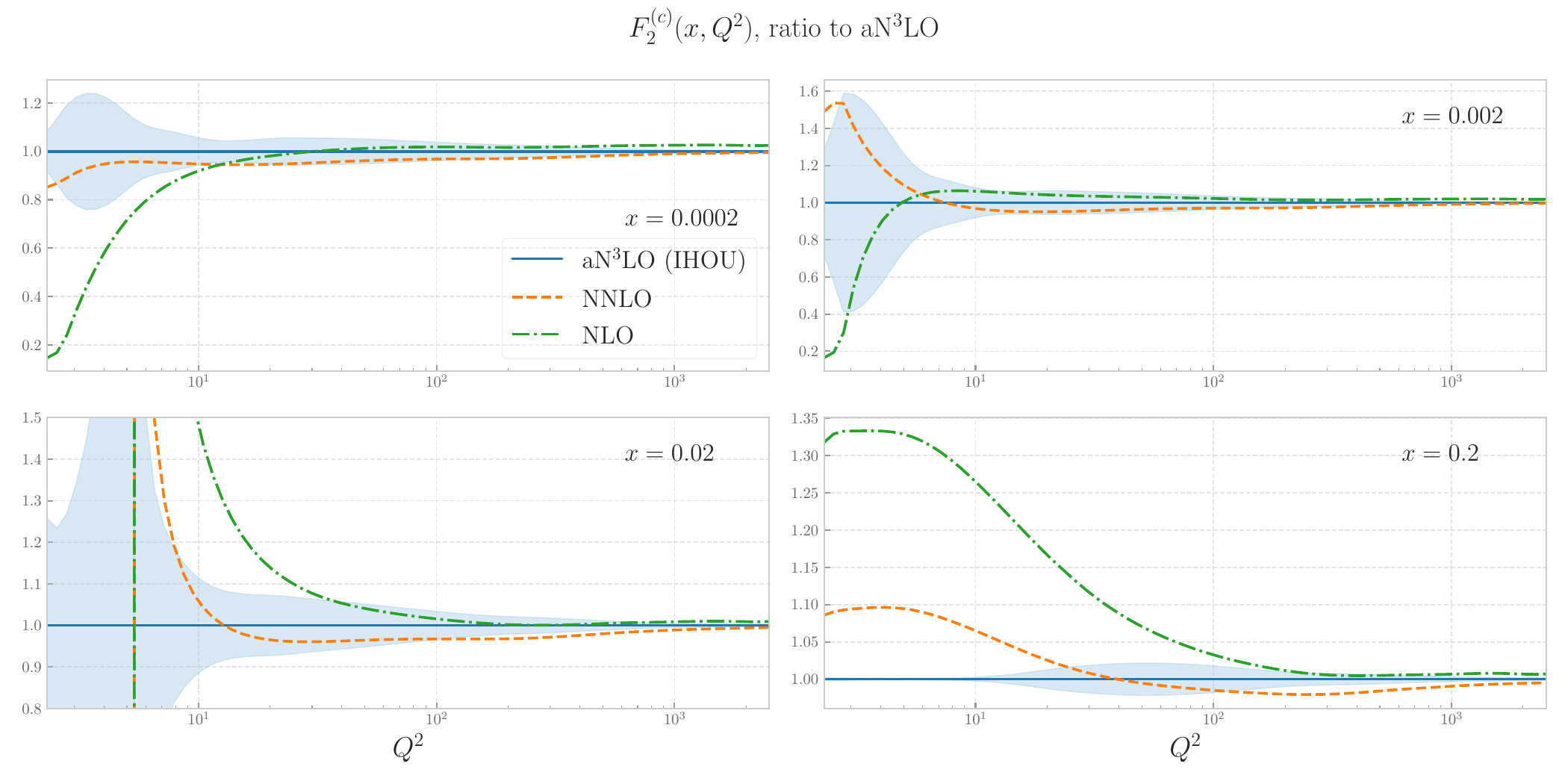}
    \caption{Same as Fig.~\ref{fig:f2_charm_n3lo}, now comparing  the
      FONLL-A (used at NLO $\mathcal{O}\lp \alpha_s\rp$),
      FONLL-C (used at NNLO $\mathcal{O}\lp \alpha_s^2\rp$), 
      and FONLL-E (used at N$^3$LO $\mathcal{O}\lp \alpha_s^3\rp$), all
      shown as a ratio to FONLL-E. The FONLL-E result
      includes the IHOU on the heavy quark coefficient functions
      Eq.~(\ref{eq:covihouC}). 
    }
    \label{fig:f2_charm_pto} 
\end{figure}

The  N$^3$LO DIS coefficients functions described in the previous 
section enable the extension to 
$\mathcal{O}\lp \alpha_s^3\rp$ of the FONLL~\cite{Cacciari:1998it}
general-mass variable flavor number
scheme for DIS~\cite{Forte:2010ta,Ball:2015dpa,Ball:2015tna}.
The goal of the  FONLL strategy is results that 
are accurate and reliable for all values of $Q^2$
from the production threshold $Q^2 \sim m_h^2$ to the asymptotic
limit $Q^2 \gg m_h^2$.

Assuming a single heavy quark, calculations performed in a decoupling scheme 
with $n_\ell$ light quarks  retain the full dependence on the heavy
quark mass and include the contribution of heavy quarks at a fixed
perturbative order (massive scheme, henceforth). Calculations performed in a
scheme in which the
heavy quark is treated as massless (massless scheme, henceforth),
and endowed with a PDF that
satisfies perturbative matching conditions,
resums logarithms of $Q^2/m_h^2$ to all orders through the running of
the coupling and the evolution of PDFs, but does not include terms that
are suppressed as powers of $\frac{m_h^2}{Q^2}$ and thus become
relevant when  $Q^2\gsim m_h^2$. The FONLL prescription matches the two calculations by defining
\be
\label{eq:fonll_schematic}
F^{{\rm FONLL}}_i = F_{i}^{(n)}(x, Q^2,m_h^2) + F_{i}^{(n+1)}(x, Q) - F_{i}^{(n,0)}(x, \ln(Q^2/m_h^2)) \, ,
\qquad \ i = {2,L,3} \, ,
\ee
where $F_{i}^{(n)}$ denotes the massive computation in which the $(n+1)$-th (heavy)
flavor decouples, $ F_{i}^{(n+1)}$ the one in which it is treated
as massless,
and $F_{i}^{(n,0)}$ is the asymptotic large-$Q^2$ limit of the
massive scheme calculation, which depends only logarithmically on
the heavy quark mass. 
This construction reduces to the decoupling calculation for $Q^2 \approx m_h^2$
and to the massless one for $Q^2 \gg m_h^2$.

The FONLL prescription of Eq.~(\ref{eq:fonll_schematic}) was
implemented in Ref.~\cite{Forte:2010ta,Ball:2015dpa,Ball:2015tna} for DIS to NNLO, by expressing all terms on the right-hand side in terms of $\alpha_s$ and PDFs all
defined in the massless scheme. This has the advantage of providing an
expression that can used with externally provided PDFs, that are
typically available only in a single
factorization scheme for each value of the scale $Q$.

However, the recent  {\sc\small EKO}
code~\cite{Candido:2022tld} 
allows, at any given scale, the coexistence of PDFs defined in
schemes with a different 
number of massless flavors. Furthermore, the
recent {\sc\small YADISM} program~\cite{Candido:2024rkr} implements DIS
coefficient functions corresponding to all three contributions on the
right-hand side of Eq.~(\ref{eq:fonll_schematic}).
It is then possible to implement the FONLL prescription
Eq.~(\ref{eq:fonll_schematic}) by simply combining expressions
computed in different schemes~\cite{num_fonll}. This formalism is especially
advantageous at higher perturbative orders, where the analytic
expressions relating PDFs in different scheme grow in complexity, and
also above
bottom threshold, where the iteration of Eq.~(\ref{eq:fonll_schematic}) 
on charm and bottom heavy quarks leads to
coexisting $n_f=3,\,4,\,5$ PDFs, while the method of
Ref.~\cite{Forte:2010ta,Ball:2015dpa,Ball:2015tna} would require
re-expressing the massive scheme PDFs into massless scheme PDFs twice. 

In the FONLL method, Eq.~(\ref{eq:fonll_schematic}), the first two terms
on the right-hand side may be computed at different perturbative
orders, provided one ensures that the third term correctly
includes only their common contributions. In
Ref.~\cite{Forte:2010ta} some natural choices were discussed, based on
the observation that in the massive scheme, the heavy quark
contributes to the structure functions only at $\mathcal{O}\lp
\alpha_s\rp$ and beyond, while in the massless scheme 
it already contributes at $\mathcal{O}\lp \alpha_s^0\rp$. Hence 
natural choices are to combine both the massive and
massless  contributions at $\mathcal{O}\lp \alpha_s\rp$ (FONLL-A), or
else the massive contribution at
$\mathcal{O}\lp \alpha_s^2\rp$ and the massless  
contribution at $\mathcal{O}\lp \alpha_s\rp$, i.e.\ both at
second nontrivial order (FONLL-B). The corresponding two options at the next
order are called FONLL-C and -D.

Here, we will consider FONLL-E, in which both the massless  and
massive contributions are determined at  $\mathcal{O}\lp\alpha_s^3\rp$. 
The charm structure function $F^{(c)}_{2}(x,Q^2)$, computed in this scheme,
is displayed in Fig.~\ref{fig:f2_charm_n3lo}  as a function of $Q^2$
for four values of $x$ (with $m_c=1.51$~GeV), and compared to the
massive and massless scheme results, with the IHOU on the massive
coefficient function shown for the first two cases. The structure
functions are computed using the NNPDF4.0 aN$^3$LO PDF set (to be
discussed in Sect.~\ref{sec:results} below) which satisfies aN$^3$LO evolution
equations, as is necessary for consistency with the massless scheme
result at high scale.
It is clear that the FONLL results interpolate between the massive and
massless calculations as the scale grows. 
The $Q^2$ value at which either of the massive or massless results
dominate depend strongly on $x$.
Except for the lowest $Q^2$ values, the IHOUs associated with the 
calculation remain moderate.

The perturbative convergence of the charm structure function is assessed in
Fig.~\ref{fig:f2_charm_pto}, where we compare the FONLL-A, FONLL-C and
FONLL-E results, all shown as a ratio to FONLL-E, the latter also
including the IHOU 
as in Fig.~\ref{fig:f2_charm_n3lo}.
Clearly, convergence is faster at higher scales due to asymptotic
freedom, and it appears that the perturbative expansion has
essentially converged for $Q^2\gtrsim 10$~GeV$^2$. On the other hand, the
impact of aN$^3$LO at low scale is sizable,  up to 50\% for small
$Q^2$ and  $x = 2\times 10^{-3}$. The IHOUs are correspondingly
sizable at low scale, and in fact always larger than the difference
between the NNLO and aN$^3$LO results except at the highest $x$
values and the lowest scales, implying that for the charm
structure function aN$^3$LO may be more accurate, but possibly 
not more precise than NNLO.

\begin{figure}[!t]
  \centering
  \includegraphics[width=\textwidth]{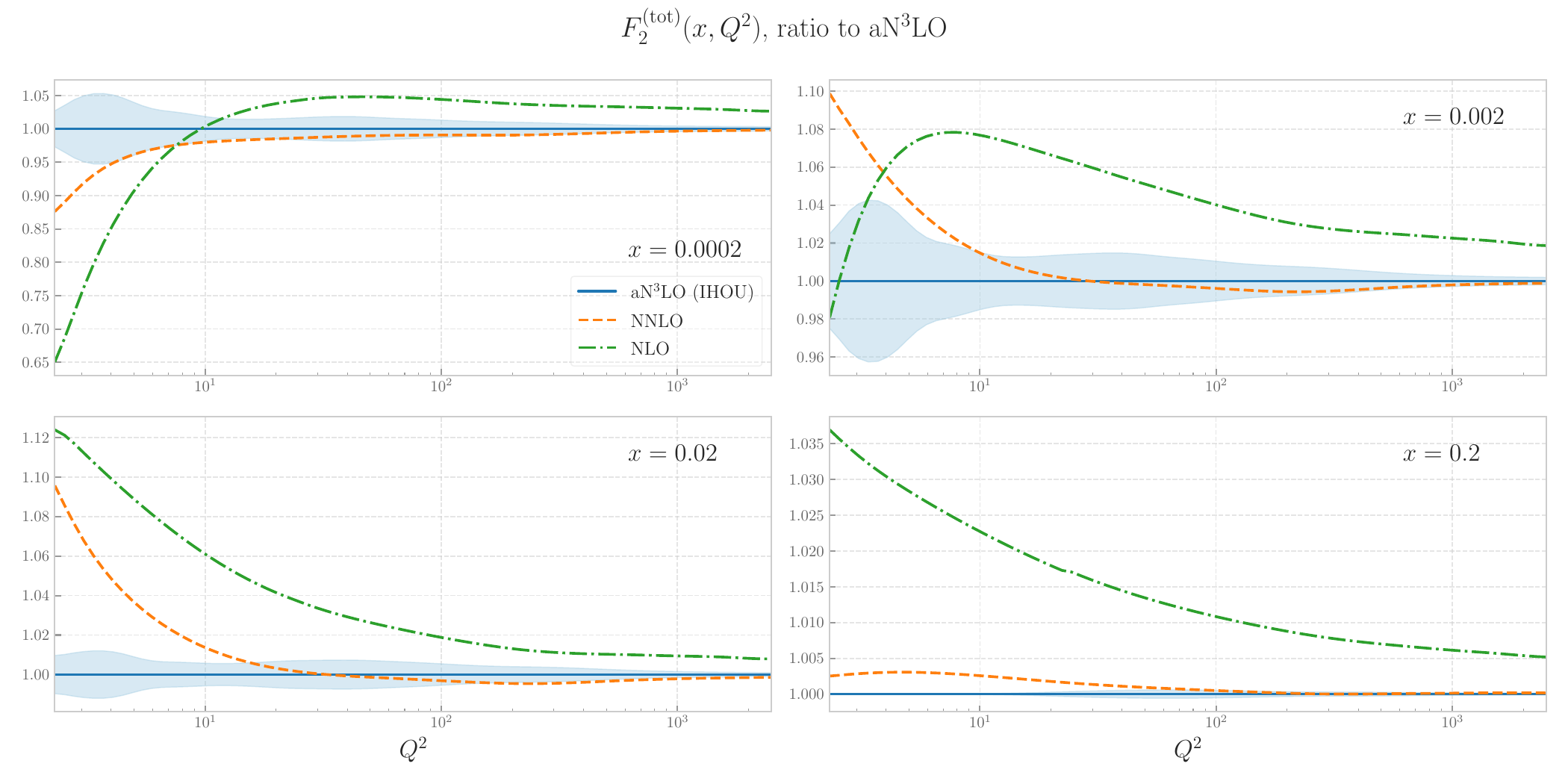}
  \caption{Same as Fig.~\ref{fig:f2_charm_pto} for the inclusive structure
    function $F^{(\text{tot})}_{2}(x,Q^2)$. Note the different scale
    on the $y$ axis.
   }
  \label{fig:f2_total_pto} 
\end{figure}

An analogous study of perturbative convergence of the 
inclusive structure function is shown  in Fig.~\ref{fig:f2_total_pto}
(note the different scale on the $y$ axis). Interestingly, the effect
of the aN$^3$LO corrections changes sign when going from $x=2\times 10^{-4}$
to larger values of $x$.
In general, N$^3$LO corrections 
are smaller at the inclusive level: specifically,
aN$^3$LO corrections to the inclusive
structure function are below 
2\% for $Q^2\gsim 10$ GeV$^{2}$, and at most of order  10\% around the
charm mass scale. The  impact of the IHOUs on the heavy coefficient
is further reduced  due to the fact that charm contributes at most one
quarter of the total structure function. Consequently, the  aN$^3$LO correction to the NNLO
result is now larger than the IHOU in a significant kinematic
region. This, together with the fact that aN$^3$LO corrections are comparable
or larger than typical experimental uncertainties on structure
function data,
motivates their inclusion  in a
global PDF determination.

\subsection{N$^3$LO corrections to hadronic processes}
\label{sec:n3lo_hadronic_coeff}

\begin{figure}[!t]
    \centering
    \includegraphics[width=.80\textwidth]{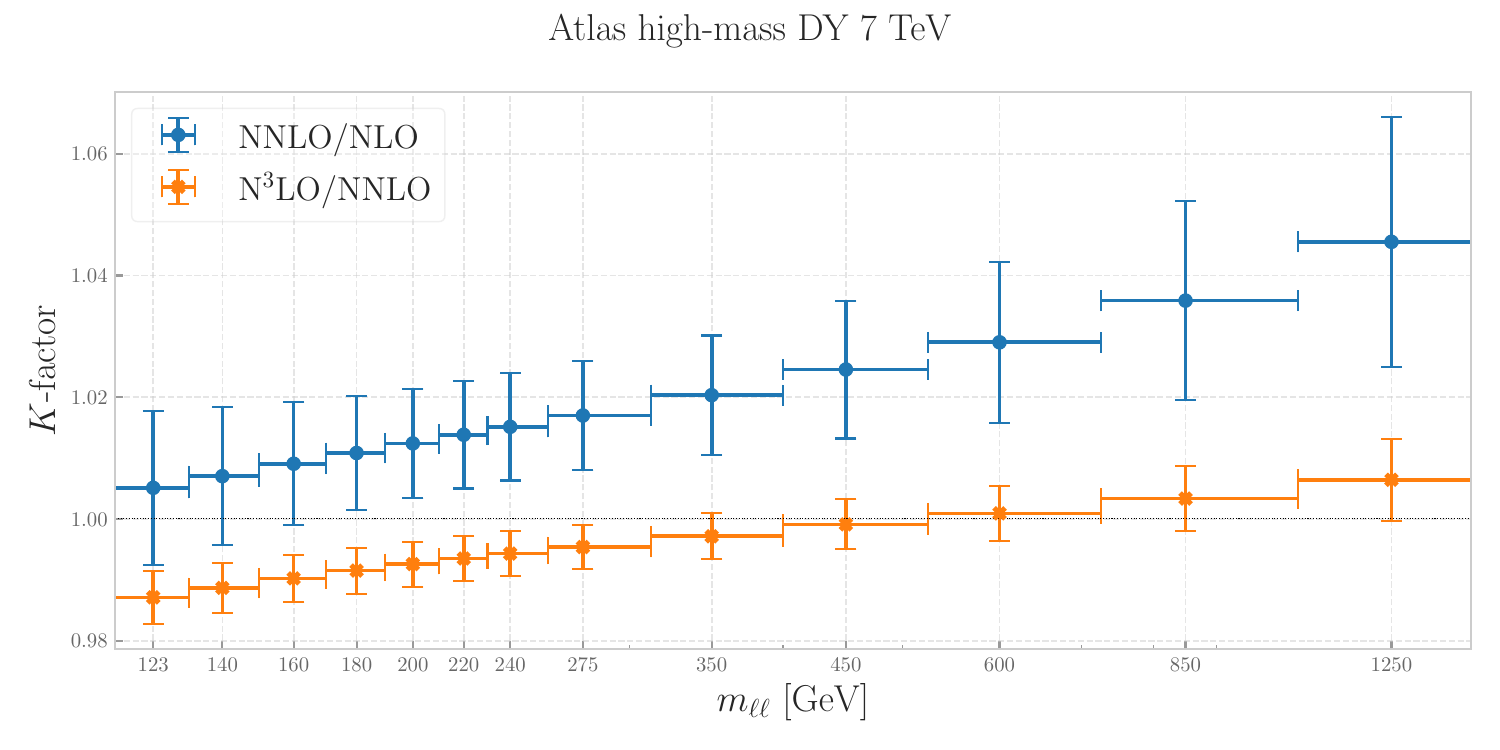}
    \caption{Ratio of the total NC Drell-Yan cross-section as a
      function of the NNLO/NLO  and N$^3$LO/NNLO
      calculations for the inclusive NC Drell-Yan cross-section
      in bins of $m_{\ell \ell}$, the invariant mass of the final-state
      dilepton pair, using the {\sc\small n3loxs} code, integrated over all
      other kinematic variables.
      The  $m_{\ell \ell}$ binning is chosen to be that
      of the ATLAS 7 TeV high-mass DY measurement~\cite{Aad:2013iua}.
      The same PDF set is used in the numerator and denominator,
      namely NNPDF4.0 NNLO (for NNLO/NLO) and aN$^3$LO (for
      N$^3$LO/NNLO). The vertical bands show the MHOU on the
      $K$-factors computed from scale variations.
    }    
    \label{fig:kf_ATLASZHIGHMASS49FB}
   \end{figure}

N$^3$LO corrections  to the total cross-section 
for inclusive neutral- (NC) and charged-current (CC) Drell-Yan
production~\cite{Duhr:2020sdp,Duhr:2021vwj} are available 
through the {\sc\small n3loxs} public code~\cite{Baglio:2022wzu}, both
for on-shell $W$ and $Z$ and as a function of the dilepton invariant
mass  $m_{\ell \ell}$.
Differential distributions at the level of leptonic observables for
the same processes have also been computed~\cite{Chen:2021vtu,Chen:2022lwc},
but are not publicly available. 
No  N$^3$LO calculations are available for other processes included in
the NNPDF4.0 dataset.

The ratio of the NC total cross-section evaluated at two subsequent
perturbative orders 
 with a fixed set of PDFs, chosen as NNPDF4.0 NNLO when comparing NNLO
 to NLO results, and   aN$^3$LO when comparing N$^3$LO to NNLO
 results,
 is shown in
 Fig.~\ref{fig:kf_ATLASZHIGHMASS49FB}. 
Results are shown in the high-mass region, as a function of  $m_{\ell
  \ell}$, with the same binning as the ATLAS 7~TeV
measurement~\cite{Aad:2013iua}. Perturbative convergence is apparent,
with the  N$^3$LO/NNLO ratio  
closer to unity and smoother than its NNLO/NLO counterpart:
while NNLO corrections range between $+0.5\%$ and $+4\%$, at N$^3$LO
they are reduced to  $-1.2\%$ and $+0.5\%$.

Total cross-section data are obtained by extrapolating measurements
performed in a fiducial region. Whereas  for NC Drell-Yan production
in the central rapidity region and for dilepton invariant masses
around the $Z$-peak, the  N$^3$LO/NNLO cross-section ratio
depends only mildly on
the dilepton rapidity $y_{\ell\ell}$~\cite{Chen:2021vtu,Chen:2022lwc},
it is unclear whether this is the case also off-peak or at very large
and very small rapidities. Hence, the inclusion of N$^3$LO corrections for
hadronic processes is, at present, not fully reliable. We have
consequently not included them in our default determination, but only
in a dedicated variant, with the goal of assessing their impact.

The datasets for which N$^3$LO corrections have been included in this variant
are listed in Table~\ref{tab:DY-kf}. We include
the high-mass NC cross-section,  the $Z$
rapidity distribution in the central  rapidity region for on-shell
$Z$-production, and the  total $W$ and $Z$ 
cross-sections. For all these processes the N$^3$LO cross-section is
determined by multiplying the NNLO result by a $K$-factor determined using
a fixed underlying PDF set, namely the aN$^3$LO NNPDF4.0 PDF set to
be discussed in Sect.~\ref{sec:results} below. Specifically, for the
rapidity distribution we take the same  fixed $K$-factor as for the
total cross-section. We do not include off-shell or
double-differential rapidity
distributions (specifically from CMS), off-forward rapidity
distributions (specifically from LHCb) and   low-mass total
cross-sections, for all of which the approximation of assuming the
$K$-factor to be independent of rapidity and/or amenable to
fiducial extrapolation is even less reliable. The datasets are
labeled as in Table~2.4 of Ref.~\cite{NNPDF:2021njg} \footnote{The
number of datapoints for the rapidity distributions differs from the
numbers in this table because here we only include $Z$ distributions}.

\begin{table}[!t]
  \centering
  \footnotesize
    \renewcommand{\arraystretch}{1.60}
\begin{tabularx}{\textwidth}{Xccccc}
    \toprule
    Dataset
    & Ref.
    & $n_{\rm dat}$
    & Kin$_1$
    & Kin$_2$~[GeV]
    & $C$-factor N$^3$LO/NNLO
    \\
    \midrule
    ATLAS high-mass DY 7 TeV
    & \cite{Aad:2013iua}
    & 13
    & $|\eta_\ell|\leq 2.1$
    & $116\le m_{\ell\ell}\le 1500$
    & $\d \sigma / d m_{\ell \ell}$
    \\
    ATLAS $Z$ 7 TeV ($\mathcal{L}=35$~pb$^{-1}$)
    & \cite{Aad:2011dm}
    & 8
    & $|\eta_\ell,y_Z|\leq 3.2$
    & $Q=m_Z$
    & $\d \sigma / d m_{\ell \ell}~ (66 < m_{\ell \ell} < 150)$
    \\
    ATLAS $Z$ 7 TeV ($\mathcal{L}=4.6$~fb$^{-1}$) CC
    & \cite{Aaboud:2016btc}
    & 24
    & $|\eta_\ell,y_Z|\leq 2.5,3.6$
    & $Q=m_Z$ 
    & $\d \sigma / d m_{\ell \ell}~(46 < m_{\ell \ell} < 116)$
    \\
    \midrule
    ATLAS $\sigma_{W,Z}^{\rm tot}$ 13 TeV
    & \cite{Aad:2016naf}
    & 3
    & ---
    & $Q=m_W,m_Z$
    & $\sigma$
    \\
    \bottomrule
  \end{tabularx}
 \\
    \vspace{0.3cm}
    \caption{The LHC NC DY  production datasets in NNPDF4.0 for which
      an N$^3$LO $K$-factor has been included in a variant of
      the default aN$^3$LO PDF determination (see
      Sect.~\ref{sec:results}). 
      For each dataset we indicate
      the published references, the number of datapoints and the
      kinematic variables.
    }
    \label{tab:DY-kf}
\end{table}

Despite the fact that we are not yet able to 
determine reliably N$^3$LO corrections for 
currently available LHC measurements, we wish to include
the full NNPDF4.0 dataset  in our aN$^3$LO PDF
determination. To this purpose, we endow all data for which   N$^3$LO
are not included with an extra uncertainty that accounts for the
missing N$^3$LO terms. This is estimated using the methodology of
Refs.~\cite{NNPDF:2019vjt,NNPDF:2019ubu}, recently used in
Ref.~\cite{NNPDF:2024dpb}  to produce a variant of the NNPDF4.0 PDF
sets that includes MHOUs. 

Thus, when not including  N$^3$LO corrections to the hard
cross-section,  the theory prediction is evaluated by combining
aN$^3$LO evolution with the NNLO cross-sections. 
The prediction is then  supplemented with a  theory covariance matrix,
computed varying 
the renormalization scale $\mu_R$ 
using a three-point
prescription~\cite{NNPDF:2019vjt,NNPDF:2019ubu}:
\be\label{eq:covihouNNLO}
\text{cov}^{\rm NNLO}_{mn} = \frac{1}{2} \left(\Delta_m(+)\Delta_n(+)+\Delta_m(-)\Delta_n(-)\right),
\ee
analogous to Eq.~(\ref{eq:covihouC}), but now with $\Delta_m(\pm)$ the
shift in the prediction for the $m$-th data point obtained by
replacing the coefficient functions with those obtained by performing
upper or lower renormalization
scale variation using the methodology of Ref.~\cite{NNPDF:2019ubu}
(as implemented and
discussed in Ref.~\cite{NNPDF:2024dpb}, Eq.~(2.9)).
This MHOU covariance
matrix is then added to the IHOU covariance matrix as a further
term on the right-hand side of Eq.~(\ref{eq:covihou}).

\begin{figure}[!t]
  \centering
  \includegraphics[width=.7\textwidth]{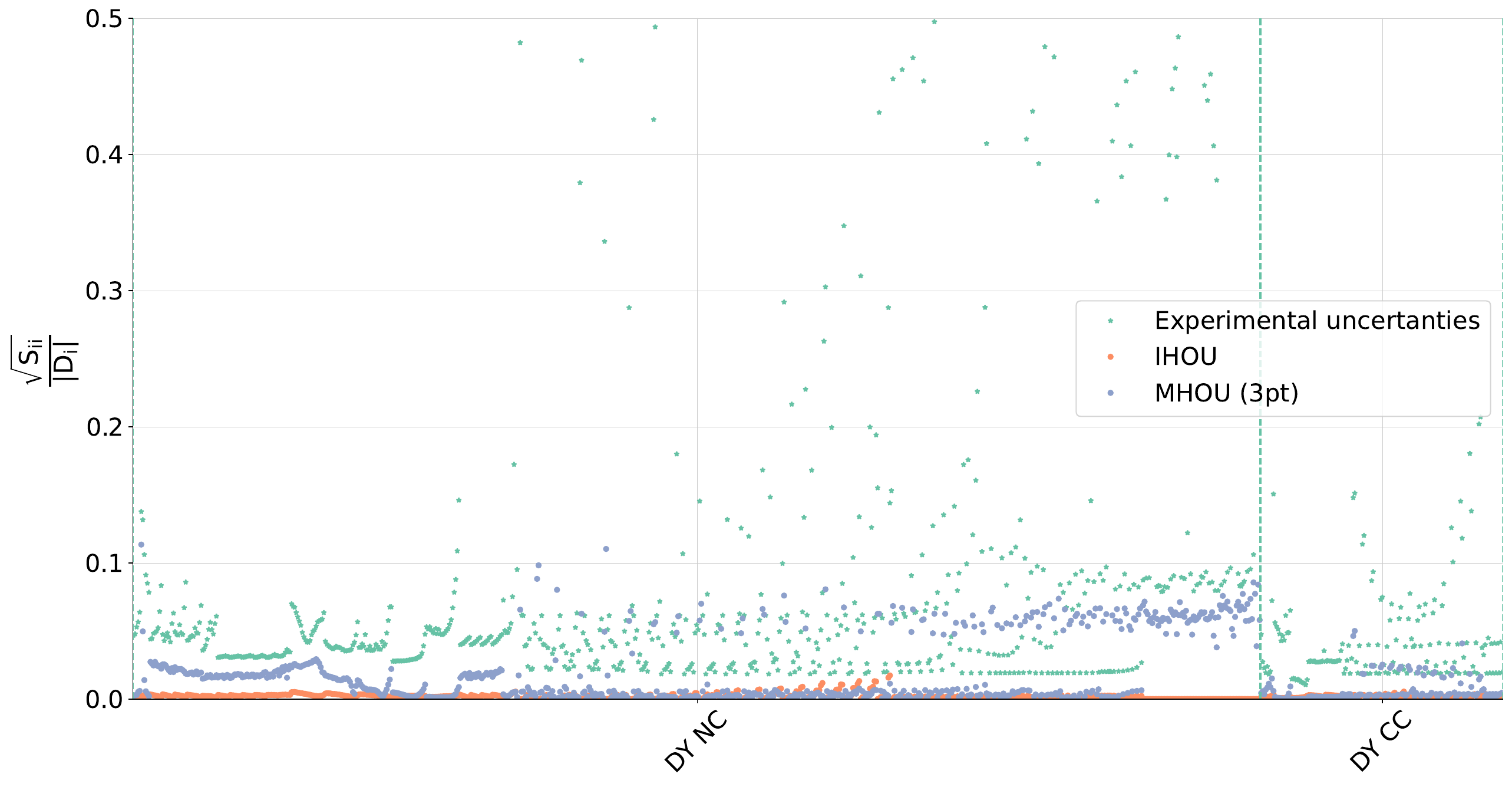} \\
  \includegraphics[width=.7\textwidth]{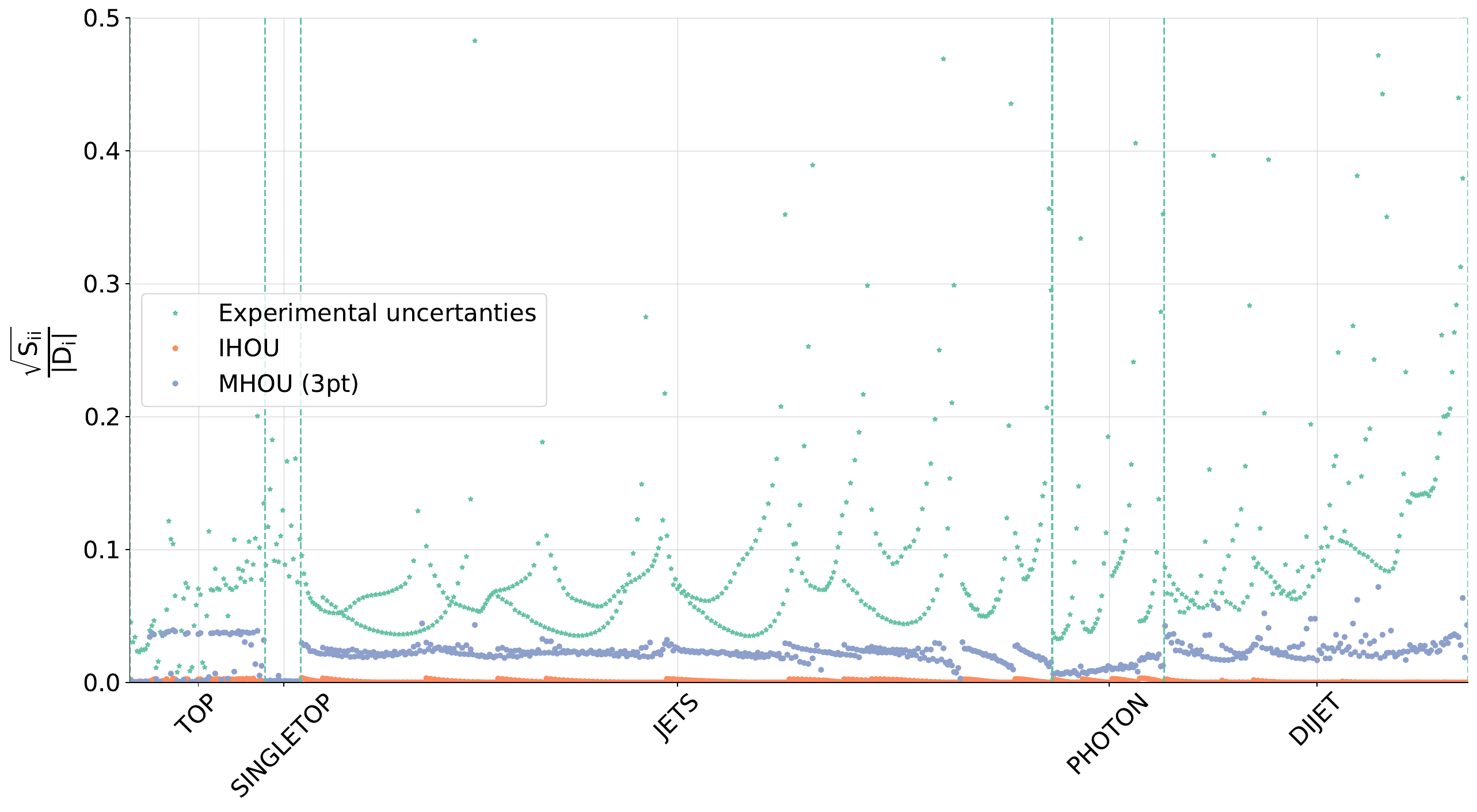}
  \caption{Same as Fig.~\ref{fig:dis_ihou}
    now comparing the IHOU from
    Eq.~(\ref{eq:covihou}) with the MHOU from
    Eq.~(\ref{eq:covihouNNLO}) due to the missing N$^3$LO correction
    to the matrix element. Results are shown for all hadronic data in
    the NNPDF4.0 dataset: specifically Drell-Yan (top) and top pair,
    single top, single-inclusive jet, prompt photon and dijet
    production (bottom). 
  }    
  \label{fig:hadronic_mhou}
 \end{figure}

The impact of this uncertainty is shown in
Fig.~\ref{fig:hadronic_mhou}, where we show for all hadronic datasets
the square root of the 
diagonal entries of the MHOU covariance matrix Eq.~(\ref{eq:covihouNNLO}), 
compared to those of the IHOU covariance matrix Eq.~(\ref{eq:covihou}),
and to the experimental uncertainties, all normalized to the central
theory prediction. The MHOU is generally larger than the IHOU,
indicating that the missing N$^3$LO terms in the hard cross-sections are
larger than the IHOU uncertainty in N$^3$LO perturbative evolution.
The experimental uncertainties are generally larger still.

In addition to the NNPDF4.0 aN$^3$LO baseline PDF set obtained in this
manner, we will also produce a NNPDF4.0 MHOU aN$^3$LO set, in analogy
to the NLO and NNLO MHOU sets recently presented in
Ref.~\cite{NNPDF:2024dpb}. For this set, MHOUs on both perturbative
evolution and on the hard matrix elements are included  using the
methodology of Refs.~\cite{NNPDF:2019vjt,NNPDF:2019ubu} with a theory
covariance matrix determined performing combined correlated
renormalization and factorization scale variations with a  7-point
prescription, as discussed in detail in
Ref.~\cite{NNPDF:2024dpb}. In this case, we simply perform scale
variation on the expressions at the order at which they are being
computed, namely aN$^3$LO for 
anomalous dimensions and DIS coefficient functions and NNLO for
hadronic processes. The scale variation then is automatically larger
and suitable deweights processes for which N$^3$LO corrections are
not available. The possibility of simultaneously including in a PDF
determination processes for which theory predictions are only
available at different perturbative orders is an advantage of the
inclusion of MHOUs in the PDF determination, as already pointed out
in Refs.~\cite{Faura:2020oom,Amoroso:2022eow}.

\section{NNPDF4.0 at aN$^3$LO}
\label{sec:results}

We now present the aN$^3$LO NNPDF4.0 PDF sets.
They have been obtained by using the dataset and methodology discussed
in~\cite{NNPDF:2021njg} and used for the construction of the LO,
NLO, and NNLO NNPDF4.0 presented there, now extended to aN$^3$LO.
The  aN$^3$LO results are obtained using the approximate N$^3$LO splitting
functions of Sect.~\ref{sec:dglap}, the exact massless and approximate massive
N$^3$LO coefficient functions of Sect.~\ref{sec:n3lo_dis}, and NNLO
hadronic cross-sections supplemented by an extra uncertainty as per
Sect.~\ref{sec:n3lo_hadronic_coeff}. 

Theoretical predictions are obtained using
the new theory pipeline of Ref.~\cite{Barontini:2023vmr}, which
relies on the {\sc\small EKO} evolution 
code~\cite{Candido:2022tld} and on the {\sc\small YADISM} DIS
module~\cite{Candido:2024rkr}. As  discussed in
Sect.~\ref{sec:fonll_d}, this pipeline in particular includes a new
FONLL implementation, that differs from the previous one by subleading
terms. A further small difference in comparison to Ref.~\cite{NNPDF:2021njg}
is the correction of a few minor bugs in the data implementation.
The overall impact of all these changes was assessed in Appendix~A of
Ref.~\cite{NNPDF:2024djq}, and was found to be very limited, so that
the new and old implementations can be considered equivalent,
and the PDF sets presented here can be considered the extension
to aN$^3$LO of the NNPDF4.0 PDF sets of Ref.~\cite{NNPDF:2021njg}.

In addition to the default NNPDF4.0 aN$^3$LO PDF determination, we also
present an aN$^3$LO PDF determination that includes MHOUs on all the
theory predictions used in the PDF determination. This is constructed
using the same methodology recently used to produce the
NNPDF4.0MHOU NNLO PDF set in Ref.~\cite{NNPDF:2024dpb}. In order to be
able to discuss perturbative convergence and the impact of MHOUs we
will also present a NNPDF4.0MHOU NLO PDF set constructed using the
same methodology, and exactly the same dataset as the default NNPDF4.0
NLO PDF set (which differs from the NNPDF4.0 NNLO dataset).

We finally construct two variants of the aN$^3$LO PDF sets (both
with and without MHOUs) with  modified N$^3$LO theory. In a first
variant, we replace our own approximation to the N$^3$LO anomalous dimensions,
discussed in Sect.~\ref{sec:dglap}, with that of
Refs.~\cite{Falcioni:2023luc,Falcioni:2023vqq,Moch:2023tdj}. In
the second variant, we will also include  N$^3$LO corrections for the processes
listed in Tab.~\ref{tab:DY-kf}, as discussed in
Sect.~\ref{sec:n3lo_hadronic_coeff}.

We first assess the fit quality, then present the PDFs and their
uncertainties, and study perturbative convergence and the
effect on it of the inclusion of MHOUs. We then specifically study the
impact of aN$^3$LO corrections on intrinsic charm. We then turn to
the variants, and finally compare our results to the recent MSHT20
aN$^3$LO PDFs.

\subsection{Fit quality}
\label{sec:fit_settings}

\begin{figure}[!t]
  \centering    
  \includegraphics[width=.75\textwidth]{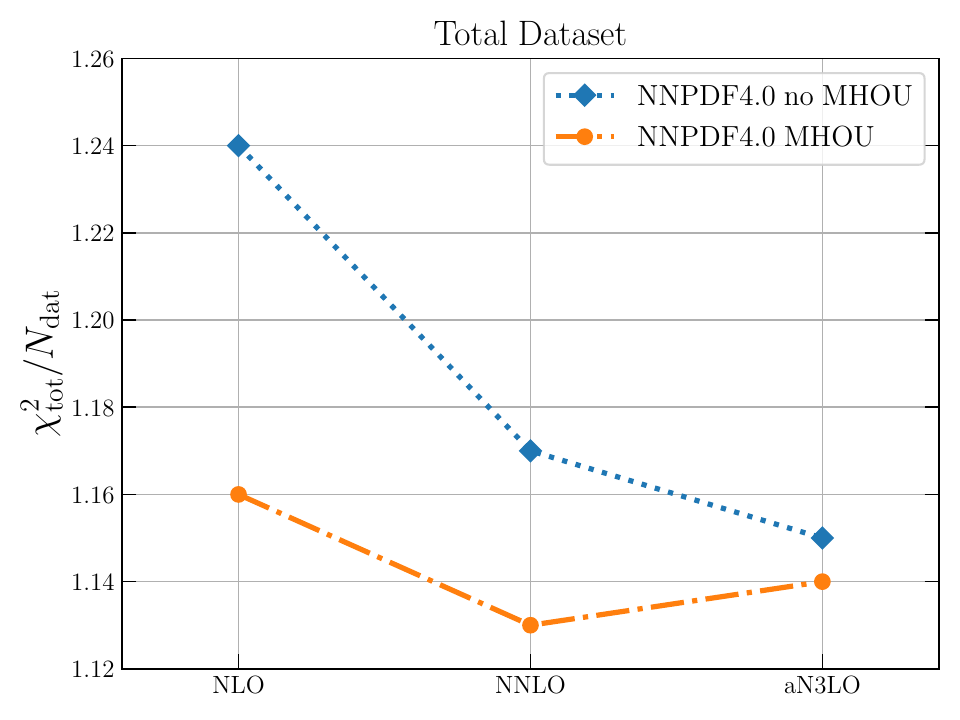}
  \caption{The values of the total $\chi^2$ per data point in the
    NNPDF4.0 NLO, NNLO, and aN$^3$LO fits without and with MHOUs.}
  \label{fig:chi2_n3lo_summary} 
\end{figure}

Tables~\ref{tab:chi2_TOTAL}-\ref{tab:chi2_OTHER} display the number
of data points and the $\chi^2$ per data point obtained in the NLO, NNLO, and
aN$^3$LO NNPDF4.0 fits with and without MHOUs. In Table~\ref{tab:chi2_TOTAL}
the datasets are grouped according to the process categorization used
in Ref.~\cite{NNPDF:2024dpb}. Results for individual datasets are displayed in
Table~\ref{tab:chi2_DIS}, (NC and CC DIS), in Table~\ref{tab:chi2_DY} (NC and
CC DY), and in Table~\ref{tab:chi2_OTHER} (top pairs, single-inclusive jets,
dijets, isolated photons, and single top). The naming of the datasets follows
Ref.~\cite{NNPDF:2021njg}. The value of the total $\chi^2$ per data point is
also shown as a function of the perturbative order in
Fig.~\ref{fig:chi2_n3lo_summary}. 

\begin{table}[!t]
  \scriptsize
  \centering
  \renewcommand{\arraystretch}{1.4}
  \begin{tabularx}{\textwidth}{Xrccrccrcc}
  \toprule
  & \multicolumn{3}{c}{NLO}
  & \multicolumn{3}{c}{NNLO}
  & \multicolumn{3}{c}{aN$^3$LO} \\
  Dataset
  & $N_{\rm dat}$
  & no MHOU
  & MHOU
  & $N_{\rm dat}$
  & no MHOU
  & MHOU 
  & $N_{\rm dat}$
  & no MHOU
  & MHOU \\
  \midrule
  DIS NC
  & 1980 & 1.30 & 1.22
  & 2100 & 1.22 & 1.20
  & 2100 & 1.22 & 1.20 \\
  DIS CC
  &  988 & 0.92 & 0.87
  &  989 & 0.90 & 0.90 
  &  989 & 0.91 & 0.92 \\
  DY NC
  &  667 & 1.49 & 1.32
  &  736 & 1.20 & 1.15
  &  736 & 1.17 & 1.16 \\
  DY CC
  &  193 & 1.31 & 1.27
  &  157 & 1.45 & 1.37
  &  157 & 1.37 & 1.36 \\
  Top pairs
  &   64 & 1.90 & 1.24
  &   64 & 1.27 & 1.43
  &   64 & 1.23 & 1.41 \\
  Single-inclusive jets
  &  356 & 0.86 & 0.82
  &  356 & 0.94 & 0.81
  &  356 & 0.84 & 0.83 \\
  Dijets
  &  144 & 1.55 & 1.81
  &  144 & 2.01 & 1.71
  &  144 & 1.78 & 1.67 \\
  Prompt photons 
  &   53 & 0.58 & 0.47
  &   53 & 0.76 & 0.67
  &   53 & 0.72 & 0.68 \\
  Single top
  &   17 & 0.35 & 0.34
  &   17 & 0.36 & 0.38
  &   17 & 0.35 & 0.36 \\
  \midrule
  Total
  & 4462 & 1.24 & 1.16
  & 4616 & 1.17 & 1.13
  & 4616 & 1.15 & 1.14 \\
\bottomrule
\end{tabularx}

  \vspace{0.3cm}
  \caption{The number of data points and the $\chi^2$ per data point obtained
    in the NLO, NNLO, and aN$^3$LO NNPDF4.0 fits without and with MHOUs,
    see text for details. The datasets are grouped according to
    the same process categorization as that used in Ref.~\cite{NNPDF:2024dpb}.}
  \label{tab:chi2_TOTAL}
\end{table}

\begin{table}[!t]
  \scriptsize
  \centering
  \renewcommand{\arraystretch}{1.4}
  \begin{tabularx}{\textwidth}{Xrccrccrcc}
  \toprule
  & \multicolumn{3}{c}{NLO}
  & \multicolumn{3}{c}{NNLO}
  & \multicolumn{3}{c}{aN$^3$LO} \\
  Dataset
  & $N_{\rm dat}$
  & no MHOU
  & MHOU
  & $N_{\rm dat}$
  & no MHOU
  & MHOU 
  & $N_{\rm dat}$
  & no MHOU
  & MHOU \\
  \midrule
  NMC $F_2^d/F_2^p$
  &  121 & 0.87 & 0.86
  &  121 & 0.87 & 0.88
  &  121 & 0.88 & 0.88 \\
  NMC $\sigma^{{\rm NC},p}$
  &  203 & 1.82 & 1.30
  &  204 & 1.57 & 1.33
  &  204 & 1.57 & 1.36 \\
  SLAC $F_2^p$
  &   33 & 1.64 & 0.74
  &   33 & 0.91 & 0.68
  &   33 & 0.93 & 0.72 \\
  SLAC $F_2^d$
  &   34 & 0.90 & 0.68
  &   34 & 0.61 & 0.54
  &   34 & 0.62 & 0.58 \\
  BCDMS $F_2^p$
  &  333 & 1.62 & 1.24
  &  333 & 1.40 & 1.29
  &  333 & 1.39 & 1.40 \\
  BCDMS $F_2^d$
  &  248 & 1.05 & 1.00
  &  248 & 1.01 & 0.99
  &  248 & 1.04 & 1.03 \\
  HERA I+II $\sigma_{\rm NC}\ e^-p$
  &  159 & 1.44 & 1.40
  &  159 & 1.40 & 1.39
  &  159 & 1.45 & 1.40 \\
  HERA I+II $\sigma_{\rm NC}\ e^+p$ ($E_p=460$~GeV)
  &  192 & 1.12 & 1.05
  &  204 & 1.09 & 1.04
  &  204 & 1.07 & 1.05 \\
  HERA I+II $\sigma_{\rm NC}\ e^+p$ ($E_p=575$~GeV)
  &  236 & 0.85 & 0.84
  &  254 & 0.93 & 0.88
  &  254 & 0.87 & 0.88 \\
  HERA I+II $\sigma_{\rm NC}\ e^+p$ ($E_p=820$~GeV)
  &   54 & 1.15 & 0.85
  &   70 & 1.12 & 0.95
  &   70 & 0.96 & 0.86 \\
  HERA I+II $\sigma_{\rm NC}\ e^+p$ ($E_p=920$~GeV)
  &  317 & 1.30 & 1.21
  &  377 & 1.31 & 1.25
  &  377 & 1.27 & 1.24 \\
  HERA I+II $\sigma_{\rm NC}^{c}$
  &   24 & 2.18 & 1.40
  &   37 & 1.96 & 1.75
  &   37 & 1.86 & 1.57 \\
  HERA I+II $\sigma_{\rm NC}^{b}$
  &   26 & 1.42 & 1.05
  &   26 & 1.44 & 1.11
  &   26 & 1.26 & 1.07 \\
  \midrule
  CHORUS $\sigma_{CC}^{\nu}$
  &  416 & 0.96 & 0.95 
  &  416 & 0.96 & 0.97
  &  416 & 0.97 & 0.98 \\
  CHORUS $\sigma_{CC}^{\bar{\nu}}$
  &  416 & 0.90 & 0.88
  &  416 & 0.88 & 0.87
  &  416 & 0.88 & 0.88 \\
  NuTeV $\sigma_{CC}^{\nu}$ (dimuon)
  &   39 & 0.24 & 0.22
  &   39 & 0.33 & 0.33
  &   39 & 1.27 & 1.28 \\
  NuTeV $\sigma_{CC}^{\bar{\nu}}$ (dimuon)
  &   36 & 0.43 & 0.35
  &   37 & 0.56 & 0.64
  &   37 & 0.63 & 0.59 \\
  HERA I+II $\sigma_{\rm CC}\ e^-p$
  &   42 & 1.34 & 1.19
  &   42 & 1.25 & 1.29
  &   42 & 1.29 & 1.34 \\
  HERA I+II $\sigma_{\rm CC}\ e^+p$
  &   39 & 1.26 & 1.22
  &   39 & 1.23 & 1.25
  &   39 & 1.27 & 1.28 \\
\bottomrule
\end{tabularx}

  \vspace{0.3cm}
  \caption{Same as Table~\ref{tab:chi2_TOTAL} for DIS NC (top) and DIS CC
  (bottom) datasets.}
  \label{tab:chi2_DIS}
\end{table}

\begin{table}[!t]
  \scriptsize
  \centering
  \renewcommand{\arraystretch}{1.4}
  \begin{tabularx}{\textwidth}{Xrccrccrcc}
  \toprule
  & \multicolumn{3}{c}{NLO}
  & \multicolumn{3}{c}{NNLO}
  & \multicolumn{3}{c}{aN$^3$LO} \\
  Dataset
  & $N_{\rm dat}$
  & no MHOU
  & MHOU
  & $N_{\rm dat}$
  & no MHOU
  & MHOU 
  & $N_{\rm dat}$
  & no MHOU
  & MHOU \\
  \midrule
  E866 $\sigma^d/2\sigma^p$ (NuSea)
  & 15 & 0.59 & 0.47
  & 15 & 0.52 & 0.51
  & 15 & 0.53 & 0.51 \\
  E866 $\sigma^p$ (NuSea)
  & 89 & 1.33 & 0.86
  & 89 & 1.59 & 1.00
  & 89 & 1.08 & 1.03 \\
  E605 $\sigma^d/2\sigma^p$ (NuSea)
  & 85 & 0.43 & 0.42
  & 85 & 0.46 & 0.45
  & 85 & 0.45 & 0.45 \\
  E906 $\sigma^d/2\sigma^p$ (SeaQuest)
  &  6 & 1.47 & 0.74
  &  6 & 0.97 & 0.90
  &  6 & 0.82 & 0.88 \\
  CDF $Z$ differential
  & 28 & 1.23 & 1.24
  & 28 & 1.23 & 1.18
  & 28 & 1.23 & 1.22 \\
  D0 $Z$ differential
  & 28 & 0.69 & 0.71
  & 28 & 0.64 & 0.64
  & 28 & 0.64 & 0.63 \\
  ATLAS low-mass DY 7 TeV
  &  4 & 0.69 & 0.66
  &  6 & 0.88 & 0.78
  &  6 & 0.78 & 0.76 \\
  ATLAS high-mass DY 7 TeV
  &  5 & 1.74 & 1.66
  &  5 & 1.60 & 1.67
  &  5 & 1.64 & 1.68 \\
  ATLAS $Z$ 7 TeV ($\mathcal{L}=35$~pb$^{-1}$)
  &  8 & 0.67 & 0.44
  &  8 & 0.58 & 0.57
  &  8 & 0.56 & 0.61 \\
  ATLAS $Z$ 7 TeV ($\mathcal{L}=4.6$~fb$^{-1}$) CC
  & 16 & 3.82 & 2.99
  & 24 & 1.80 & 1.68
  & 24 & 1.66 & 1.69 \\
  ATLAS $Z$ 7 TeV ($\mathcal{L}=4.6$~fb$^{-1}$) CF
  & 15 & 1.77 & 1.22
  & 15 & 1.07 & 1.02
  & 15 & 1.02 & 0.99 \\
  ATLAS low-mass DY 2D 8 TeV
  & 47 & 1.38 & 0.94
  & 60 & 1.23 & 1.08
  & 60 & 1.17 & 1.13 \\
  ATLAS high-mass DY 2D 8 TeV
  & 48 & 1.52 & 1.38
  & 48 & 1.11 & 1.08
  & 48 & 1.09 & 1.09 \\
  ATLAS $\sigma_{Z}^{\rm tot}$ 13 TeV
  & 1 & 0.12 & 0.41
  & 1 & 0.24 & 0.60
  & 1 & 0.24 & 0.66 \\
  ATLAS $Z$ $p_T$ 8 TeV ($p_T,m_{\ell\ell}$)
  & 41 & 1.08 & 0.91
  & 44 & 0.91 & 0.91
  & 44 & 0.89 & 0.89 \\
  ATLAS $Z$ $p_T$ 8 TeV ($p_T,y_Z$)
  & 28 & 0.87 & 0.52
  & 48 & 0.90 & 0.70
  & 48 & 0.77 & 0.68 \\
  CMS DY 2D 7 TeV
  &  88 & 1.29 & 1.11
  & 110 & 1.34 & 1.32
  & 110 & 1.34 & 1.36 \\
  CMS $Z$ $p_T$ 8 TeV
  &  28 & 1.66 & 1.47
  &  28 & 1.40 & 1.41
  &  28 & 1.35 & 1.39 \\
  LHCb $Z\to ee$ 7 TeV
  &   9 & 1.47 & 1.18
  &   9 & 1.65 & 1.53
  &   9 & 1.48 & 1.46 \\
  LHCb $Z \to \mu$ 7 TeV
  &  15 & 1.03 & 0.87
  &  15 & 0.80 & 0.73
  &  15 & 0.77 & 0.73 \\
  LHCb $Z\to ee$ 8 TeV
  &  17 & 1.58 & 1.38
  &  17 & 1.24 & 1.26
  &  17 & 1.31 & 1.27 \\
  LHCb $Z\to \mu$ 8 TeV
  &  15 & 1.25 & 1.06
  &  15 & 1.44 & 1.59
  &  15 & 1.60 & 1.60 \\
  LHCb $Z\to ee$ 13 TeV
  &  15 & 1.68 & 1.60
  &  15 & 1.72 & 1.80
  &  15 & 1.78 & 1.76 \\
  LHCb $Z\to \mu\mu$ 13 TeV
  &  16 & 1.10 & 1.11
  &  16 & 0.94 & 0.99
  &  16 & 0.99 & 0.94 \\
  \midrule
  D0 $W$ muon asymmetry
  &   8 & 2.42 & 2.17
  &   9 & 1.86 & 1.95
  &   9 & 2.07 & 2.03 \\
  ATLAS $W$ 7 TeV ($\mathcal{L}=35$~pb$^{-1}$)
  &  22 & 1.20 & 1.13
  &  22 & 1.11 & 1.12
  &  22 & 1.09 & 1.12 \\
  ATLAS $W$ 7 TeV ($\mathcal{L}=4.6$~fb$^{-1}$)
  &  22 & 2.18 & 2.13
  &  22 & 2.08 & 2.16
  &  22 & 2.16 & 2.10 \\
  ATLAS $\sigma_{W}^{\rm tot}$ 13 TeV
  &   2 & 0.16 & 0.54 
  &   2 & 1.21 & 1.60
  &   2 & 1.38 & 1.67 \\
  ATLAS $W^+$+jet 8 TeV
  &  15 & 0.26 & 0.28
  &  15 & 0.79 & 0.79
  &  15 & 0.73 & 0.73 \\
  ATLAS $W^-$+jet 8 TeV
  &  15 & 0.98 & 1.27
  &  15 & 1.49 & 1.45
  &  15 & 1.41 & 1.41 \\
  CMS $W$ electron asymmetry 7 TeV
  &  11 & 0.92 & 1.03
  &  11 & 0.84 & 0.85
  &  11 & 0.82 & 0.86 \\
  CMS $W$ muon asymmetry 7 TeV
  &  11 & 2.03 & 1.77
  &  11 & 1.71 & 1.73
  &  11 & 1.70 & 1.71 \\
  CMS $W$ rapidity 8 TeV
  &  22 & 0.93 & 0.74
  &  22 & 1.33 & 1.03
  &  22 & 1.11 & 1.08 \\
  LHCb $W \to \mu$ 7 TeV
  &  14 & 1.63 & 1.26 
  &  14 & 2.78 & 1.99
  &  14 & 2.12 & 2.03 \\
  LHCb $W \to \mu$ 8 TeV
  &  14 & 0.60 & 0.44
  &  14 & 0.97 & 0.92
  &  14 & 0.80 & 0.84 \\
\bottomrule
\end{tabularx}

  \vspace{0.3cm}
  \caption{Same as Table~\ref{tab:chi2_TOTAL} for DY NC (top) and DY CC
  (bottom) datasets.}
  \label{tab:chi2_DY}
\end{table}

\begin{table}[!t]
  \scriptsize
  \centering
  \renewcommand{\arraystretch}{1.4}
  \begin{tabularx}{\textwidth}{Xrccrccrcc}
  \toprule
  & \multicolumn{3}{c}{NLO}
  & \multicolumn{3}{c}{NNLO}
  & \multicolumn{3}{c}{aN$^3$LO} \\
  Dataset
  & $N_{\rm dat}$
  & no MHOU
  & MHOU
  & $N_{\rm dat}$
  & no MHOU
  & MHOU 
  & $N_{\rm dat}$
  & no MHOU
  & MHOU \\
  \midrule
  ATLAS $\sigma_{tt}^{\rm tot}$ 7 TeV
  & 1 & 10.4 & 0.96
  & 1 & 4.50 & 2.40 
  & 1 & 2.78 & 2.05 \\
  ATLAS $\sigma_{tt}^{\rm tot}$ 8 TeV
  & 1 & 1.74 & 0.59
  & 1 & 0.02 & 0.03
  & 1 & 0.04 & 0.08 \\
  ATLAS $\sigma_{tt}^{\rm tot}$ 13 TeV ($\mathcal{L}$=139~fb$^{-1}$)
  & 1 & 3.82 & 0.96
  & 1 & 0.49 & 0.41
  & 1 & 0.51 & 0.44 \\
  ATLAS $t\bar{t}~\ell$+jets 8 TeV ($1/\sigma d\sigma/dy_t$)
  & 4 & 4.16 & 1.79
  & 4 & 3.13 & 3.70
  & 4 & 2.98 & 3.64 \\
  ATLAS $t\bar{t}~\ell$+jets 8 TeV ($1/\sigma d\sigma/dy_{t\bar t}$)
  & 4 & 8.93 & 3.73
  & 4 & 4.50 & 5.80
  & 4 & 4.26 & 4.92 \\
  ATLAS $t\bar{t}~2\ell$ 8 TeV ($1/\sigma d\sigma/dy_{t\bar t}$)
  & 4 & 1.94 & 1.76
  & 4 & 1.60 & 1.86
  & 4 & 1.66 & 1.80 \\
  CMS $\sigma_{tt}^{\rm tot}$ 5 TeV
  & 1 & 0.61 & 0.73
  & 1 & 0.02 & 0.01
  & 1 & 0.03 & 0.02 \\
  CMS $\sigma_{tt}^{\rm tot}$ 7 TeV
  & 1 & 5.27 & 1.30
  & 1 & 1.01 & 0.50
  & 1 & 0.60 & 0.34 \\
  CMS $\sigma_{tt}^{\rm tot}$ 8 TeV
  & 1 & 3.50 & 0.85
  & 1 & 0.26 & 0.17
  & 1 & 0.21 & 0.10 \\
  CMS $\sigma_{tt}^{\rm tot}$ 13 TeV
  & 1 & 0.75 & 0.26
  & 1 & 0.06 & 0.01
  & 1 & 0.04 & 0.05 \\
  CMS $t\bar{t}~\ell$+jets 8 TeV ($1/\sigma d\sigma/dy_{t\bar t}$)
  & 9 & 1.87 & 1.59
  & 9 & 1.21 & 1.59
  & 9 & 1.31 & 1.52 \\
  CMS $t\bar{t}$ 2D $2\ell$ 8 TeV ($1/\sigma d\sigma/dy_tdm_{t\bar t}$)
  & 15 & 2.03 & 1.89
  & 15 & 1.30 & 1.25
  & 15 & 1.28 & 1.37 \\
  CMS $t\bar{t}~2\ell$ 13 TeV ($d\sigma/dy_t$)
  & 10 & 0.78 & 0.69
  & 10 & 0.51 & 0.59
  & 10 & 0.55 & 0.60 \\
  CMS $t\bar{t}~\ell$+jet 13 TeV ($d\sigma/dy_t$)
  & 11 & 0.66 & 0.25
  & 11 & 0.60 & 0.66
  & 11 & 0.52 & 0.71 \\ 
  \midrule
  ATLAS incl. jets 8~TeV, $R=0.6$
  & 171 & 0.67 & 0.74
  & 171 & 0.68 & 0.64
  & 171 & 0.68 & 0.64 \\
  CMS incl. jets 8 TeV
  & 185 & 0.95 & 0.83
  & 185 & 1.19 & 0.95
  & 185 & 0.97 & 0.99 \\
  \midrule
  ATLAS dijets 7 TeV, $R=0.6$
  & 90 & 1.47 & 1.72
  & 90 & 2.14 & 1.69
  & 90 & 1.76 & 1.63 \\
  CMS dijets 7 TeV
  & 54 & 1.57 & 2.01
  & 54 & 1.79 & 1.74
  & 54 & 1.84 & 1.78 \\
  \midrule
  ATLAS isolated $\gamma$ prod. 13 TeV
  & 53 & 0.57 & 0.47
  & 53 & 0.76 & 0.67
  & 53 & 0.72 & 0.68 \\
  \midrule
  ATLAS single~$t$ $R_{t}$ 7 TeV
  & 1 & 0.43 & 0.29
  & 1 & 0.50 & 0.57
  & 1 & 0.51 & 0.58 \\
  ATLAS single~$t$ $R_{t}$ 13 TeV
  & 1 & 0.04 & 0.03
  & 1 & 0.06 & 0.07
  & 1 & 0.06 & 0.07 \\
  ATLAS single~$t$ 7 TeV ($1/\sigma d\sigma/dy_t$)
  & 3 & 0.83 & 0.84
  & 3 & 0.96 & 0.94
  & 3 & 0.97 & 0.97 \\
  ATLAS single~$t$ 7 TeV ($1/\sigma d\sigma/dy_{\bar t}$)
  & 3 & 0.06 & 0.06
  & 3 & 0.06 & 0.06
  & 3 & 0.06 & 0.06 \\
  ATLAS single~$t$ 8 TeV ($1/\sigma d\sigma/dy_t$)
  & 3 & 0.38 & 0.31
  & 3 & 0.25 & 0.26
  & 3 & 0.22 & 0.24 \\
  ATLAS single~$t$ 8 TeV ($1/\sigma d\sigma/dy_{\bar t}$)
  & 3 & 0.19 & 0.21
  & 3 & 0.19 & 0.19
  & 3 & 0.20 & 0.20 \\
  CMS single~$t$ $\sigma_{t}+\sigma_{\bar{t}}$ 7 TeV
  & 1 & 0.89 & 0.88
  & 1 & 0.74 & 0.84
  & 1 & 0.39 & 0.43 \\
  CMS single~$t$ $R_{t}$ 8 TeV
  & 1 & 0.15 & 0.08
  & 1 & 0.18 & 0.20
  & 1 & 0.18 & 0.21 \\
  CMS single~$t$ $R_{t}$ 13 TeV
  & 1 & 0.33 & 0.27
  & 1 & 0.36 & 0.38
  & 1 & 0.36 & 0.38 \\
\bottomrule
\end{tabularx}

  \vspace{0.3cm}
  \caption{Same as Table~\ref{tab:chi2_TOTAL} for (from top to bottom) top pair,
    single-inclusive jet, isolated photon and single top production datasets.}
  \label{tab:chi2_OTHER}
\end{table}

The NLO and NNLO results without MHOUs are obtained  using the NLO and NNLO
NNPDF4.0 PDF sets~\cite{NNPDF:2021njg}. The NNLO result with MHOUs is
obtained using the NNPDF4.0MHOU NNLO set from Ref.~\cite{NNPDF:2024dpb},
while, as already mentioned, the  NNPDF4.0MHOU NLO presented here for
the first time uses an identical methodology to NNPDF4.0MHOU
NNLO~\cite{NNPDF:2024dpb}, but the same dataset as NNPDF4.0
NLO~\cite{NNPDF:2021njg}. Hence, the datasets with and without MHOU are always
the same, but the NLO and NNLO datasets are not the same but rather follow
Ref.~\cite{NNPDF:2021njg}. The N$^3$LO dataset is the same as NNLO. 
In all cases, the theoretical predictions entering the computation of the
$\chi^2$ are obtained with the new theory pipeline. The covariance matrix,
whenever needed, is computed as described in Sect.~4.1 of
Ref.~\cite{NNPDF:2024dpb}. The N$^3$LO predictions are computed with the
aforementioned aN$^3$LO PDF sets. These are based on the same datasets and
kinematic cuts as the NNPDF4.0 NNLO PDF sets, use the theoretical predictions
discussed in Sects.~\ref{sec:dglap}-\ref{sec:n3lo_coefffun}, and are
supplemented with a IHOU covariance matrix as discussed in
Sects.~\ref{sec:ihou}-\ref{sec:n3lo_dis} and a MHOU for hadronic processes for
which N$^3$LO hard cross-sections are not available as discussed in
Sect.~\ref{sec:n3lo_hadronic_coeff}.

Table~\ref{tab:chi2_TOTAL} and Fig.~\ref{fig:chi2_n3lo_summary}
show that without MHOUs fit quality improves as the perturbative
order increases. Note that this also happens when going from NNLO to
N$^3$LO, despite the fact that N$^3$LO corrections are only partially included,
with hadronic matrix elements still computed at NNLO. This shows that
the impact of N$^3$LO corrections to  evolution and DIS coefficient
functions is significant enough to affect fit quality in a way that is
qualitatively compatible with what one would expect when adding an
extra perturbative order to the improvement already seen when going
from NLO to NNLO.

On the other hand, when MHOUs are included, fit quality
becomes independent of perturbative order within uncertainties (note
that, with $N_{\rm dat}=4462$, $\sigma_{\chi^2}=0.03$). This suggests
that the MHOU covariance matrix estimated through scale variation is
correctly reproducing the observed shift between perturbative orders,
i.e. the true MHOU. Note that if true this also means that at
aN$^3$LO the missing N$^3$LO corrections to  hadronic processes are
correctly accounted for by the corresponding MHOU which is always
included. Also, at aN$^3$LO
the fit quality is the same within uncertainties irrespective of whether MHOUs 
are included or not. This strongly suggests that inclusion of higher
order terms in perturbative evolution and DIS coefficient function
would not lead to further improvements, i.e. that in this respect, with
experimental uncertainties,  current methodology and   current dataset
the perturbative expansion has converged.

\subsection{Parton distributions}
\label{sec:PDFs}

\begin{figure}[!p]
  \centering
  \includegraphics[width=0.45\textwidth]{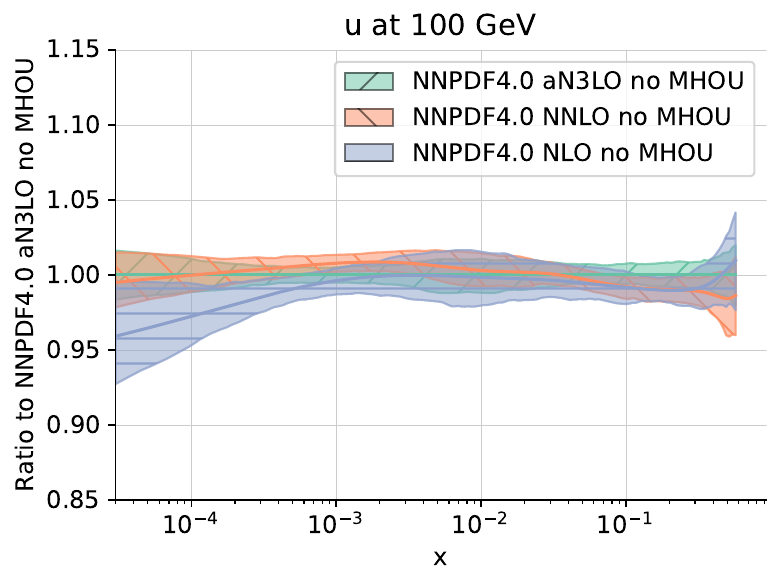}
  \includegraphics[width=0.45\textwidth]{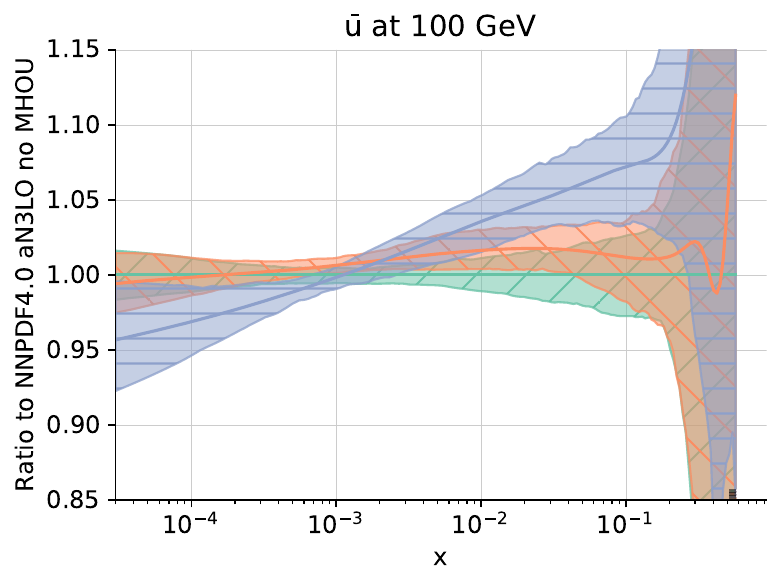}
  \includegraphics[width=0.45\textwidth]{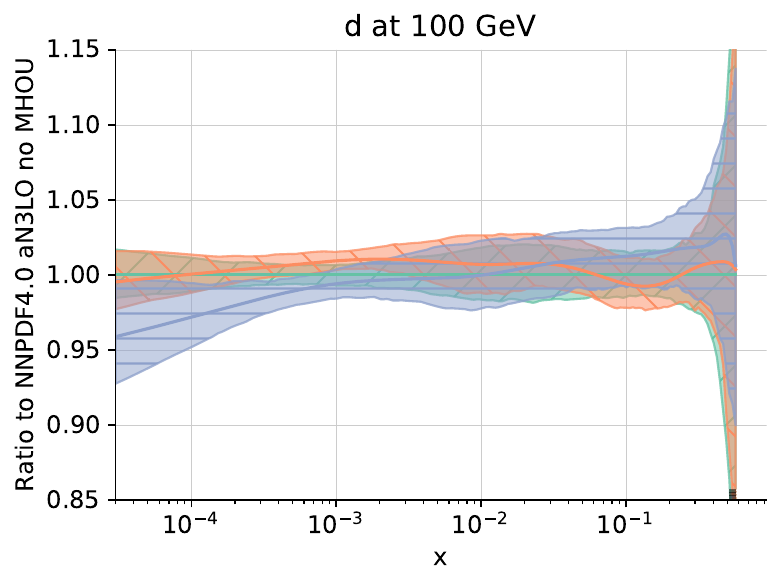}
  \includegraphics[width=0.45\textwidth]{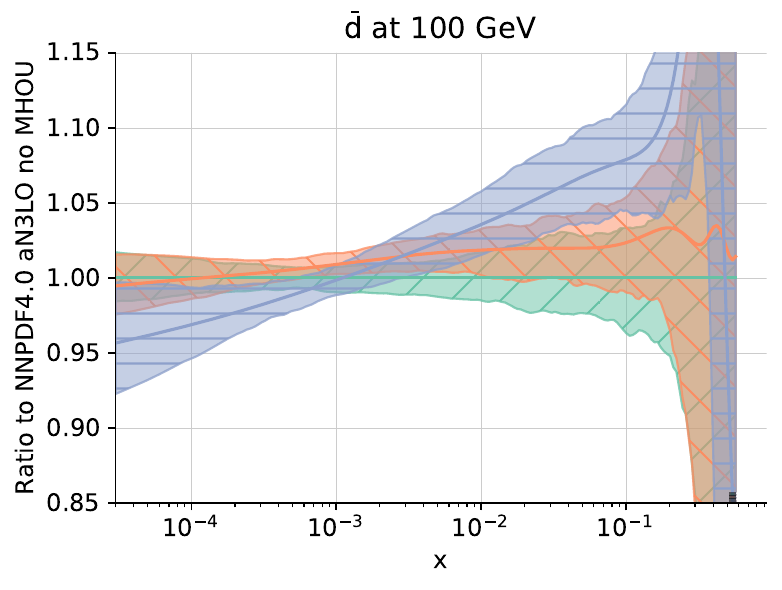}
  \includegraphics[width=0.45\textwidth]{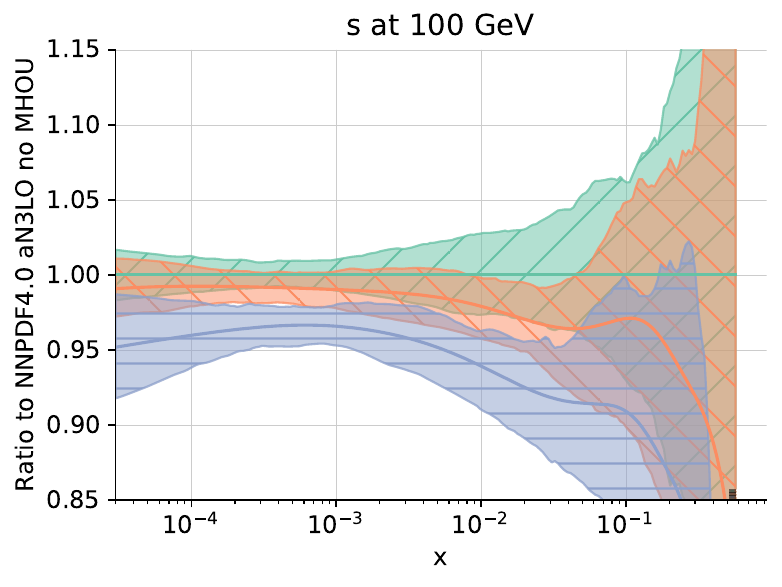}
  \includegraphics[width=0.45\textwidth]{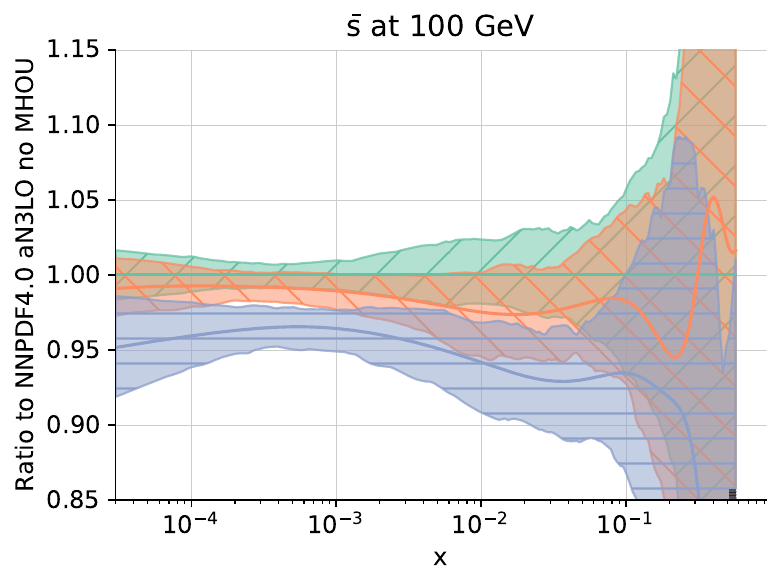}
  \includegraphics[width=0.45\textwidth]{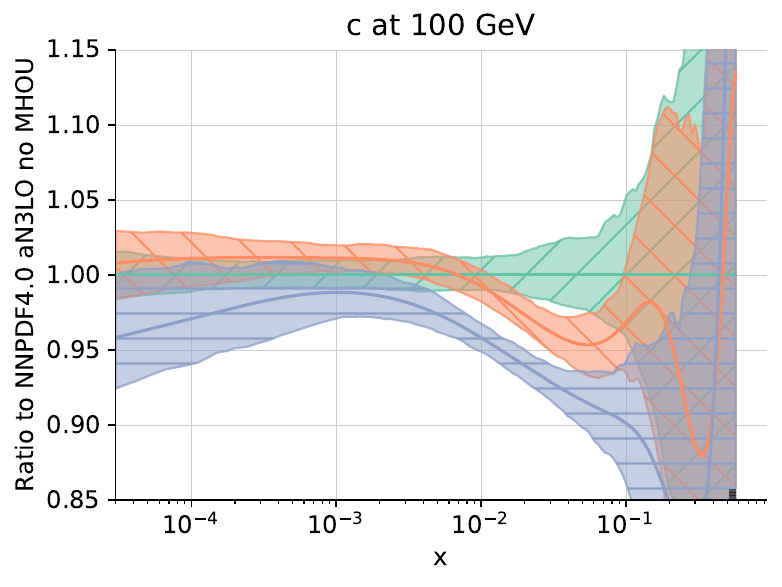}
  \includegraphics[width=0.45\textwidth]{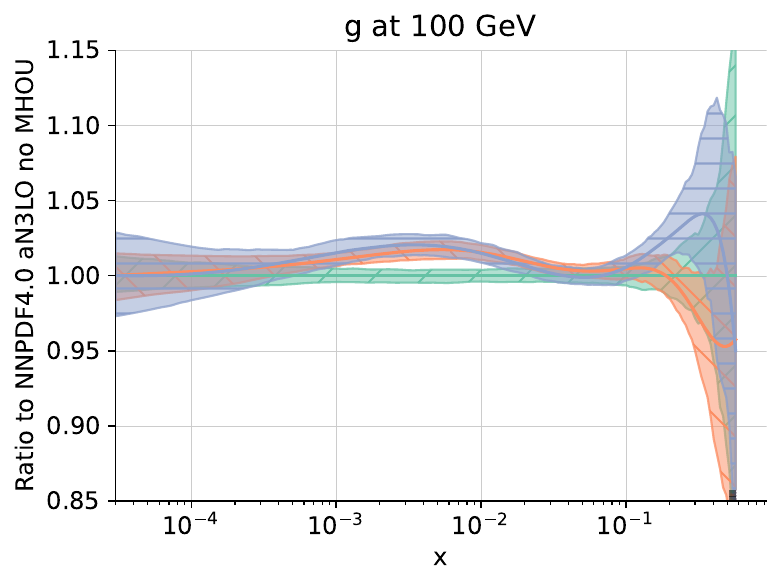}
  \caption{The NLO, NNLO and aN$^3$LO NNPDF4.0 PDFs at $Q=100$~GeV. We display
    the up, antiup, down, antidown, strange, antistrange, charm and gluon PDFs
    normalized to the aN$^3$LO result. Error bands correspond to one sigma
    PDF uncertainties, not including MHOUs on the theory predictions
    used in the fit.}
  \label{fig:pdfs_noMHOU_log}
\end{figure}

\begin{figure}[!p]
  \centering
  \includegraphics[width=0.45\textwidth]{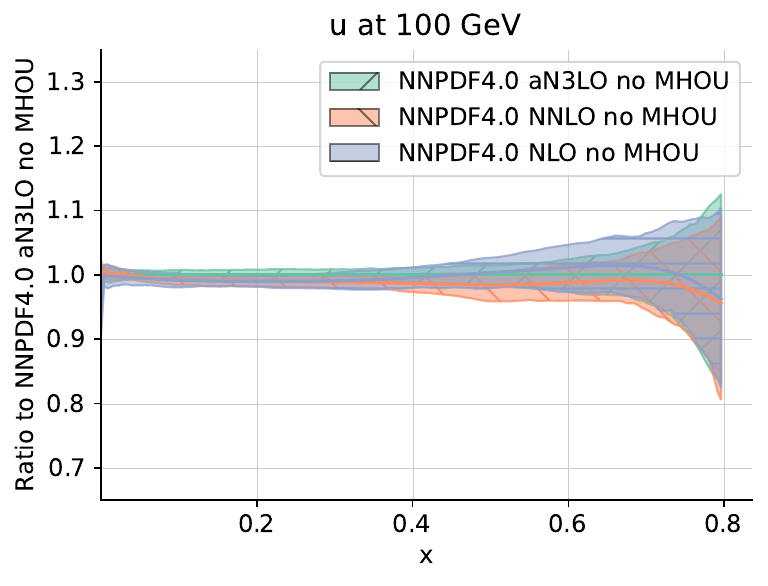}
  \includegraphics[width=0.45\textwidth]{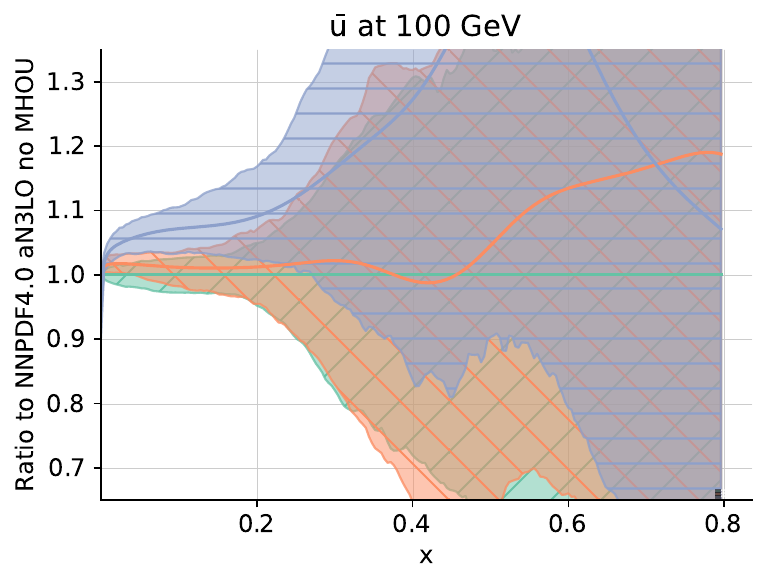}
  \includegraphics[width=0.45\textwidth]{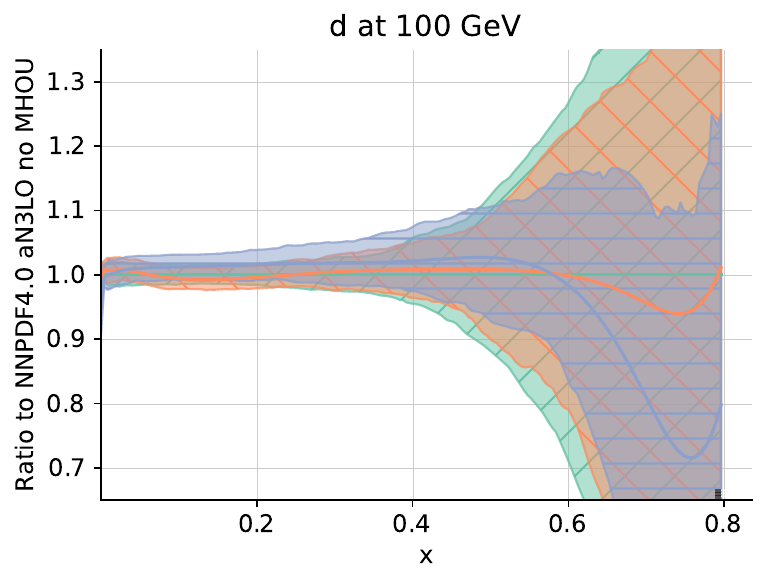}
  \includegraphics[width=0.45\textwidth]{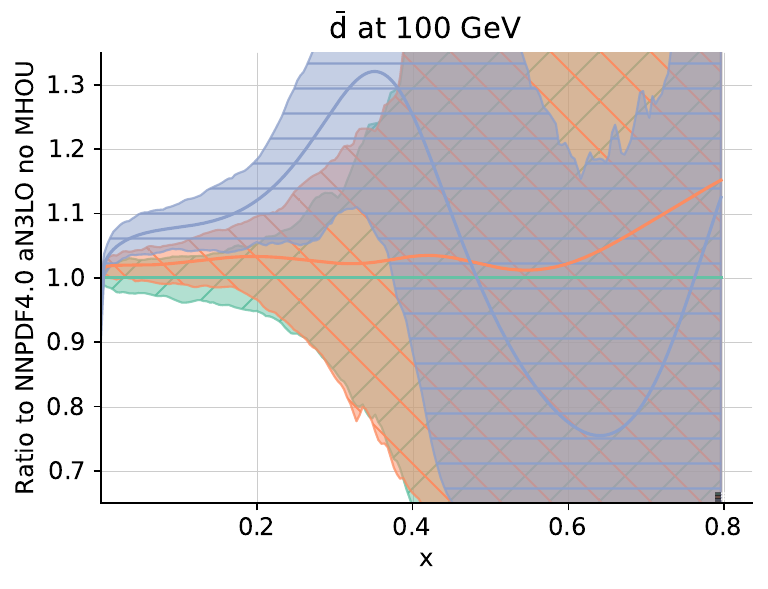}
  \includegraphics[width=0.45\textwidth]{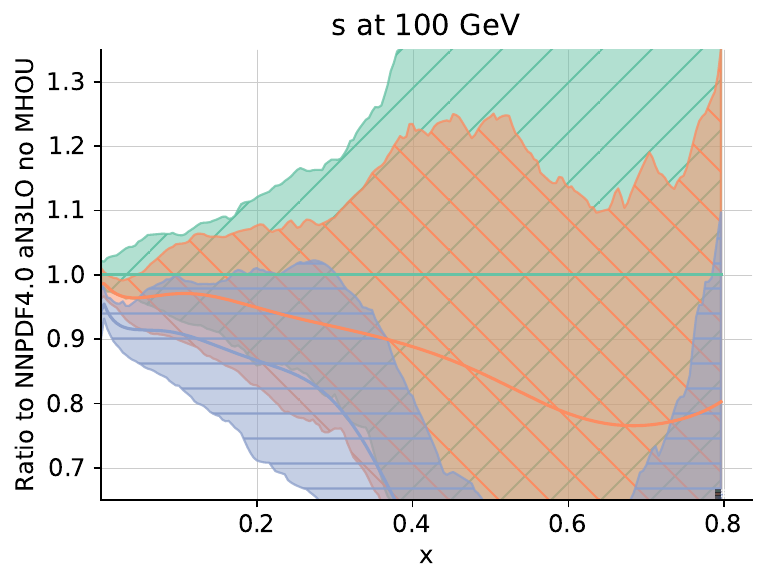}
  \includegraphics[width=0.45\textwidth]{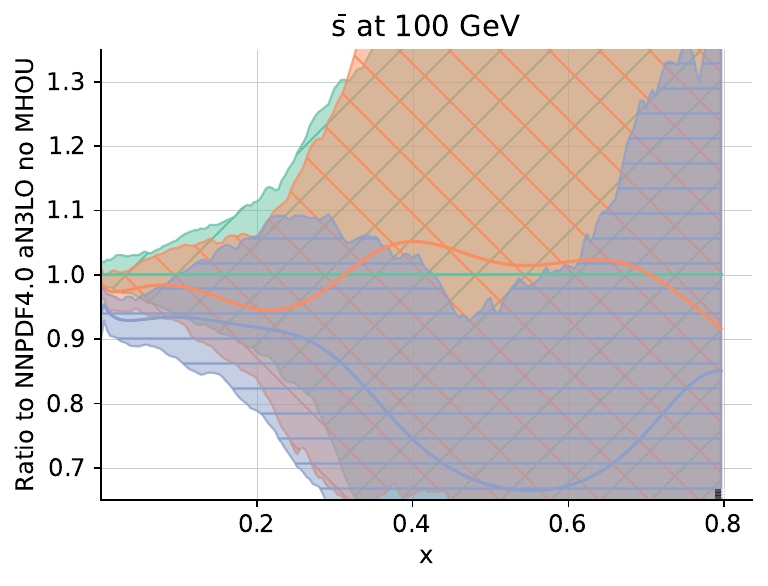}
  \includegraphics[width=0.45\textwidth]{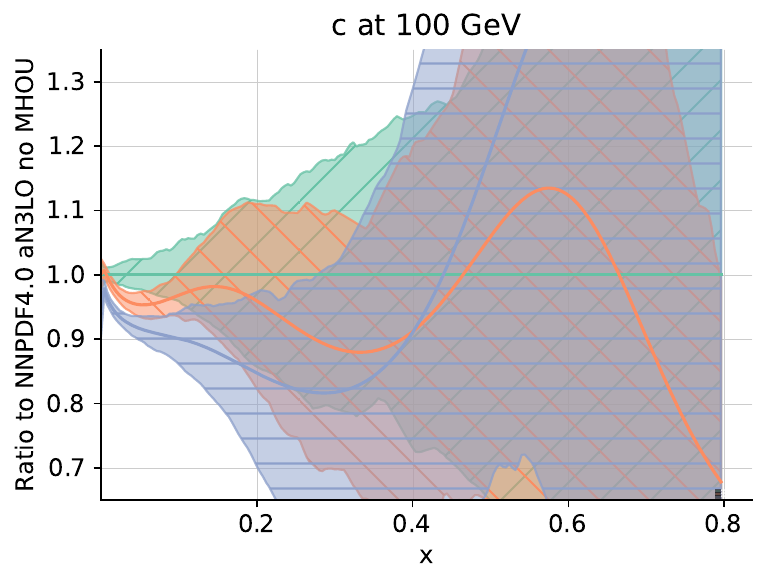}
  \includegraphics[width=0.45\textwidth]{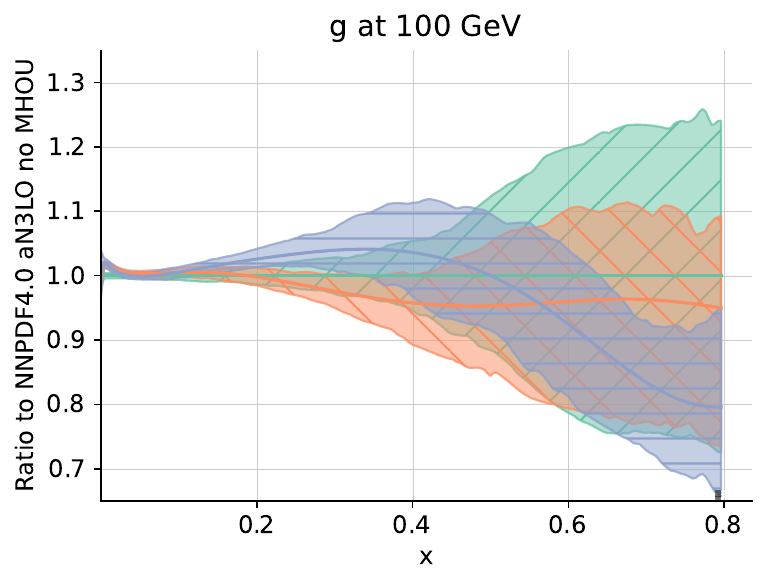}
  \caption{Same as Fig.~\ref{fig:pdfs_noMHOU_log} in linear scale.}
  \label{fig:pdfs_noMHOU_lin}
\end{figure}

\begin{figure}[!p]
  \centering
  \includegraphics[width=0.45\textwidth]{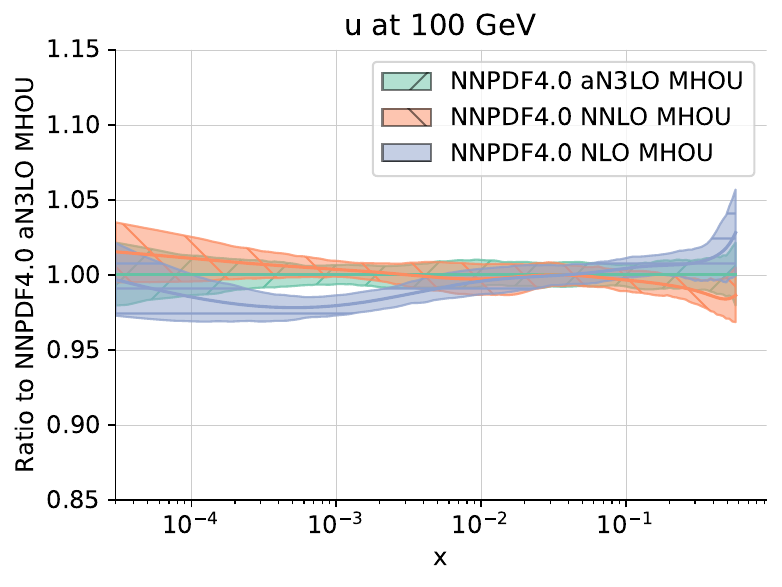}
  \includegraphics[width=0.45\textwidth]{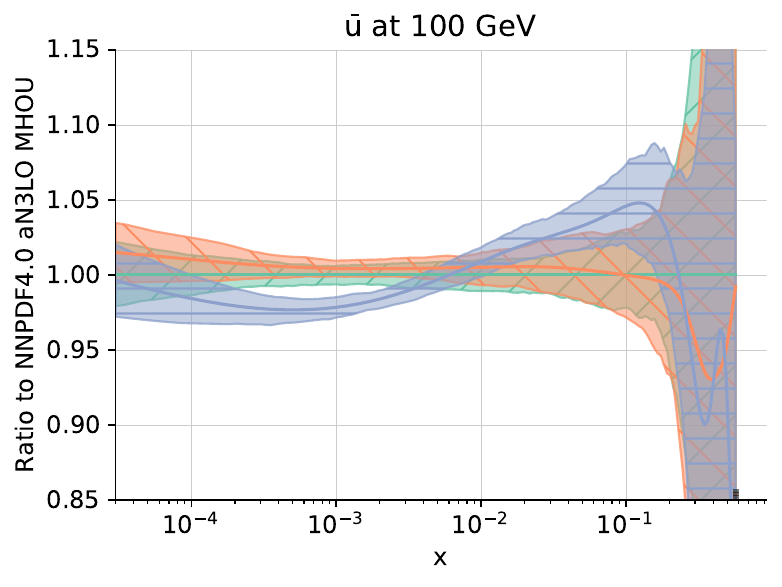}
  \includegraphics[width=0.45\textwidth]{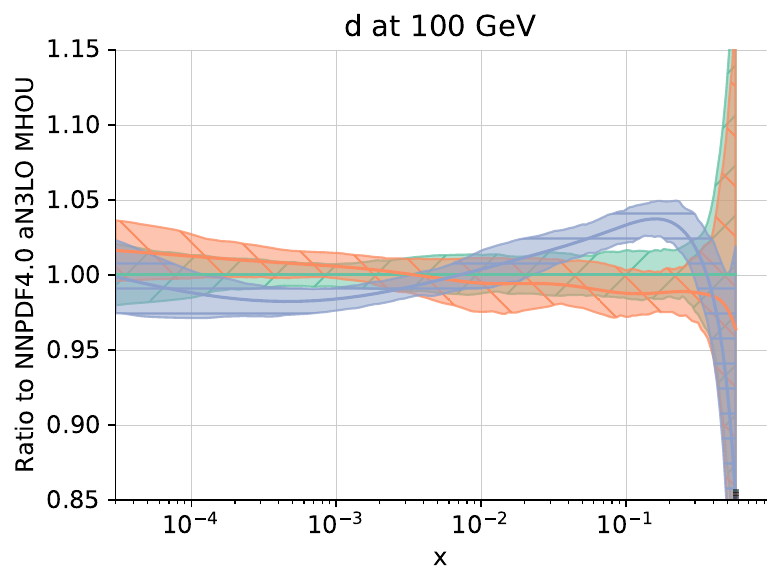}
  \includegraphics[width=0.45\textwidth]{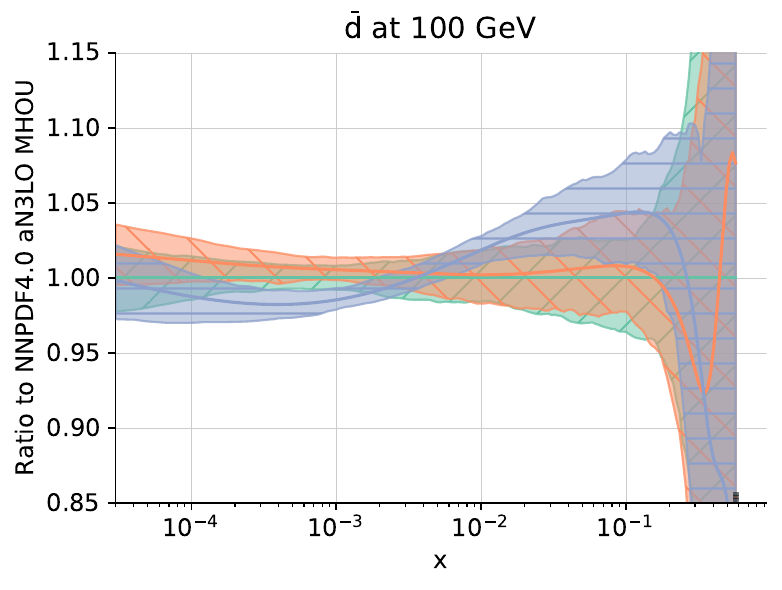}
  \includegraphics[width=0.45\textwidth]{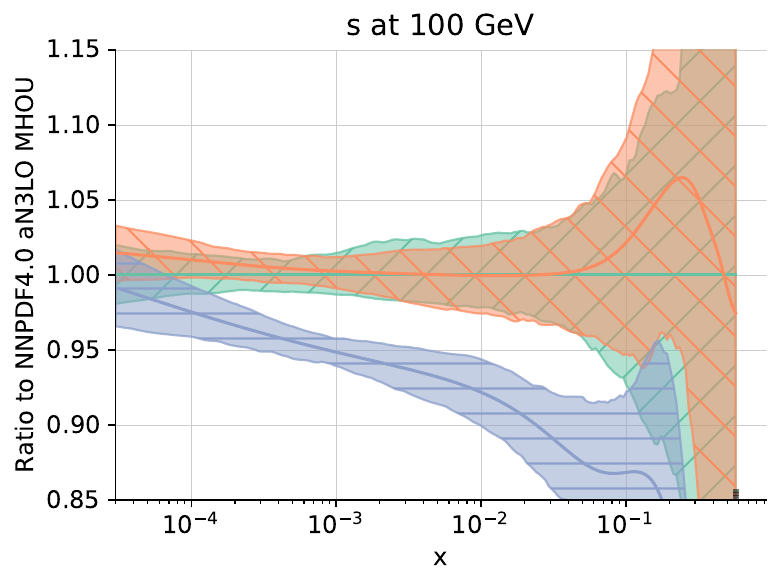}
  \includegraphics[width=0.45\textwidth]{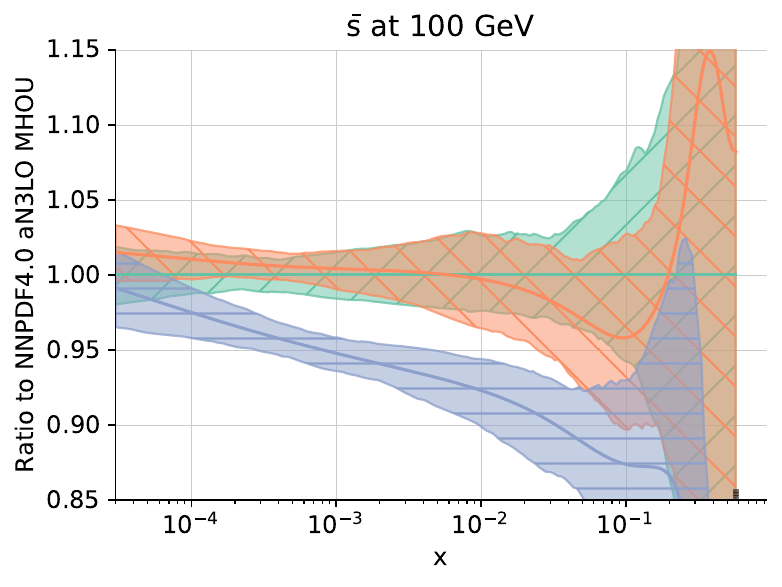}
  \includegraphics[width=0.45\textwidth]{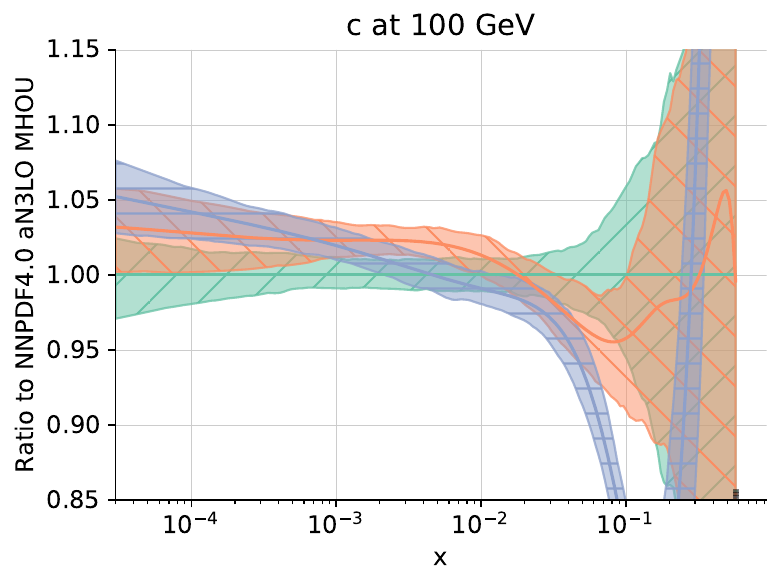}
  \includegraphics[width=0.45\textwidth]{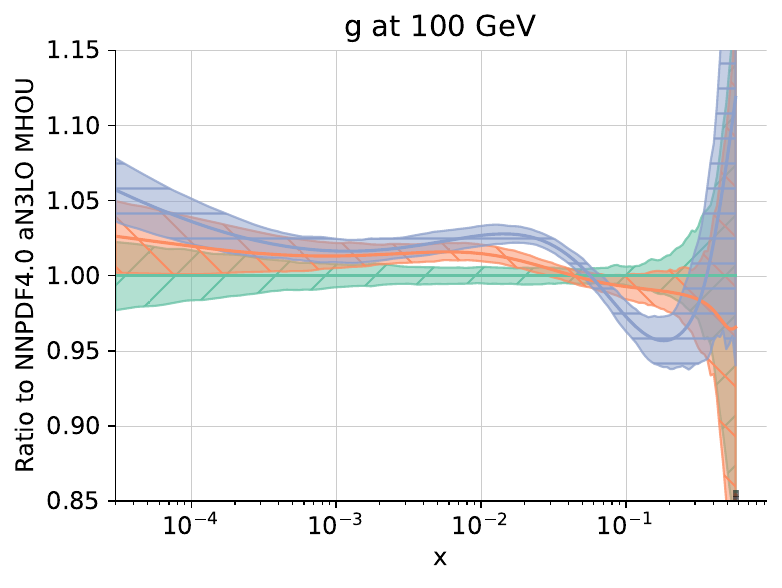}
  \caption{Same as Fig.~\ref{fig:pdfs_noMHOU_log} for NNPDF4.0MHOU PDF
    sets. Error bands correspond to one sigma
    PDF uncertainties also including MHOUs on the theory predictions
    used in the fit.}
  \label{fig:pdfs_MHOU_log}
\end{figure}

\begin{figure}[!p]
  \centering
  \includegraphics[width=0.45\textwidth]{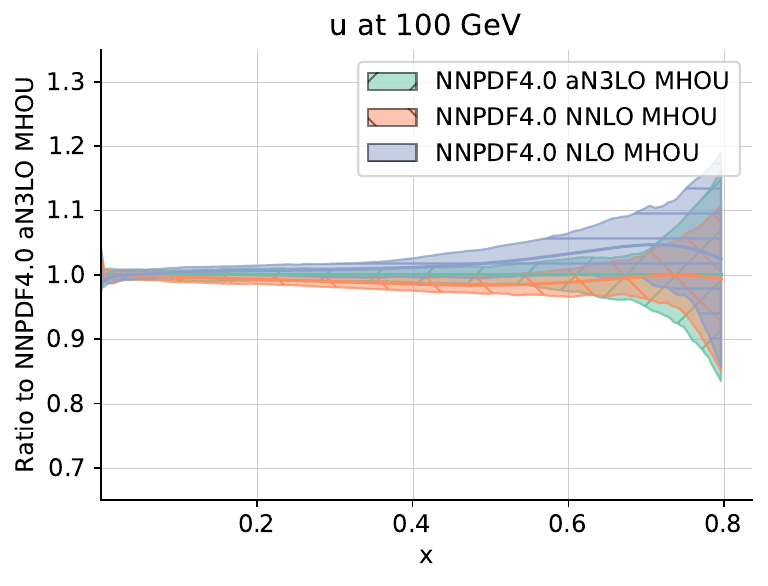}
  \includegraphics[width=0.45\textwidth]{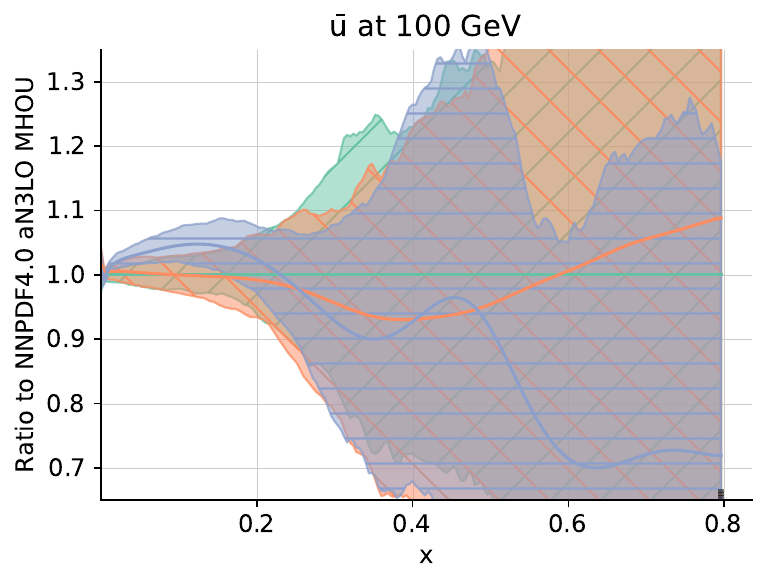}
  \includegraphics[width=0.45\textwidth]{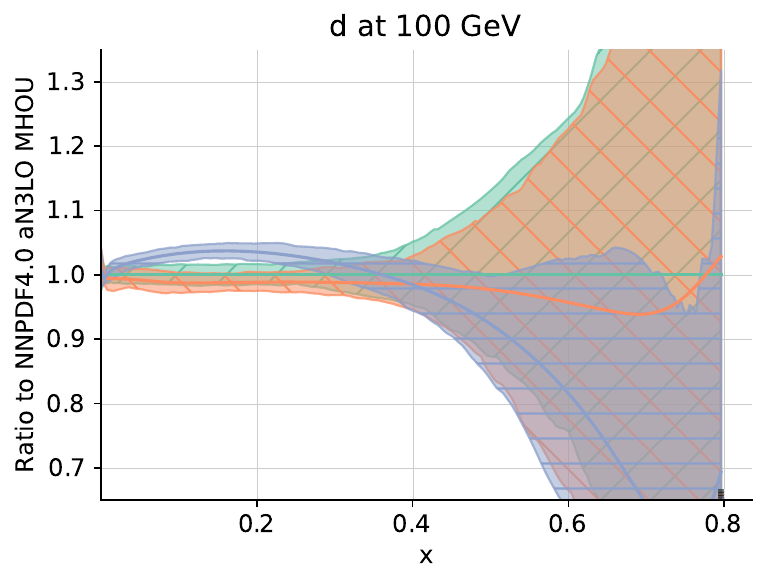}
  \includegraphics[width=0.45\textwidth]{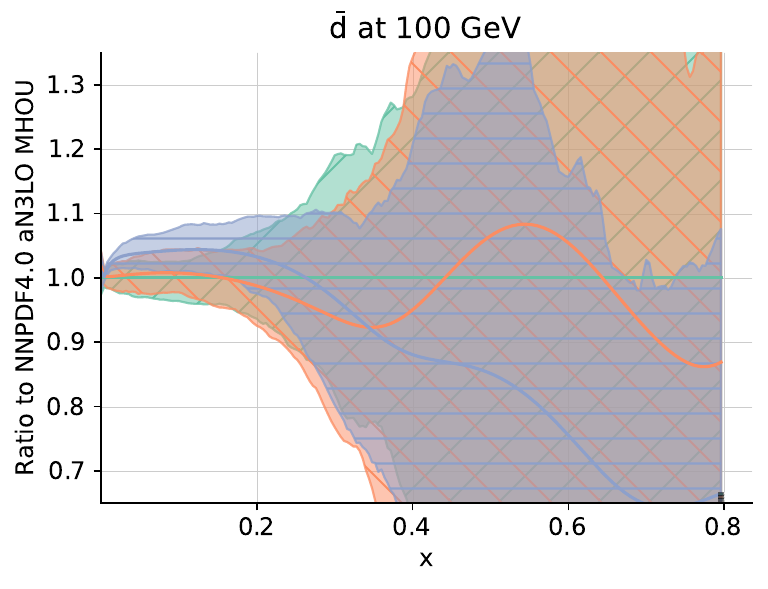}
  \includegraphics[width=0.45\textwidth]{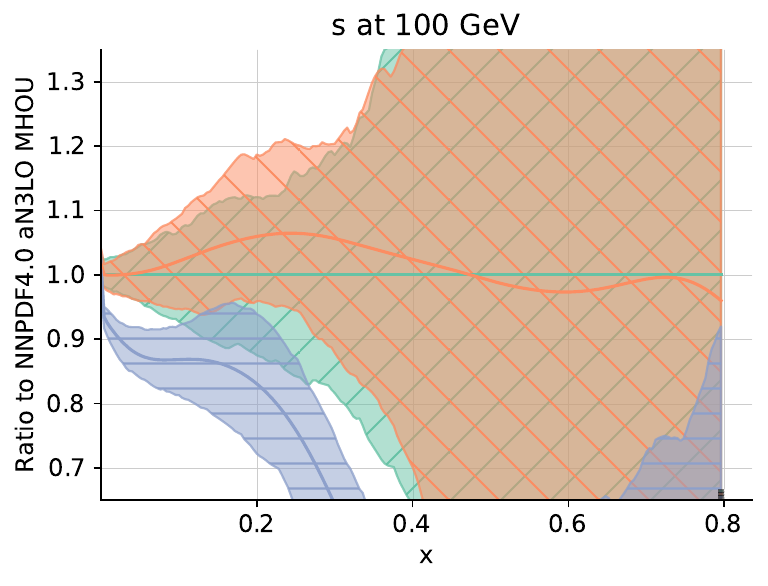}
  \includegraphics[width=0.45\textwidth]{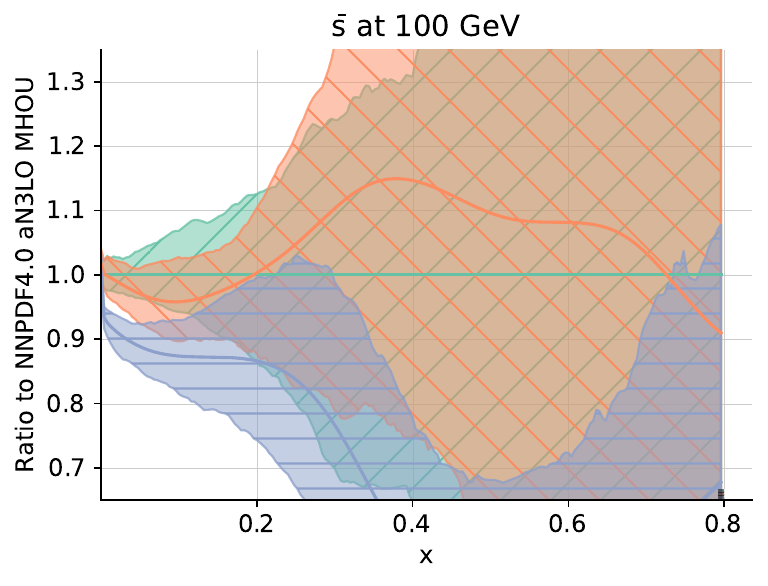}
  \includegraphics[width=0.45\textwidth]{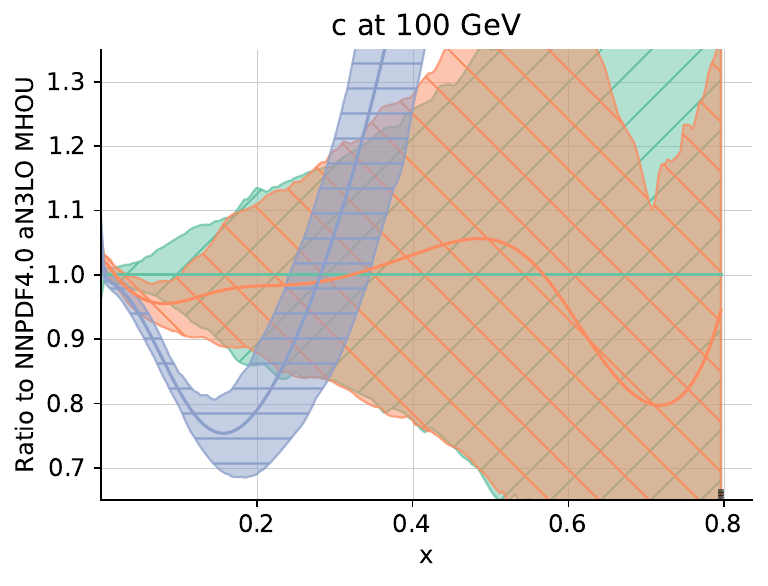}
  \includegraphics[width=0.45\textwidth]{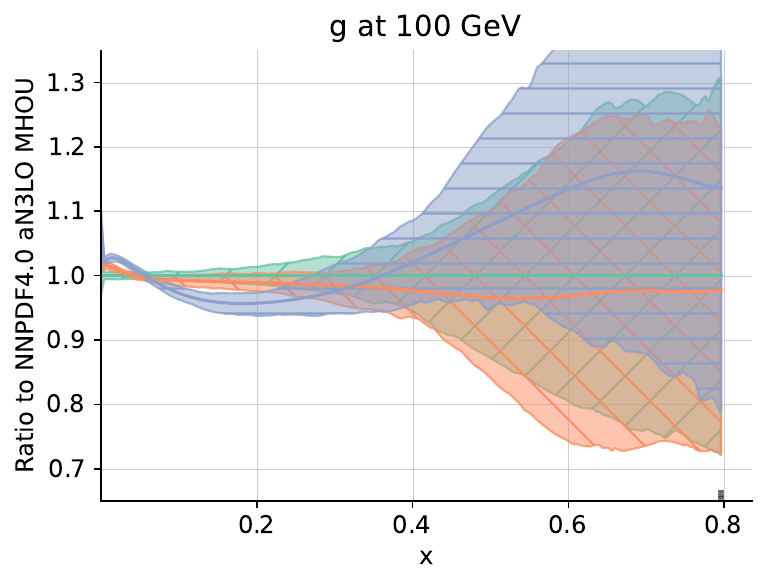}
  \caption{Same as Fig.~\ref{fig:pdfs_MHOU_log} in linear scale.}
  \label{fig:pdfs_MHOU_lin}
\end{figure}

We now examine the NNPDF4.0 aN$^3$LO parton distributions. We compare the NLO,
NNLO and aN$^3$LO NNPDF4.0 PDFs, obtained without and with inclusion of MHOUs,
in Figs.~\ref{fig:pdfs_noMHOU_log}-\ref{fig:pdfs_noMHOU_lin}
and in Figs.~\ref{fig:pdfs_MHOU_log}-\ref{fig:pdfs_MHOU_lin},
respectively. Specifically, we show the up, antiup, down, antidown, strange,
antistrange, charm and gluon PDFs at $Q=100$~GeV, normalized to the aN$^3$LO
result, as a function of $x$ in logarithmic and linear scale. Error bands
correspond to one sigma PDF uncertainties, which do (MHOU sets) or do
not (no MHOU sets) include
MHOUs on all theory predictions used in  the fit. The PDF sets, with
and without 
MHOUs, are the same used to compute the values of the $\chi^2$
in Tables~\ref{tab:chi2_TOTAL}-\ref{tab:chi2_OTHER}. 

The excellent perturbative convergence seen in the fit quality is also manifest
at the level of PDFs. In particular, the NNLO PDFs
are either very close to or indistinguishable from their aN$^3$LO counterparts.
Inclusion of MHOUs further improves the consistency between NNLO and aN$^3$LO
PDFs, which lie almost on top of each other. This means that the NNLO
PDFs are made more accurate by the inclusion of MHOUs, and that the
aN$^3$LO PDFs have converged, in the sense discussed above.
Exceptions to this stability  are the
charm and gluon PDFs, for which aN$^3$LO corrections have a sizable impact.
In the case of charm, they lead to  an enhancement of the central value of
about 4\% for $x\sim 0.05$; in the case of gluon, to a suppression of
about 2-3\% for $x\sim 0.005$. In both cases, inclusion of MHOUs leads to an
increase in PDF uncertainties by about 1-2\%. This makes the NNLO and
aN$^3$LO charm PDFs with MHOUs compatible within uncertainties, and the
NNLO and aN$^3$LO gluon PDFs with MHOU almost compatible. 

\begin{figure}[!p]
  \centering
  \includegraphics[width=0.45\textwidth]{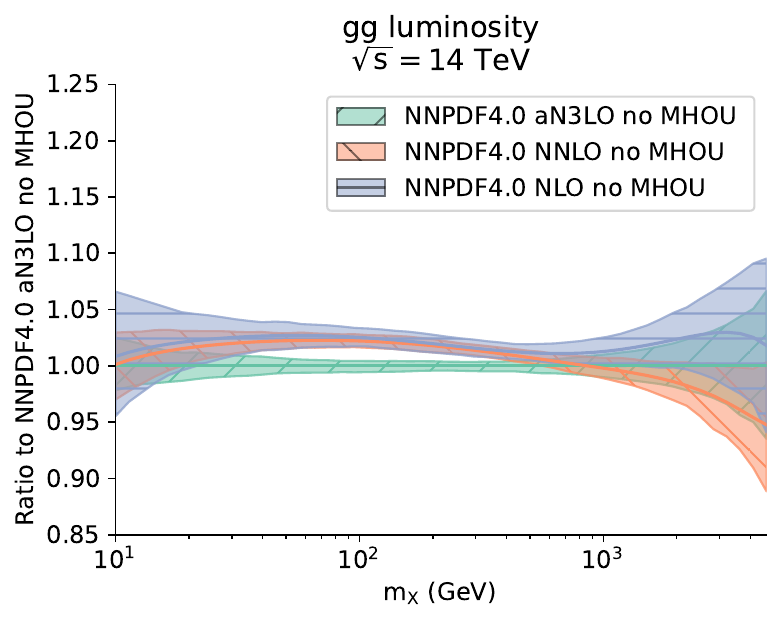}
  \includegraphics[width=0.45\textwidth]{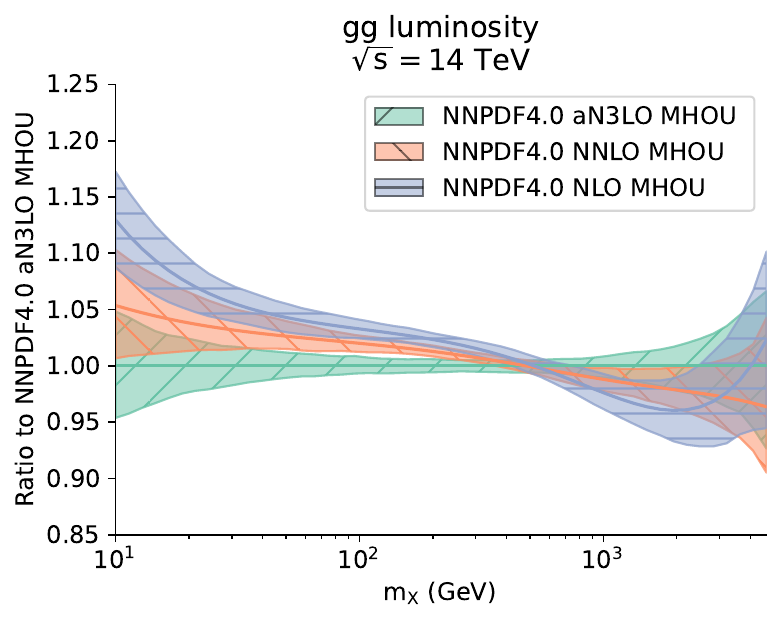}
  \includegraphics[width=0.45\textwidth]{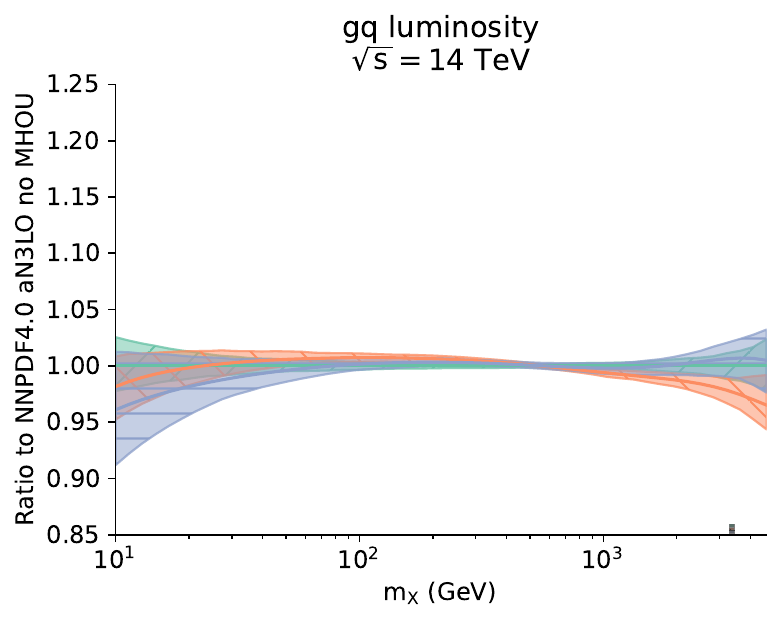}
  \includegraphics[width=0.45\textwidth]{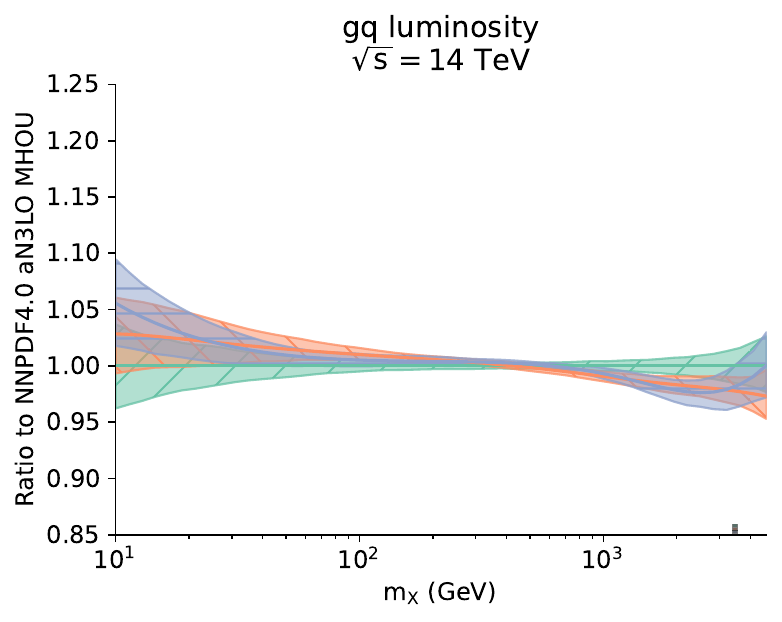}
  \includegraphics[width=0.45\textwidth]{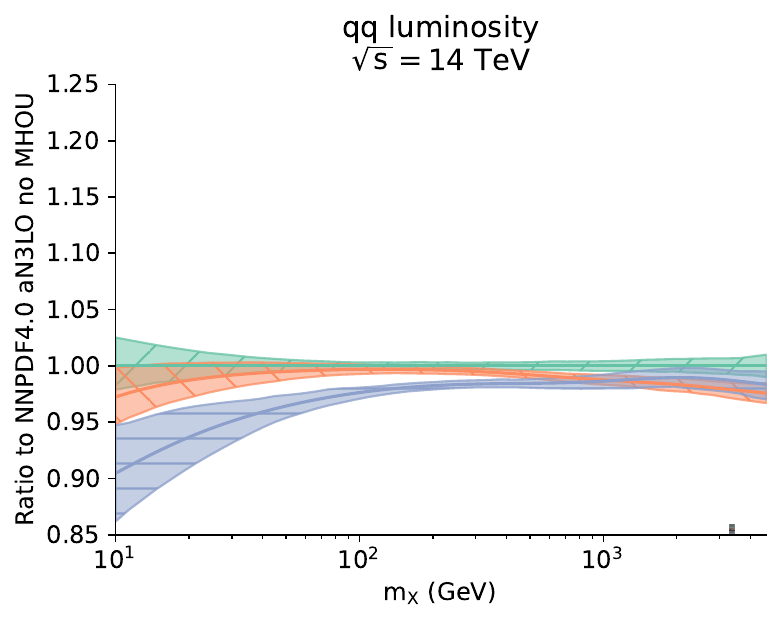}
  \includegraphics[width=0.45\textwidth]{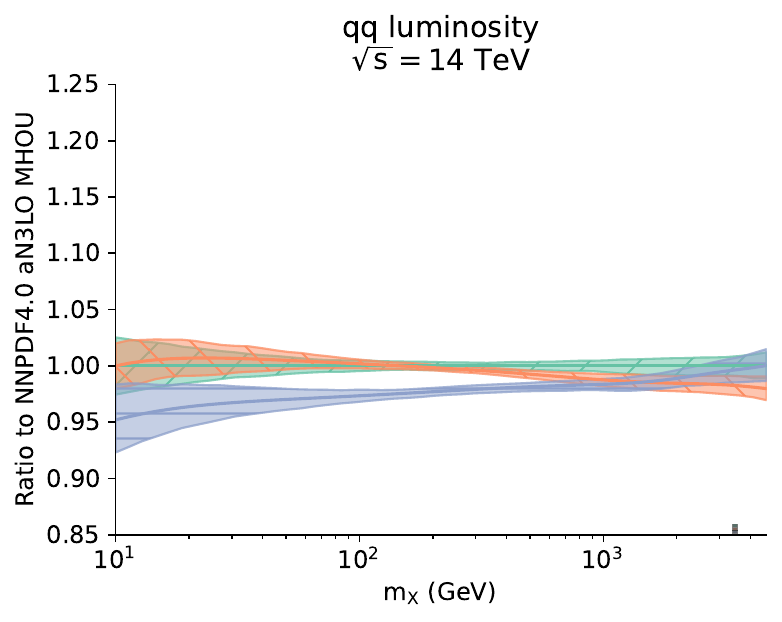}
  \includegraphics[width=0.45\textwidth]{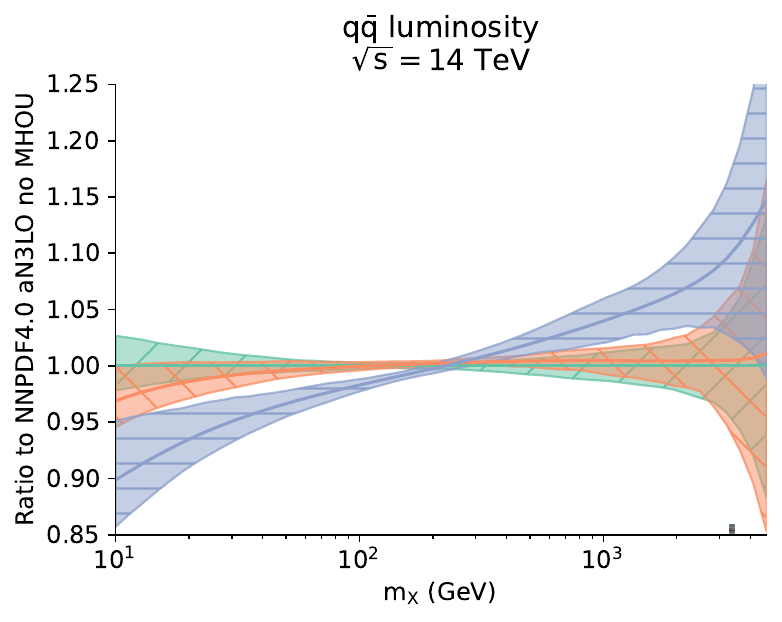}
  \includegraphics[width=0.45\textwidth]{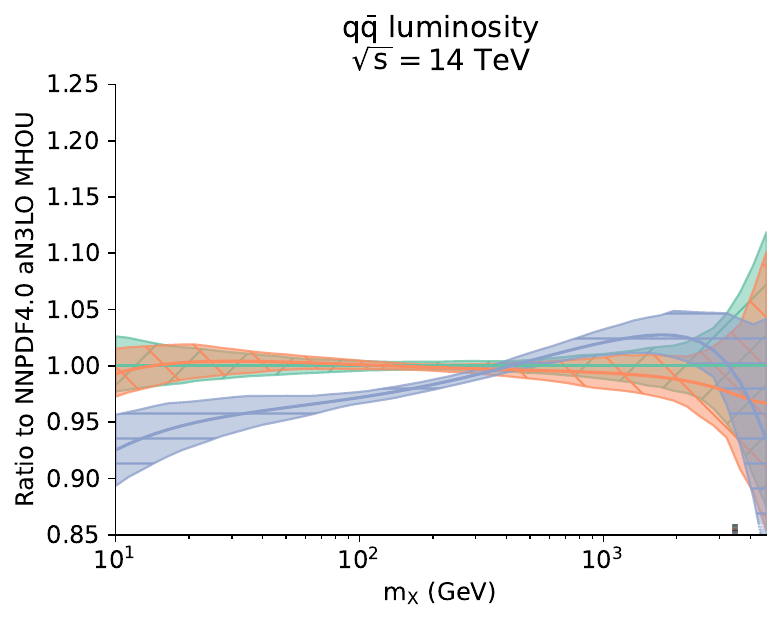}
  \caption{The gluon-gluon, gluon-quark, quark-quark, and quark-antiquark
    parton luminosities as a function of $m_X$ at $\sqrt{s}=14$~TeV,
    computed with NLO, NNLO and aN$^3$LO NNPDF4.0 PDFs without MHOUs (left) and
    with MHOUs (right), all shown as a ratio to the respective
    aN$^3$LO results. Uncertainties are as in Figs.~\ref{fig:pdfs_noMHOU_log}-\ref{fig:pdfs_MHOU_log}}
  \label{fig:lumis}
\end{figure}

Figure~\ref{fig:lumis} presents a comparison similar to that of
Figs.~\ref{fig:pdfs_noMHOU_log}-\ref{fig:pdfs_MHOU_log} for the gluon-gluon,
gluon-quark, quark-quark, and quark-antiquark parton luminosities.
These are shown integrated in rapidity as a function of the invariant mass of
the final state $m_X$ for a center-of-mass energy $\sqrt{s}=14$~TeV. Their
definition follows Eqs.~(1)-(4) of Ref.~\cite{Mangano:2016jyj}.

As already observed for PDFs, perturbative convergence is excellent, and
improves upon inclusion of MHOUs. The NNLO and aN$^3$LO results are
compatible within uncertainties for the gluon-quark, quark-quark, and
quark-antiquark luminosities. Some differences are seen for the
gluon-gluon luminosity, consistent with the differences seen in
the gluon PDF. Specifically, the  aN$^3$LO corrections lead to  a
suppression of the gluon-gluon luminosity of 2-3\% for
$m_X\sim 100$~GeV. This effect is somewhat compensated by an increase in
uncertainty of about 1\% upon inclusion of MHOUs. Indeed,
the NNLO and aN$^3$LO gluon-gluon luminosities for $m_X\sim 100$~GeV
differ by about  $2.5\sigma$ without MHOU, but become almost
compatible within uncertainties when MHOUs are included.

All in all, these results show that  aN$^3$LO corrections are generally small,
except for the gluon PDF, and that at aN$^3$LO the perturbative expansion 
has all but converged, with NNLO and aN$^3$LO PDFs very close to each
other, especially upon inclusion of MHOUs. They also show that MHOUs
generally improve the accuracy of PDFs, though at aN$^3$LO they
have a very small
impact. The phenomenological consequences of this state of
affairs will be further discussed in Sect.~\ref{sec:pheno}.

\subsection{PDF uncertainties}
\label{sec:uncertainties}

\begin{figure}[!p]
  \centering
  \includegraphics[width=0.45\textwidth]{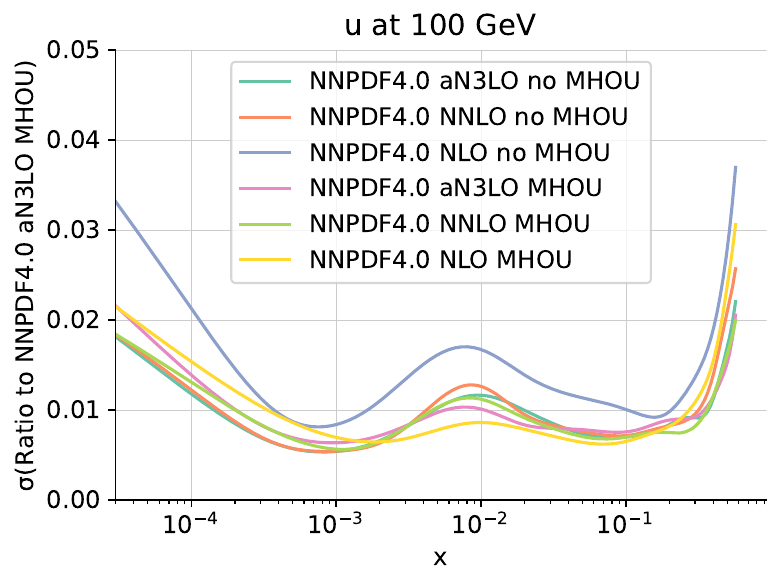}
  \includegraphics[width=0.45\textwidth]{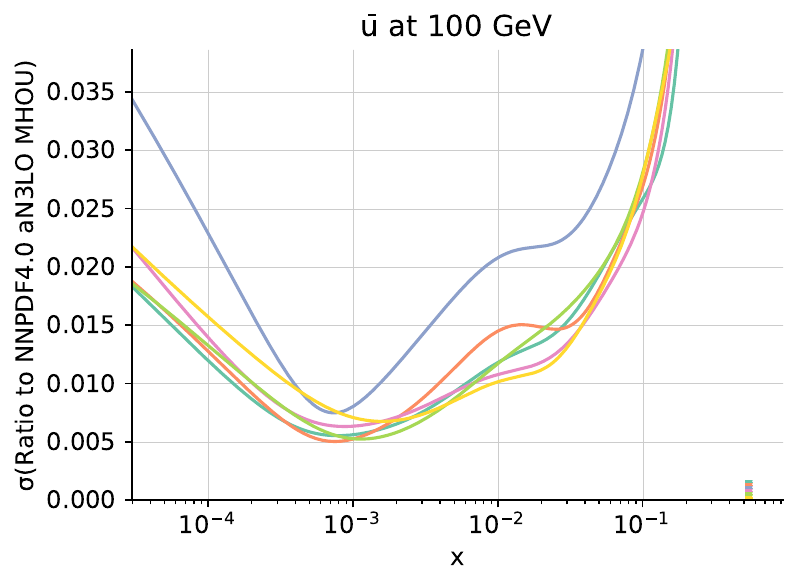}\\
  \includegraphics[width=0.45\textwidth]{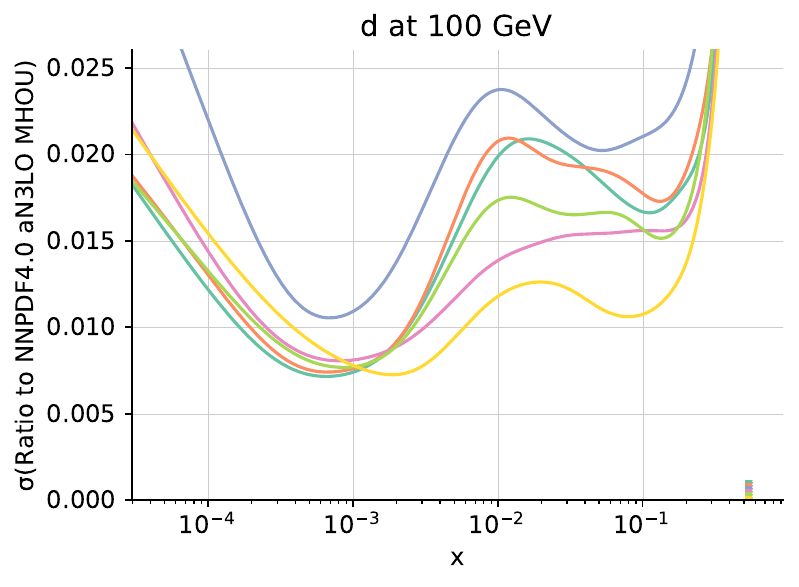}
  \includegraphics[width=0.45\textwidth]{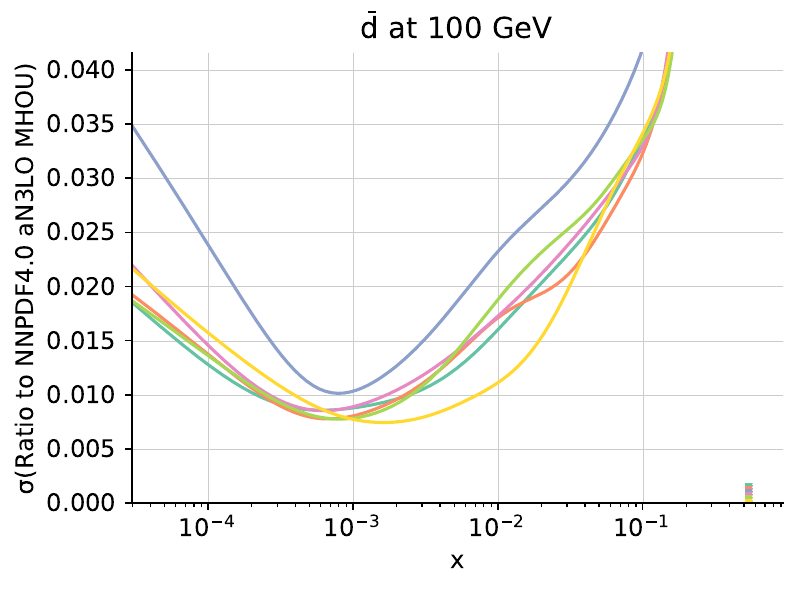}\\
  \includegraphics[width=0.45\textwidth]{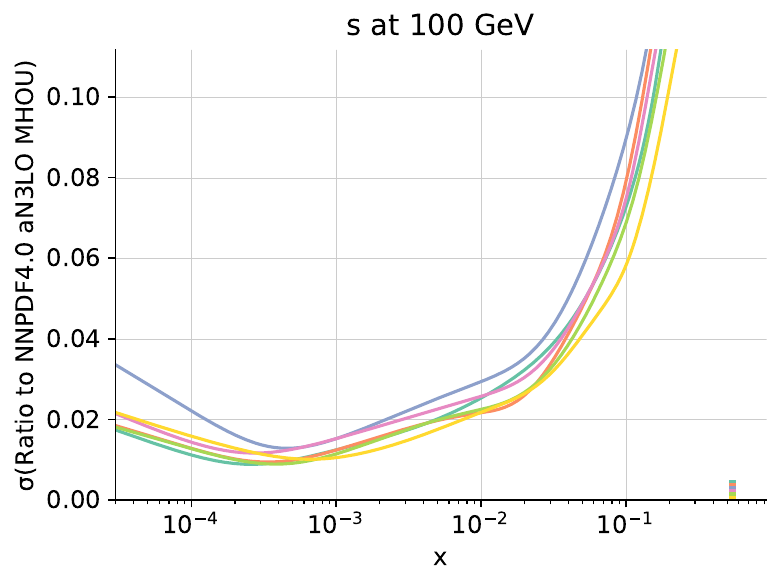}
  \includegraphics[width=0.45\textwidth]{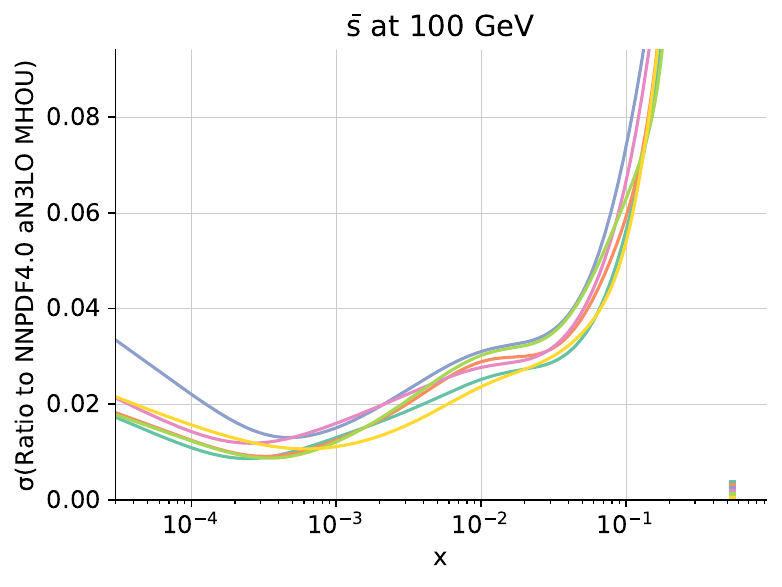}\\
  \includegraphics[width=0.45\textwidth]{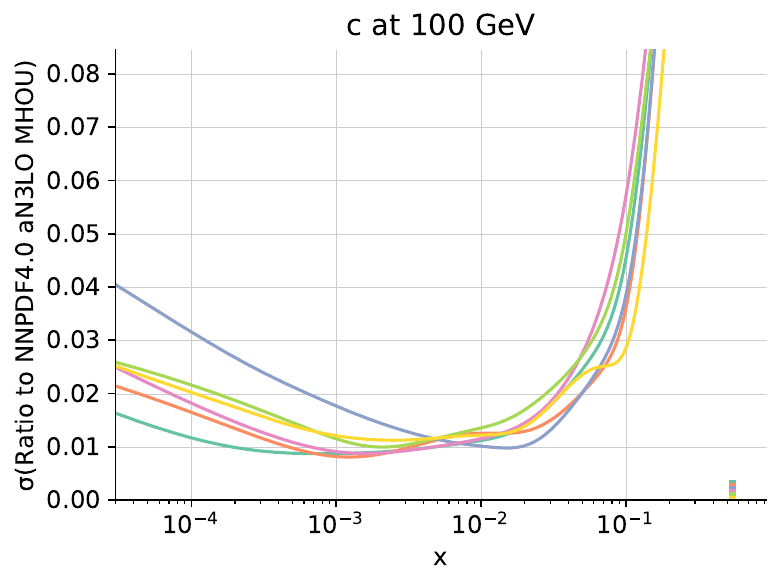}
  \includegraphics[width=0.45\textwidth]{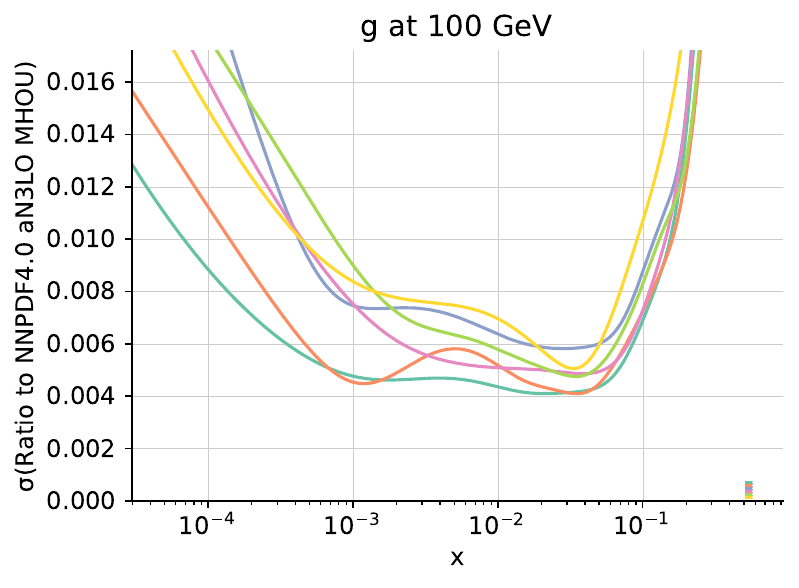}\\
  \caption{Relative one sigma uncertainties for the PDFs shown in
    Figs.~\ref{fig:pdfs_noMHOU_log}-\ref{fig:pdfs_MHOU_log}. All uncertainties
    are normalized to the central value of the NNPDF4.0 aN$^3$LO set with
    MHOUs.}
  \label{fig:pdf_uncs}
\end{figure}

We now take a closer look at  PDF uncertainties.
In Fig.~\ref{fig:pdf_uncs} we display one sigma uncertainties for the NNPDF4.0
NLO, NNLO, and aN$^3$LO PDFs with and without MHOUs at $Q=100$~GeV. All
uncertainties are normalized to the central value of the NNPDF4.0 aN$^3$LO PDF
set with MHOUs. The NLO uncertainty is generally the
largest of all in the absence of MHOUs, and for quark distributions
the smallest once MHOUs are included. All other uncertainties, at NNLO
and aN$^3$LO, with and without MHOUs, are quite similar to each other,
especially for quark PDFs. The fact that upon inclusion of an extra
source of uncertainty, namely the MHOU, 
PDF uncertainties are reduced (at NLO) or unchanged (at NNLO and
aN$^3$LO) may look counter-intuitive. However, as already pointed out  in
Refs.~\cite{Ball:2018twp,Ball:2020xqw,NNPDF:2024dpb}, this can be
understood to be a consequence of the increased compatibility of the
data due to inclusion of MHOUs and of
higher-order perturbative corrections.

\begin{figure}[!p]
  \centering
  \includegraphics[width=0.45\textwidth]{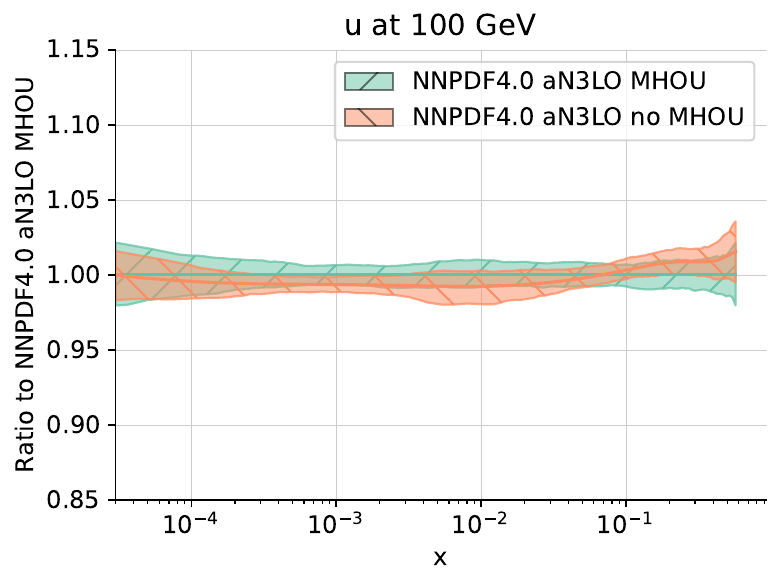}
  \includegraphics[width=0.45\textwidth]{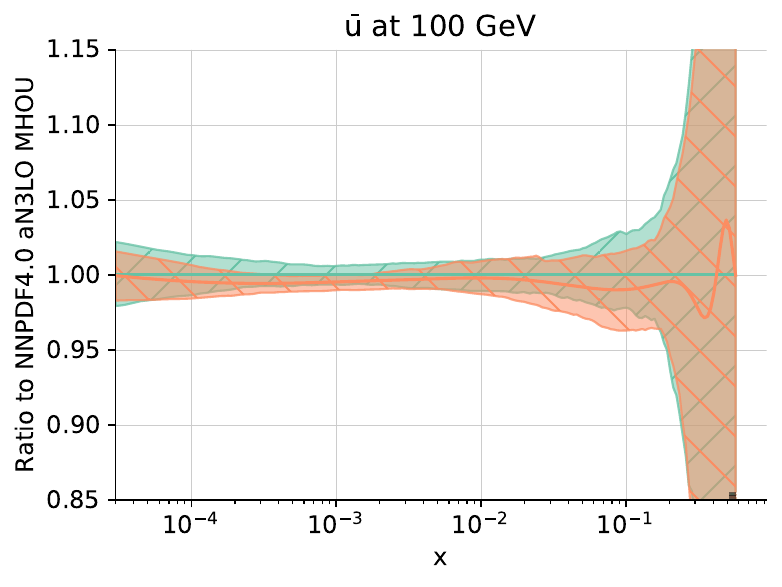}\\
  \includegraphics[width=0.45\textwidth]{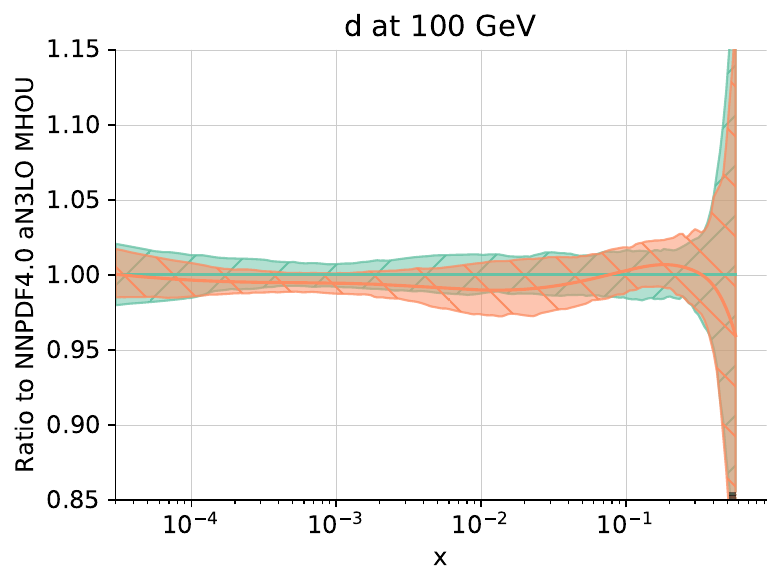}
  \includegraphics[width=0.45\textwidth]{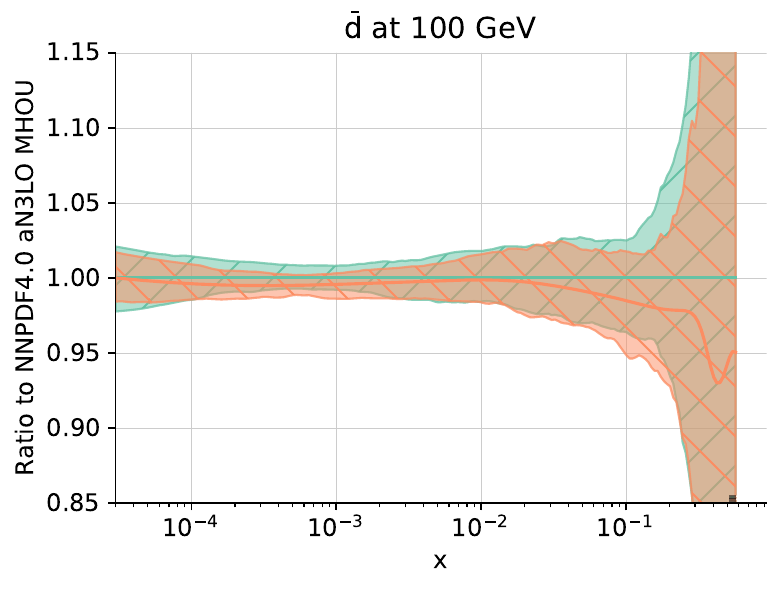}\\ 
  \includegraphics[width=0.45\textwidth]{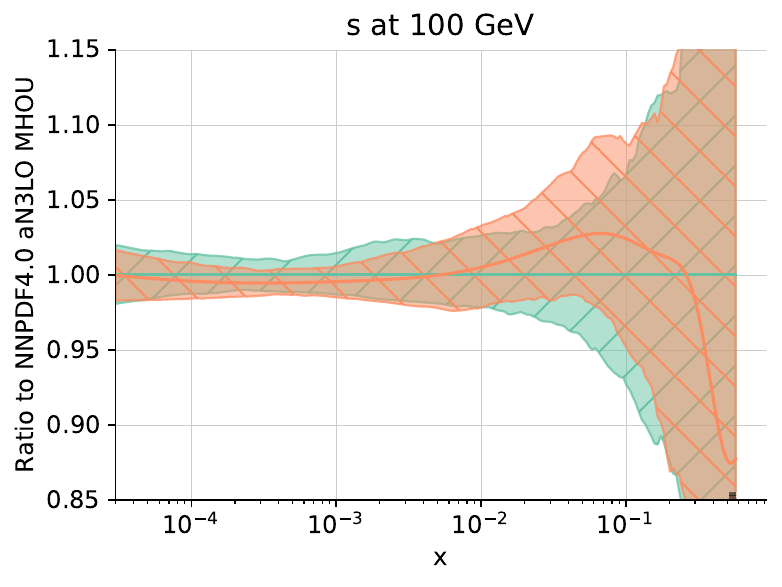}
  \includegraphics[width=0.45\textwidth]{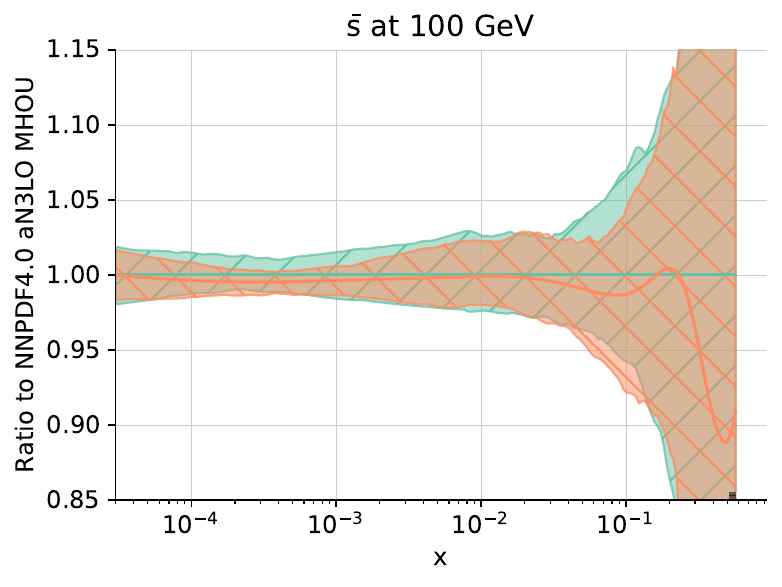}\\
  \includegraphics[width=0.45\textwidth]{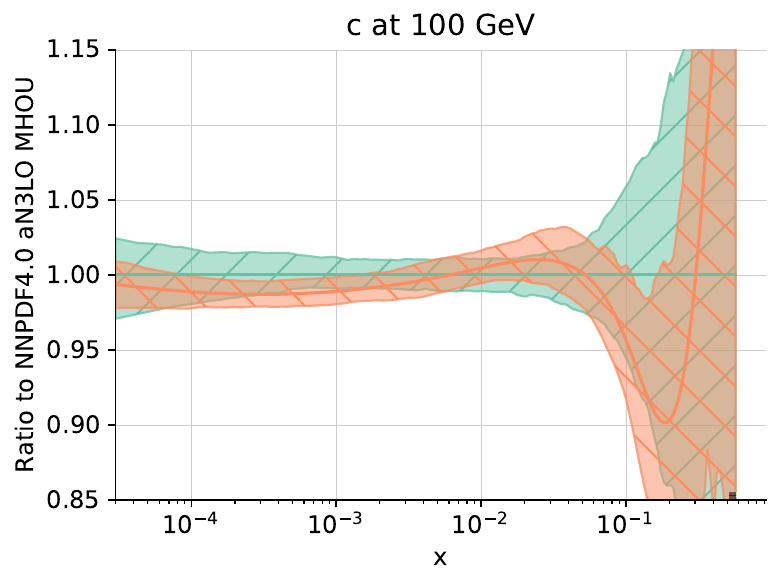}
  \includegraphics[width=0.45\textwidth]{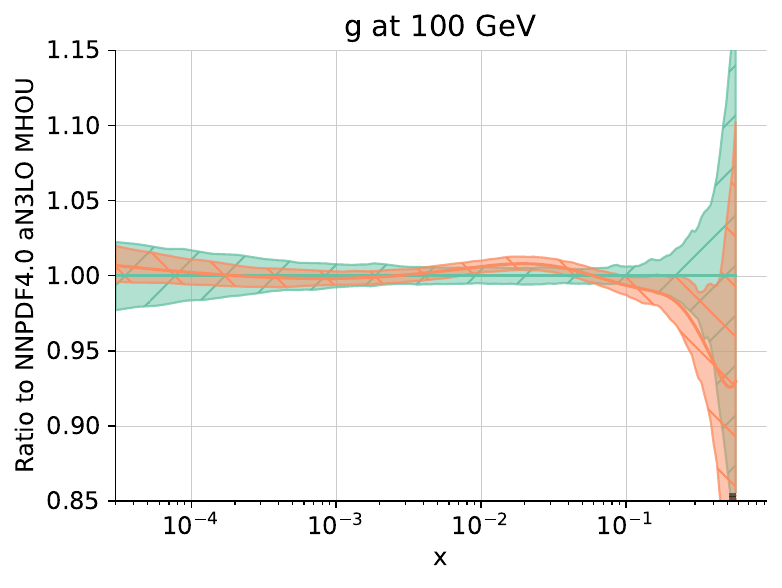}\\
  \caption{Same as
    Figs.~\ref{fig:pdfs_noMHOU_log}-\ref{fig:pdfs_MHOU_log}, now comparing NNPDF4.0 aN$^3$LO PDFs without and with
    MHOUs.}
  \label{fig:NNLO_vs_N3LO_pdfs}
\end{figure}

\begin{figure}[!t]
  \centering
  \includegraphics[width=0.45\textwidth]{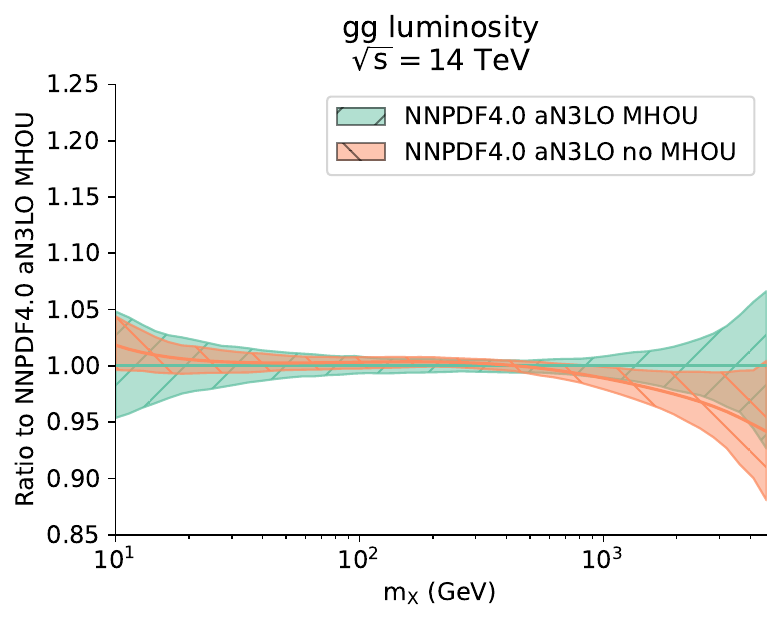}
  \includegraphics[width=0.45\textwidth]{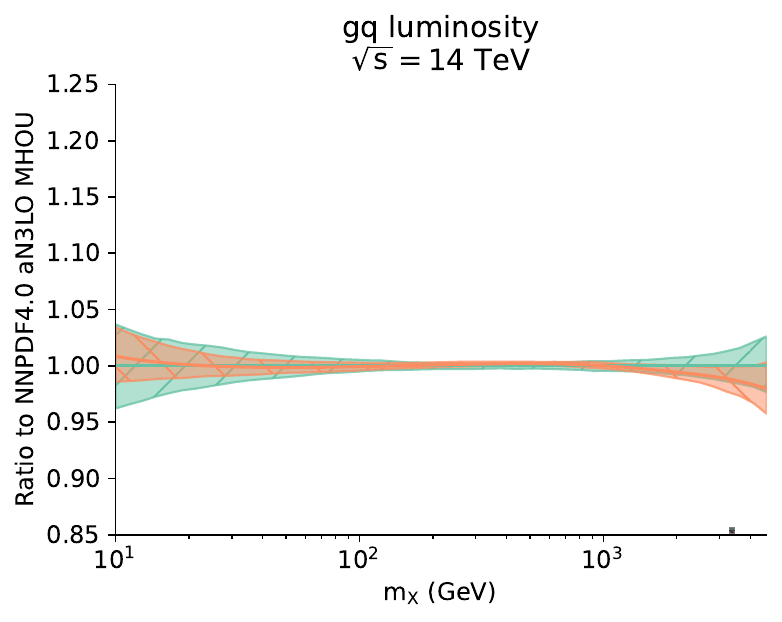}\\
  \includegraphics[width=0.45\textwidth]{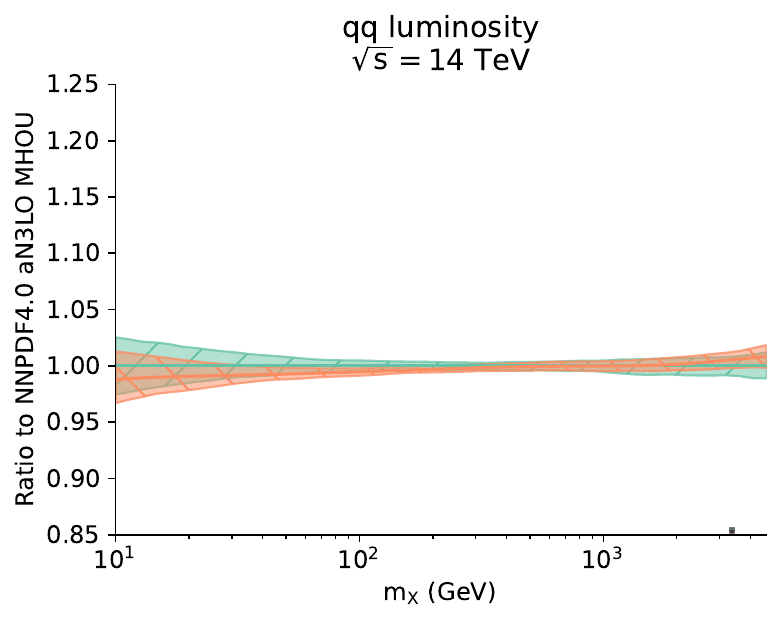}
  \includegraphics[width=0.45\textwidth]{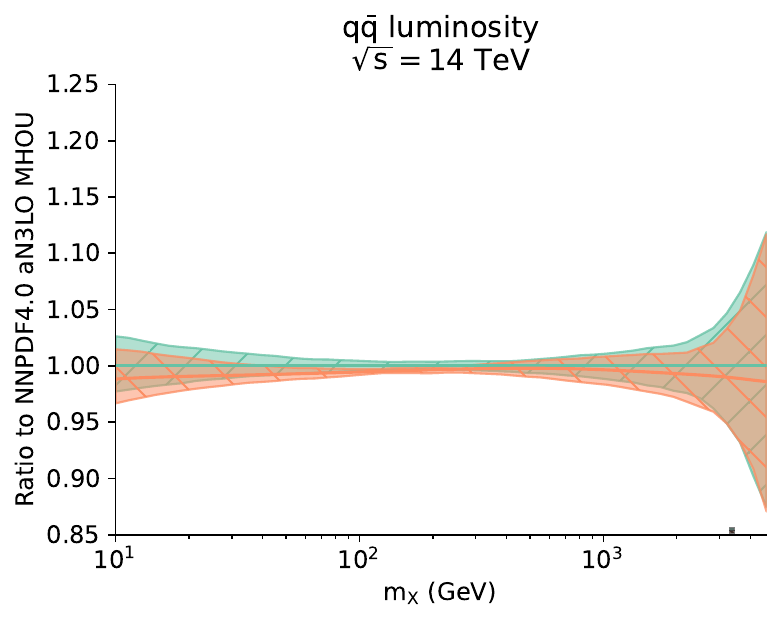}
  \caption{Same as
    Fig.~\ref{fig:lumis}, now comparing NNPDF4.0 aN$^3$LO PDFs without and with
    MHOUs.}
  \label{fig:NNLO_vs_N3LO_lumis}
\end{figure}

The impact of MHOUs on NLO and NNLO PDFs was extensively assessed in
Ref.~\cite{NNPDF:2024dpb}. In a similar vein, here we focus on the
impact of MHOUs on aN$^3$LO PDFs. To this purpose, in
Fig.~\ref{fig:NNLO_vs_N3LO_pdfs} we compare the NNPDF4.0 aN$^3$LO PDFs
with and without MHOUs. The related comparison for parton luminosities
is presented in Fig.~\ref{fig:NNLO_vs_N3LO_lumis}. Again, 
aN$^3$LO PDFs and  luminosities with and without MHOU are very 
compatible with each other. This evidence reinforces the expectation that
perturbative corrections beyond N$^3$LO will not alter PDFs significantly,
at least with current data and methodology.

In analogy with Ref.~\cite{NNPDF:2024dpb}, we also compare the
$\phi$ estimator introduced in Ref.~\cite{NNPDF:2014otw} (see Eq.~(4.6) there).
The estimator gives the ratio of the average correlated PDF uncertainty to the
data uncertainty. As such, it provides an estimate of the consistency of the
data: consistent data are combined by the underlying theory and lead to an
uncertainty in the prediction which is smaller than that of the original data.
The value of $\phi$ obtained in the NLO, NNLO, and aN$^3$LO NNPDF4.0 fits
with and without MHOUs (as in Table~\ref{tab:chi2_TOTAL}) is reported in
Table~\ref{tab:phi_TOTAL}. It is clear that $\phi$ converges to very similar
values with the increase of the perturbative order and/or with inclusion of
MHOUs for both the total dataset and for most of the data categories.
This fact is further quantitative evidence of the excellent perturbative
convergence of the PDF uncertainties.

\begin{table}[!t]
  \scriptsize
  \centering
  \renewcommand{\arraystretch}{1.4}
  \begin{tabularx}{\textwidth}{Xcccccc}
  \toprule
  & \multicolumn{2}{c}{NLO}
  & \multicolumn{2}{c}{NNLO}
  & \multicolumn{2}{c}{N$^3$LO} \\
  Dataset
  & no MHOU
  & MHOU
  & no MHOU
  & MHOU 
  & no MHOU
  & MHOU \\
  \midrule
  DIS NC
  & 0.14 & 0.13 
  & 0.15 & 0.13
  & 0.13 & 0.13 \\
  DIS CC
  & 0.11 & 0.11 
  & 0.12 & 0.12 
  & 0.12 & 0.12 \\
  DY NC
  & 0.19 & 0.17
  & 0.18 & 0.17
  & 0.17 & 0.18  \\
  DY CC
  & 0.33 & 0.27
  & 0.35 & 0.32
  & 0.31 & 0.32 \\
  Top pairs
  & 0.18 & 0.17
  & 0.17 & 0.17
  & 0.16 & 0.19 \\
  Single-inclusive jets
  & 0.13 & 0.13
  & 0.13 & 0.13
  & 0.13 & 0.13 \\
  Dijets
  & 0.10 & 0.10
  & 0.11 & 0.10
  & 0.10 & 0.10 \\
  Prompt photons 
  & 0.06 & 0.07
  & 0.06 & 0.06
  & 0.05 & 0.05 \\
  Single top
  & 0.04 & 0.04
  & 0.04 & 0.04
  & 0.04 & 0.04 \\
  \midrule
  Total
  & 0.18 & 0.15
  & 0.16 & 0.15
  & 0.15 & 0.15 \\
\bottomrule
\end{tabularx}

  \caption{The $\phi$ uncertainty estimator for NNPDF4.0 PDFs at NLO, NNLO and
    aN$^3$LO without and with MHOUs for the process categories as in
    Table~\ref{tab:chi2_TOTAL}.}
  \label{tab:phi_TOTAL}
\end{table}

\subsection{Implications for intrinsic charm}
\label{sec:ic}

The availability of the aN$^3$LO PDFs discussed in
Sects.~\ref{sec:PDFs}-\ref{sec:uncertainties} allows us to revisit and
consolidate our recent results on intrinsic charm. Specifically, based on the
NNPDF4.0 NNLO PDF determination, we have found evidence for intrinsic
charm~\cite{Ball:2022qks} and an indication for a non-vanishing valence charm
component~\cite{NNPDF:2023tyk}. In these analyses, the dominant source of theory
uncertainty was estimated to come from the matching conditions that are used in
order to obtain PDFs in a three-flavor charm decoupling scheme from
high-scale data, while MHOUs were assumed to be subdominant. The
uncertainty in the matching conditions was in turn estimated by comparing
results obtained using NNLO matching and the best available aN$^3$LO matching
conditions, both applied to NNLO PDFs.

It is now possible to improve these results on three counts. First, we
can now fully include MHOUs. Second, we can consistently combine
aN$^3$LO matching conditions and aN$^3$LO PDFs, and perform a consistent
comparison of NNLO and  aN$^3$LO results. Finally, knowledge of
aN$^3$LO matching conditions themselves is now improved thanks to
recent results~\cite{Ablinger:2022wbb} that were not available at the
time of the analysis of Ref.~\cite{Ball:2022qks}. We will specifically
discuss the determination of the total intrinsic charm component and we
do not consider the valence component,
because effects of MHOUs and of the flavor scheme transformation are
already very small at  NNLO~\cite{NNPDF:2023tyk}.

\begin{figure}[!t]
  \centering
  \includegraphics[width=0.45\textwidth]{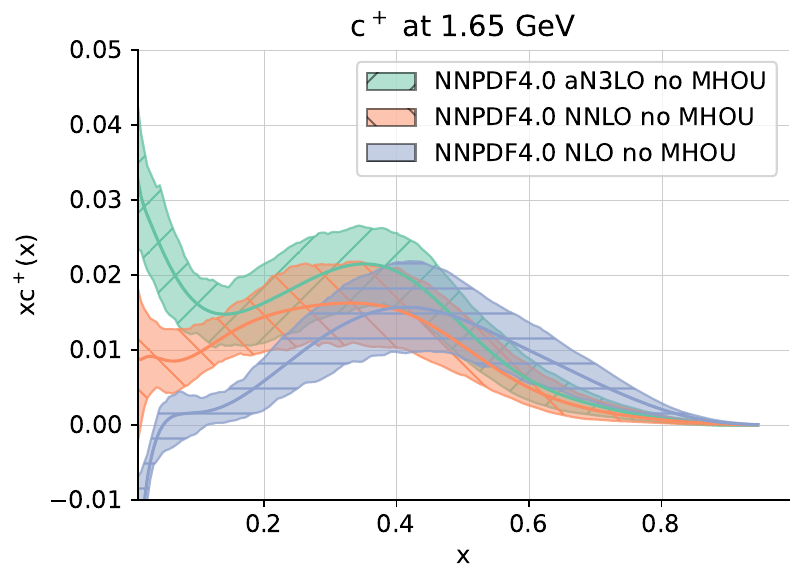}
  \includegraphics[width=0.45\textwidth]{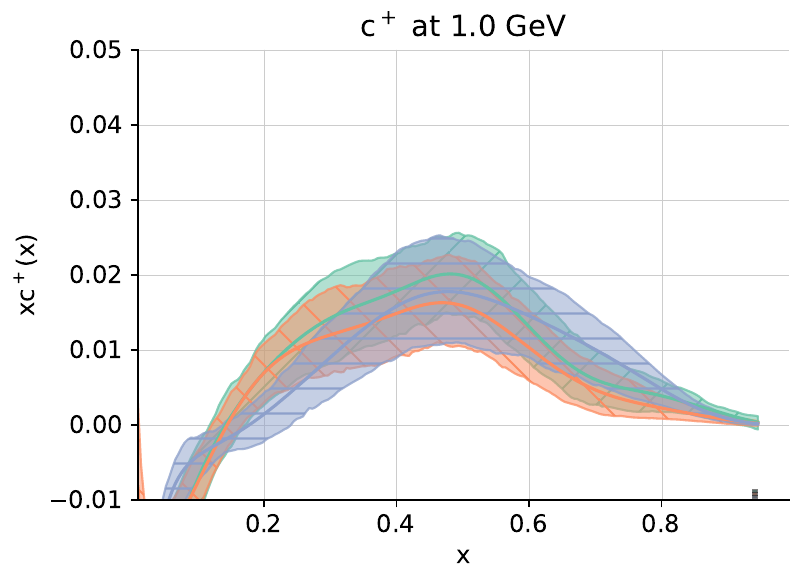}\\
  \includegraphics[width=0.45\textwidth]{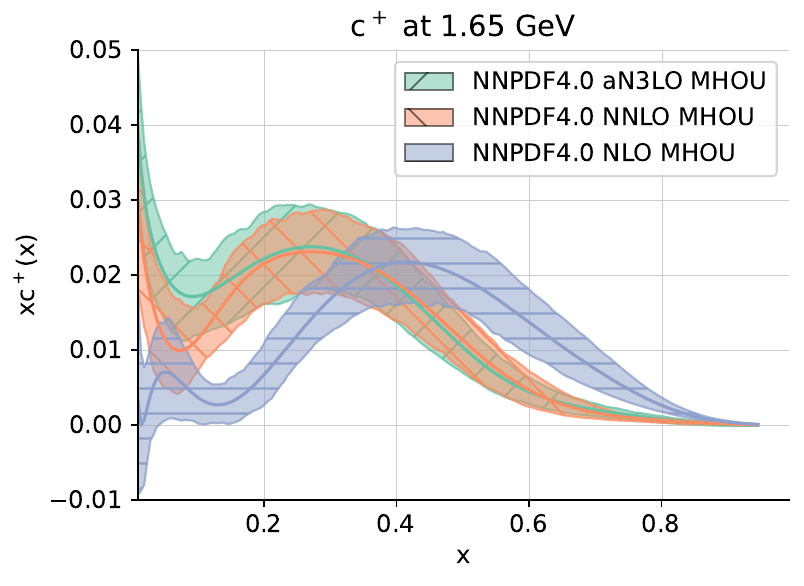}
  \includegraphics[width=0.45\textwidth]{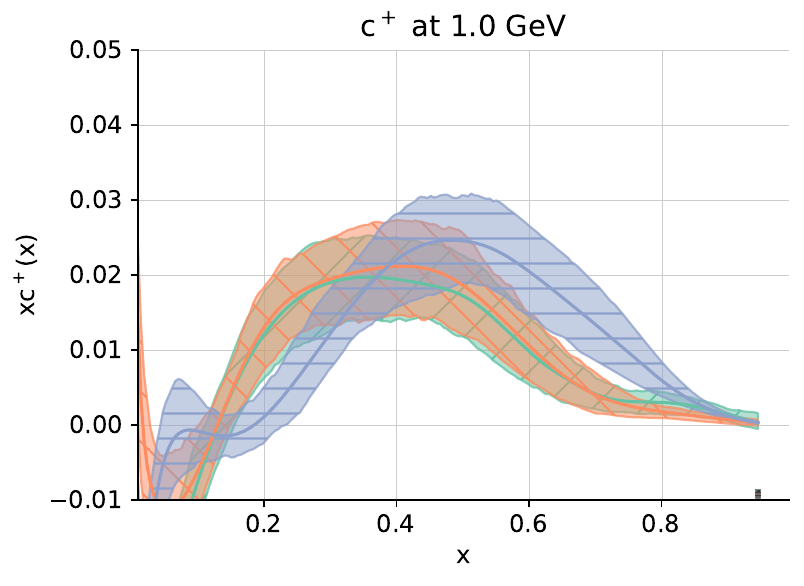}\\ 
  \caption{The total charm PDF, $xc^+(x,Q^2)$, in the 4FNS at $Q=1.65$~GeV
    (left) and 3FNS (right), as obtained from the NNPDF4.0 NLO, NNLO, and
    aN$^3$LO fits without (top) and with (bottom) MHOUs. Error bands correspond
    to one sigma PDF uncertainties. Note that in the 3FNS the charm
    PDF does not depend on scale.}
  \label{fig:PDFs-n3lo-q1p65gev-Charm-perturbative-convergence} 
\end{figure}

To this purpose, in
Fig.~\ref{fig:PDFs-n3lo-q1p65gev-Charm-perturbative-convergence} we
show the total charm PDF, $xc^+(x,Q^2)$, in the 4FNS at $Q=1.65$~GeV and in the
3FNS, as obtained from using  NNPDF4.0 NLO, NNLO and aN$^3$LO without and with
MHOUs. Note that in the 3FNS the charm PDF does not depend on scale.
Error bands correspond to one sigma PDF uncertainties. The 4FNS
results share the general features discussed in Sect.~\ref{sec:PDFs}:
the perturbative expansion converges nicely, with the aN$^3$LO result
very close to the NNLO. The convergence is further improved by the
inclusion of MHOUs, which move the NNLO yet closer to the
aN$^3$LO. The 3FNS result is especially remarkable: whereas the
combination of aN$^3$LO matching with NNLO PDFs, used in
Ref.~\cite{Ball:2022qks} to conservatively estimate MHOUs, was somewhat
unstable, now results display complete stability, and in particular
the NNLO and aN$^3$LO results completely coincide.

\begin{figure}[!t]
  \centering
  \includegraphics[width=0.45\textwidth]{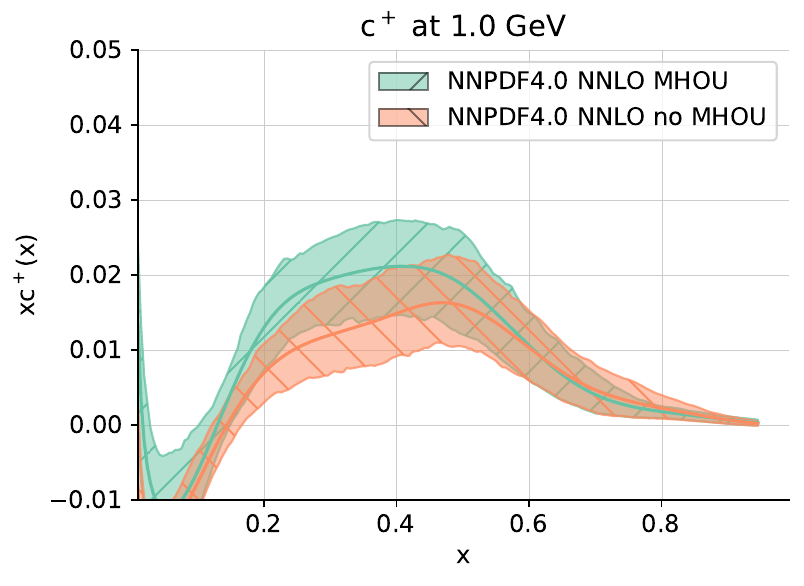}
  \includegraphics[width=0.45\textwidth]{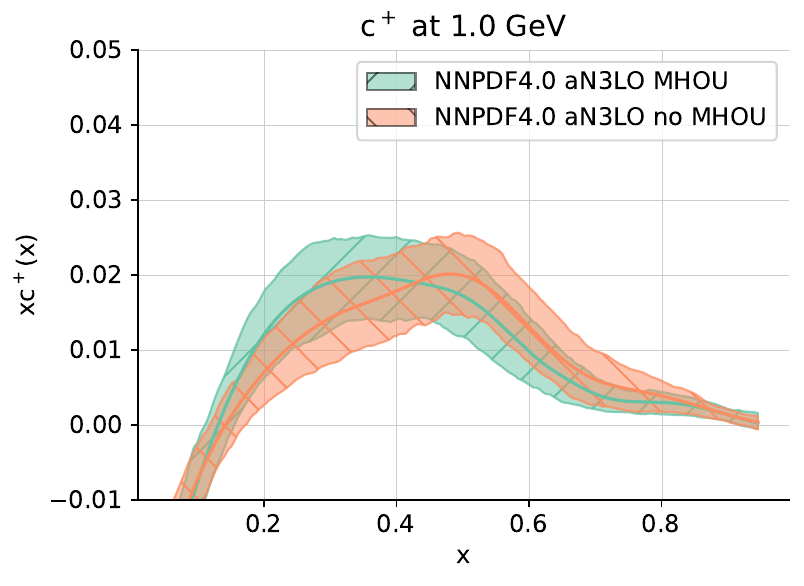}
  \caption{Same as
    Fig.~\ref{fig:PDFs-n3lo-q1p65gev-Charm-perturbative-convergence}, now
    comparing the total charm PDF in the 3FNS with and without
    MHOUs, respectively at NNLO (left) and aN$^3$LO (right).}
  \label{fig:PDFs-n3lo-q1p65gev-Charm-impactMHOU} 
\end{figure}

In order to assess the impact of MHOUs more clearly, in
Fig.~\ref{fig:PDFs-n3lo-q1p65gev-Charm-impactMHOU} we compare the total charm
PDF in the 3FNS with and without MHOUs, respectively at NNLO
and aN$^3$LO. At NNLO MHOUs have a small but non-negligible impact
on central values, with almost unchanged uncertainty, but at aN$^3$LO
they have essentially no impact, confirming the perturbative
convergence of the result.

\begin{figure}[!t]
  \centering
  \includegraphics[width=0.55\textwidth]{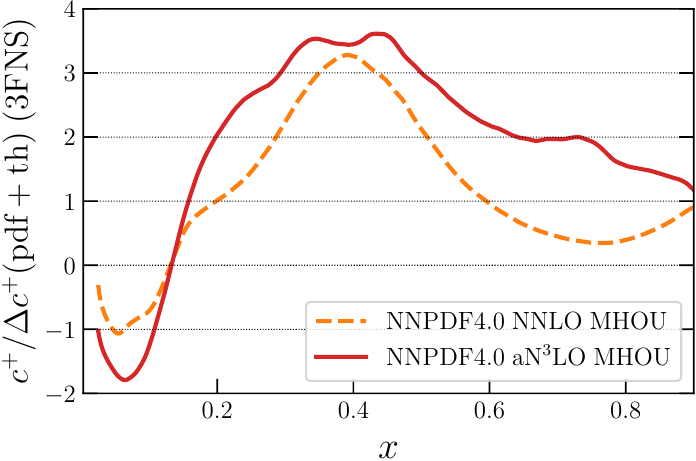}
  \caption{The pull (central value divided by total uncertainty)
    for the total charm PDF in the 3FNS obtained
    in the NNPDF4.0 NNLO and aN$^3$LO fits with MHOUs.}
  \label{fig:IC-Significance-N3LO} 
\end{figure}

We thus proceed to a final re-assessment of the significance of
intrinsic charm through the pull, defined as the central value divided
by total uncertainty, using NNPDF4.0MHOU NNLO and aN$^3$LO PDFs.
We estimate the total uncertainty by adding in quadrature to the PDF uncertainty
(which already includes the MHOU from the theory predictions used in the fit) a 
further  theory uncertainty, taken equal to the difference between the
central value at given perturbative order, and that at the previous perturbative
order (so at NNLO from the difference to NLO, and so on).
This now also includes the MHOU due to change in the 
matching from 4FNS to 3FNS, but also the shift in the 4FNS result that
is in principle already accounted for by the MHOU. Also, it 
conservatively assumes that the shift between the current order and
the next is equal to that from the previous order, rather than
smaller. Results obtained with this conservative error
estimate are shown in Fig.~\ref{fig:IC-Significance-N3LO}.
It is clear that the
significance of intrinsic charm is increased somewhat when going from
NNLO to aN$^3$LO. It is now also somewhat increased already at NNLO in
comparison to the result of Ref.~\cite{Ball:2022qks}, despite the more
conservative uncertainty estimate, thanks to the  increased accuracy of MHOU
PDFs and the consistent and improved treatment of matching aN$^3$LO
conditions. Indeed, local significance at the peak is now more than
three sigma for the default fit.

\subsection{Dependence on the treatment of aN$^3$LO corrections.}
\label{app:splitting_bench}
\label{sec:N3LOKfact}

\begin{figure}[!p]
  \centering
  \includegraphics[width=0.45\textwidth]{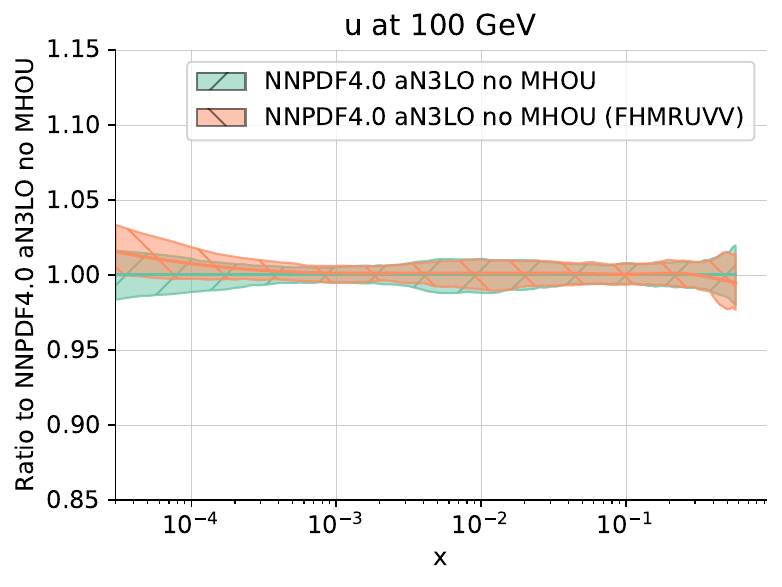}
  \includegraphics[width=0.45\textwidth]{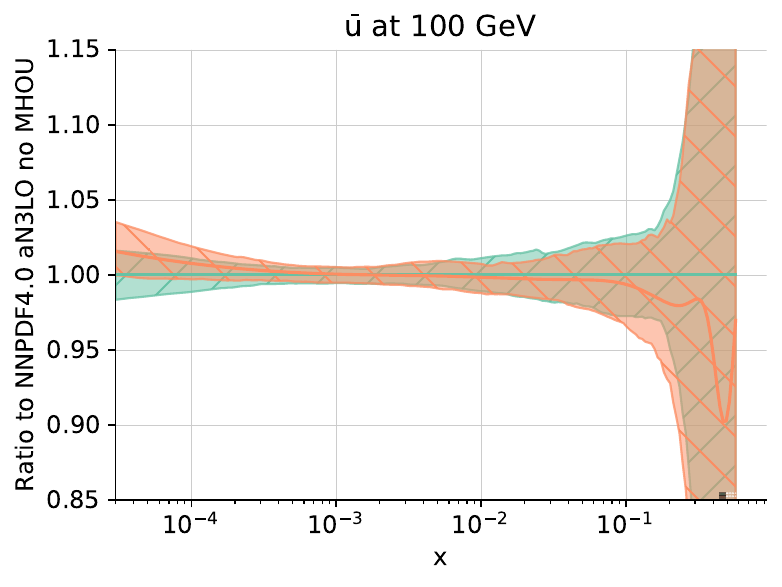}
  \includegraphics[width=0.45\textwidth]{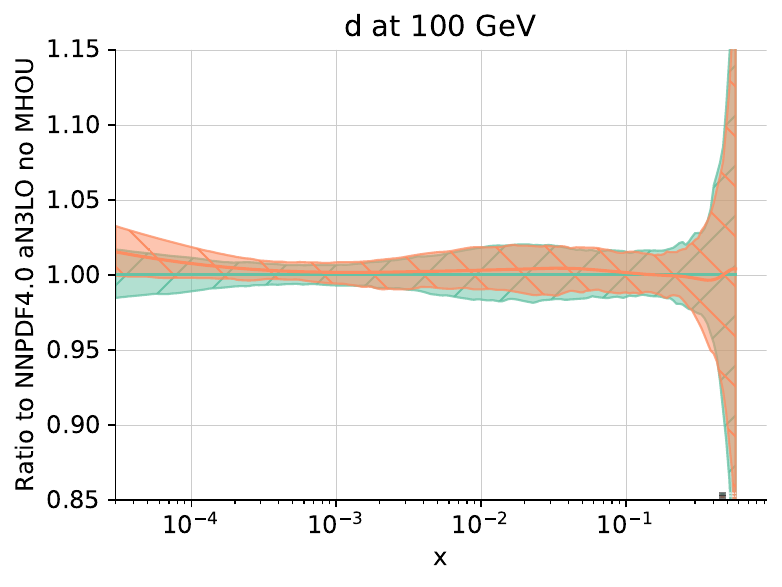}
  \includegraphics[width=0.45\textwidth]{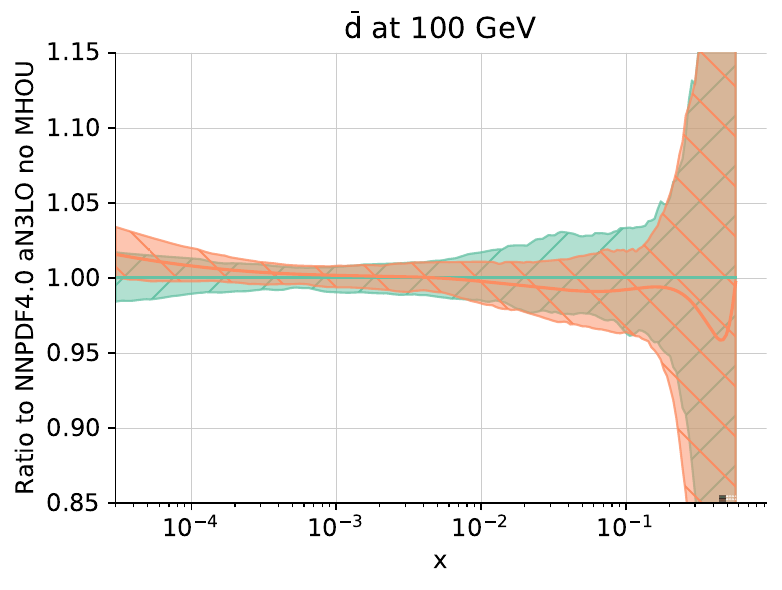}
  \includegraphics[width=0.45\textwidth]{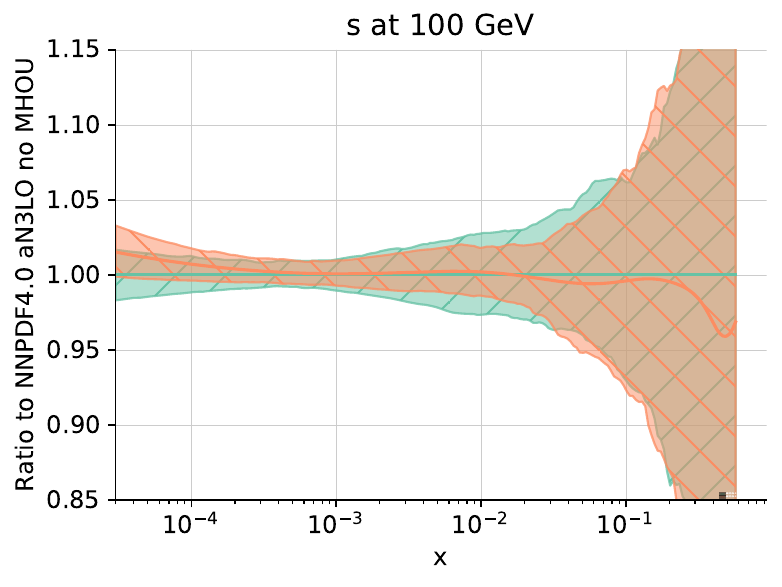}
  \includegraphics[width=0.45\textwidth]{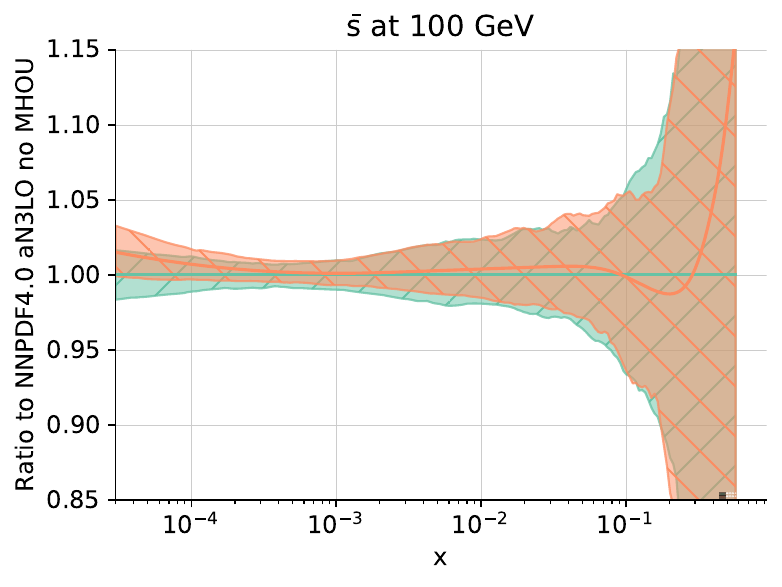}
  \includegraphics[width=0.45\textwidth]{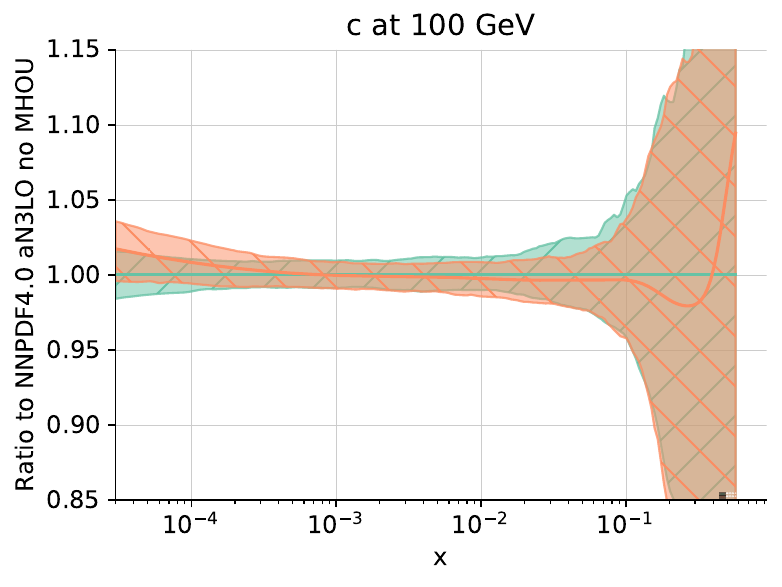}
  \includegraphics[width=0.45\textwidth]{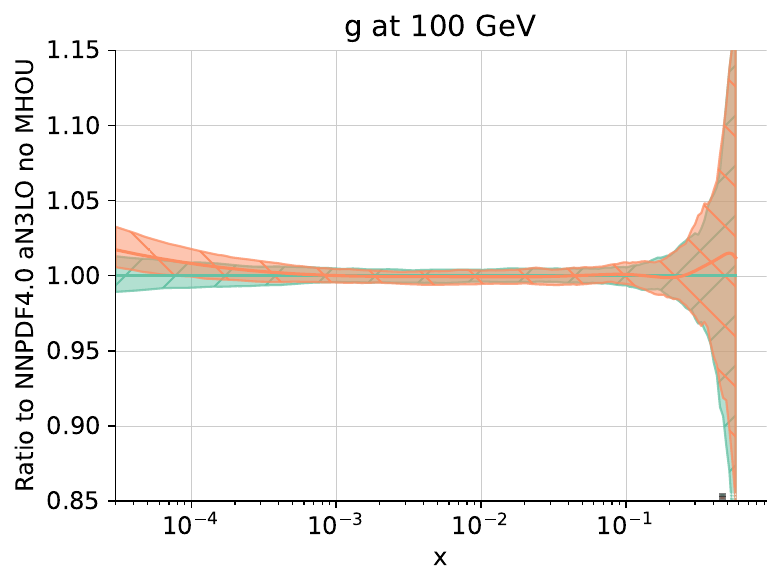}
  \caption{Same as
    Fig.~\ref{fig:pdfs_noMHOU_log}, now
    comparing the  NNPDF4.0 aN$^3$LO PDFs and PDFs
    obtained using the  FHMRUVV approximation to N$^3$LO perturbative
    evolution.}
  \label{fig:PDFs-n3lo-q100gev-ratios-FHMRUVV} 
\end{figure}

\begin{figure}[!p]
  \centering
  \includegraphics[width=0.45\textwidth]{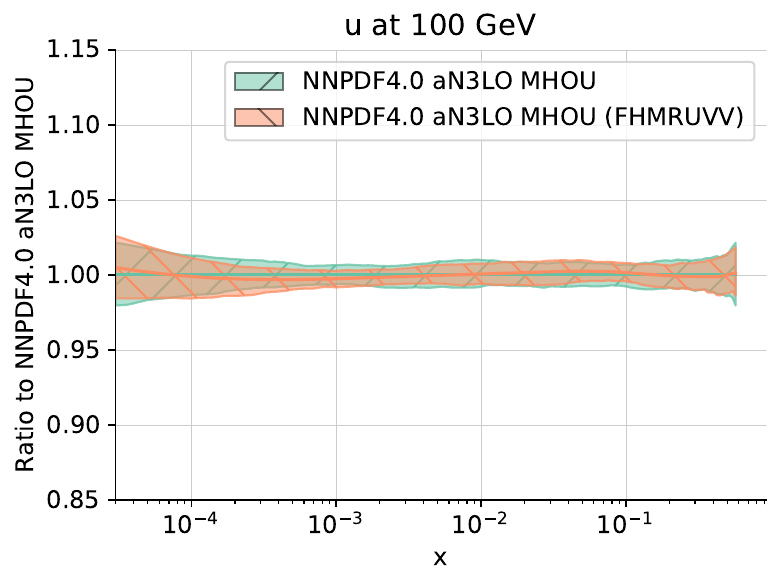}
  \includegraphics[width=0.45\textwidth]{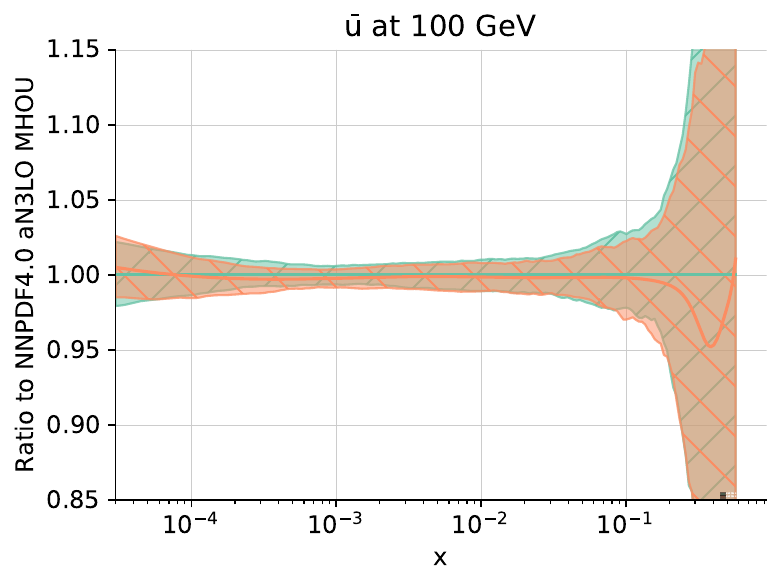}
  \includegraphics[width=0.45\textwidth]{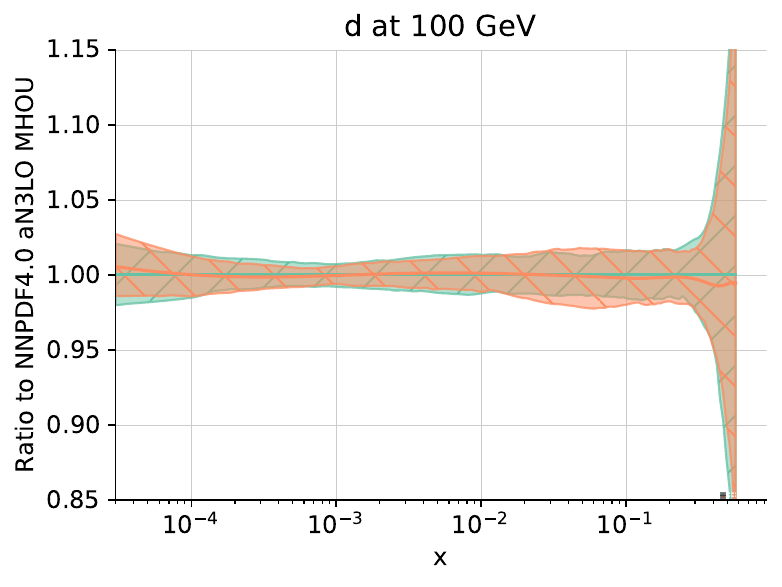}
  \includegraphics[width=0.45\textwidth]{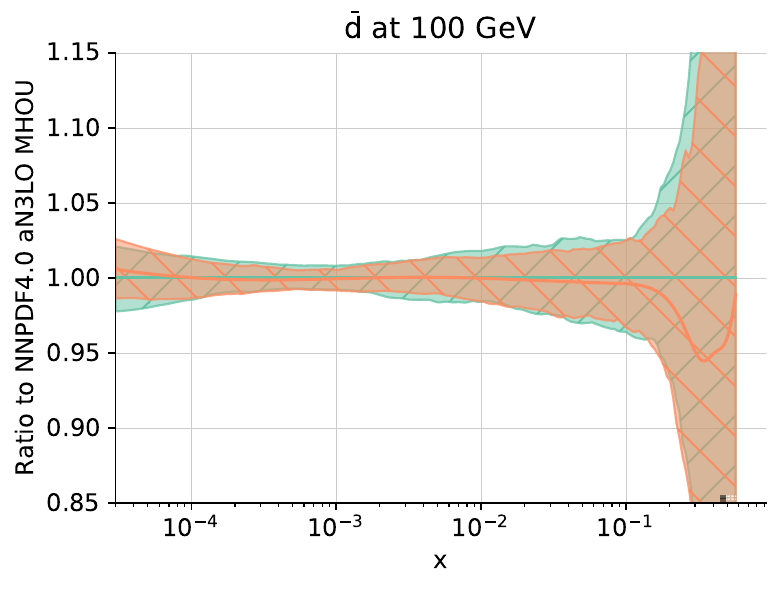}
  \includegraphics[width=0.45\textwidth]{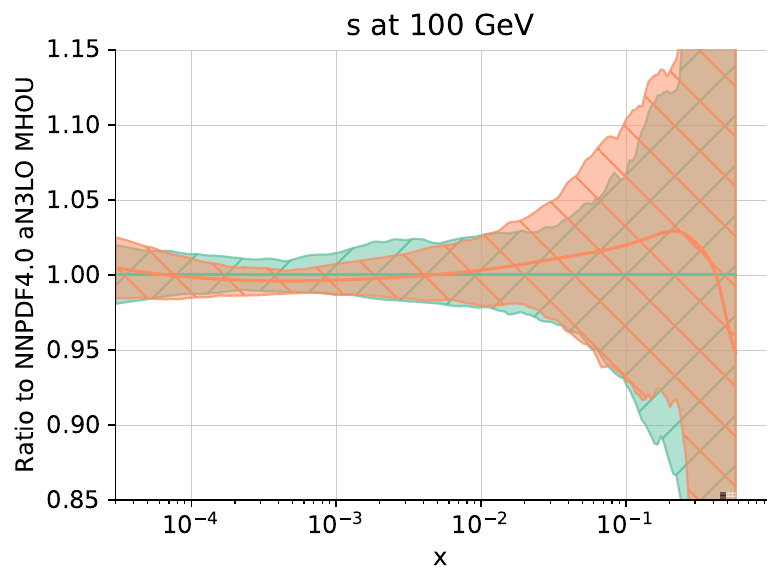}
  \includegraphics[width=0.45\textwidth]{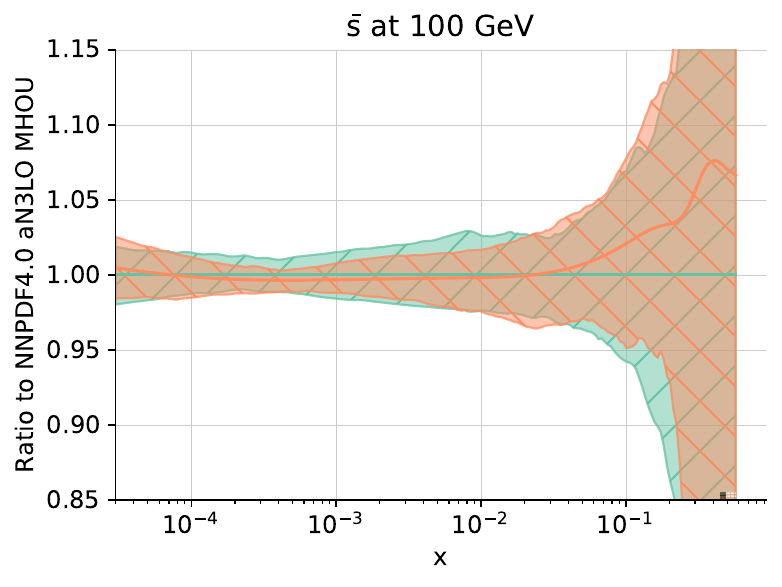}
  \includegraphics[width=0.45\textwidth]{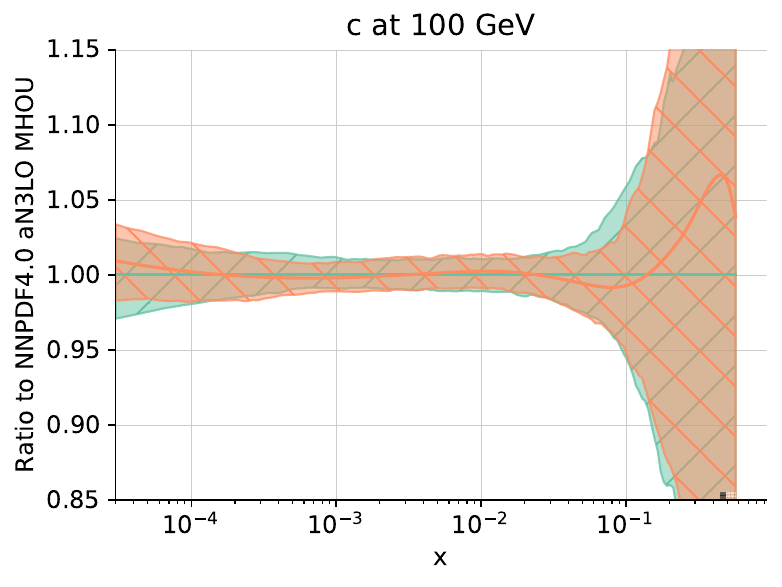}
  \includegraphics[width=0.45\textwidth]{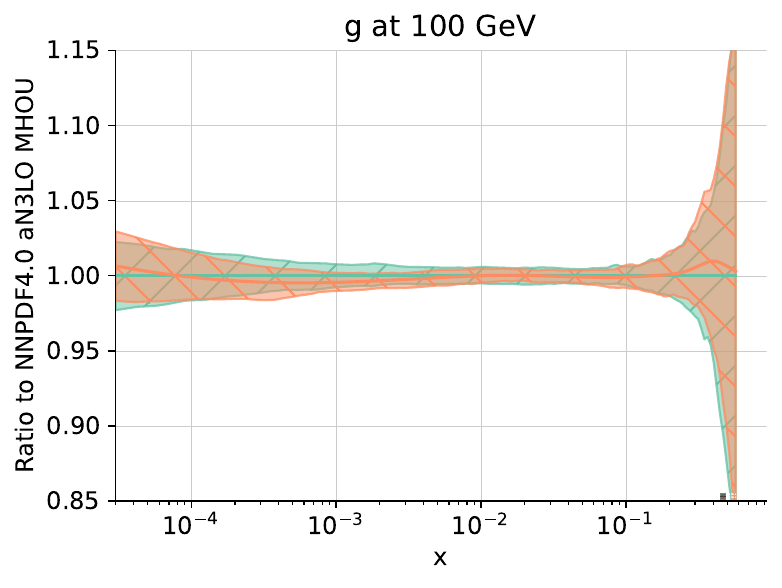}
  \caption{Same as Fig.~\ref{fig:PDFs-n3lo-q100gev-ratios-FHMRUVV} for aN$^3$LO
  PDF sets with MHOUs.}
  \label{fig:PDFs-n3lo-q100gev-ratios-FHMRUVV_MHOU} 
\end{figure}

We now discuss two variants in the treatment of aN$^3$LO corrections:
a different approximation to perturbative evolution, and a different treatment
of hadronic cross-sections.

The aN$^3$LO PDF sets presented in Sect.~\ref{sec:PDFs} are based on
our approximation to the full set of N$^3$LO splitting functions
presented in Sect.~\ref{sec:dglap}. This approximation was  compared in
Sect.~\ref{sec:n3lo_comp} to that recently presented in
Refs.~\cite{Moch:2017uml,Falcioni:2023luc,Falcioni:2023vqq,Moch:2023tdj} (FHMRUVV).
In order to fully assess the impact of this different approximation to N$^3$LO
perturbative evolution we have repeated the  aN$^3$LO PDF determination,
without and with MHOUs, but now using  the FHMRUVV approximation to N$^3$LO
perturbative evolution, with everything else
unchanged. PDFs obtained using the FHMRUVV approximation vs our own are
compared in Fig.~\ref{fig:PDFs-n3lo-q100gev-ratios-FHMRUVV}
without MHOUs and in Fig.~\ref{fig:PDFs-n3lo-q100gev-ratios-FHMRUVV_MHOU} with
MHOUs. PDFs are displayed at $Q=100$~GeV, normalized to 
the central value of our default result. Differences turn out to be
completely negligible, as might have been expected given the good
agreement at the level of splitting functions seen in Fig.~\ref{fig:splitting-functions-mhst}.

\begin{figure}[!p]
  \centering
  \includegraphics[width=0.45\textwidth]{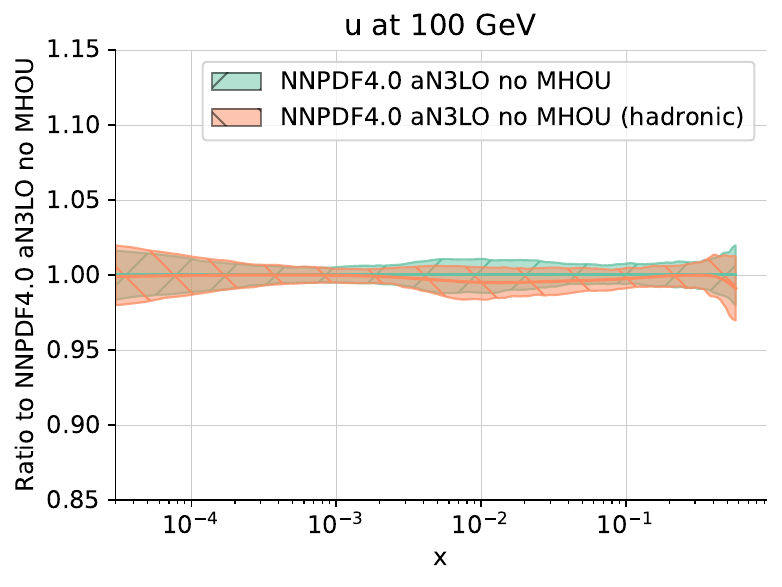}
  \includegraphics[width=0.45\textwidth]{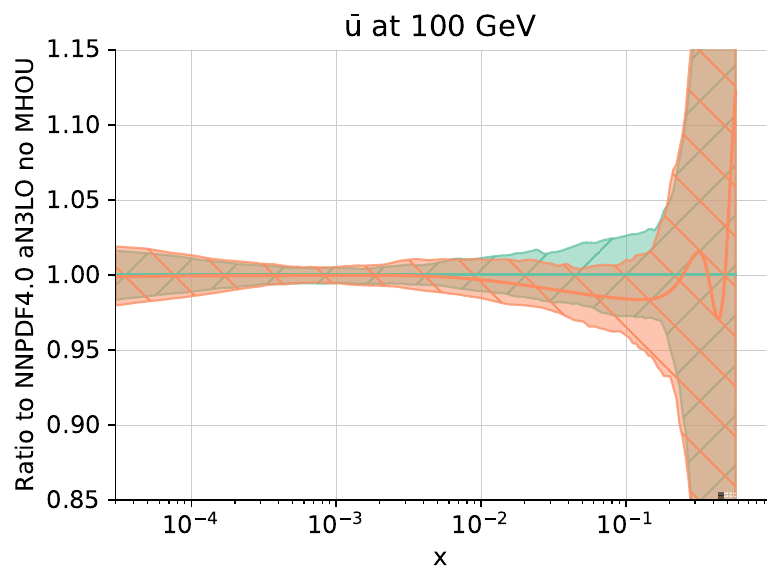}
  \includegraphics[width=0.45\textwidth]{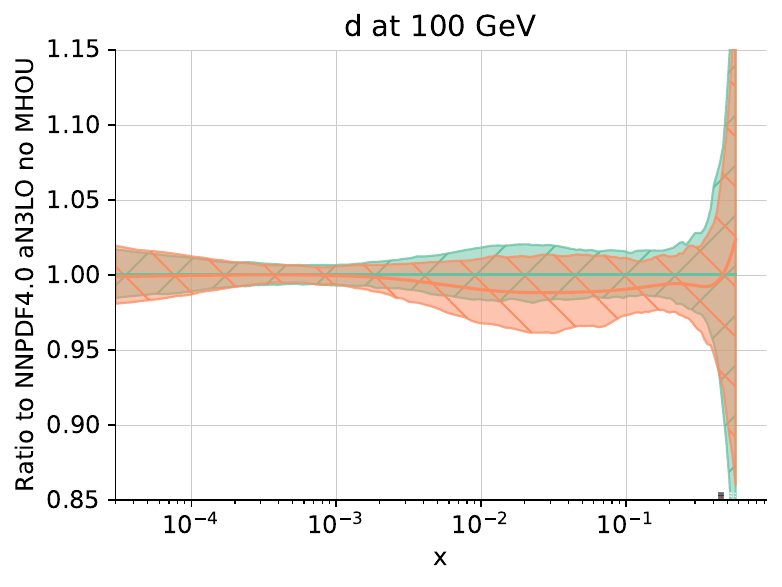}
  \includegraphics[width=0.45\textwidth]{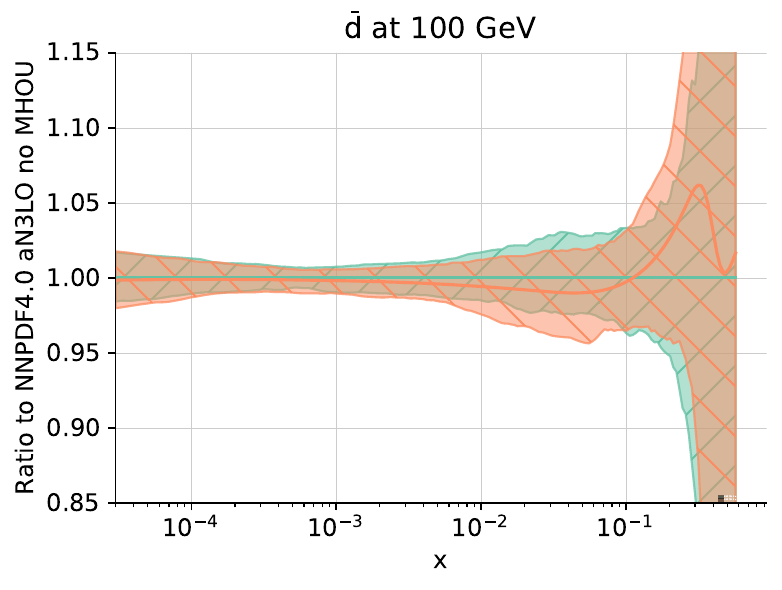}
  \includegraphics[width=0.45\textwidth]{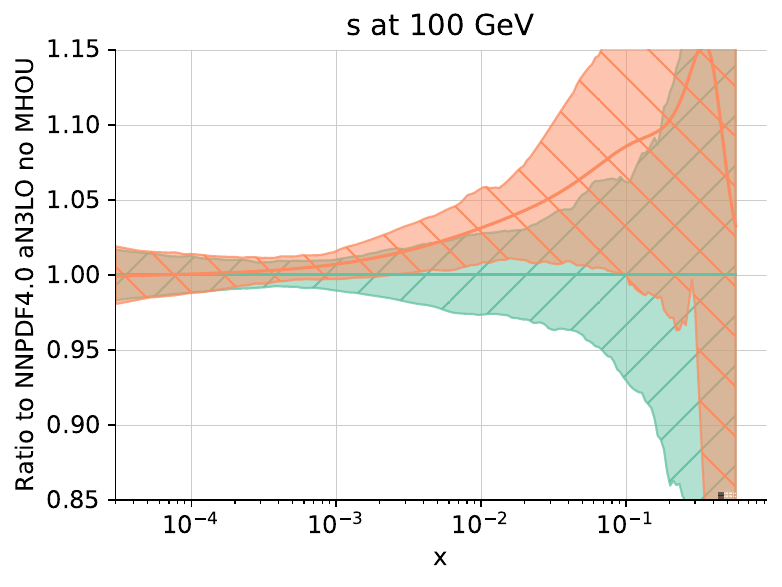}
  \includegraphics[width=0.45\textwidth]{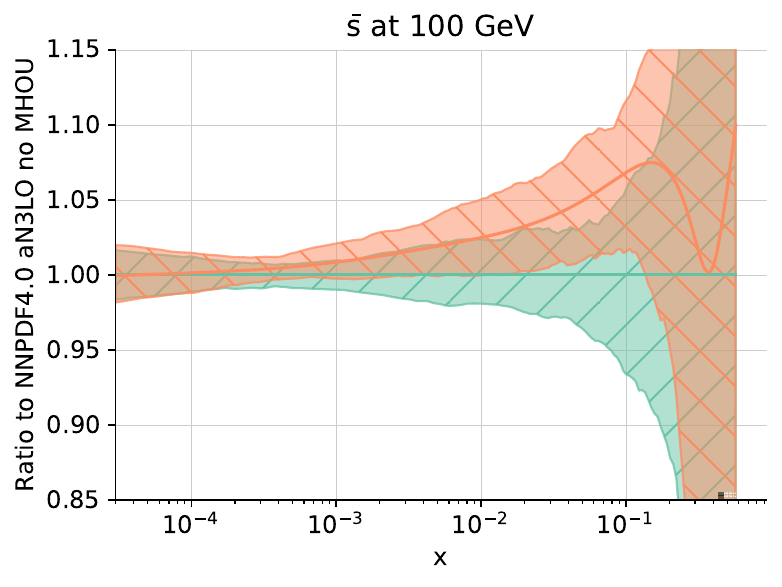}
  \includegraphics[width=0.45\textwidth]{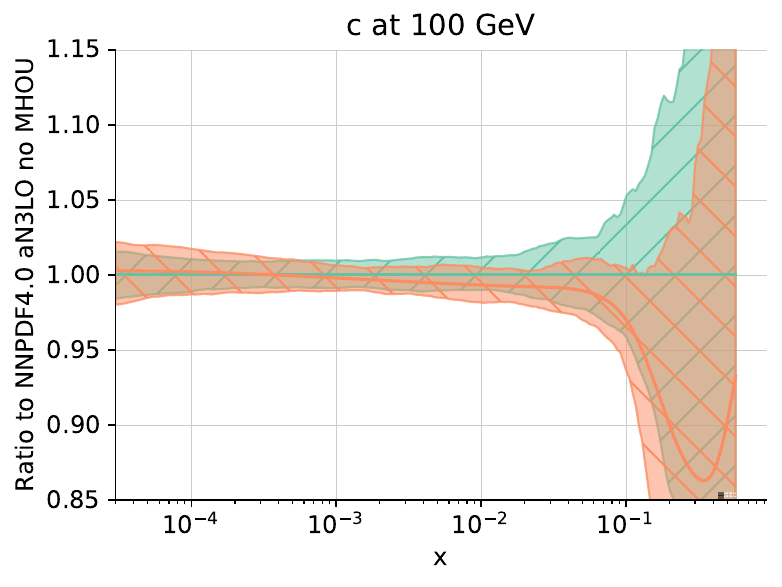}
  \includegraphics[width=0.45\textwidth]{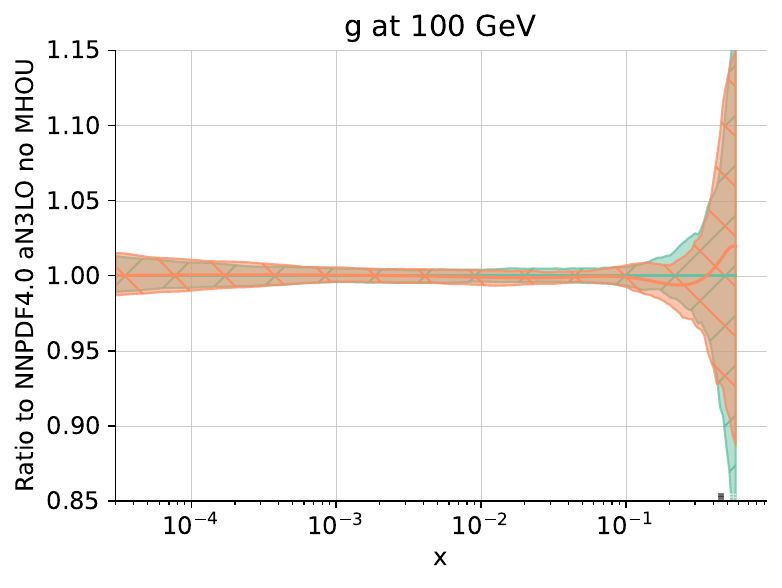}
  \caption{Same as
    Fig.~\ref{fig:pdfs_noMHOU_log}, now
    comparing the NNPDF4.0 aN$^3$LO baseline PDF set without
    MHOUs to a variant obtained with inclusion of N$^3$LO corrections
    for the hadronic processes of Tab.~\ref{tab:DY-kf}.}
  \label{fig:hadronic_noMHOU} 
\end{figure}

\begin{figure}[!p]
  \centering
  \includegraphics[width=0.45\textwidth]{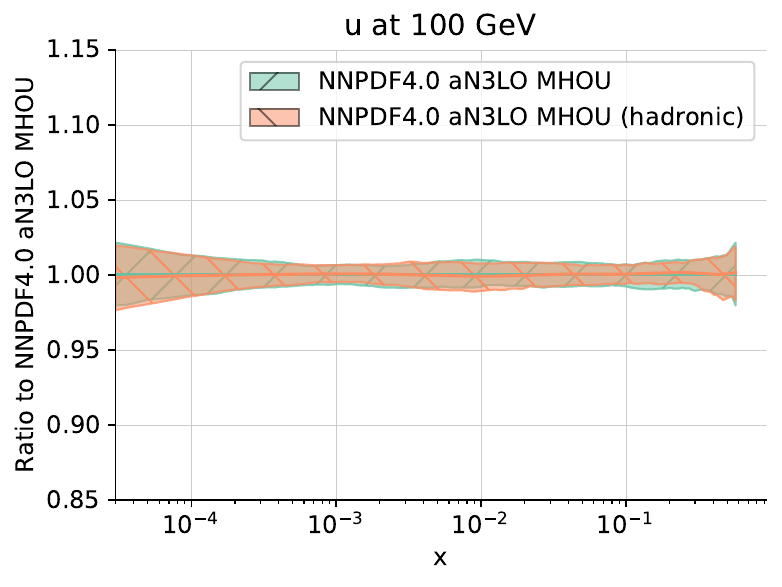}
  \includegraphics[width=0.45\textwidth]{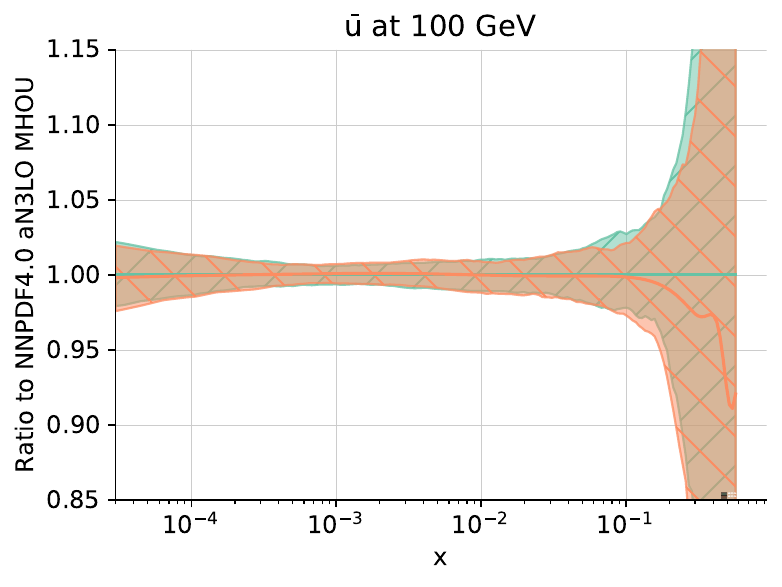}
  \includegraphics[width=0.45\textwidth]{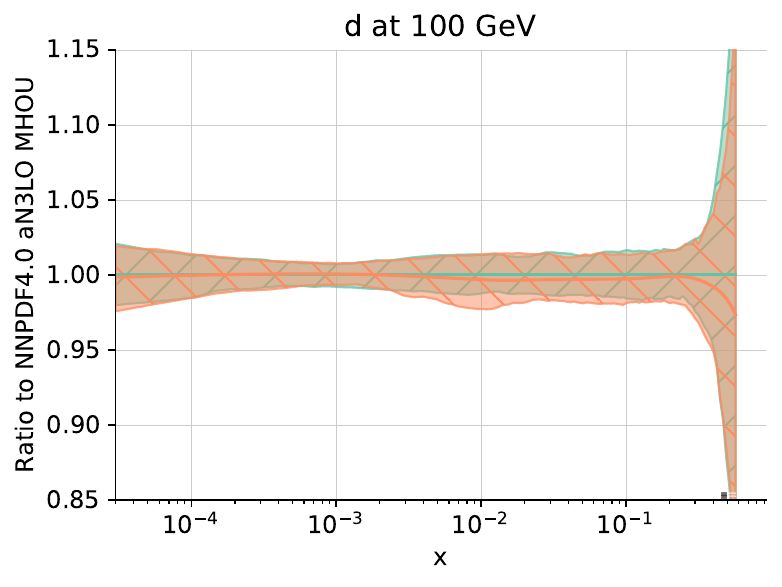}
  \includegraphics[width=0.45\textwidth]{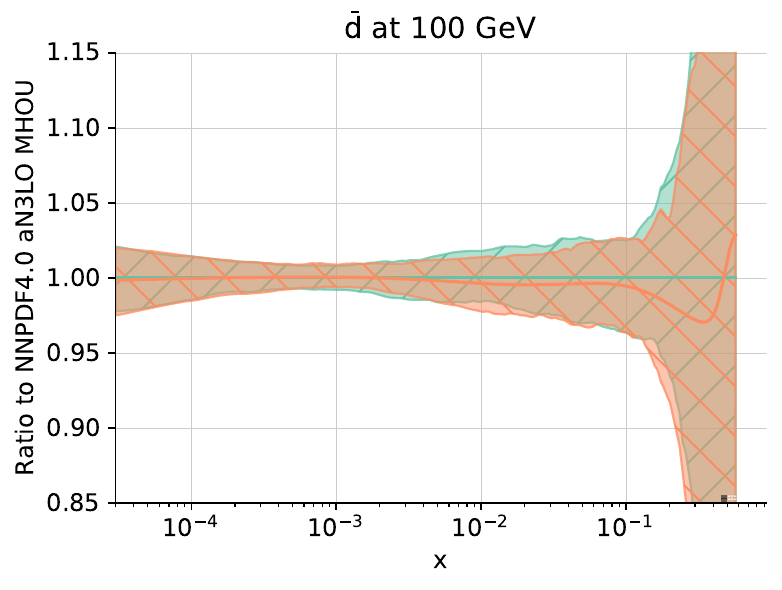}
  \includegraphics[width=0.45\textwidth]{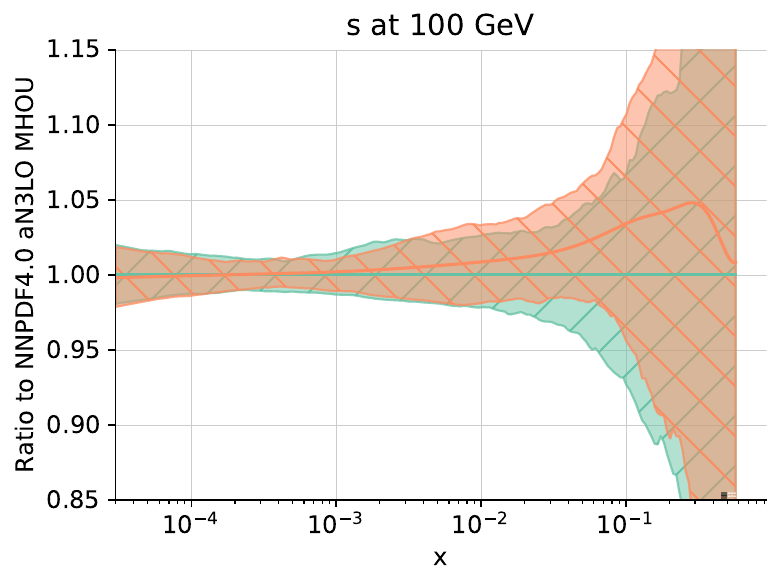}
  \includegraphics[width=0.45\textwidth]{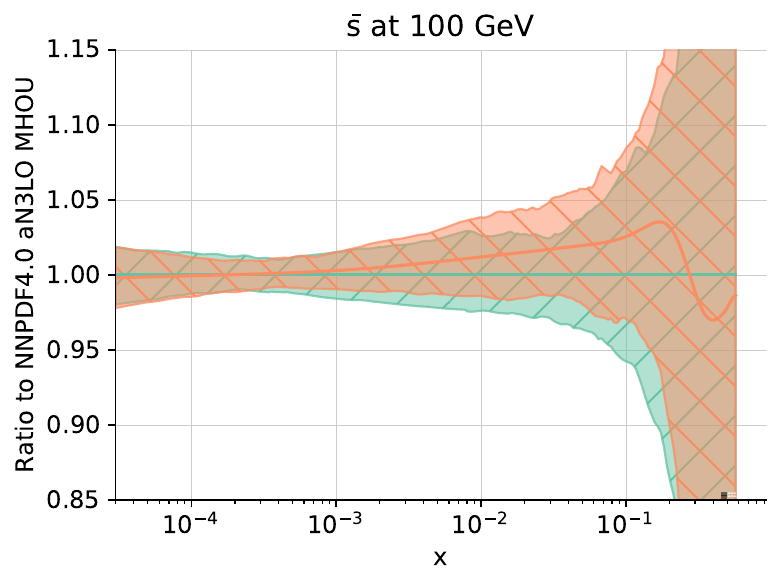}
  \includegraphics[width=0.45\textwidth]{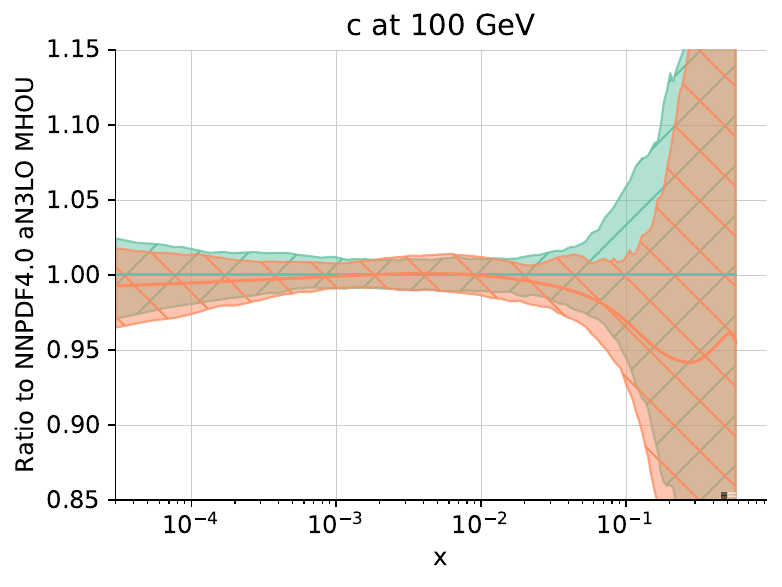}
  \includegraphics[width=0.45\textwidth]{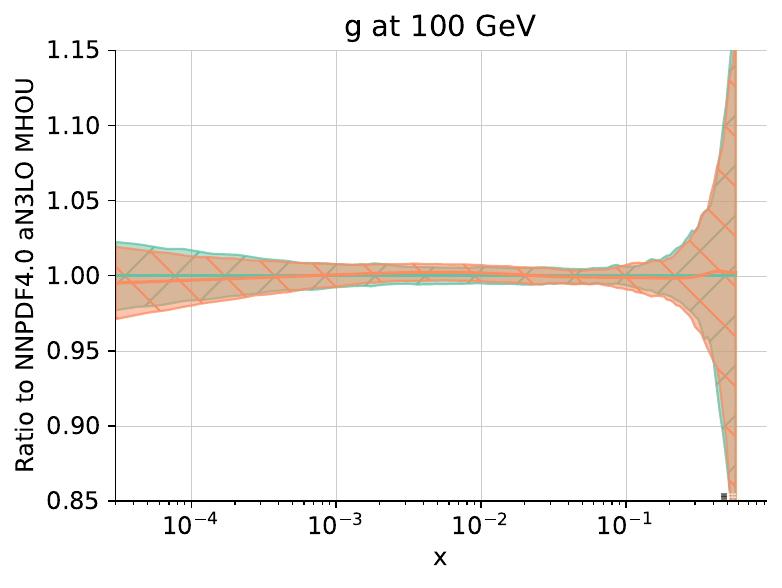}
  \caption{Same as Fig.~\ref{fig:hadronic_noMHOU} for aN$^3$LO
  PDF sets with MHOUs.}
  \label{fig:hadronic_MHOU} 
\end{figure}

As discussed in Sect.~\ref{sec:n3lo_hadronic_coeff} N$^3$LO
corrections to hadronic hard cross-sections are not included in our
default aN$^3$LO PDF determination. We can however assess the impact of the
inclusion of all the publicly available corrections for all relevant
data in the NNPDF4.0 dataset, listed in Tab.~\ref{tab:DY-kf} and
discussed in Sect.~\ref{sec:n3lo_hadronic_coeff}, using the
methodology discussed in that Section. To this purpose, we have
repeated the aN$^3$LO PDF determination, with and without  
MHOUs, now with N$^3$LO corrections for these datasets included. Note that for
these processes NNLO corrections are currently already included
through $K$-factors, hence this requires combining NNLO and N$^3$LO
$K$-factors, with ensuing loss of accuracy, which is one of the
reasons why these corrections are not included in our default aN$^3$LO
determination.  MHOUs on these N$^3$LO predictions can be  determined by
scale variation as usual, given that  renormalization scale variation
at  N$^3$LO
only requires 
knowledge of the NNLO result.

  \begin{table}[!t]
    \scriptsize
    \centering
    \renewcommand{\arraystretch}{1.4}
    \begin{tabularx}{\textwidth}{Xrccrcc}
  \toprule
  & \multicolumn{3}{c}{default}
  & \multicolumn{3}{c}{with N$^3$LO $K$-factors} \\
  Dataset
  & $N_{\rm dat}$
  & no MHOU
  & MHOU 
  & $N_{\rm dat}$
  & no MHOU
  & MHOU \\
  \midrule
  ATLAS high-mass DY 7 TeV
  &  5 & 1.64 & 1.68
  &  5 & 1.63 & 1.56 \\
  ATLAS $Z$ 7 TeV ($\mathcal{L}=35$~pb$^{-1}$)
  &  8 & 0.56 & 0.61 
  &  8 & 0.47 & 0.52 \\
  ATLAS $Z$ 7 TeV ($\mathcal{L}=4.6$~fb$^{-1}$) CC
  & 24 & 1.66 & 1.69 
  & 24 & 1.90 & 1.59 \\
  ATLAS $\sigma_{Z}^{\rm tot}$ 13 TeV
  & 1 & 0.24 & 0.66
  & 1 & 0.06 & 0.00 \\
  ATLAS $\sigma_{W}^{\rm tot}$ 13 TeV
  &   2 & 1.38 & 1.67 
  &   2 & 1.33 & 1.59 \\
\bottomrule
\end{tabularx}

    \caption{\small The number of data points and the $\chi^2$ per
      data point for the datasets of Table~\ref{tab:DY-kf}  comparing
      the default fits (same as Table~\ref{tab:chi2_DY}) to fits in
      which N$^3$LO corrections are included following the methodology
      of Sect.~\ref{sec:n3lo_hadronic_coeff}, 
      in both cases with and without MHOUs.
    }
    \label{tab:DY_chi2_kfactor}
  \end{table}
  The  fit quality for the datasets of Table~\ref{tab:DY-kf} both in the
default determinations (with and without MHOU, same as Table~\ref{tab:chi2_DY})
and after the inclusion of N$^3$LO corrections (also with and without
MHOU) is shown in Table~\ref{tab:DY_chi2_kfactor}.
The ensuing PDFs are compared to the default NNPDF4.0 aN$^3$LO in
Fig.~\ref{fig:hadronic_noMHOU} 
without MHOUs and in Fig.~\ref{fig:hadronic_MHOU} with MHOUs. In both cases,
PDFs are displayed at $Q=100$~GeV and are normalized to the central value of
the corresponding default NNPDF4.0 aN$^3$LO set. 
The impact of these N$^3$LO corrections on fit quality is very
moderate, though in all cases but one (and in all cases with MHOUs) it
leads to improved agreement. At the level of PDFs, however, the impact
is only (barely) visible in the PDFs
without MHOUs, and even in this case it only significantly impacts the
strange and antistrange PDFs, for which it  leads to an enhancement of
4-5\% for $x\sim 0.1$, though well within the PDF uncertainty. Even this
small effect is absent in the PDFs with MHOUs. We conclude that at
present available N$^3$LO corrections for hadronic processes
have no effect on PDF determination. On the other hand, the
improvement in fit quality is reassuring, and suggests that a more
significant effect might be seen once N$^3$LO corrections become
available for a wider set of processes.

\subsection{Comparison with MSHT20}
\label{sec:msht20}

We compare the NNPDF4.0 aN$^3$LO PDF set to the only other existing aN$^3$LO
PDF set, MSHT20 aN$^3$LO~\cite{McGowan:2022nag}. As already discussed
in Sect.~\ref{sec:n3lo_comp}, MSHT20 aN$^3$LO PDFs are determined by fitting
to the data the nuisance parameters that parametrize the IHOU
uncertainty on a prior approximation to splitting functions.
It follows that the ensuing central
value is partly determined by the data, and the IHOU is entirely
data-driven. When comparing NNPDF4.0 and MSHT20 aN$^3$LO PDF sets it should of
course be borne in mind that the sets already differ at NNLO due to
differences in dataset and methodology. The NNLO MSHT20 and NNPDF4.0
PDF sets were compared in Fig.~21 and the corresponding parton
luminosity in Fig.~60 of Ref.~\cite{NNPDF:2021njg}, while a detailed
benchmarking was presented in Ref.~\cite{PDF4LHCWorkingGroup:2022cjn}.

\begin{figure}[!p]
  \centering
  \includegraphics[width=0.45\textwidth]{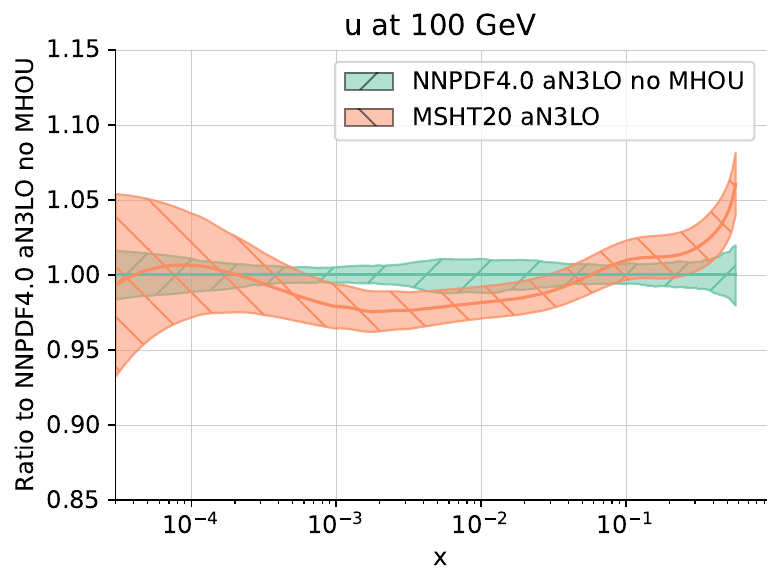}
  \includegraphics[width=0.45\textwidth]{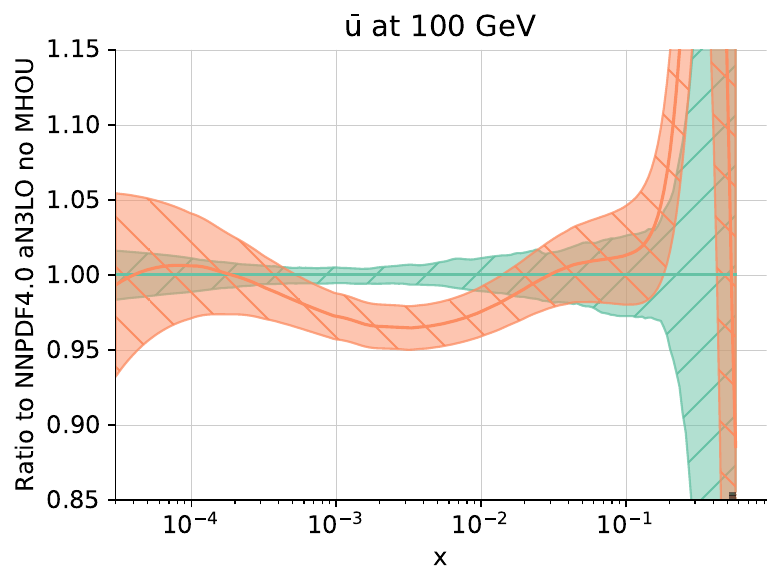}\\
  \includegraphics[width=0.45\textwidth]{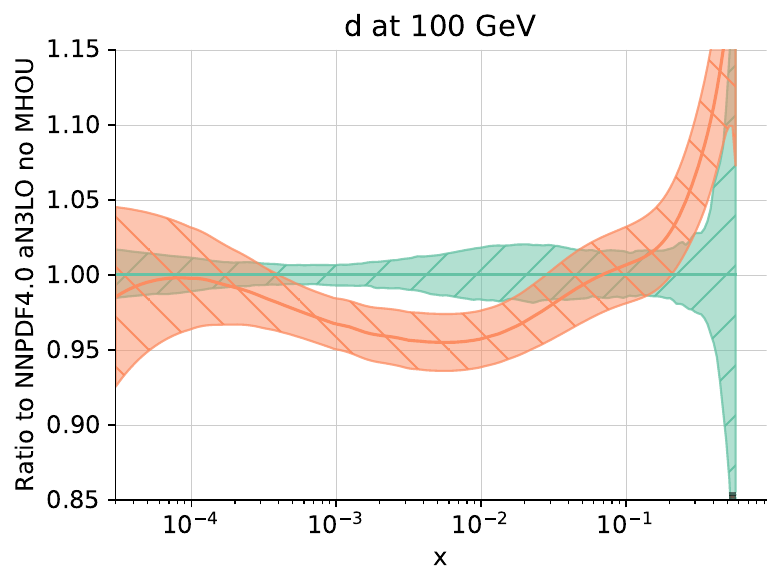}
  \includegraphics[width=0.45\textwidth]{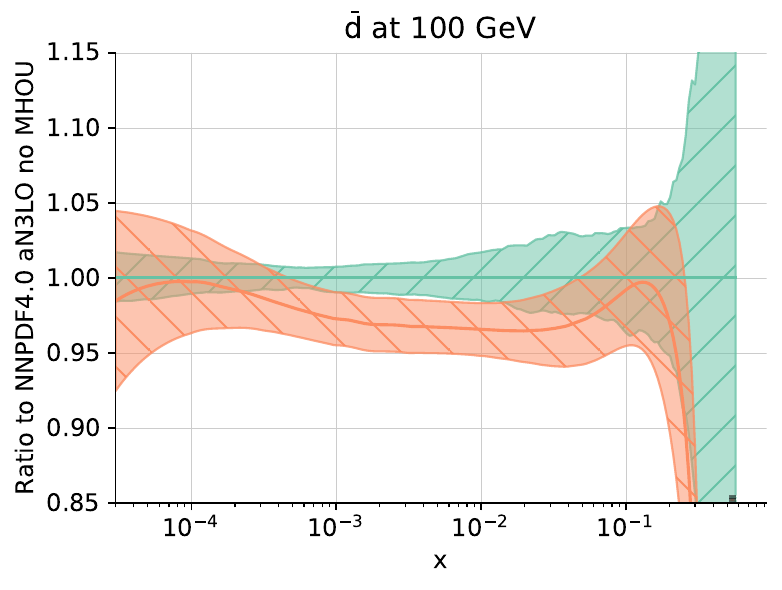}\\
  \includegraphics[width=0.45\textwidth]{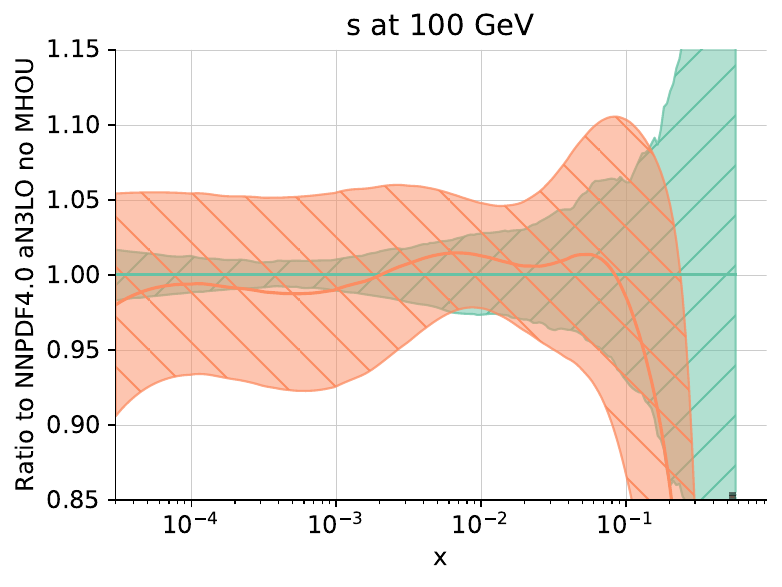}
  \includegraphics[width=0.45\textwidth]{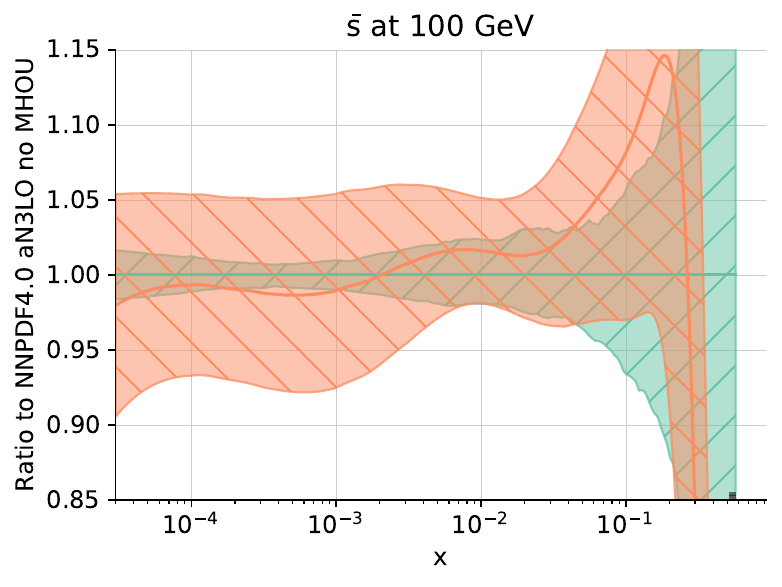}\\
  \includegraphics[width=0.45\textwidth]{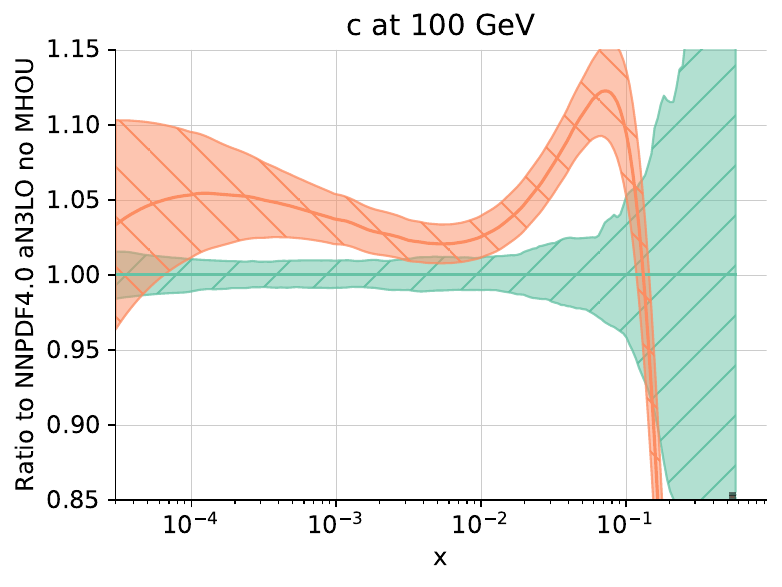}
  \includegraphics[width=0.45\textwidth]{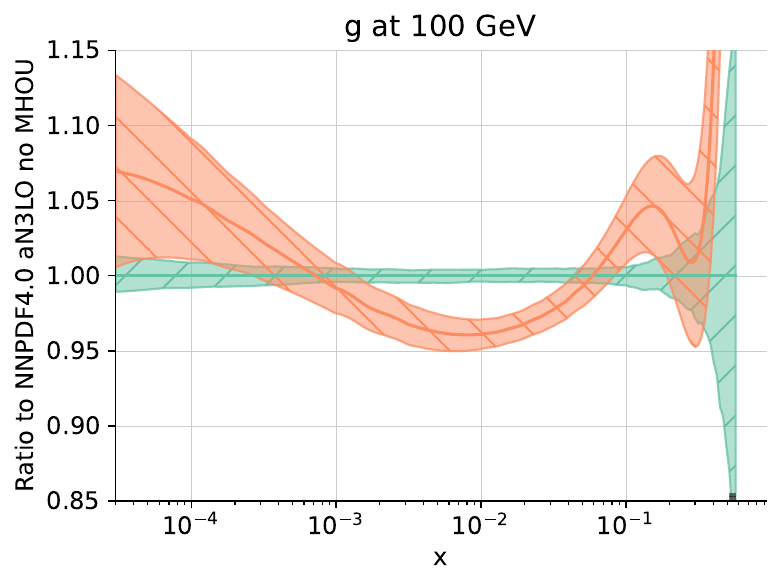}\\
  \caption{Same as
    Fig.~\ref{fig:pdfs_noMHOU_log}, now
    comparing the NNPDF4.0 aN$^3$LO baseline PDF set without
    MHOUs to  the MSHT20 set recommended as baseline in
    Ref.~\cite{McGowan:2022nag}.}
  \label{fig:PDFs_MSHT20}
\end{figure}

\begin{figure}[!t]
  \centering
  \includegraphics[width=0.45\textwidth]{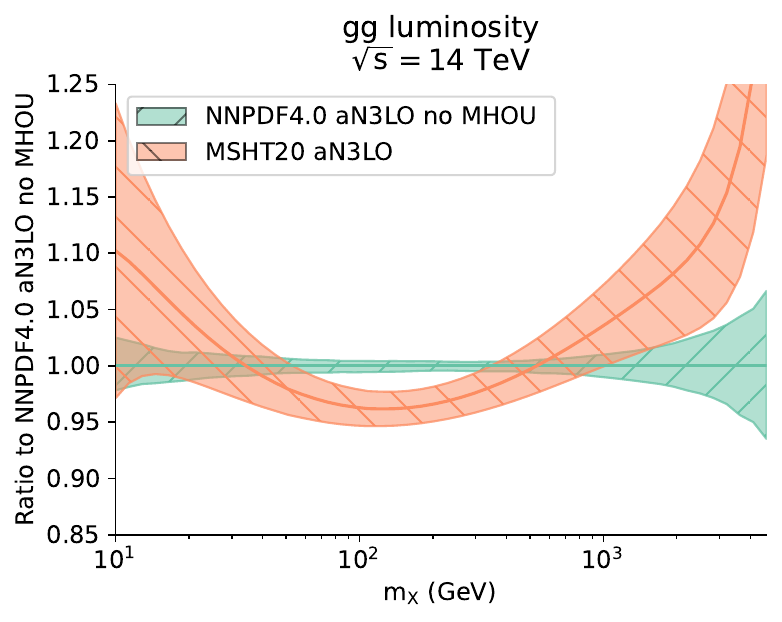}
  \includegraphics[width=0.45\textwidth]{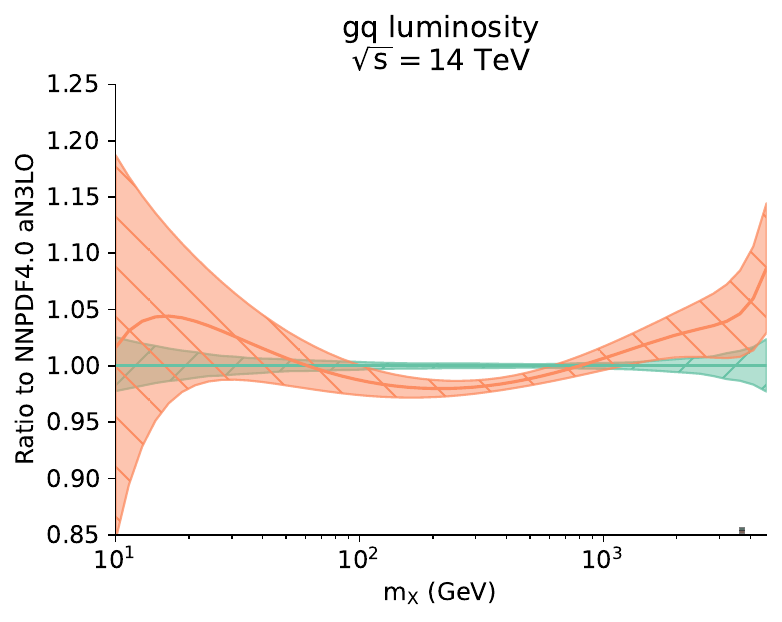}\\
  \includegraphics[width=0.45\textwidth]{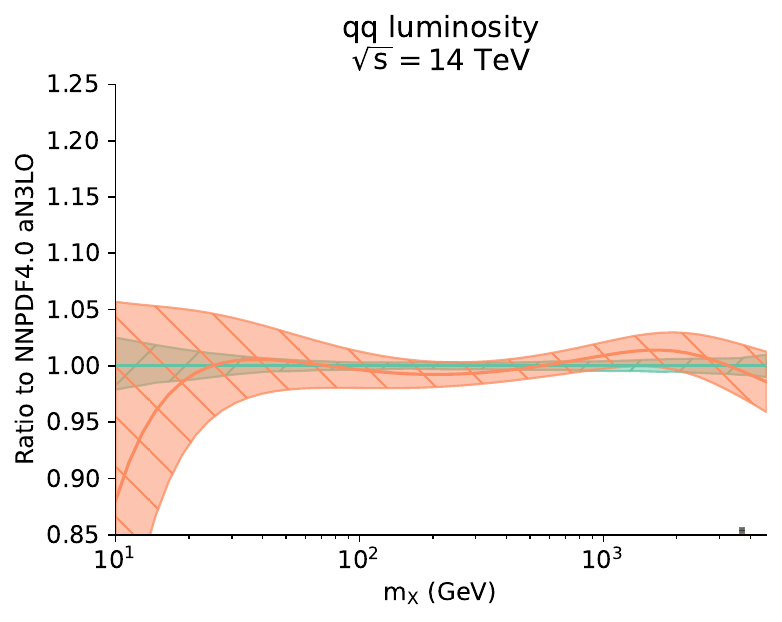}
  \includegraphics[width=0.45\textwidth]{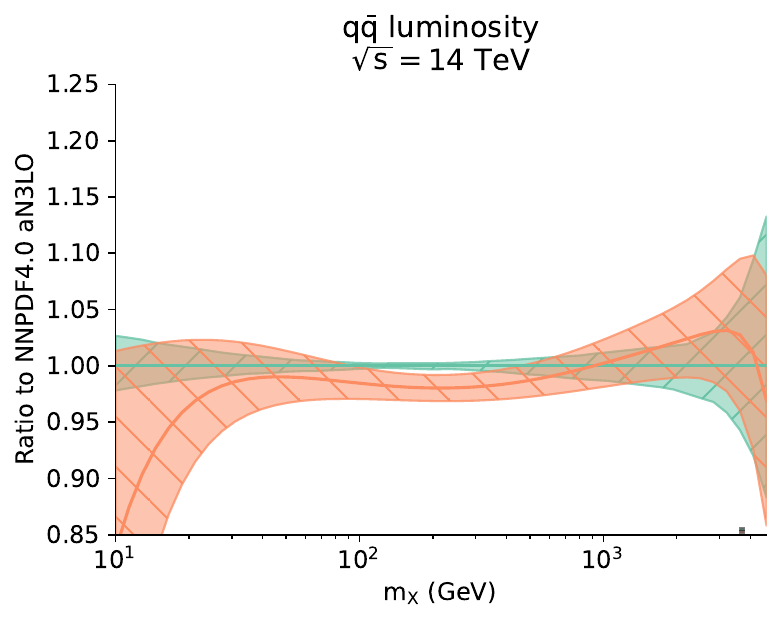}\\
  \caption{Same as Fig.~\ref{fig:PDFs_MSHT20} for parton luminosities as in
    Fig.~\ref{fig:lumis}.}
  \label{fig:lumis_MSHT20}
\end{figure}

The comparison of the aN$^3$LO sets is presented in
Fig.~\ref{fig:PDFs_MSHT20}, where we show the  NNPDF4.0
no MHOU set and the  MSHT20 set recommended as baseline in
Ref.~\cite{McGowan:2022nag} at $Q=100$~GeV, normalized to the NNPDF4.0 central
value. All error bands are  one sigma
uncertainties.  The dominant differences between the PDF sets are the
same as already observed at NNLO, with  the largest difference
observed for the charm PDF, which is independently parametrized in
NNPDF4.0, but not in MSHT20, where it is determined by perturbative
matching conditions. However, the differences, while remaining qualitatively
similar, are slightly reduced (by 1-2\%) when moving from NNLO to aN$^3$LO.
Exceptions are the charm and especially the gluon PDF, which differ more at
aN$^3$LO. Specifically, the gluon PDF,
while reasonably compatible for 
$x\lesssim 0.07$ at NNLO, disagrees  at aN$^3$LO, with the
MSHT20 result suppressed by 3-4\% in the region
$10^{-3}\lesssim x \lesssim 10^{-1}$, with a PDF uncertainty of
1-2\%. This suppression of the MSHT20 gluon can likely be
traced to the behavior of the $P_{gq}$  splitting function seen in Fig.~\ref{fig:splitting-functions-mhst}.

\begin{figure}[!p]
  \centering
  \includegraphics[width=0.45\textwidth]{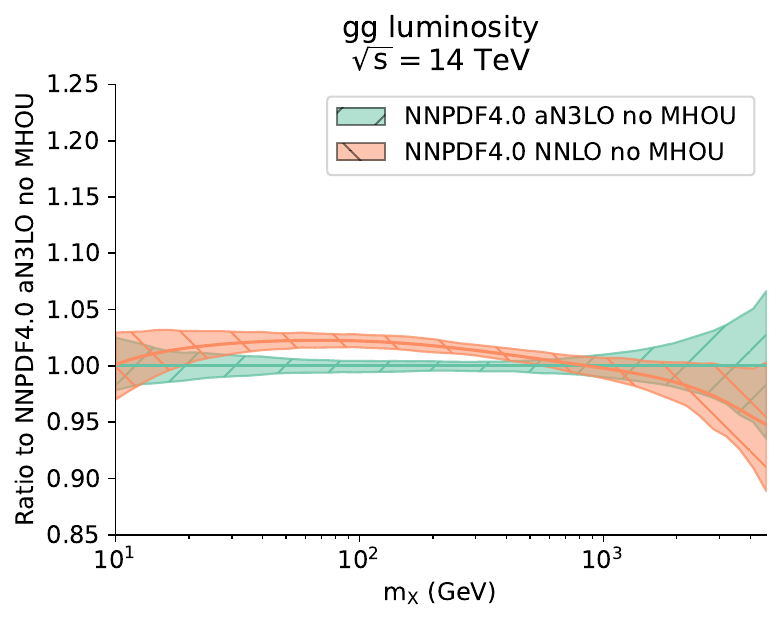}
  \includegraphics[width=0.45\textwidth]{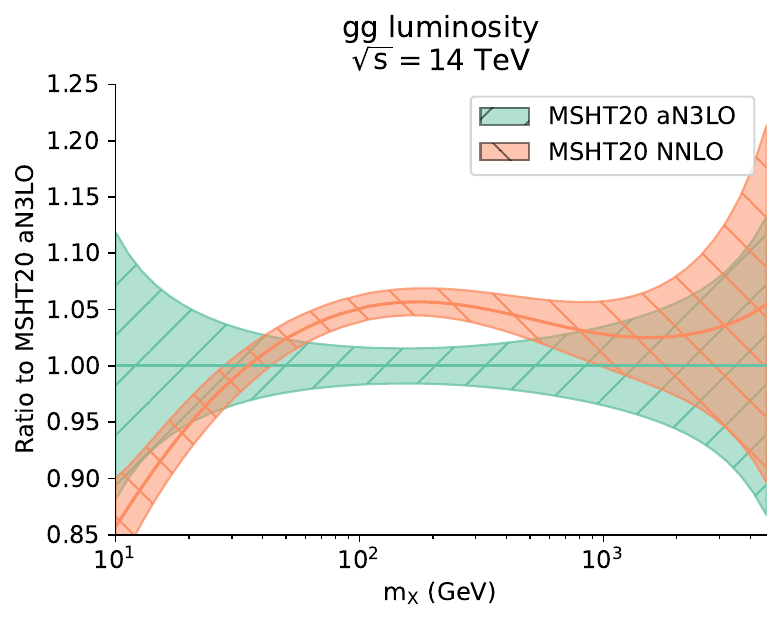}\\
  \includegraphics[width=0.45\textwidth]{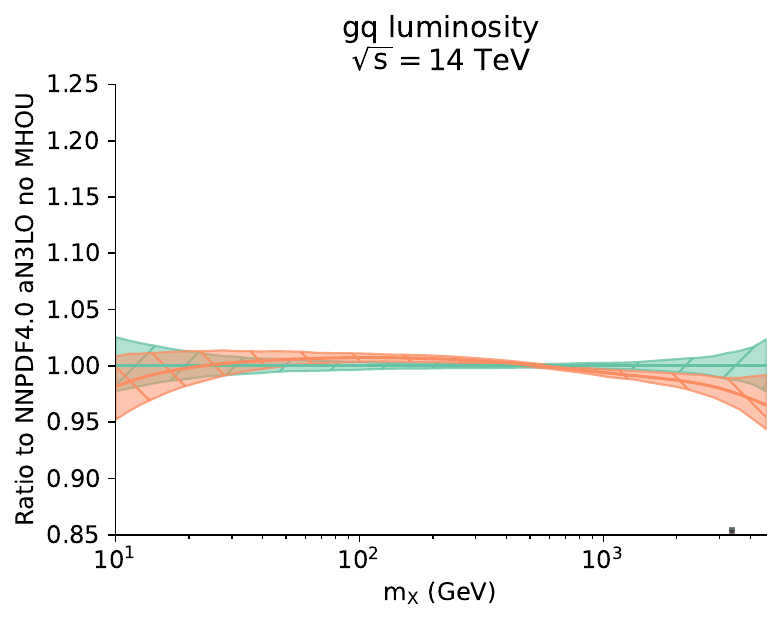}
  \includegraphics[width=0.45\textwidth]{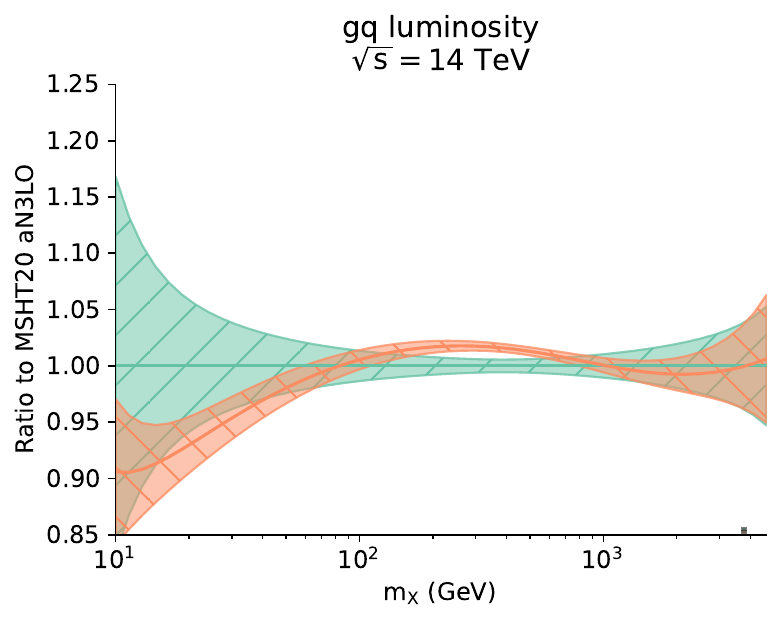}\\
  \includegraphics[width=0.45\textwidth]{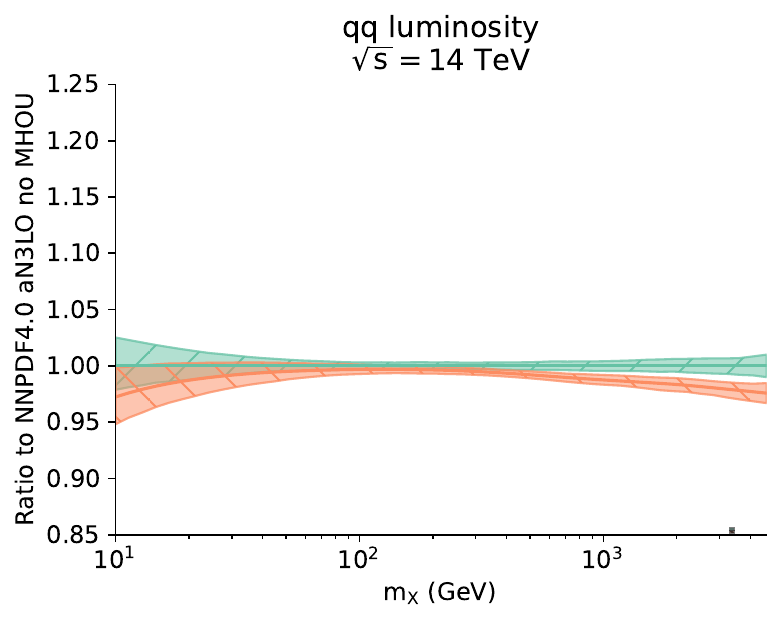}
  \includegraphics[width=0.45\textwidth]{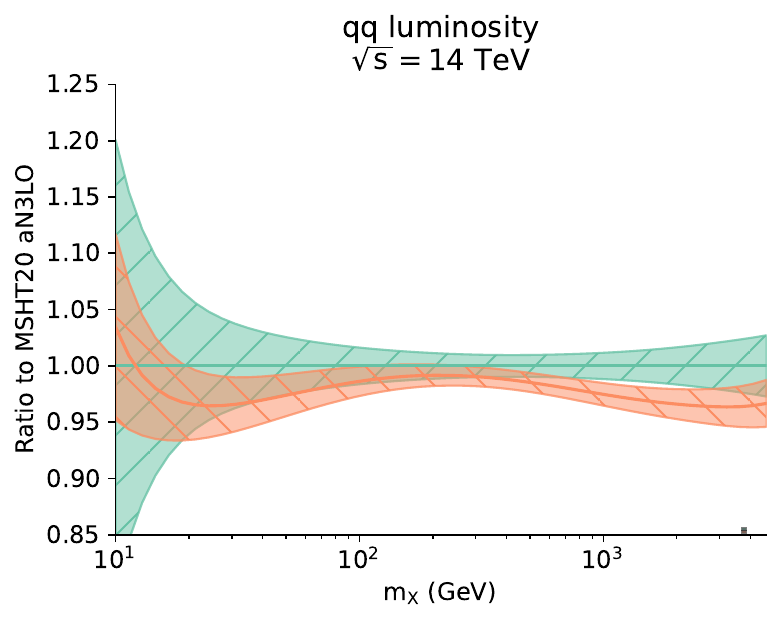}\\
  \includegraphics[width=0.45\textwidth]{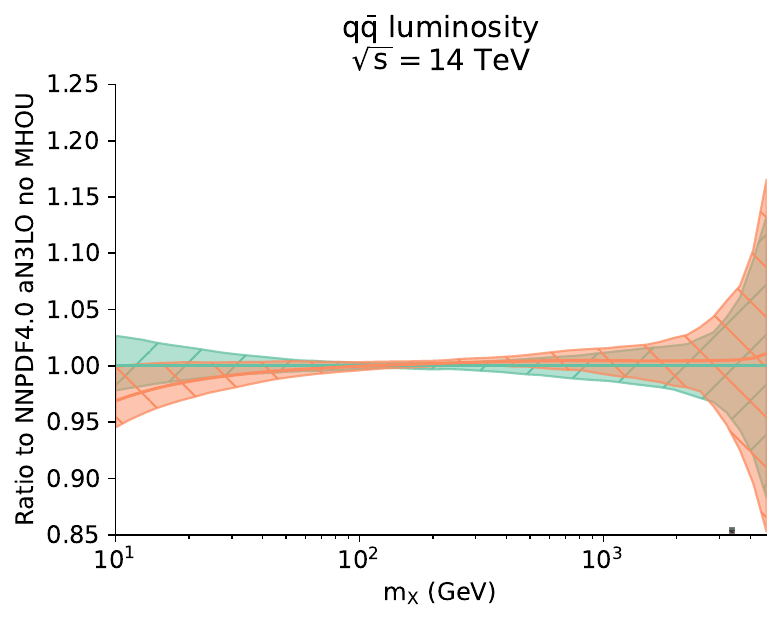}
  \includegraphics[width=0.45\textwidth]{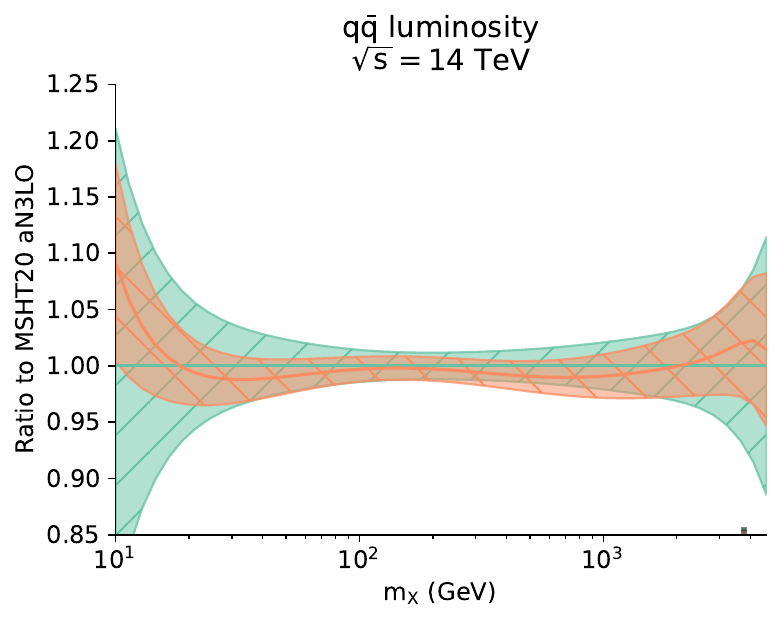}\\
  \caption{Same as Fig.~\ref{fig:lumis_MSHT20}, now comparing aN$^3$LO and NNLO
    parton luminosities, separately for the NNPDF4.0 (left) and MSHT20
    (right) PDF sets, normalized to the aN$^3$LO result.}
  \label{fig:lumis_NLO_vs_N3LO}
\end{figure}

Parton luminosities are compared in
Fig.~\ref{fig:lumis_MSHT20}. Again the pattern is similar to that seen
at NNLO, but now with a considerable  suppression of the gluon-gluon and
gluon-quark luminosities in the $M_X\sim100$~GeV region that can be
traced to the behavior of the gluon PDF seen in Fig.~\ref{fig:PDFs_MSHT20}.
The quark-quark luminosity remains similar in MSHT20 and NNPDF4.0 both at NNLO
and aN$^3$LO. The impact of these effects on the computation of
precision LHC cross-sections will be addressed in Sect.~\ref{sec:pheno}.

In order to understand better the comparative impact of aN$^3$LO corrections,
we compare for each set the NNLO and aN$^3$LO luminosities. Results are shown
in Fig.~\ref{fig:lumis_NLO_vs_N3LO}, normalized to the
aN$^3$LO result. The qualitative impact of the aN$^3$LO corrections on
either set is similar, but with a  stronger aN$^3$LO suppression of gluon
luminosities for MSHT20. In particular the gluon-gluon luminosity is
suppressed for $10^2\lesssim m_X\lesssim 10^3$~GeV  by about 3\% in NNPDF4.0
and 6\% in MSHT20 and the gluon-quark luminosity is suppressed in the same
region by about 1\% in NNPDF4.0 and 3\% in MSHT20. 
In the case of the gluon-gluon luminosity the differences between
NNLO and aN$^3$LO  are larger than the respective PDF uncertainties
(that do not include MHOUs in either case). As already mentioned in
Sect.~\ref{sec:n3lo_comp}, a dedicated benchmark of aN$^3$LO results
is ongoing and will be presented in Ref.~\cite{lh24}.

\section{LHC phenomenology at aN$^3$LO accuracy}
\label{sec:pheno}

We  present a first assessment of
the implications of aN$^3$LO PDFs
for LHC phenomenology, by looking at processes for which N$^3$LO
results are publicly available, namely the Drell-Yan and Higgs total
inclusive cross-sections.
We  present predictions at NLO, NNLO, and aN$^3$LO using both 
NNPDF4.0 and MSHT20 PDFs, consistently matching the perturbative order
of the PDF and matrix element. At N$^3$LO  we also show results
obtained  with the currently common approximation
of using NNLO PDFs with aN$^3$LO matrix elements.

At each perturbative order, the uncertainty on the cross-section is
determined by adding in quadrature the PDF uncertainty to the MHOU on
the hard matrix element determined
 performing 7-point renormalization and factorization scale
 variation and taking the envelope of the results. This is the procedure
 that is most commonly used for the estimation of the total
 uncertainty on hadron collider processes, and we follow it here for
 ease of comparison with available results. In a more refined treatment, 
 MHOUs on the hard cross-section can be included through a theory covariance
matrix for the hard cross-section itself, like the MHOUs and IHOUs on
the PDF. This would then make it possible to keep track of the correlation
between these different sources of
uncertainty~\cite{Harland-Lang:2018bxd,Ball:2021icz,Kassabov:2022orn}.

Here we  plot 
results with a total uncertainty obtained combining these
uncertainties in quadrature (both with and
without MHOUs in the fit), and we also tabulate this total uncertainty (without
MHOUs in the fit) along with the PDF uncertainty both with and without MHOUs.   
Also, in order to assess the impact of the use of aN$^3$LO PDFs,
we plot N$^3$LO results obtained using NNLO  and aN$^3$LO PDFs, we
tabulate the shift  between the N$^3$LO
prediction obtained using NNLO and aN$^3$LO PDFs, and we compare it to previous
estimate of this expected shift based on the differences between NNLO
and NLO PDFs. Indeed, predictions for processes computed at N$^3$LO
accuracy are commonly obtained using NNLO PDFs, with an extra
uncertainty assigned to the result dues to this mismatch in
perturbative order between the PDF and the matrix element. A commonly
used prescription in order to estimate this
uncertainty~\cite{Anastasiou:2016cez,Baglio:2022wzu} is to take it
equal to
    \begin{equation}
      \label{eq:PDFimpact_xsec_approx}
\Delta^{\rm app}_{\rm NNLO} \equiv \frac{1}{2}\Bigg| \frac{\sigma^{\rm NNLO}_{\rm NNLO-PDF}
      - \sigma^{\rm NNLO}_{\rm NLO-PDF}}{\sigma^{\rm NNLO}_{\rm
          NNLO-PDF}}\Bigg|,
    \end{equation}
    namely to assume that the same percentage shift, computed at one less
    perturbative order, would be twice as large.
This prescription can now be compared to the exact result
        \begin{equation}
     \label{eq:PDFimpact_xsec_exact}
      \Delta^{\rm exact}_{\rm NNLO} \equiv \Bigg| \frac{\sigma^{\rm N^3LO}_{\rm N^3LO-PDF}
      - \sigma^{\rm N^3LO}_{\rm NNLO-PDF}}{\sigma^{\rm N^3LO}_{\rm N^3LO-PDF}}\Bigg| \, .
        \end{equation}

\subsection{Inclusive Drell-Yan production}
\label{sec:gaugeboson}

\begin{figure}[!p]
  \centering
  \includegraphics[width=0.47\linewidth]{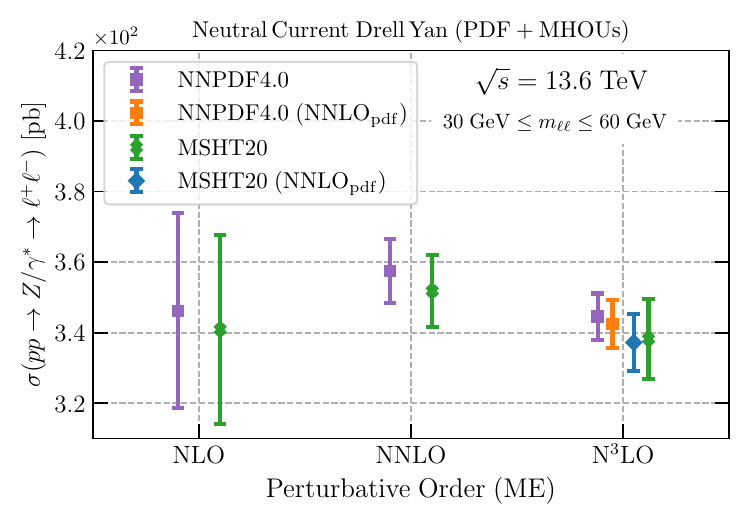}
  \includegraphics[width=0.47\linewidth]{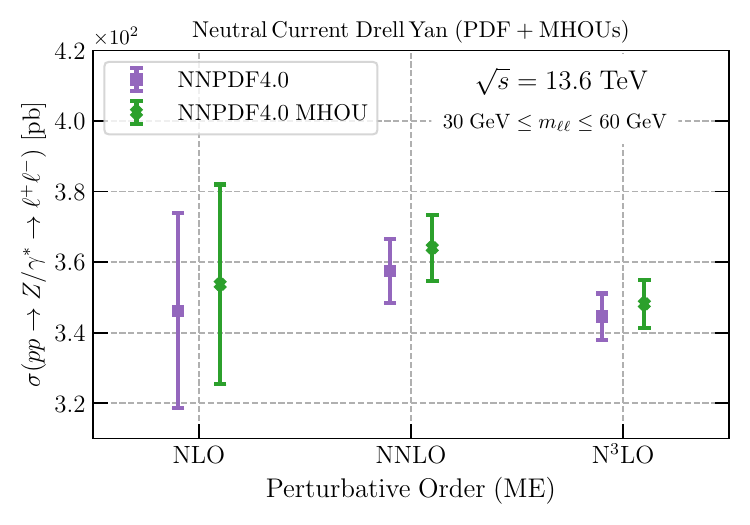}
  \includegraphics[width=0.47\linewidth]{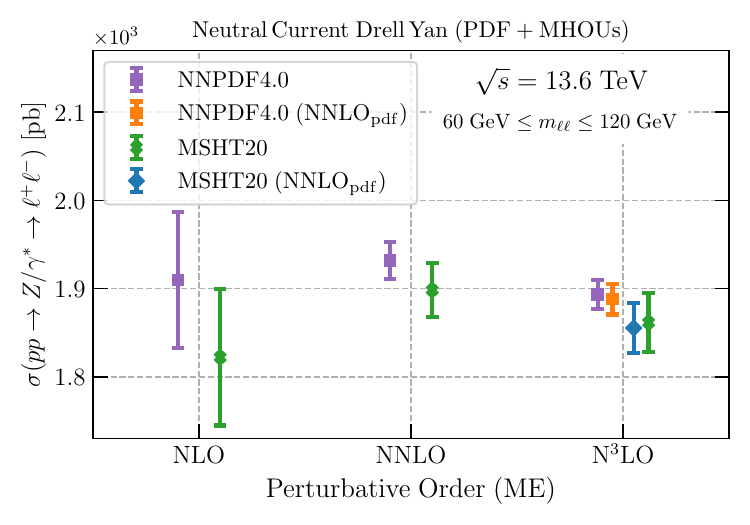}
  \includegraphics[width=0.47\linewidth]{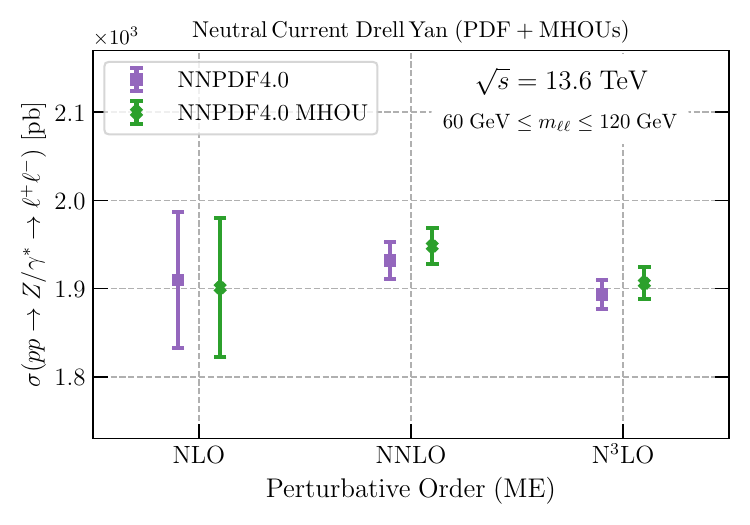}
  \includegraphics[width=0.47\linewidth]{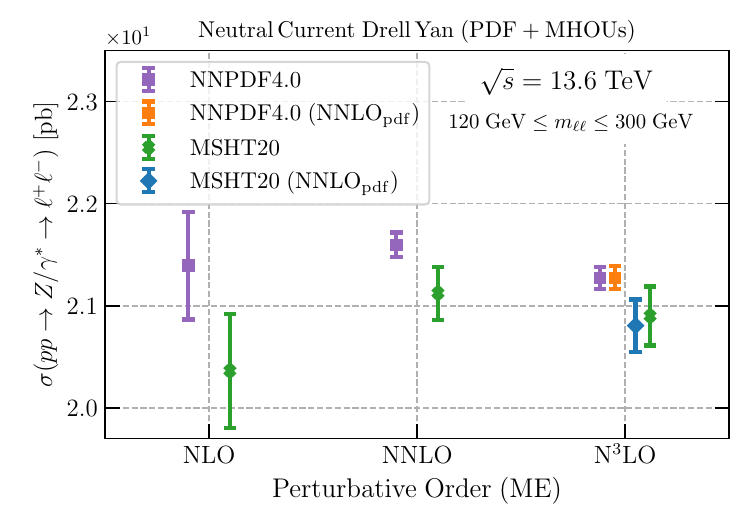}
  \includegraphics[width=0.47\linewidth]{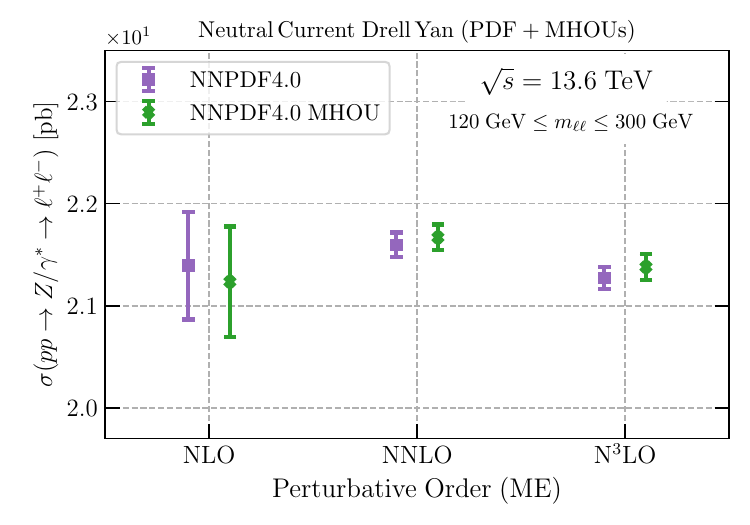}
  \includegraphics[width=0.47\linewidth]{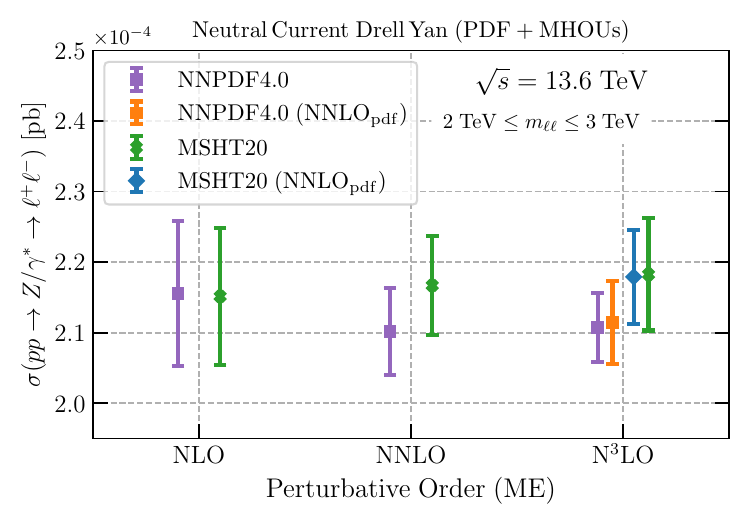}
  \includegraphics[width=0.47\linewidth]{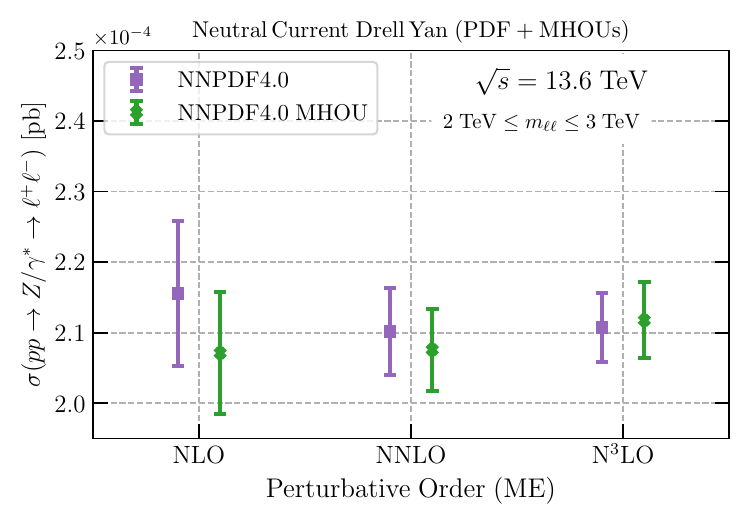}
  \caption{\small The inclusive neutral-current Drell-Yan
    production cross-section, $pp\to \gamma^*/Z \to \ell^+\ell^-$,
    for different ranges of the dilepton invariant mass $Q=m_{\ell\ell}$, from
  low to high invariant masses (top to bottom). Results are shown (left)
  comparing NLO, NNLO  and aN$^3$LO with matched perturbative order in the
  matrix element and PDF, and also at aN$^3$LO with NNLO PDFs using
  NNPDF4.0 and MSHT20 PDFs and
  (right) at aN$^3$LO  with PDFs without and with MHOUs.
}
  \label{fig:nc-dy-pheno} 
\end{figure}

\begin{figure}[!p]
  \centering
  \includegraphics[width=0.49\linewidth]{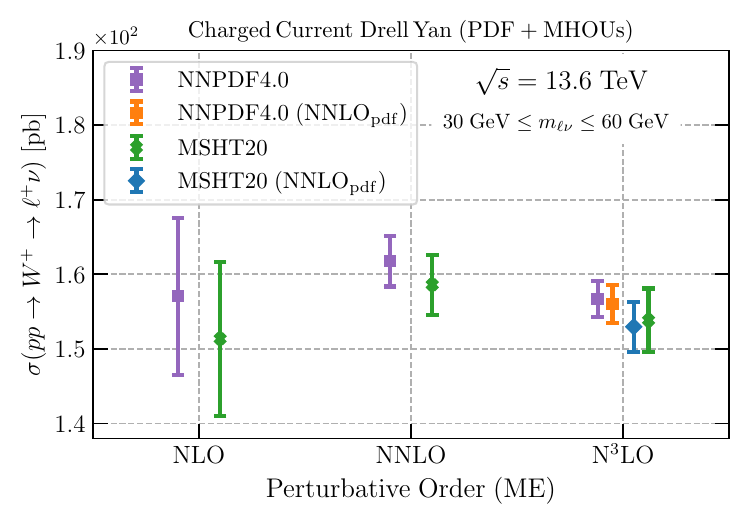}
  \includegraphics[width=0.49\linewidth]{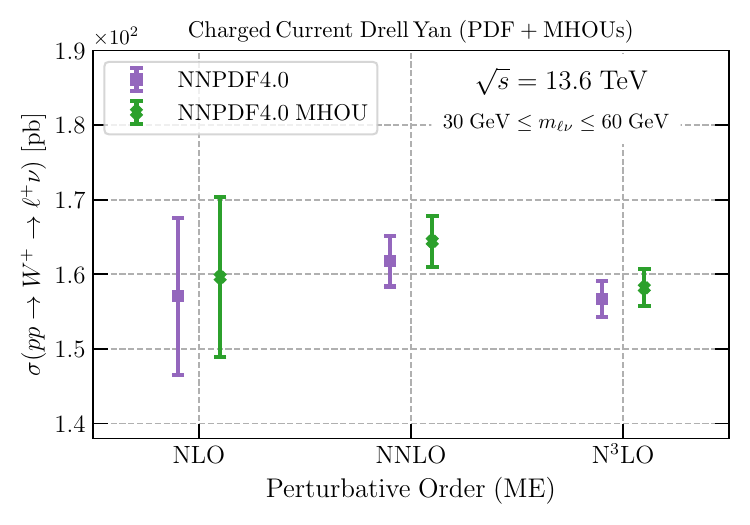}
  \includegraphics[width=0.49\linewidth]{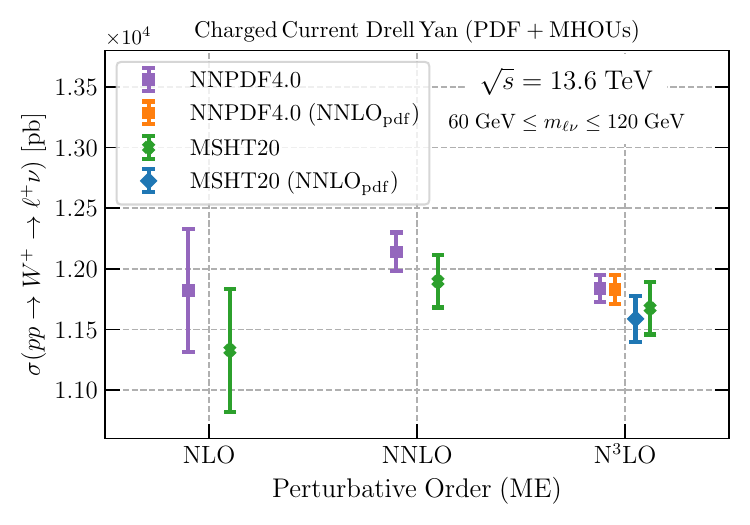}
  \includegraphics[width=0.49\linewidth]{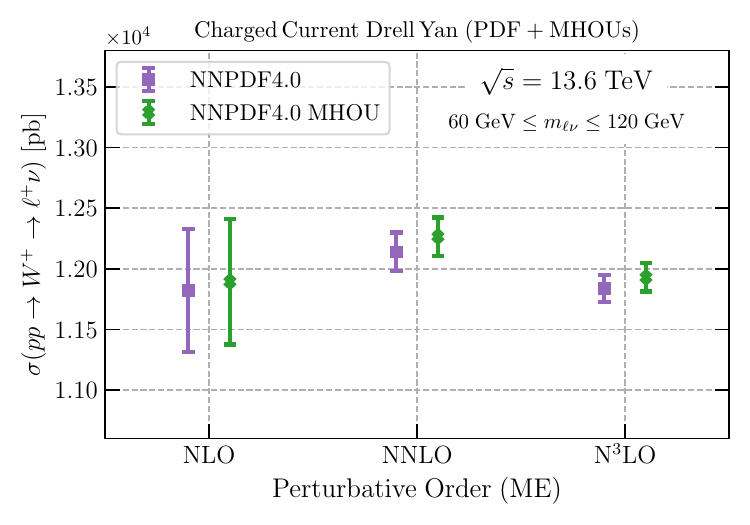}
  \includegraphics[width=0.49\linewidth]{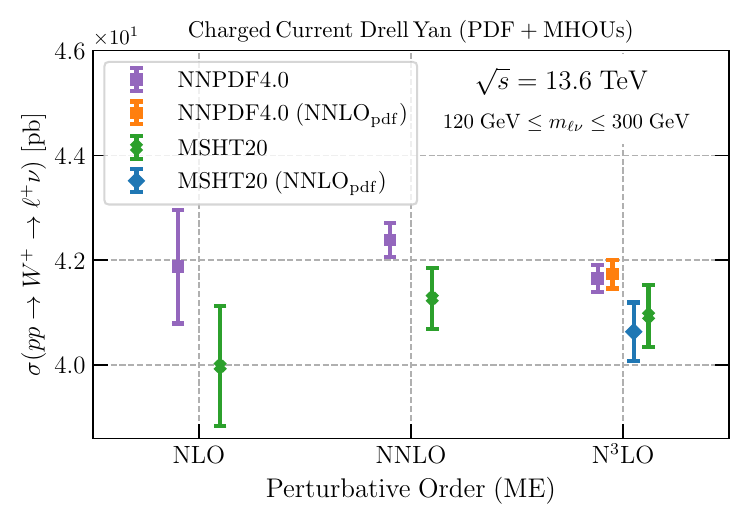}
  \includegraphics[width=0.49\linewidth]{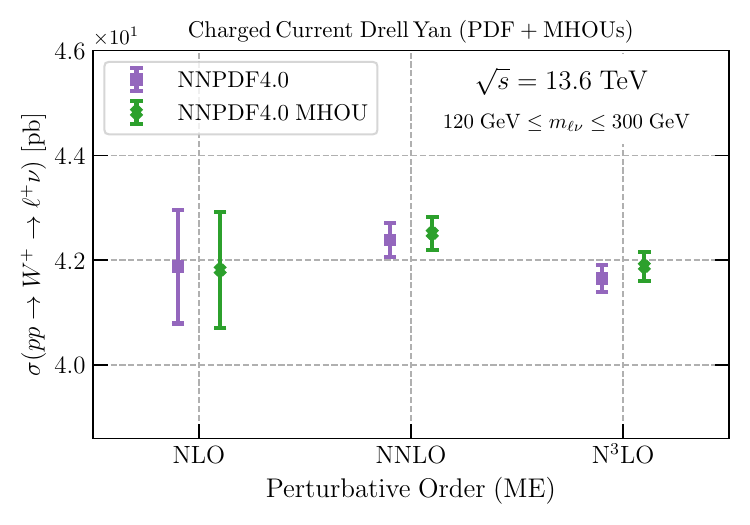}
  \includegraphics[width=0.49\linewidth]{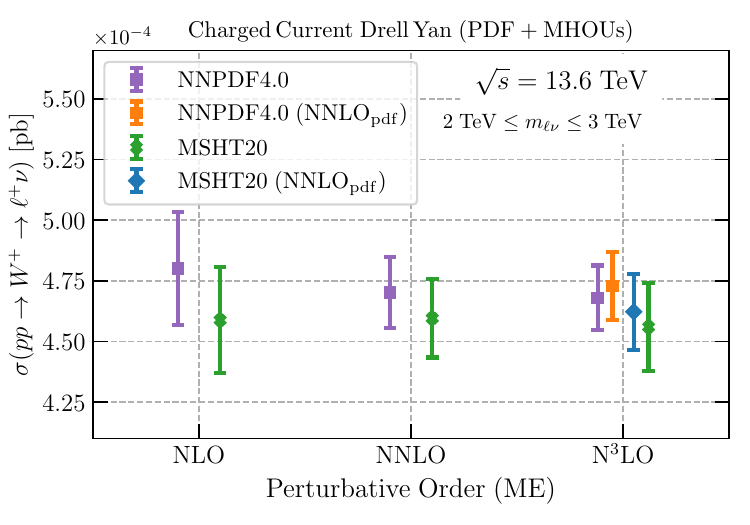}
  \includegraphics[width=0.49\linewidth]{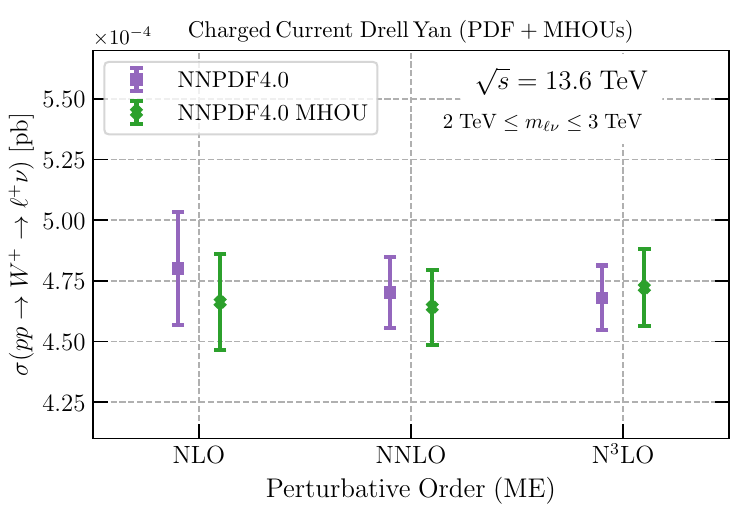}
  \caption{\small Same as Fig.~\ref{fig:nc-dy-pheno} for the inclusive charged-current Drell-Yan
    production cross-section, $pp\to W^+ \to \ell^+\nu_{\ell}$.
}
  \label{fig:ccp-dy-pheno} 
\end{figure}

\begin{figure}[!p]
  \centering
  \includegraphics[width=0.49\linewidth]{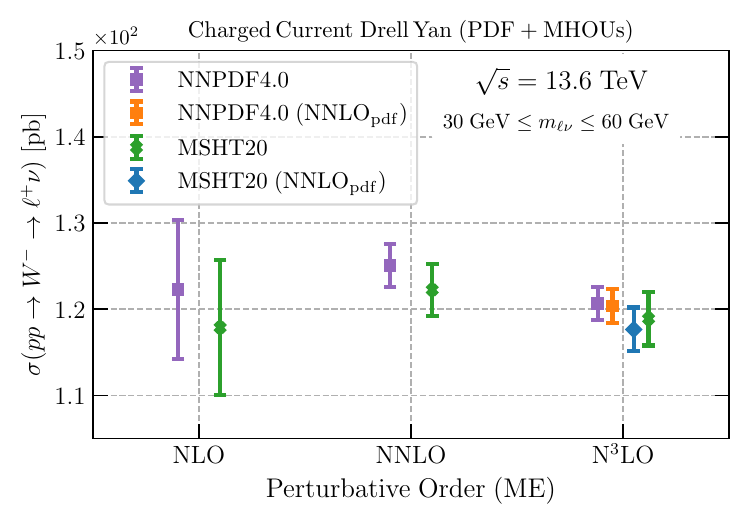}
  \includegraphics[width=0.49\linewidth]{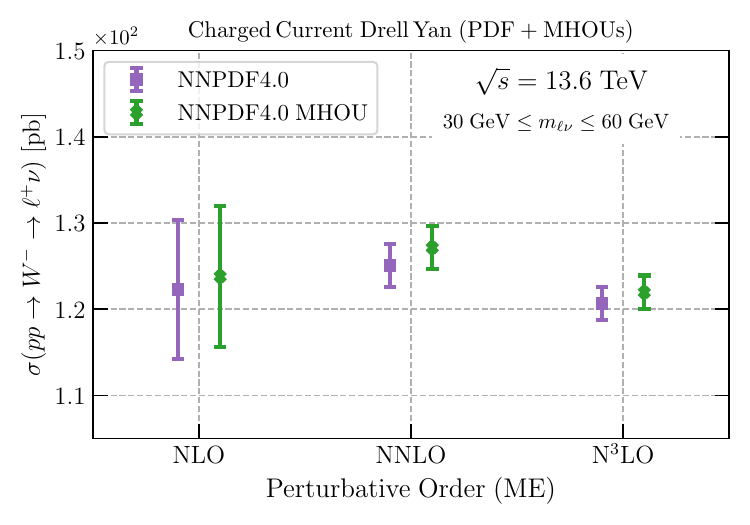}
  \includegraphics[width=0.49\linewidth]{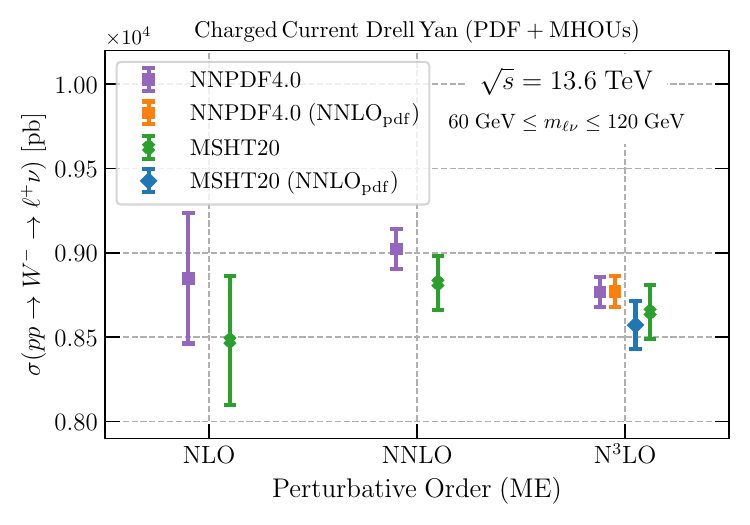}
  \includegraphics[width=0.49\linewidth]{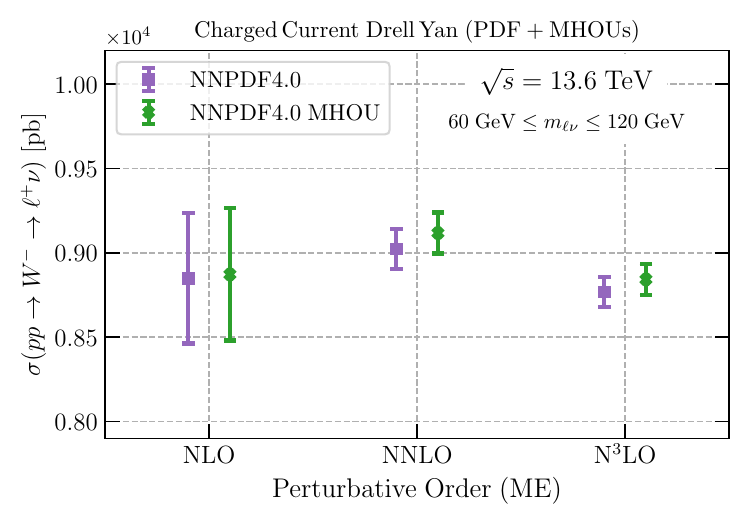}
  \includegraphics[width=0.49\linewidth]{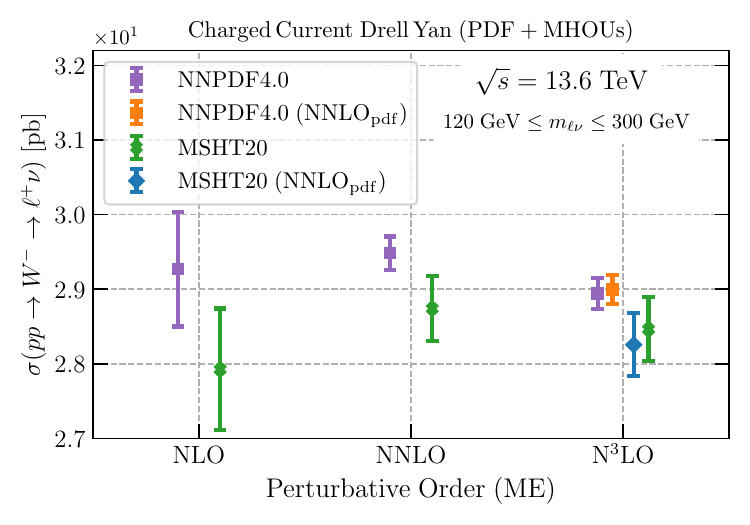}
  \includegraphics[width=0.49\linewidth]{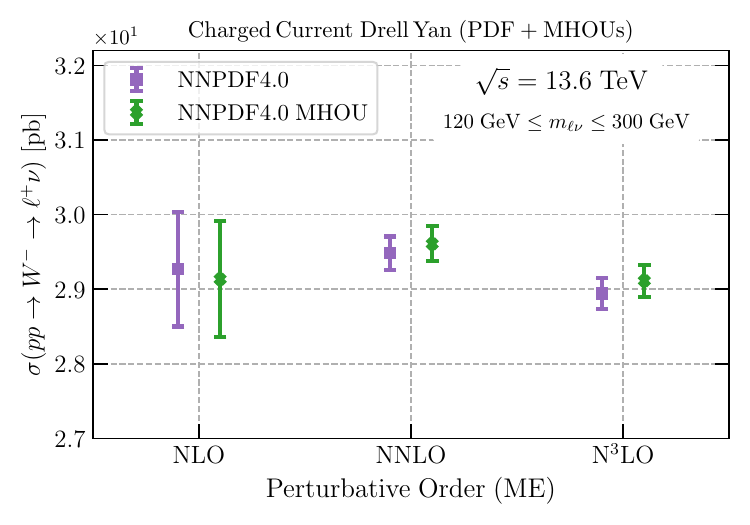}
  \includegraphics[width=0.49\linewidth]{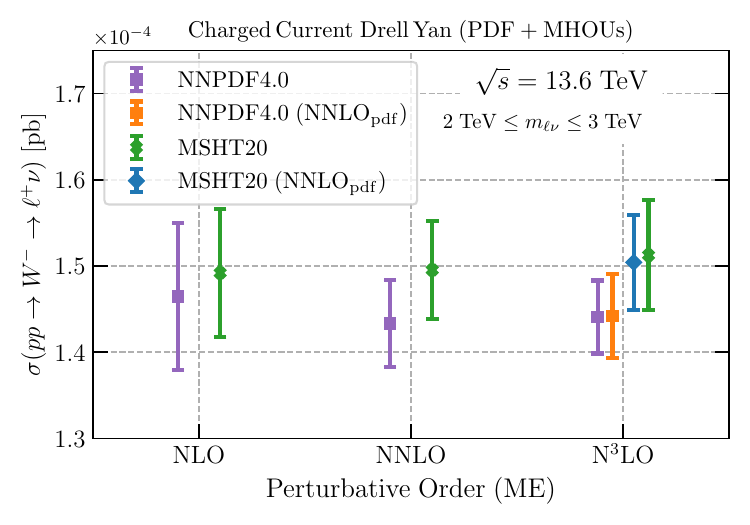}
  \includegraphics[width=0.49\linewidth]{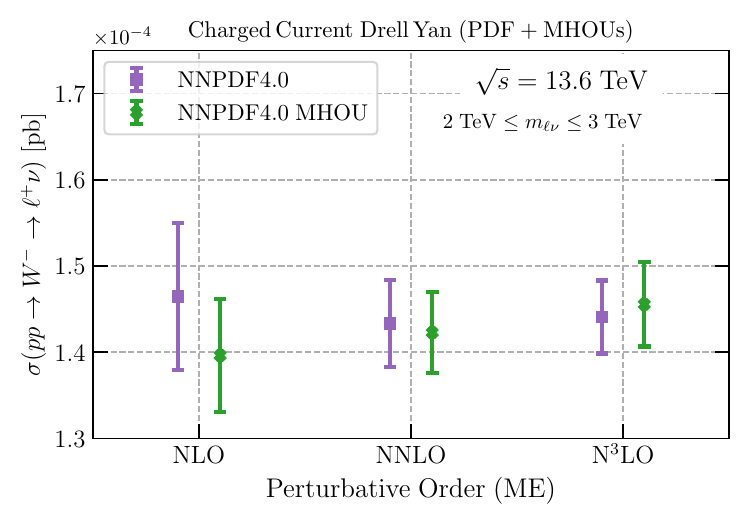}
  \caption{\small Same as Fig.~\ref{fig:nc-dy-pheno} for the inclusive charged-current Drell-Yan
    production cross-section, $pp\to W^- \to \ell^-\bar{\nu}_{\ell}$.
}
  \label{fig:ccm-dy-pheno} 
\end{figure}

We show results for inclusive charged-current and neutral-current
gauge boson production cross-sections 
followed by their decays into the dilepton final state.
Cross-sections are evaluated using the {\sc\small n3loxs} code~\cite{Baglio:2022wzu}
for different ranges in the  final-state dilepton
invariant mass, $Q=m_{\ell\ell}$ for neutral-current and $Q=m_{\ell \nu}$ for charged-current
scattering. Fig.~\ref{fig:nc-dy-pheno} displays
the inclusive neutral-current Drell-Yan
cross-section $pp\to \gamma^*/Z \to \ell^+\ell^-$ and
Figs.~\ref{fig:ccp-dy-pheno}-\ref{fig:ccm-dy-pheno} the
charged-current cross-sections  $pp\to W^\pm \to \ell^\pm\nu_{\ell}$.
We consider one low-mass bin ($30~{\rm GeV}\le Q \le 60~{\rm GeV}$),
the mass peak bin ($60~{\rm GeV}\le Q \le 120~{\rm GeV}$), and two high-mass
bins ($120~{\rm GeV}\le Q \le 300~{\rm GeV}$ and $2~{\rm
  TeV}\le Q\le 3~{\rm TeV}$), relevant for high-mass new physics searches~\cite{Ball:2022qtp}.
In all cases, we compare the NLO, NNLO, and aN$^3$LO predictions using
NNPDF4.0 and MSHT20 PDFs
determinations, with the same perturbative order in matrix element and
PDFs, and also the  aN$^3$LO result with NNLO PDFs, and then we
compare the aN$^3$LO with NNPDF4.0 aN$^3$LO PDFs with and without MHOUs.
The values of cross-sections and uncertainties are collected in
Table~\ref{tab:DY_unc}.

\begin{table}[!t]
  \scriptsize
  \centering
  \renewcommand{\arraystretch}{1.4}
    \begin{tabularx}{\textwidth}{Xccccccccccc}
    \toprule
 \multirow{2}{*}{Process} & 
  \multicolumn{6}{c}{NNPDF4.0}
  & \multicolumn{5}{c}{MSHT20} \\
 &  $\sigma$ (pb) & $\delta_{\rm th}$
  &   $\delta^{\rm no MHOU}_{\rm PDF}$  &
  $\delta^{\rm MHOU}_{\rm PDF}$ &   $\Delta^{\rm app}_{\rm NNLO}$
  & $\Delta^{\rm exact}_{\rm NNLO}$   
 &  $\sigma$ (pb) & $\delta_{\rm th} \sigma$
  &   $\delta_{\rm PDF}$&   $\Delta^{\rm app}_{\rm
    NNLO}~ $   &$\Delta^{\rm exact}_{\rm NNLO}$    \\
  \midrule
  $W^+$ (p)  &   $1.2 \times 10^{4}$  &    1.0 & 0.5 & 0.5   &  1.1     &
  0.1    &  $1.2 \times 10^{4}$    &    1.9 & 1.7  & 2.3   & 0.8  \\
  $W^-$ (p)  &   $8.8 \times 10^{3}$  &   1.0  &  0.5& 0.5  &  1.1    & 0.1
  &   $8.7 \times 10^{3}$    &   1.9  &  1.6 &   2.1 & 0.0   \\
  $Z$ (p)  &  $1.9 \times 10^{3}$   &   0.9 & 0.4 & 0.5   &    1.1   & 0.3
  &
  $1.9 \times 10^{3}$     &    1.8 & 1.6 &   2.6  & 0.3  \\
  $W^+$ (hm)  &   $4.7 \times 10^{-4}$  &  2.8   & 2.8 &  3.3   & 3.2     &   1.1
  &  $4.6 \times 10^{-4}$     &  4.0   & 3.9  &  2.0  & 1.3  \\
  $W^-$ (hm)  & $1.4 \times 10^{-4}$    &  2.9  & 2.9 &  3.3   &  3.3    &  0.1    &
  $1.5 \times 10^{-4}$    &   4.2  & 4.2  &  2.0  & 0.6  \\
  $Z$ (hm)  &  $2.1 \times 10^{-4}$   &    2.3 & 2.3 &   2.5  &   3.4   &  0.3
  & $2.2 \times 10^{-4}$      &  3.6 & 3.6    & 2.7   & 0.2  \\
\bottomrule
\end{tabularx}

  \vspace{0.3cm}
  \caption{The N$^3$LO cross-sections and uncertainties for the
    inclusive gauge boson production processes displayed in
    Figs.~\ref{fig:nc-dy-pheno}-\ref{fig:ccm-dy-pheno} and evaluated using the
    NNPDF4.0 and MSHT20 aN$^3$LO PDFs.
    We show the percentage total theory uncertainty $\delta_{\rm th}$,
    obtained  adding in quadrature the 7-point scale variation
    MHOUs and the PDF uncertainty $\delta_{\rm PDF}$ (not including
    MHOUs in the fit), which is also separately provided.
    In the case of NNPDF4.0 the value of $\delta_{\rm PDF}$  with
    MHOUs in the fit is also listed. All uncertainties are
    expressed as percentage   of the cross-section. We finally show
    the error $\Delta^{\rm exact}_{\rm NNLO}$ 
    Eq.~(\ref{eq:PDFimpact_xsec_exact})
    due to using NNLO PDFs at N$^3$LO, and the 
    estimate of this error $\Delta^{\rm app }_{\rm NNLO}$
    Eq.~(\ref{eq:PDFimpact_xsec_approx}), also expressed as a percentage. }
  \label{tab:DY_unc}
\end{table}

In general we observe a
good perturbative convergence,
with predictions at two subsequent orders in agreement within
uncertainties, and generally improved convergence upon including MHOUs
on the PDF.
Predictions based on NNPDF4.0 and MSHT20 are always  consistent with
each other within uncertainties.
We can draw three main conclusions from
Figs.~\ref{fig:nc-dy-pheno}-\ref{fig:ccm-dy-pheno} and Tab.~\ref{tab:DY_unc}. First,  in many cases
differences between the NNLO and N$^3$LO predictions tend to be
reduced when using consistently the appropriate PDFs at each order,
rather than NNLO PDFs with  N$^3$LO matrix elements
(though in some cases the results are unchanged).
For instance, for the two lowest $m_{\ell\ell}$ bins for NC production
aN$^3$LO PDFs drive upwards the N$^3$LO prediction,  making it
closer to the NNLO result. Second, the difference between PDFs with and
without MHOUs, while moderate, remains non-negligible even at
N$^3$LO, where it starts being comparable to the overall uncertainty,
and thus it must be included in precision calculations. Third, the
impact of using aN$^3$LO instead of NNLO PDFs is actually smaller than
the guess based on the estimate Eq.~(\ref{eq:PDFimpact_xsec_approx}).

\subsection{Inclusive Higgs production}
\label{sec:higgsproduction}

\begin{figure}[!t]
  \centering
  \includegraphics[width=0.49\linewidth]{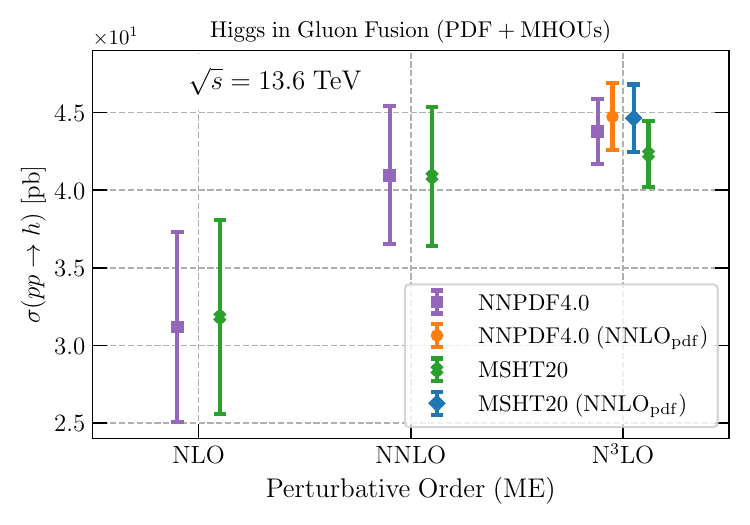}
  \includegraphics[width=0.49\linewidth]{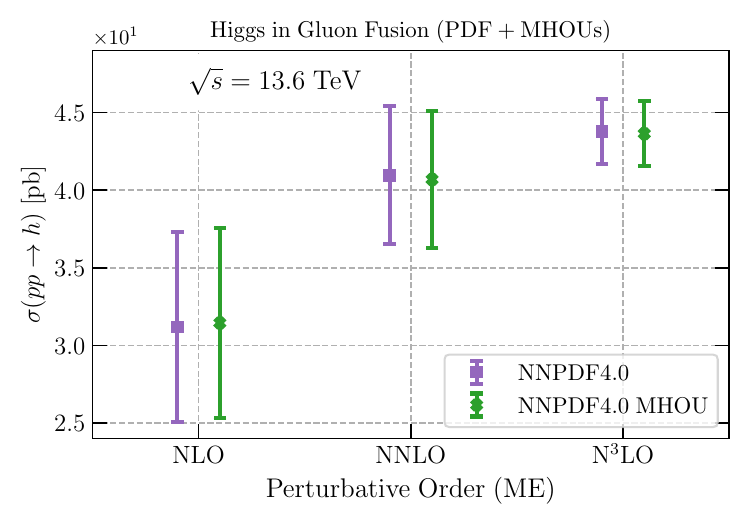}
  \includegraphics[width=0.49\linewidth]{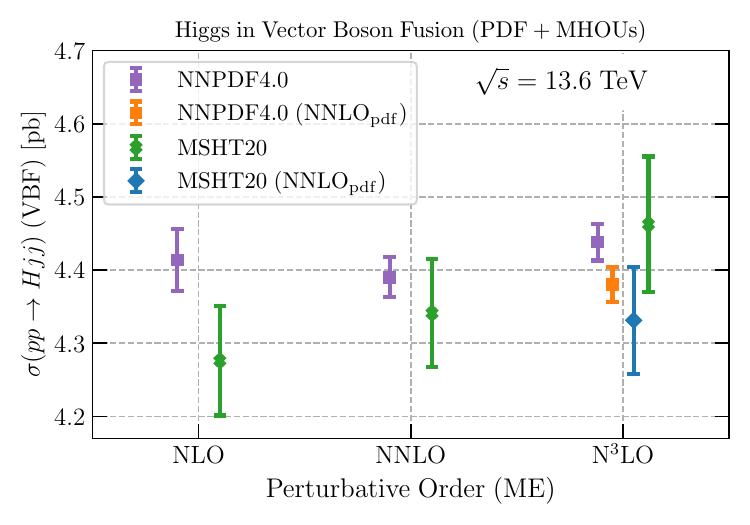}
  \includegraphics[width=0.49\linewidth]{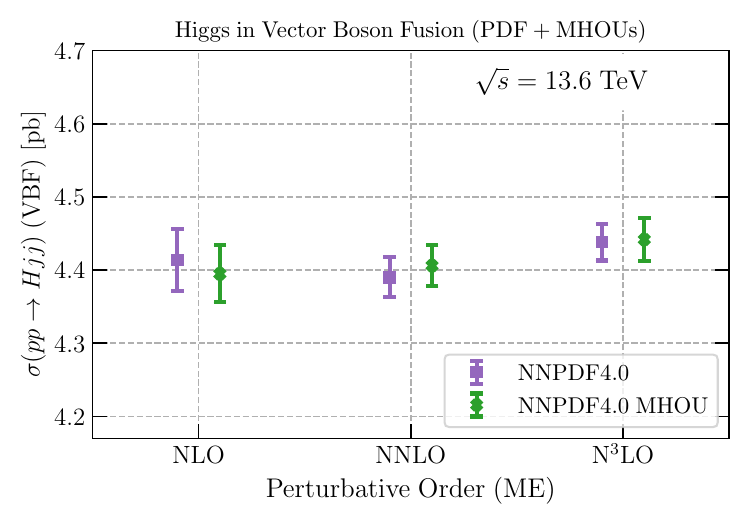}
  \caption{\small Same as  Fig.~\ref{fig:nc-dy-pheno} for
 Higgs production in gluon-fusion and  via vector-boson fusion.
}
  \label{fig:higgs-pheno-1} 
\end{figure}

\begin{figure}[!t]
  \centering
  \includegraphics[width=0.49\linewidth]{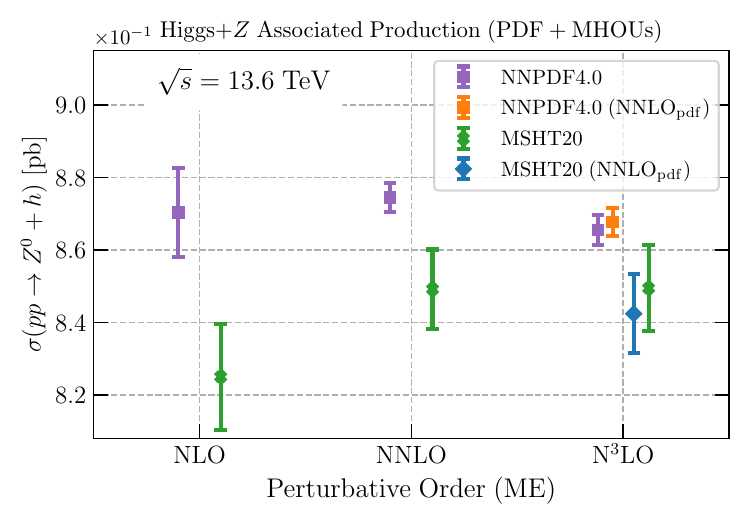}
  \includegraphics[width=0.49\linewidth]{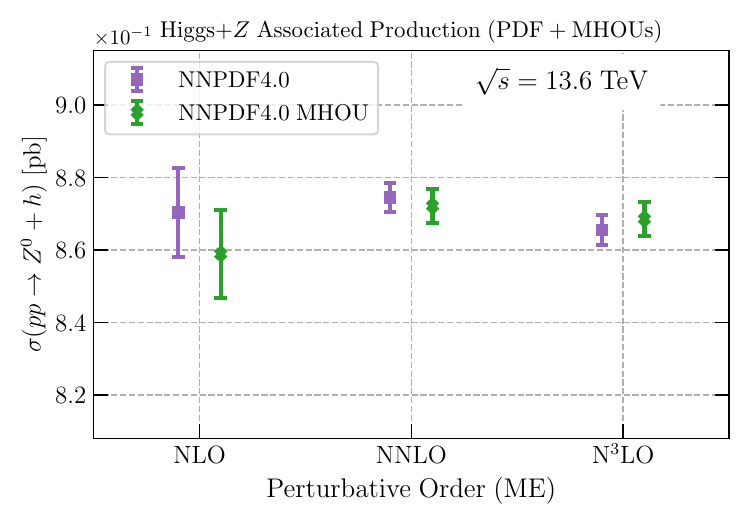}
  \includegraphics[width=0.49\linewidth]{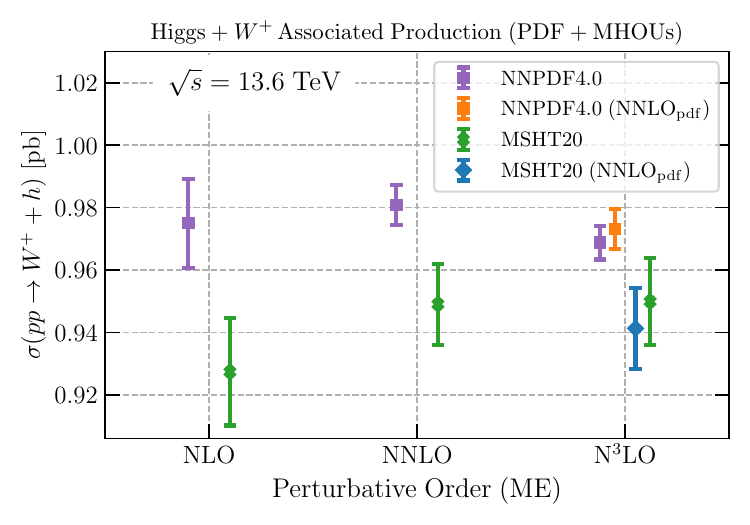}
  \includegraphics[width=0.49\linewidth]{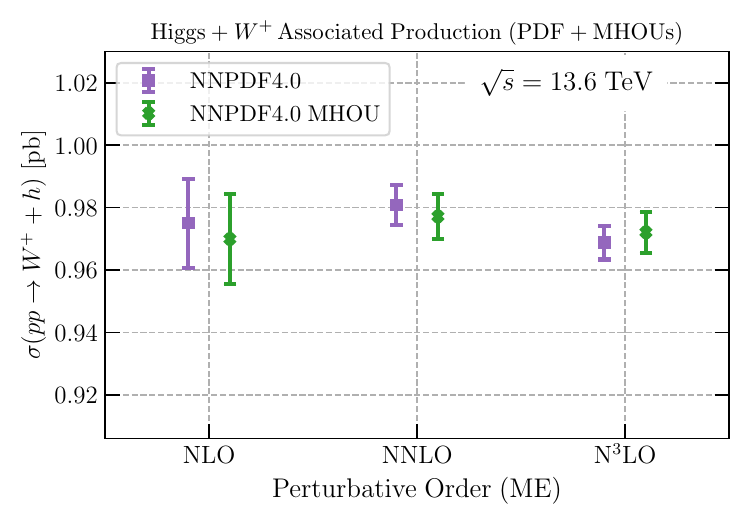}
  \includegraphics[width=0.49\linewidth]{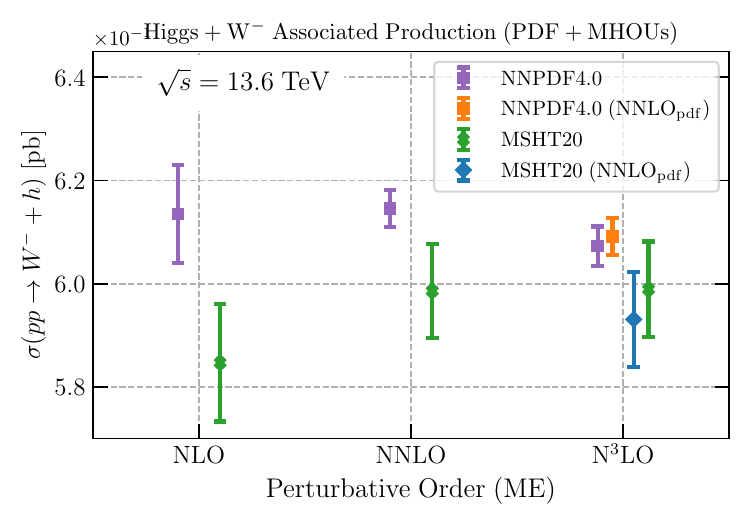}
  \includegraphics[width=0.49\linewidth]{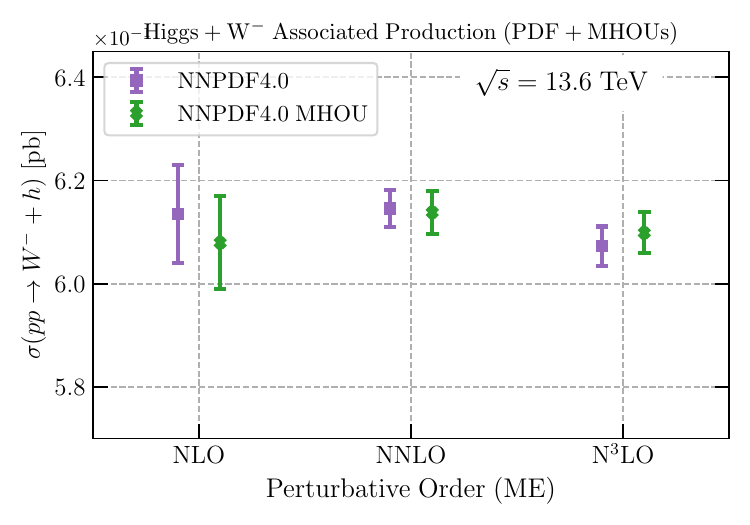}
  \caption{\small Same as  Fig.~\ref{fig:nc-dy-pheno} for
 Higgs production  in association
 with $W^+$ and $Z$ gauge bosons: from top to bottom, $Zh$, $W^+h$, and $W^-h$.
}
  \label{fig:higgs-pheno-2} 
\end{figure}

We now consider 
Higgs production in gluon fusion, in associated production
with vector bosons, and via vector-boson fusion (VBF). Predictions are
obtained using the {\sc\small ggHiggs} code~\cite{Bonvini:2014jma} for gluon
fusion, {\sc\small n3loxs} for associate production, and    {\sc\small
  proVBFH} code~\cite{Dreyer:2018qbw} for VBF.
Results are shown in Fig.~\ref{fig:higgs-pheno-1} and Table~\ref{tab:Higgs_unc}
for  gluon-fusion and VBF, and Fig.~\ref{fig:higgs-pheno-2} for associate
production with $W^+$ and $Z$.

\begin{table}[!t]
  \scriptsize
  \centering
  \renewcommand{\arraystretch}{1.4}
    \begin{tabularx}{\textwidth}{Xccccccccccc}
    \toprule
 \multirow{2}{*}{Process} & 
  \multicolumn{6}{c}{NNPDF4.0}
  & \multicolumn{5}{c}{MSHT20} \\
 &  $\sigma$ (pb) & $\delta_{\rm th}$
  &   $\delta^{\rm no MHOU}_{\rm PDF}$  &   $\delta^{\rm
    MHOU}_{\rm PDF}$ &   $\Delta^{\rm app}_{\rm NNLO}$
  & $\Delta^{\rm exact }_{\rm NNLO} $   
 &  $\sigma$ (pb) & $\delta_{\rm th} \sigma\,  $
  &   $\delta_{\rm PDF}~ $&   $\Delta^{\rm app}_{\rm
    NNLO}~ $   &$\Delta^{\rm exact }_{\rm NNLO} $    \\
  \midrule
  $gg\to h$  &  43.8   &    4.8  & 0.6  & 0.7   & 0.2     &   2.2   &    42.3  &  5.1   &  1.7    & 1.4    & 5.3  \\
  $h$ VBF  &   4.44  &   0.6    & 0.5  & 0.6    &  0.2    &  1.3    &    4.46    &    2.1      & 2.0   & 1.3   & 2.9  \\
  $hW^+$  &   0.97  &    0.6    &  0.5  & 0.6   &  0.2    &   0.5   &  0.95    &   1.5        & 1.4   &   0.8  &0.9   \\
  $hW^-$  &      0.61    &  0.6    & 0.6  &  0.6  & 0.2     & 0.3      &0.60      & 1.6     & 1.5  &   0.9  &  1.0 \\
  $hZ$  &     0.87  &  0.5    &  0.4 &   0.5   &  0.1    &  0.3    &    0.85    &    1.4    &  1.4   &  1.1  &   0.8 \\
\bottomrule
\end{tabularx}

  \vspace{0.3cm}
  \caption{Same as Table~\ref{tab:DY_unc}  for  the Higgs production processes
    displayed in Figs.~\ref{fig:higgs-pheno-1}-\ref{fig:higgs-pheno-2}}
  \label{tab:Higgs_unc}
\end{table}

Here too we observe generally good perturbative convergence, even for
gluon fusion, that notoriously has a very slowly converging expansion.
Also in this case, there is generally better agreement between
NNPDF4.0 and MSHT20 as the perturbative order increases, except for
gluon fusion where the agreement is similar at all orders. Indeed, in all
cases MSHT20 and NNPDF4.0 results agree within uncertainties at
aN$^3$LO, while they do not at NLO for VBF, nor at NLO and NNLO
for associated production. 
The impact
of using aN$^3$LO PDFs  instead of NNLO PDFs at N$^3$LO for NNPDF4.0 is
very moderate for gluon fusion, somewhat more significant for associated
production, and more significant for VBF, in which it is
comparable to the PDF uncertainty.  For MSHT20 instead a significant change
from using aN$^3$LO instead of NNLO PDFs
is also observed for gluon fusion, where
suppression of the cross-sections is seen when replacing NNLO with
aN$^3$LO PDFs. This follows from the behavior of the gluon luminosity
seen in Fig.~\ref{fig:lumis_NLO_vs_N3LO}.
The impact of MHOUs on the PDFs is generally quite small on the scale
of the PDF uncertainty at all perturbative orders, and
essentially absent for gluon fusion. For associated production it
marginally improves perturbative convergence. Interestingly, for
NNPDF4.0, for all
Higgs production processes considered, and especially for gluon fusion,
 the estimate Eq.~(\ref{eq:PDFimpact_xsec_approx}) is a substantial
 underestimate of the actual error which is made using NNLO PDFs at
 N$^3$LO. This follows from the fact that   (see Fig.~\ref{fig:lumis})
 for $m_X\sim 100$~GeV 
 the NNLO  gluon-gluon luminosity is actually closer to the NLO than to
 the aN$^3$LO, which in turn appears to be an accidental consequence
 of the behavior of the gluon PDF for $x\sim 10^{-2}$. 

\clearpage

\section{Summary and outlook}
\label{sec:summary}

We have presented the first aN$^3$LO PDF sets within the NNPDF framework, by
constructing a full set of approximate N$^3$LO
splitting functions based on available partial
results and known limits, approximate massive DIS coefficient functions,
and extending to this order the FONLL general-mass scheme for DIS
coefficient functions. We now summarize the new PDF sets that we are
releasing, our main conclusions on their features, and our plans
for future developments.

The NNPDF4.0 aN$^3$LO PDF sets are available via the {\sc\small LHAPDF6}
interface, 
\begin{center}
{\bf \url{http://lhapdf.hepforge.org/}~} .
\end{center}
Specifically, we provide an  aN$^3$LO  NNPDF4.0 set 
\begin{flushleft}
  \tt NNPDF40\_an3lo\_as\_01180 \\
\end{flushleft}
that supplements the LO, NLO and NNLO sets of
Ref.~\cite{NNPDF:2021njg}.\\ We also provide
NLO and  aN$^3$LO NNPDF4.0MHOU sets
\begin{flushleft}
  \tt NNPDF40\_nlo\_as\_01180\_mhou \\
  \tt NNPDF40\_an3lo\_as\_01180\_mhou \\
\end{flushleft}
that supplement the NNLO NNPDF4.0MHOU PDF set of
Ref.~\cite{NNPDF:2024dpb}. These sets include in the PDF uncertainty the MHOU
on the processes used for PDF determination, but in all other respects
(including the dataset) follow the default sets to which they can be directly compared.

For both aN$^3$LO sets, we also release the corresponding sets including $\alpha_s$ variations,
\begin{flushleft}
  \tt NNPDF40\_an3lo\_as\_01180\_pdfas \\
  \tt NNPDF40\_an3lo\_as\_01180\_mhou\_pdfas \\
\end{flushleft}
in which replicas 101 and 102 correspond to fits with
$\alpha_s(m_Z)=0.117$ and 0.119
respectively, in order to evaluate the combined PDF+$\alpha_s$ uncertainties
following the prescription of~\cite{Butterworth:2015oua,Demartin:2010er,PDF4LHCWorkingGroup:2022cjn}.

All these sets are delivered as ensembles of  $N_{\rm rep}=100$ Monte Carlo replicas.\\
However, we also make available
Hessian variants following~\cite{Carrazza:2015aoa,Carrazza:2016htc}
and denoted 
\begin{flushleft}
  \tt NNPDF40\_an3lo\_as\_01180\_hessian \\
  \tt NNPDF40\_an3lo\_as\_01180\_mhou\_hessian \\
  \tt NNPDF40\_an3lo\_as\_01180\_pdfas\_hessian \\
  \tt NNPDF40\_an3lo\_as\_01180\_mhou\_pdfas\_hessian \\
\end{flushleft}
each set comprising $N_{\rm eig}=50$ eigenvectors.

All these sets are also available on
the NNPDF Collaboration website,
\begin{center}
{\bf \url{https://nnpdf.mi.infn.it/nnpdf4-0-n3lo/}~} .
\end{center}
where we also give the PDF sets discussed in
Sect.~\ref{sec:N3LOKfact} based on variant treatments of the aN$^3$LO corrections.
In addition to the {\sc\small LHAPDF} grids themselves,
all the  results obtained in this work are reproducible
by means of the open-source NNPDF code~\cite{NNPDF:2021uiq} and the related suite
of theory tools.

We have provided a first assessment of these PDF sets
by comparing them to their NLO and NNLO counterparts with and without MHOUs.
Our main conclusions are the following
\begin{itemize}
  \item For all PDFs good perturbative convergence is observed, with
    differences decreasing as the perturbative order increases, and the
    aN$^3$LO result always compatible with the NNLO within
    uncertainties.
    \item For quark PDFs the difference between NNLO and aN$^3$LO
      results is extremely small, suggesting that with current data
      and methodology the effect of yet higher orders is negligible.
      \item For the gluon PDF a more significant shift is observed between
        NNLO and N$^3$LO, thus making the inclusion of  N$^3$LO
        important for precision phenomenology.
        \item The inclusion of MHOUs improves perturbative
          convergence, mostly by shifting central values at each order
          towards the higher-order result, by an amount that decreases
          with increasing perturbative order.
          \item Upon inclusion of MHOUs the fit quality becomes all
            but independent of perturbative  order, and PDF
            uncertainties generally decrease (or remain unchanged) due
            to the improved data compatibility.
          \item The effect of MHOUs at  N$^3$LO is very small for
            quarks but not negligible for the gluon PDF.
          \item Evidence for intrinsic charm is somewhat
            increased already at NNLO by
              the inclusion of MHOUs, and somewhat increased again when
              going from NNLO to N$^3$LO.
              \item The impact of N$^3$LO corrections on the total
                cross-section for Higgs in gluon fusion is very small
                on the scale of the PDF uncertainty.
\end{itemize}
All in all, these results underline the importance of the inclusion of
N$^3$LO corrections and MHOUs for precision phenomenology at
sub-percent accuracy.

Future NNPDF  releases will include by default  MHOUs, will be at all
orders up to aN$^3$LO, and will include a photon PDF. Specifically, we
aim to extend to aN$^3$LO with MHOUs our recent construction of
NNPDF4.0QED PDFs~\cite{NNPDF:2024djq}.
Indeed, aN$^3$LO
PDFs including a photon PDF (such as those recently released by
MSHT20~\cite{Cridge:2023ryv}) will be a necessary ingredient for 
theory predictions based
on state-of-the art QCD and electroweak (EW) corrections. In fact, we
are working towards the
consistent inclusion of combined QCD$\times$EW corrections also in the theory
predictions used for PDF determination.

Another important line of future development involves the all-order
resummation of potentially large perturbative contributions
in the large $x$ and small $x$ regions~\cite{Bonvini:2015ira,Ball:2017otu}.
This will involve matching
resummed and fixed-order cross-sections and (at small $x$) perturbative
evolution in the new streamlined NNPDF theory
pipeline. Such resummed PDFs will be instrumental for precision phenomenology:
specifically at small $x$, 
forward neutrino production at the LHC and
scattering processes for high-energy astroparticle physics, and at
large $x$, searches for new physics in high-mass final states at the
LHC and future hadron colliders.

\begin{center}
\rule{5cm}{.1pt}
\end{center}
\bigskip

\subsection*{Acknowledgments}

We thank  James McGowan, Thomas Cridge, Lucian Harland-Lang,
and Robert Thorne for discussions on  MSHT20 PDFs.
We are grateful to Sven Moch and Joshua Davies for discussion and
for communications concerning  their N$^3$LO results..

R.D.B, L.D.D, and R.S. are supported by the U.K. Science and Technology
Facility Council (STFC) consolidated grants ST/T000600/1 and ST/X000494/1.
F.H. is supported by the Academy
of Finland project 358090 and is funded as a part of the Center of Excellence
in Quark Matter of the Academy of Finland, project 346326. E.R.N. is
supported by the Italian Ministry of University and Research (MUR) through the
``Rita Levi-Montalcini'' Program. M.U. and Z.K. are supported by the
European Research Council under the European Union's Horizon 2020 research and
innovation Programme (grant agreement n.950246), and partially by the STFC
consolidated grant ST/T000694/1 and ST/X000664/1. J.R. is partially supported
by NWO, the Dutch Research Council. C.S. is supported by the German Research
Foundation (DFG) under reference number DE 623/6-2.

\appendix

\section{Explicit expressions of anomalous dimensions}
\label{app:splitting_asy}

We provide here explicit expressions for the $\gamma_{{\rm ns}\,
  \pm,\,N\to \infty}^{(3)}(N)$ and $\gamma_{{\rm ns}\, \pm,\,N\to
  0}^{(3)}(N)$ anomalous dimension discussed in Sect.~\ref{sec:non_singlet}
and the $\gamma_{{\rm ns}\, \pm,\,N\to \infty}^{(3)}(N)$,  $\gamma_{{\rm ns}\,
  \pm,\,N\to 0}^{(3)}(N)$ and $\gamma_{{\rm ns}\, \pm,\,N\to
  1}^{(3)}(N)$ discussed in Sect.~\ref{sec:singlet}.

\paragraph{$\gamma_{{\rm ns},\pm}^{(3)}(N)$:}

  \begin{align}
  \begin{split}
    \gamma_{{\rm ns},\pm, N\to 0}^{(3)}(N) & =
      -\frac{252.84}{N^7} + \frac{1580.25 -126.42 n_f}{N^6} \\
      & +\frac{-5806.8 +752.198 n_f -18.963 n_f^2}{N^5} \\
      & +\frac{14899.9 -2253.11 n_f +99.1605 n_f^2 -0.790123 n_f^3}{N^4} \\
      & +\frac{- 28546.4 +5247.18 n_f -226.441 n_f^2 +2.89712 n_f^3}{N^3} \\   
      & +\frac{50759.7 -8769.15 n_f +395.605 n_f^2 -3.16049 n_f^3}{N^2}
      \label{eq:Pqqns_to0}
  \end{split}
  \end{align}
      
  \begin{align}
  \begin{split}
    \gamma_{{\rm ns},\pm N\to \infty}^{(3)}(N) & =
      \left ( +20702.4 -5171.92 n_f +195.577 n_f^2 +3.27234 n_f^3 \right ) S_1(N) \\
      & -23393.8+5550.04 n_f -193.855 n_f^2 -3.01498 n_f^3 \\
      & + \left ( 16950.9 - 2741.83 n_f + 26.6886 n_f^2 \right ) \frac{S_1(N)}{N} \\
      & +\frac{ +11126.6 -3248.4 n_f +180.432 n_f^2  + 0.526749 n_f^3 }{N}
      \label{eq:Pqqns_toinf}
  \end{split}
  \end{align}

\paragraph{$\gamma_{gg}^{(3)}(N)$:}

  \begin{align}
  \begin{split}
  \gamma_{gg,N\to 0}^{(3)}(N) & =
    \frac{- 103680 + 20005.9 n_f -568.889 n_f^2}{N^7} \\
    & + \frac{- 17280 -19449.7 n_f +1725.63 n_f^2}{N^6} \\
    & + \frac{- 627979 +80274.1 n_f -2196.54 n_f^2 + 4.74074 n_f^3}{N^5}
  \label{eq:Pgg_to0}
  \end{split}
  \end{align}

  \be
  \gamma_{gg,N\to 1}^{(3)}(N) = -\frac{49851.7}{(N-1)^4} + \frac{213824 + 1992.77 n_f }{(N-1)^3}
  \label{eq:Pgg_to1}
  \ee

  \begin{align}
  \begin{split}
  \gamma_{gg,N\to \infty}^{(3)}(N) & =
    \left ( +40880.3 -11714.2 n_f  +440.049 n_f^2  + 7.36278 n_f^3 \right ) S_1(N) \\
    & -68587.9  +18144 n_f -423.811 n_f^2  -0.906722 n_f^3 \\
    & - \left ( -85814.1+13880.5 n_f -135.111 n_f^2 \right ) \frac{S_1(N)}{N} \\
    & + \frac{- 54482.8 + 4341.13 n_f +21.3333n_f^2}{N}
  \label{eq:Pgg_toinf}
  \end{split}
  \end{align}

\paragraph{$\gamma_{gq}^{(3)}(N)$:}

  \begin{align}
    \begin{split}
    \gamma_{gq,N\to 0}^{(3)}(N) & =
    \frac{- 37609.9 + 5309.63 n_f}{N^7} \\
    & + \frac{ -35065.7 + 221.235 n_f}{N^6} \\
    & + \frac{ -175455 + 9092.91 n_f + 778.535 n_f^2}{N^5}
  \label{eq:Pgq_to0}
  \end{split}
  \end{align}

  \be  
  \gamma_{gq,N\to 1}^{(3)}(N) = -\frac{22156.3}{(N-1)^4} + \frac{95032.9 + 885.674 n_f}{(N-1)^3}
  \label{eq:Pgq_to1}
  \ee

  \begin{align}
    \begin{split}
    \gamma_{gq,N\to \infty}^{(3)}(N) & =
    \left ( -13.4431 +0.548697 n_f \right ) L_{5,0}(N) \\
    & + \left ( -375.398 +34.4947 n_f  -0.877915 n_f^2 \right )L_{4,0}(N)
  \label{eq:Pgq_toinf}
  \end{split}
  \end{align}

\paragraph{$\gamma_{qg}^{(3)}(N)$:}

  \begin{align}
    \begin{split}
    \gamma_{qg,N\to 0}^{(3)}(N) & =
    \frac{14103.7 n_f -1991.11 n_f^2}{N^7} \\
    & +\frac{2588.84 n_f +2069.33 n_f^2}{N^6} \\
    & + \frac{68802.3 n_f -7229.38 n_f^2 -99.1605 n_f^3}{N^5}
  \label{eq:Pqg_to0}
  \end{split}
  \end{align}

  \be  
  \gamma_{qg,N\to 1}^{(3)}(N) = -\frac{7871.52 n_f}{(N-1)^3}
  \label{eq:Pqg_to1}
  \ee

  \begin{align}
    \begin{split}
    \gamma_{qg,N\to \infty}^{(3)}(N) & =
    \left ( -1.85185 n_f +0.411523 n_f^2 \right ) L_{5,0}(N) \\
    & + \left ( -35.6878 n_f +3.51166 n_f^2 + 0.0823045 n_f^3 \right ) L_{4,0}(N) \\
    & + \left ( -2.88066 n_f -0.823045 n_f^2 \right ) L_{5,1}(N) \\
    & + \left (  +40.5114 n_f -5.54184 n_f^2  -0.164609 n_f^3 \right ) L_{4,1}(N)   
  \label{eq:Pqg_toinf}
  \end{split}
  \end{align}

\paragraph{$\gamma_{qq,{\rm ps}}^{(3)}(N)$:}

  \begin{align}
    \begin{split}
      \gamma_{qq,{\rm ps},N\to 0}^{(3)}(N) & =
      \frac{5404.44 n_f -568.889 n_f^2}{N^7} \\  
      & +\frac{3425.98 n_f +455.111 n_f^2}{N^6} \\
      & +\frac{20515.2 n_f -1856.79 n_f^2 + 4.74074 n_f^3}{N^5}
  \label{eq:Pqqps_to0}
  \end{split}
  \end{align}

  \be  
  \gamma_{qq,{\rm ps},N\to 1}^{(3)}(N) = -\frac{3498.45 n_f}{(N-1)^3},
  \label{eq:Pqqps_to1}
  \ee

  \begin{align}
    \begin{split}
      \gamma_{qq,{\rm ps},N\to \infty}^{(3)}(N) & =
      \left ( +56.4609 n_f -3.6214 n_f^2 \right )  L_{4,1}(N) \\
      & + \left ( +247.551 n_f -40.5597 n_f^2 + 1.58025 n_f^3 \right )  L_{3,1}(N) \\
      & +13.1687 n_f  L_{4,2}(N) \\
      & + \left ( +199.111 n_f -13.6955 n_f^2 \right )  L_{3,2}(N).
  \label{eq:Pqqps_toinf}
  \end{split}
  \end{align}

The functions $L_{k,j}(N)$ are defined as the Mellin transform of $(1-x)^j\ln^k(1-x)$: 

\begin{align}
  L_{k,0}(N) & = \mathcal{M}\left[\ln^k(1-x)\right](N) =(-1)^k k! \frac{S_{1_k, \dots, 1_1}(N)}{N} \\
  L_{k,1}(N) & = \mathcal{M}\left[(1-x) \ln^k(1-x)\right] = L_{k,0}(N) - L_{k,0}(N+1) \\
  L_{k,2}(N) & =\mathcal{M}\left[(1-x)^2 \ln^k(1-x)\right] = L_{k,0}(N) - 2 L_{k,0}(N+1)+  L_{k,0}(N+2)
  \label{eq:mellin_lm1kmk}
\end{align}
with the multi-index harmonics of weight-$k$ defined recursively as
\be
  S_{1_k,\dots,1_1}(N) = \sum_{j=1}^{N} \frac{S_{1_{k-1}, \dots, 1_1}(j)}{j} \ 
  \label{eq:multi_idx_harmonics}
\ee
and the termination condition
\be
S_\emptyset = 1.
  \label{eq:harmterm}
\ee

\bibliography{nnpdf40n3lo}

\end{document}